\documentclass[prd,aps,floats,floatfix,eqsecnum,nofootinbib]{revtex4}
\usepackage{amsmath,amssymb,verbatim,epsfig,graphicx,rotating}
\usepackage{psfrag}
\usepackage{multirow}
\newcommand{\be}{\begin{equation}}
\newcommand{\ee}{\end{equation}}
\newcommand{\bea}{\begin{eqnarray}}
\newcommand{\eea}{\end{eqnarray}}
\newcommand{\vx}{\ensuremath{\vec{x}}}
\newcommand{\vk}{\ensuremath{\vec{k}}}
\newcommand{\vq}{\ensuremath{\vec{q}}}
\newcommand{\bds}[1]{\boldsymbol{#1}}
\newcommand{\avg}[1]{\langle #1 \rangle}
\begin{document}
\title{The Effective Theory of Inflation in the Standard Model of the 
Universe

and the CMB+LSS data analysis}
\author{\bf D. Boyanovsky $^{(a,b,c)}$}\email{boyan@pitt.edu}
\author{\bf C. Destri $^{(d,c)}$} \email{Claudio.Destri@mib.infn.it}
\author{\bf H. J. de Vega $^{(b,c)}$} \email{devega@lpthe.jussieu.fr}
\author{\bf N. G. Sanchez $^{(c)}$} \email{Norma.Sanchez@obspm.fr}
\affiliation{$^{(a)}$ Department of Physics and
Astronomy, University of Pittsburgh, Pittsburgh, Pennsylvania 15260,
USA.\\
$^{(b)}$ LPTHE, Laboratoire Associ\'e au CNRS UMR 7589,\\
Universit\'e Pierre et Marie Curie (Paris VI) et Denis Diderot 
(Paris VII), \\ Tour 24, 5 \`eme. \'etage, 4, Place Jussieu, 
75252 Paris, Cedex 05, France.\\ 
$^{(c)}$ Observatoire de Paris, LERMA, Laboratoire Associ\'e au 
CNRS UMR 8112, \\61, Avenue de l'Observatoire, 75014 Paris, France.\\
$^{(d)}$ Dipartimento di Fisica G. Occhialini, Universit\`a
Milano-Bicocca and INFN, sezione di Milano-Bicocca, Piazza della Scienza 3,
20126 Milano, Italia\\}

\date{\today}

\begin{abstract}
Inflation is today a part of the Standard Model of the Universe supported by
the cosmic microwave background (CMB) and large scale structure (LSS) datasets.
Inflation solves the horizon and flatness problems and 
naturally generates  density fluctuations that seed LSS and CMB anisotropies, 
and tensor perturbations (primordial gravitational waves). Inflation theory is 
based on a scalar field  $ \varphi $ (the inflaton) whose potential 
is fairly flat leading to a slow-roll evolution. This review focuses on the
following new aspects of inflation. We present the
effective theory of inflation \`a la Ginsburg-Landau in which
the inflaton potential is a polynomial in the field $ \varphi $ and has
the universal form $ V(\varphi) = N \; M^4 \;
w(\varphi/[\sqrt{N}\; M_{Pl}]) $, where $ w = {\cal O}(1) , \;
M \ll M_{Pl} $ is the scale of inflation and  $ N \sim 60 $ is the number 
of efolds since the cosmologically relevant modes exit the horizon till 
inflation ends.
The slow-roll expansion becomes a systematic $ 1/N $ expansion and
the inflaton couplings become {\bf naturally small} as powers of the ratio
$ (M / M_{Pl})^2 $. The spectral index and the ratio of tensor/scalar
fluctuations are $ n_s - 1 = {\cal O}(1/N), \; r = {\cal O}(1/N) $ while
the running index turns to be $ d n_s/d \ln k =  {\cal O}(1/N^2) $
and therefore can be neglected. The energy scale of inflation $ M \sim 0.7 
\times 10^{16}$ GeV is
completely determined by the amplitude of the scalar adiabatic
fluctuations. A complete analytic study plus the 
Monte Carlo Markov Chains (MCMC) analysis of the available
CMB+LSS data (including WMAP5) with fourth degree trinomial potentials 
showed: (a) the {\bf spontaneous breaking} of the
$ \varphi \to - \varphi $ symmetry of the inflaton potential. (b)
a {\bf lower bound} for $ r $ in new inflation:
$ r > 0.023 \; (95\% \; {\rm CL}) $ and $ r > 0.046 \;  (68\% \;
{\rm CL}) $. (c) The preferred inflation potential is a {\bf double
well}, even function of the field with a moderate quartic coupling
yielding as most probable values: $ n_s \simeq 0.964 ,\; r\simeq
0.051 $. This value for $ r $ is within reach of forthcoming CMB
observations. The present data in the effective theory of
inflation clearly {\bf prefer new inflation}. Study of higher degree 
inflaton potentials show that terms of degree higher than four do not 
affect the fit in a significant way. In addition, horizon exit happens for 
$ \varphi/[\sqrt{N} \; M_{Pl}] \sim 0.9 $ making higher order terms 
in the potential $ w $ negligible. We summarize the physical effects of 
{\bf generic} initial conditions (different from Bunch-Davies) on the 
scalar and tensor perturbations during slow-roll and
introduce the transfer function $ D(k) $ which encodes the observable 
initial conditions effects on the power spectra. 
These effects are more prominent in the \emph{low}
CMB multipoles: a change in the initial conditions during slow roll can
account for the observed CMB quadrupole suppression.
Slow-roll inflation is generically preceded by a
short {\bf fast-roll} stage. Bunch-Davies initial conditions are the 
natural initial conditions for the fast-roll perturbations. During 
fast-roll, the potential in the wave equations of curvature and tensor 
perturbations is purely attractive and leads to a suppression of the 
curvature and tensor CMB quadrupoles. A MCMC analysis of 
the WMAP+SDSS data including fast-roll shows that the quadrupole 
mode exits the horizon about 0.2 efold before fast-roll ends and its 
amplitude gets suppressed. In addition, fast-roll fixes the initial 
inflation redshift to be $ z_{init} = 0.9 \times 10^{56} $ and
the {\bf total number} of efolds of inflation to be $ N_{tot} \simeq 64 $.
Fast-roll fits the TT, the TE and the EE
modes well reproducing the quadrupole supression. A thorough study of the 
quantum loop corrections reveals that they are very small and controlled by
powers of $(H /M_{Pl})^2 \sim {10}^{-9} $, a conclusion that validates the 
reliability of the
effective theory of inflation. The present review shows how powerful is
the Ginsburg-Landau effective theory of inflation in predicting
observables that are being or will soon be contrasted to observations.
\end{abstract}

\maketitle

\tableofcontents
\section{Introduction to the Effective Theory of Inflation}

\subsection{Overview and present status of inflation}

Inflation was introduced to solve several outstanding problems of the
standard Big Bang model \cite{guthsato} and has now become an important 
part of the standard cosmology. It provides a natural mechanism
for the generation of scalar density fluctuations that seed large scale
structure, thus explaining the origin of the temperature anisotropies in
the cosmic microwave background (CMB), and for the generation of tensor
perturbations (primordial gravitational waves) 
\cite{kt,libros,mass,hu,fluc}.

\medskip

\begin{figure}[h]
\includegraphics[height=9.cm,width=12.cm]{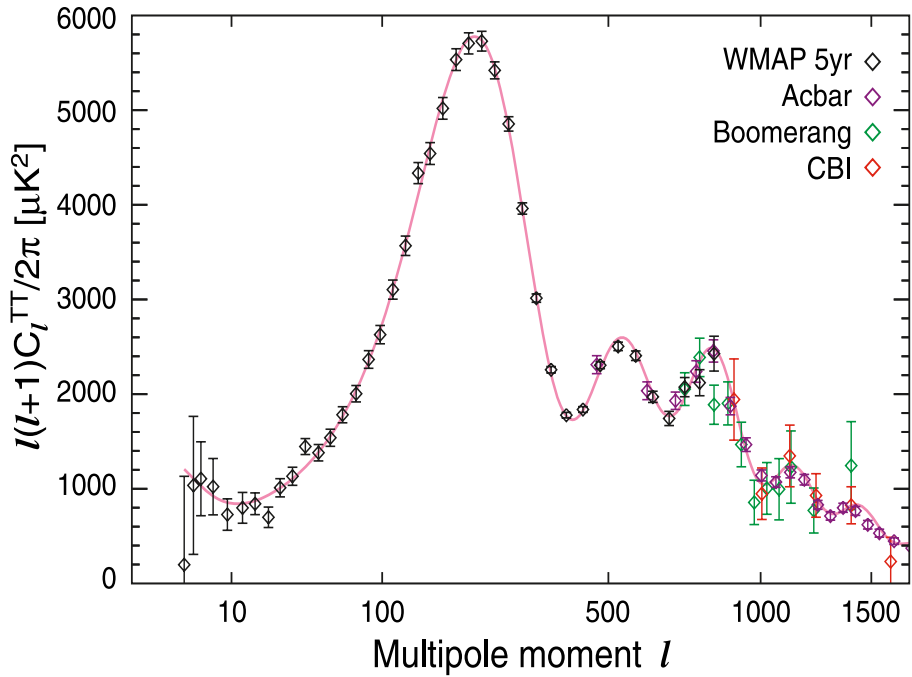}
\caption{Acoustic oscillations from WMAP 5 years data set plus other 
CMB data. Theory and observations nicely agree except for the lowest 
multipoles: the quadrupole CMB suppression. See sects. \ref{supcua} and 
\ref{fastroll} for discussions on this.} 
\label{acus}
\end{figure}

A distinct aspect of inflationary perturbations is that they are 
generated by quantum fluctuations of the scalar field(s) that drive 
inflation. After their wavelength becomes larger than the Hubble radius, 
these fluctuations are amplified and grow, becoming classical and 
decoupling from causal microphysical processes. Upon re-entering the 
horizon, during the radiation and matter dominated eras, these classical 
perturbations seed the inhomogeneities which generate structure upon 
gravitational collapse \cite{libros,hu,fluc}. 
A great diversity of inflationary models predict fairly generic features: a
gaussian, nearly scale invariant spectrum of (mostly) adiabatic scalar 
and tensor primordial fluctuations, which provide an
excellent fit to the highly precise wealth of data provided by the
Wilkinson Microwave Anisotropy Probe (WMAP)\cite{WMAP1,WMAP3,WMAP5}
making the inflationary paradigm fairly robust.
Precision CMB data reveal peaks and valleys in the temperature
fluctuations resulting from  acoustic oscillations in the
electron-photon fluid at recombination. These are depicted in fig. 
\ref{acus} where up to five peaks can be seen.

Baryon acoustic oscillations driven by primordial fluctuations
produce a peak in the galaxy correlations at 
$ \sim 109 \; h^{-1} $ Mpc (comoving
sound horizon) \cite{eis}. This peak is the real-space version of the 
acoustic oscillations in momentum (or $l$) space and are confirmed 
by LSS data \cite{eis}.

\begin{figure}[h]
\includegraphics[height=9.cm,width=12.cm]{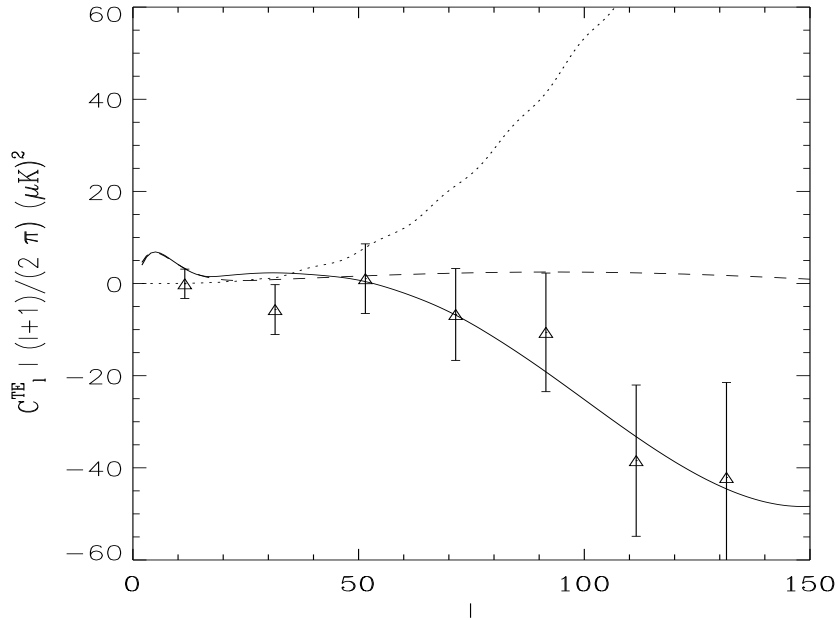}
\caption{Temperature-Polarization angular power spectrum.
The large-angle TE power spectrum predicted in primordial adiabatic 
models (solid), primordial isocurvature models (dashed) and by defects
such as cosmic strings (dotted). The WMAP TE data (Kogut et al. \cite{WMAP1})
are shown for comparison, in bins of $ \Delta l = 10 $. Superhorizon 
adiabatic modes from inflation fit the data while subhorizon sources of 
TE power go in directions opposite to the data.}
\label{TE}
\end{figure}

Perhaps the most striking validation of inflation
as a mechanism for generating \emph{superhorizon} fluctuations
is the anticorrelation peak in the temperature-polarization (TE) angular
power spectrum at $ l \sim 150 $ corresponding to superhorizon
scales \cite{WMAP1} and depicted in fig. \ref{TE}. 
The observed TE power spectrum can only 
be generated by fluctuations that exited the horizon during inflation
and re-entered the horizon later, when the expansion of the universe
decelerates.

\begin{figure}[h]
\includegraphics[height=9.cm,width=12.cm]{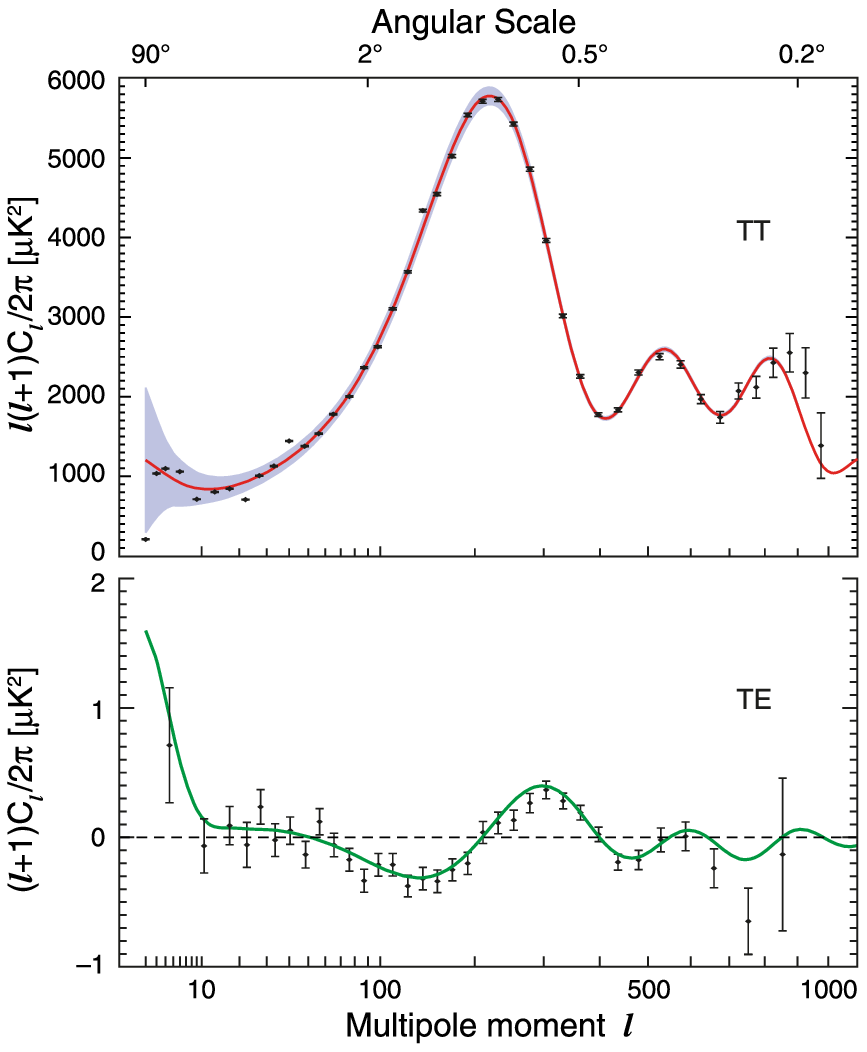}
\caption{The temperature (TT) and temperature-polarization correlation 
(TE) power spectra based on the 5 year WMAP data \cite{WMAP5}.}
\label{TEwmap5}
\end{figure}

The confirmation of many of the robust predictions of
inflation by current high precision observations places inflationary
cosmology on solid grounds.

\medskip

Amongst the wide variety of inflationary scenarios, single field slow-roll
models provide an appealing, simple and fairly generic description of
inflation. Its simplest implementation is based on a scalar field (the
inflaton) whose homogeneous expectation value drives the dynamics of the
scale factor, plus small quantum fluctuations. The inflaton potential is
fairly flat during inflation and it dominates the universe energy during 
inflation. This flatness not only leads to a slowly varying Hubble 
parameter, hence ensuring a sufficient number of efolds of inflation,
but also provides an explanation for the gaussianity of the fluctuations as
well as for the (almost) scale invariance of their power spectrum. A flat
potential precludes large non-linearities in the dynamics of the
\emph{fluctuations} of the scalar field.

\medskip

The current WMAP data are validating the single field slow-roll scenario 
\cite{WMAP1,WMAP3,WMAP5}. Furthermore, because the potential is flat the
scalar field is almost {\bf massless}, and modes cross the horizon with an
amplitude proportional to the Hubble parameter. This fact combined with a
slowly varying Hubble parameter yields an almost scale invariant primordial
power spectrum.  The slow-roll approximation has been recently cast as a 
systematic $ 1/N $ expansion \cite{1sN}, where $ N \sim 60 $ is the number 
of efolds before the end of inflation when modes of cosmological relevance 
today first crossed the Hubble radius.

\medskip

The observational progress begins to discriminate among 
different inflationary models, placing stringent constraints on them. 
The upper bound on the ratio $ r $ of tensor to scalar fluctuations 
obtained by WMAP convincingly excludes the massless monomial 
$ \varphi^4 $ potential \cite{WMAP1,WMAP3,WMAP5} and hence 
{\bf strongly suggests} the presence of a {\bf mass
term} in the single field inflaton potential \cite{ciri,infwmap}.
Therefore, as a minimal single field model, one should consider a sufficiently 
general polynomial, the simplest polynomial potential bounded
from below being the fourth order trinomial potential \cite{mcmc}. 

\medskip

The observed low value of the CMB quadrupole with respect to
the $\Lambda$CDM theoretical value
has been an intriguing feature on large angular scales since 
first observed by COBE/DMR \cite{cobe}, and confirmed by the WMAP data 
\cite{WMAP3,WMAP5}.  In the best fit $\Lambda$CDM model, 
using the WMAP5 data we find that the probability that the quadrupole is as
low or lower than the observed value is just 0.031. Even if one does not 
care about the specific multipole and looks for any multipole as low or 
lower than the observed quadrupole with respect to the $\Lambda$CDM model 
value, then the probability remains smaller than 5\%. Therefore, it is 
relevant to find a cosmological explanation of the quadrupole supression 
beyond the $\Lambda$CDM model. An early fast-roll stage can explain the CMB
 quadrupole suppression as we discuss below.

\medskip

This review article focuses on the following new aspects of inflationary
cosmology:

\begin{itemize}
\item{An effective field theory description of slow-roll single
field inflation \`a la Ginsburg-Landau. In the Ginsburg-Landau framework,
the potential is a polynomial in the field starting by a constant term 
\cite{gl}. Linear terms can always be eliminated by a constant shift of 
the inflaton field. The quadratic term can have a positive or a
negative sign associated to chaotic or new inflation, respectively.
This effective Ginsburg-Landau field theory is characterized by only 
{\bf two} energy scales: the scale of inflation $ M $ and the Planck scale 
$ M_{Pl} = 2.43534 \; 10^{18}$ GeV $ \gg M $. In this context we propose a 
{\bf universal} form for the inflaton potential in slow-roll models \cite{1sN}:
\be \label{VIn}
V(\varphi) = N \; M^4 \; w(\chi)  \; ,
\ee
\noindent where $ N \sim 60 $  is the number of efolds since the 
cosmologically relevant modes exit the horizon till the end of inflation 
and $ \chi $ is a dimensionless, slowly varying field 
$$
\chi = \frac{\varphi}{\sqrt{N} \;  M_{Pl}}  \; .
$$
The slow-roll expansion becomes in this way a systematic $ 1/N $ 
expansion. The couplings in the inflaton Lagrangian become naturally small
due to suppression factors arising from eq.(\ref{VIn}) as
the ratio of the two relevant energy scales here: $ M $ and $ M_{Pl} $.
The spectral index, the ratio of tensor/scalar fluctuations, the running 
index and the amplitude of the scalar adiabatic fluctuations are 
naturally 
\be \label{predi}
n_s - 1 = {\cal O}\left(\frac1{N}\right) \quad , \quad 
r = {\cal O}\left(\frac1{N}\right) \quad , \quad
\frac{d n_s}{d \ln k} =  {\cal O}\left(\frac1{N^2}\right) \quad , \quad
|{\Delta}_{k\;ad}^{\mathcal{R}}|  \sim N \; 
\left(\frac{M}{M_{Pl}}\right)^2 \; ,
\ee
for {\bf all} inflaton potentials in the class of eq.(\ref{VIn}). Hence, 
the energy scale of inflation $ M $ is completely determined by the 
amplitude of the scalar adiabatic fluctuations 
$ |{\Delta}_{k\;ad}^{\mathcal{R}}| $ and using the WMAP5 results for it,
we find $ M \sim 10^{16}$ GeV.
The running index results $ d n_s/d \ln k \sim  10^{-4} $ and 
therefore can be neglected. Namely, within the class of models 
eq.(\ref{VIn}) one does not need to measure the ratio $ r $ in order to 
learn about the scale of inflation. Moreover, we were able to provide 
{\bf lower} bounds for $ r $ and {\bf predict its value} in the effective 
theory of inflation using the CMB+LSS data and Monte Carlo
Markov  Chains (MCMC) simulations 
\cite{mcmc}.}
\item{Besides its simplicity, the trinomial
potential  (minimal single field model in the Ginsburg-Landau spirit)
is rich enough to describe the physics of inflation and
accurately reproduce the WMAP data \cite{mcmc}. It is well motivated
within the Ginsburg-Landau approach as an effective field theory
description (see ref.\cite{gl,quir}). We provide a complete analytic
study complemented by a statistical analysis. The MCMC 
analysis of the available CMB+LSS data with the 
Ginsburg-Landau effective field theory of inflation showed \cite{mcmc}:

(i) The data strongly indicate the {\bf breaking} (whether
spontaneous or explicit) of the $ \varphi \to - \varphi $ symmetry of the
inflaton potentials both for new and for chaotic inflation. (ii)
Trinomial new inflation naturally satisfies this requirement and
provides an excellent fit to the data. (iii) Trinomial chaotic
inflation produces the best fit in a very narrow corner of the
parameter space. (iv) The chaotic symmetric trinomial potential is
almost certainly {\bf ruled out} (at more than $95\%$CL). In trinomial 
chaotic inflation the MCMC runs go towards a potential in the {\em 
boundary} of the parameter space and which resembles a spontaneously 
symmetry broken potential of new inflation. (v) The above results and 
further physical analysis here lead us to conclude that {\bf new inflation}
gives the best description of the data. (vi) We find a {\bf  lower bound}
for $ r $ within trinomial new inflation potentials: $ r > 0.023 \;
(95\% \; {\rm CL}) $ and $ r > 0.046 \;  (68\% \;   {\rm CL}) $.
(vii) The preferred new inflation potential is a double
well, even function of the field with a moderate quartic coupling 
$ y \sim 1 $,
\be\label{bini}
w(\chi) = \frac{y}{32} \left(\chi^2 - \frac8{y}\right)^{\! 2} 
= -\frac12 \; \chi^2 + \frac{y}{32} \; \chi^4 + \frac2{y}\quad   .
\ee
[see eq.(\ref{VIn})]. This new inflation model yields as most probable 
values: $ n_s \simeq 0.964 ,\; r\simeq 0.051 $. This value for $ r $ 
is within reach of forthcoming CMB observations \cite{mcmc}. 
For the best fit value $ y \simeq 1.26 $,
the inflaton field exits the horizon in the negative
concavity region $ w''(\chi) < 0 $ intrinsic to new inflation.
We find for the best fit, $ M = 0.543 \times 10^{16}$ GeV for the scale of 
inflation and $ m = 1.21 \times 10^{13}$ GeV for the inflaton mass. 
We derived explicit formulae and study in detail the spectral index
$ n_s $ of the adiabatic fluctuations, the ratio $ r $ of tensor to
scalar fluctuations and the running  index $ d n_s/d \ln k $
\cite{mcmc}. We use these analytic formulas as hard constraints on $
n_s $ and $ r $ in the MCMC analysis. Our analysis differs in this
{\bf crucial} aspect from previous MCMC studies in the literature
involving the CMB data. }

\item{The dependence of the observables ($ n_s, \; r $ and $ dn_s/d\ln k $)
on the degree of the inflaton potential $ 2 \, n $ is studied
and confronted to the WMAP and large scale structure data \cite{fi6}. This 
shows that higher degree terms ($ n>2 $) in the inflaton potential do not
affect the fit in a significant way in new inflation.
The window of consistency with the WMAP and LSS data {\bf narrows} for 
growing $ n $ in chaotic inflation. New inflation 
yields a good fit to the $ r $ and $ n_s $ data in a wide range of
field and parameter space. Small field inflation yields $ r<0.16 $
while large field inflation yields $ r>0.16 $ (for $ N=50 $). All
members of the new inflation family predict a small but negative
running $ -4 \, (n+1)\times 10^{-4} \leq  dn_s/d\ln k   \leq -2\times
10^{-4} $. (The values of $ r, \; n_s, \; dn_s/d\ln k $ for arbitrary
$ N $ follow by a simple rescaling from the $ N=50 $ values).
The reconstruction program carried out in ref.\cite{fi6} suggests quite
generally that for $ n_s $ consistent with the WMAP and LSS data
and $ r<0.1 $ the {\bf symmetry breaking scale} for new inflation is
$ |\varphi_{min}| \sim 20~M_{Pl} $ while the {\bf field scale} at Hubble
crossing is $ |\varphi_{exit}| \sim 7 \; M_{Pl} $. This corresponds to
$ \chi_{exit} \sim 0.9 $ which can make negligible the higher order terms 
in $ w(\chi_{exit}) $ for new inflation. The family of chaotic
models feature $ r \geq 0.16 $ (for $ N=50 $) and only a {\it
restricted subset} of chaotic models are consistent with the
combined WMAP bounds on $ r, \; n_s, \; dn_s/d\ln k $  with a
narrow window in field amplitude around $ |\varphi_{exit}| \sim 15~M_{Pl}
$. We conclude that a measurement of $ r<0.16 $ (for $ N =50 $)
distinctly rules out a large class of chaotic scenarios and favors
small field new inflationary models. As a general consequence, {\bf
new inflation} emerges clearly more favoured than chaotic inflation.}

\item{The dynamics of inflation is usually described by the classical 
evolution of a scalar field (the inflaton). The use of classical dynamics
 is justified by the enormous stretching of
physical lengths during inflation. When the physical wavelength of
the fluctuations become larger than the Hubble radius, these
fluctuations effectively become classical. This is probably the only
case where the time evolution itself leads to the classicalization
of fluctuations and microscopic scales near the Planck scale 
$ 10^{-32} \, {\rm cm} \lesssim \lambda = 2 \, \pi / k \lesssim
10^{-28} $ cm become macroscopic today in the range $ 1 \, {\rm
Mpc} \lesssim \lambda_{today} \lesssim 10^4 $ Mpc. This happens thanks to 
a redshift by $ \sim 10^{56} $ since the beginning of inflation for a total
number of inflation efolds $ N_{tot} \sim 64 $.

We discuss the validity of the effective theory of inflation. 
It is valid generically as long as the energy density is $ \ll M_{Pl}^4 $. 
This is true thanks to eq.(\ref{VIn}) even when the inflaton field 
$ \varphi $ takes values equal to many times $  M_{Pl} $.

We summarize the quantum loop corrections to inflationary dynamics
(see \cite{effpot,quant}). Novel phenomena emerges at the quantum
level as a consequence of the lack of kinematic thresholds, among
them the phenomenon of inflaton decay into its own quanta. A
thorough study of the effect of quantum fluctuations reveals that
these loop corrections are suppressed by powers of $
\left(H/M_{Pl}\right)^2 $ where $ H $ is the Hubble parameter during 
inflation \cite{effpot,quant}.
The amplitude of temperature fluctuations constrains the scale of
inflation with the result that $ \left(H/M_{Pl}\right)^2 \sim
10^{-9} $. Therefore, quantum loop corrections are very small
and controlled by the ratio $ \left(H/M_{Pl}\right)^2 $, a
conclusion that  validates the reliability of the classical
approximation and the effective field theory approach to
inflationary dynamics. The quantum corrections to the power spectrum
are computed and expressed in terms of $ n_s, \; r $ and $ dn_s/d \ln k $.
Trace anomalies dominate the quantum corrections to the primordial
power spectrum (see sec. \ref{trece}).} 

\item{Scalar (curvature) and tensor (gravitational wave) perturbations 
originate in quantum fluctuations during inflation. These are usually 
studied within the slow-roll approximation and with Bunch-Davies
initial conditions. We summarize the physical effects on the power 
spectrum of generic initial conditions with particular attention to 
back-reaction effects \cite{quadru1,quadru2}. We introduce a
\emph{transfer function } $ D(k) $ which encodes the effect of
generic initial conditions on the power spectra. The constraint from
renormalizability and small back reaction entails that $ D(k)
\lesssim \mu^2/k^2 $ for large $ k $ where $ \mu $ characterizes the 
asymptotic decay of the occupation number. This implies that observable
effects from initial conditions are more prominent in the \emph{low}
CMB multipoles.  The effects on high $l$-multipoles are
suppressed by a factor $ \sim 1/l^2 $ due to the large $ k $ fall off  of 
$ D(k) $. Hence, a change from the Bunch-Davies initial conditions for the 
fluctuations can naturally account for the low observed value
of the CMB quadrupole \cite{quadru1,quadru2}.}

\item{Slow-roll inflation is generically preceded by a short fast-roll 
stage during which the kinetic and the potential
energy of the inflaton field are of the same order. This fast-roll stage is
followed by the usual slow-roll regime during which the kinetic
energy is much smaller than the potential energy.  The fast-roll
stage leads to a purely attractive potential in the wave equations of
curvature and tensor perturbations. This in turn leads to a suppression 
of the quadrupole in curvature and tensor perturbations which exited the 
horizon during the fast-roll stage. Within the context of the effective 
field theory and for generic initial conditions on the inflaton field, it
is shown that a quadrupole suppression consistent with observations
is a natural consequence of the fast-roll stage 
\cite{quadru1,quadru2,quamc}. A {\bf new} parameter emerges in this way 
in the early universe model: the comoving wave number $ k_{tran} $ 
characteristic scale of this attractive potential. This mode $ k_{tran} $ 
happens to exit the horizon precisely {\bf at
the transition} from the fast-roll to the slow-roll stage. The fast-roll
stage dynamically modifies the initial power spectrum of perturbations
by a transfer function $ D(k) $. We perform a MCMC analysis of the WMAP 
and SDSS data combined with the most recent supernovae compilation \cite{SN}
and including the fast-roll stage. We find the value 
$ k_{tran} = 0.290 \; {\rm Gpc}^{-1} $ (today) and $ k_{tran}^{init} = 1.69 \; 
10^{14} \, {\rm GeV} = 14 \; m $ at the beginning of inflation. These 
values fix the redshift since the beginning of inflation till today to 
$ z_{init} = 0.915 \times 10^{56} $. From that we find the {\bf 
total number of efolds} $ N_{tot} $ during inflation to be (see sec. 
\ref{ntot64})
$$
N_{tot} \simeq 63 - \frac12 \; \log\left(\frac{H}{10^{-4} \; 
M_{Pl}}\right)  \; ,
$$
where $ H $ is the Hubble parameter at the beginning of 
inflation. The values of $ H $ and $ N_{tot} $  favoured by the current WMAP 
data within the effective theory of inflation and respecting the lower bound 
for $ N_{tot} $ that solves the horizon problem  (see sec. \ref{soluhor})
turn to be $ H \simeq 0.4 \times 10^{14} $ GeV and $ N_{tot} \simeq 64 $.
That is, the MCMC analysis of the CMB+LSS data including the early fast-roll
explanation of the CMB quadrupole suppression {\bf imposes} 
$ N_{tot} \simeq 64 $. The quadrupole mode $ k_Q = 0.238 \; 
{\rm Gpc}^{-1} $ exits the horizon earlier than $ k_{tran} $, about 0.2
efolds before the end of fast-roll. Including the fast-roll stage improves the fits 
to the TT, the TE and the EE modes, well reproducing the quadrupole 
supression. }
\end{itemize}

As we discuss in sec. \ref{fastroll}, inflation generically starts by a 
fast-roll stage where the kinetic and potential
energy of the inflaton are of the same order. This is followed by a 
slow-roll regime where the kinetic energy is much smaller than the 
potential energy. The slow-roll regime of inflation is an attractor of 
the dynamics during which
the Universe is dominated by vacuum energy. Inflation ends
when again the kinetic energy of the inflaton becomes large as the
field is rolling near the minimum of the potential. Eventually,
the energy stored in the homogeneous inflaton is transferred
explosively into the production of particles via spinodal or
parametric instabilities \cite{chalo,prehea,baacke,ramsey}. 
More precisely, non-linear phenomena eventually 
{\bf shut-off} the instabilities and {\bf stop} inflation 
\cite{chalo,chalo2,tsu}.
All these processes lead to the transition to a radiation dominated era.
This is the standard picture of the transition from
inflation to standard hot big bang cosmology.

\medskip

We formulate here inflation as an effective field theory within the 
Ginsburg-Landau spirit \cite{1sN,gl}. The theory of the second order 
phase transitions, the Ginsburg-Landau theory of superconductivity, 
the current-current Fermi theory of weak interactions, the sigma model of 
pions, nucleons (as skyrmions) and photons are all successful
effective field theories. The present review shows how powerful is
the effective theory of inflation {\bf to predict observable quantities} 
that can be or will be soon contrasted with experiments. 
There are {\bf two kind} of predictions in the effective theory of 
inflation: first, predictions on the order of magnitude of the CMB
observables valid for all inflaton potentials in the class of 
eq.(\ref{VIn}) [see eqs.(\ref{predi})]; second, precise quantitative 
predictions as those presented in secs. \ref{mcmc}-\ref{fastroll}.

The Ginsburg-Landau realization of the inflationary potential presented
in this review fits the amplitude of the CMB anisotropy remarkably well
and reveals that the
Hubble parameter, the inflaton mass and non-linear couplings are
see-saw-like, namely powers of the ratio $ (M/M_{Pl})^2 \sim 10^{-9} $ 
multiplied by further powers of $ 1/N $. Therefore, the smallness of the 
couplings is not a result of fine tuning but a {\bf natural} consequence of
the form of the potential, of the validity of the effective field theory
description and slow-roll. The quantum expansion in loops is
therefore a double expansion on $ \left(H/M_{Pl}\right)^2 $ and $
1/N $. Notice that graviton corrections are also at least of
order $ \left(H/M_{Pl}\right)^2 $ because the amplitude of tensor
modes is of order $H/M_{Pl}$ . We show in sec. \ref{potuniv}
that the form of the potential which fits the WMAP+LSS data and is 
consistent with slow-roll eqs.(\ref{VIn}) implies the small values 
for the inflaton self-couplings \cite{1sN}. 

The infrared (superhorizon) modes in the quantum loops produce large 
contributions of the order $ \sim N $. However, as shown in 
sec. \ref{trece} these large infrared contributions get multiplied by 
slow-roll factors of order  $ \sim 1/N $. As a result, the 
superhorizon contributions to physical magnitudes turn to be of
order $ N^0 $ \cite{effpot,quant} times  factors of the order of
$ \left(H/M_{Pl}\right)^2 $.

We note that the effective theory of inflation describes an evolution
spanning about 26 orders of magnitude in length scales from the beginning 
till the end of the inflationary era. This is the largest scale change 
described by a field theory so far.

It must be stressed that the energy scale of inflation,
$ M \sim 10^{16} $ GeV is the energy scale of at least two
other important physical situations: (a) the scale of Grand Unification
of strong and electroweak interactions and (b) the large energy scale 
in the see-saw formula for neutrino masses [see eq.(\ref{neum})].
This coincidence suggests a physical link between the three areas.

\medskip

Many deep problems remain to be solved in the early universe. One of 
them is the reheating problem. Namely, how the universe thermalizes 
after inflation and at what temperature. Baryogenesis provides a lower 
bound on the reheating temperature \cite{kt}. The mechanisms of 
thermalization uncovered in refs. \cite{fi4} can 
provide a starting point to understand the reheating.

The units used in this review are such that $ \hbar = c = 1 $.

\subsection{The Standard Cosmological Model}\label{uno}

The history of the Universe is a history of expansion and cooling down.

On large scales the Universe is homogeneous and isotropic and its
geometry is   described by the Friedmann-Robertson-Walker (FRW) metric
\be\label{FRWk}
 ds^2= dt^2-a^{\, 2}(t) \left[ \frac{dr^2}{1 - k \; r^2} +
r^2 \left( d\theta^2 + \sin^2 \theta \; d \phi^2
\right)\right] \; , 
\ee 
where $ r, \; \theta $ and $ \phi $ are
comoving spherical coordinates, $ t $ is the cosmic time (the
proper time of a comoving observer), $ a(t) $ the scale factor, and $
k =0, \pm 1 $ stands for the scalar curvature of three-dimensional 
spatial sections. $ k = 0, \; k > 0 $ and $ k < 0 $ describes flat, 
closed and open universe, respectively. The dynamics of the scale factor 
is completely determined by Einstein's equations and the equation of 
state. Overwhelming observational evidence indicates that the geometry 
of the Universe is spatially flat, namely $ k=0 $. Thus the FRW metric
simplifies to 
\be\label{FRW}
 ds^2= dt^2-a^2(t) \; d\vec{x}^2
\ee
or in conformal time $ \eta $
\be \label{metcon}
ds^2 = a^2(\eta)[d\eta^2 - (d\vec x)^2] \; .
\ee
where $ d\eta = dt/a(t) $. Notice that this cosmological expansion has no 
center: it happens everywhere at all spatial points $ \vx $ and it is 
identical everywhere. The scale factor grows monotonically with time.

Physical scales are stretched by the scale factor $ a(t) $ with respect to 
the time independent comoving scales
\begin{equation}\label{scale}
l_{phys}(t)= a(t) \; l_{com} \; .
\end{equation}
A physical wavelength redshifts proportional to the scale factor
[eq.(\ref{scale})], therefore its time derivative obeys the Hubble law 
$$ 
\dot{l}_{phys}(t) = H(t) \; l_{phys}(t)= \frac{l_{phys}(t)}{d_H(t)} \; .
$$ 
where $ H(t) \equiv \dot{a}(t)/a(t) $ and $ d_H(t) $ is the Hubble radius.

The redshift $ z $ at time $ t $ is defined as
\be\label{z}
z + 1 \equiv \frac{a_0}{a(t)}
\ee
where $ a_0 $ stands for the scale factor today and we choose 
$ a_0 \equiv 1 $. The farther back in time, the larger is the redshift 
and the smaller is $ a(t) $.

The temperature decreases as the universe expands as
\be \label{Ta} 
T(t) = \frac{T_0}{a(t)}  \; .
\ee 
Eq.(\ref{Ta}) applies to all particles in thermal equilibrium
as well as to massless decoupled particles (radiation).
Since the temperature decreased with time, the Universe underwent a 
succession of phase transitions towards the less symmetric phases 
\cite{tranfas}.

\medskip

The combination of data from CMB and LSS, and numerical
simulations lead to the $\Lambda$CDM or \emph{concordance model}
which has now become the standard cosmology. This impressive
convergence of observational data and theoretical and numerical results
describes a Universe that is composed of a cosmological constant, dark
matter, baryonic (atoms) matter and radiation. This model
provides the {\bf only consistent} explanation of  the
broad set of precise and independent astronomical observations over
a wide range of scales available today. Namely:

\begin{itemize}

\item{WMAP data and previous CMB data.}

\item{Light Elements Abundances.}

\item{Large Scale Structures (LSS) Observations. Baryon acoustic 
oscillations (BAO).}

\item{Acceleration of the Universe expansion:
Supernova Luminosity/Distance (SN) and Radio Galaxies.}

\item{Gravitational Lensing Observations.}

\item{Lyman $ \alpha $ Forest Observations.}

\item{Hubble Constant ($H_0$) Measurements.}

\item{Properties of Clusters of Galaxies.}

\end{itemize}

In the homogeneous and isotropic FRW universe described by eq.(\ref{FRW}), the
matter distribution must be homogeneous and isotropic, with an
energy momentum tensor having in spatial average the isotropic fluid form 
\be\label{fluido}
\langle T^{\mu}_{\nu} \rangle={\rm diag}[\rho, -p,-p,-p ] \; , 
\ee
where $ \rho, \; p $ are the energy density and pressure,
respectively. In such space-time geometry the Einstein equations of general
relativity reduce to the Friedmann equation, which determines
the evolution of the scale factor from the energy density
\be\label{fri}
\left[\frac{\dot{a}(t)}{a(t)} \right]^2 = H^2(t) = \frac{
\rho}{3 M^2_{Pl}} \,.
\end{equation}
where $ M_{Pl}= 1/\sqrt{8\pi G} = 2.43534 \times 10^{18}$ GeV
$ = 0.434\times 10^{-5}$ g.
The spatially flat Universe has today the critical density
\be\label{rhocrit}
\rho_c = 3 \; M^2_{Pl} \; H^2_0= 1.878 \; h^2 \;
10^{-29}\textrm{g}/\textrm{cm}^3 = 1.0537 \; 10^{-5} \;  
h^2 \; \textrm{GeV}/\textrm{cm}^3
\; .
\ee
where $ H_0 = 100 \; h \; \textrm{km/sec}/\textrm{Mpc} $ is the Hubble 
constant today, $ h = 0.705 \pm 0.013 $ \cite{pdg,WMAP5} and then
$ H_0 = 1.5028 \; 10^{-33} $ eV.
Notice that eq.(\ref{fri}) implies that $ a(t) $ is a monotonic
function of time.

The energy momentum tensor conservation reduces to the single
conservation equation,
\begin{equation}\label{conener}
\dot{\rho}+3 \; H(t) \; \left( \rho+p\right)  =0
\end{equation}
\noindent The two equations (\ref{fri}) and (\ref{conener}) can
be combined to yield the acceleration of the scale factor,
\begin{equation}\label{accel}
\frac{\ddot{a}}{a}= -\frac1{6 \; M^2_{Pl}}(\rho + 3 \; p)
\end{equation}
\noindent which will prove useful later.
In order to provide a close set of equations we must append an
equation of state $ p=p(\rho) $ which is typically written in the form
\begin{equation}\label{eqnofstate}
p=w(\rho) \; \rho
\end{equation}

\begin{table}
\begin{tabular}{|c|c|c|c|c|}
\hline  $\rho_c$  &  $ (2.36 \;  {\rm meV})^4 $ & & $h$   &  
$ 0.705 \pm 0.013 $ \\
\hline $ H_0 $ & $ h /[3 \; {\rm Gpc}] =h /[9.77813 \; {\rm Gyr}]$ & & 
$\Omega_{\Lambda}$ & $0.726$ \\ 
\hline  $ M_{Pl} $  & $2.43534\times 10^{18}$ GeV & &  $\Omega_M $ & $0.274$ \\ 
\hline  $ M $  & $0.543 \times 10^{16}$ GeV & &  $\Omega_r $ & 
$8.49 \; 10^{-5}$ \\ \hline  $ m $  & $1.21 \times 10^{13}$ GeV& &$ n_s $ & 
$ 0.960 \pm 0.014 $\\ \hline
\end{tabular} 
\caption{Selected Cosmological Parameters \cite{pdg,WMAP5}. $ m $ and $ M $
are given by eq.(\ref{myM}).}
\end{table}

The following are  important cosmological solutions:
\bea\label{fa}
&&  {\rm Cosmological ~ Constant } \; \Rightarrow
p=-\rho : {\bf\Lambda D}{\rm ~de~Sitter~ expansion}\Rightarrow \rho = {\rm
constant}~;~ a(t) = a(0) \; e^{Ht} ~;~H= 
\sqrt{\rho/[3 \; M^2_{Pl}]} \cr \cr
&& {\rm Radiation}\; \Rightarrow p/\rho=1/3:
{\rm {\bf RD} \; (Radiation ~ domination)}
\Rightarrow \rho(t)= \rho(t_r) \;  a^{-4}(t)~;~
a(t) = a(t_r) \; \sqrt{t/t_r} \label{RD} \\
&& {\rm Non-relativistic ~ (cold) ~ Matter} \; \Rightarrow p/\rho=0 :
{\rm {\bf MD} \;  (Matter ~ domination)}  \Rightarrow \rho(t) =
\rho(t_{eq}) \; a^{-3}(t)~;~ a(t) = a(t_{eq}) \; 
(t/t_{eq})^\frac23 \nonumber 
\eea 
where  $ t_r $ and $ t_{eq} $ are the values of cosmic time at which the 
Universe becomes radiation or matter dominated, respectively.

Notice from eqs.(\ref{accel}) and (\ref{RD}) that
accelerated expansion ($ {\ddot a}(t) > 0 $) takes place if $ p/\rho < -1/3 $. 

\medskip

The universe started by a very short accelerated inflationary stage 
dominated by the vacuum energy, lasting $ \sim 10^{-36}$ sec
ending by redshift $ z \sim 10^{29} $ and approximately described by the 
de Sitter metric. This inflationary stage was followed by decelerated 
expansion, first by the radiation dominated era and then by the matter 
dominated era. Finally, the universe entered again an accelerated phase 
dominated by the dark energy, described by a cosmological constant in the 
Standard Model of the Universe, at $ z \simeq 0.5 $.

\medskip

Particle physics at energy scales below $ \sim 200 $ GeV is on solid 
experimental footing in the framework of the standard model of strong and 
electroweak interactions.

Current theoretical ideas supported by the renormalization group
running of the couplings in the standard model of particle physics
and its supersymmetric extensions show that the strong, weak and
electromagnetic interactions are unified in a grand unified theory
(GUT) at the scale $ M_{GUT} \sim 10^{16}$ GeV. 
Furthermore, the characteristic scale at which gravity calls for a quantum
description is the Planck scale $  M_{Pl} = 1/\sqrt{8\pi G} = 2.43534 \; 10^{18}$ GeV
$ \gg M_{GUT} $.

The connection between the standard model of particle physics and
early Universe cosmology is through the semiclassical 
Einstein equations that
couple the space-time geometry to the matter-energy content. As
argued above, gravity can be studied semi-classically at energy
scales well below the Planck scale. The standard model of particle
physics is a {\em quantum field theory}, thus the space-time is
classical  but with sources that are quantum fields. Semiclassical
gravity is defined by the Einstein's equations with the
expectation value of the quantum energy-momentum tensor $ \hat{T}^{\mu
\nu} $ as the source
\begin{equation}\label{einstein}
G^{\mu \nu} = R^{\mu \nu} -\frac12 \; g^{\mu \nu} R = \frac{
\langle \hat{T}^{\mu \nu} \rangle}{M^2_{Pl}} \; .
\end{equation}
The expectation value of  $ \hat{T}^{\mu \nu} $ is taken in a
given quantum state (or density matrix) compatible with
homogeneity and isotropy which must  be translational and
rotational invariant. Such state yields an expectation value for
the energy momentum tensor with the fluid form eq.(\ref{fluido}),
and the Einstein equations (\ref{einstein}) reduce to the Friedmann
equation  (\ref{fri}).

All of the ingredients 
are now in place to understand the evolution of the early Universe.
Einstein's equations determine the evolution of the scale factor, particle physics
provides the energy momentum tensor and statistical mechanics
provides the fundamental framework to describe the thermodynamics
from the microscopic quantum field theory of the strong, 
electroweak interactions and beyond.

The sources for Einstein equations are dark energy, dark and ordinary 
matter and radiation. The standard model of particle physics describes 
ordinary matter and radiation. 

Dark energy accounts {\bf today} for $ 72 \pm 1.5\% $ of the energy of the
Universe \cite{WMAP5}. The current observed value is $ \rho_{\Lambda} =  
\Omega_{\Lambda} \; \rho_c = (2.36 \;  {\rm meV})^4 \; , 
\; 1 \;  {\rm meV} = 10^{-3}\;   {\rm eV} $ \cite{pdg,WMAP5}. 
The equation of state is $ p_{\Lambda} = - \rho_{\Lambda} $ within 
observational errors corresponding to a cosmological constant. 

The nature of the dark energy (today) is not 
yet understood. A plausible explanation of the dark energy
may be the quantum zero point energy of a light matter field in the cosmological
space-time. This has the equation of state of a cosmological constant. 
Notice that the renormalized value of the zero point energy in the cosmological
space-time is finite and may be naturally of the order of the (mass)$^4$ of the 
light field involved.

Matter accounts today for $ 28 \pm 1.5\% $ of the energy of the Universe 
\cite{WMAP5}. $ 84 \% $ of the matter is {\bf dark matter}.
Therefore,  dark matter is an essential constituent of the universe.
The nature of dark matter is still unknown but is certainly
beyond the Standard Model of strong and electroweak particle interactions 
\cite{libros,tranfas}. It is probably formed by particles in the keV mass
scale \cite{oscnos}.

\begin{figure}[h]
\begin{center}
\begin{turn}{-90}
\psfrag{Omega_{Lambda} vs. log(1+z)}{$ \frac{\rho_{\Lambda}}{\rho}$ vs. 
$ \log(1+z) $}
\psfrag{Omega_{mat} vs. log(1+z)}{$ \frac{\rho_{Mat}}{\rho} $ vs. 
$ \log(1+z) $}
\psfrag{Omega_{rad} vs. log(1+z)}{$ \frac{\rho_{rad}}{\rho} $ vs. 
$ \log(1+z) $}
\includegraphics[height=13.cm,width=8.cm]{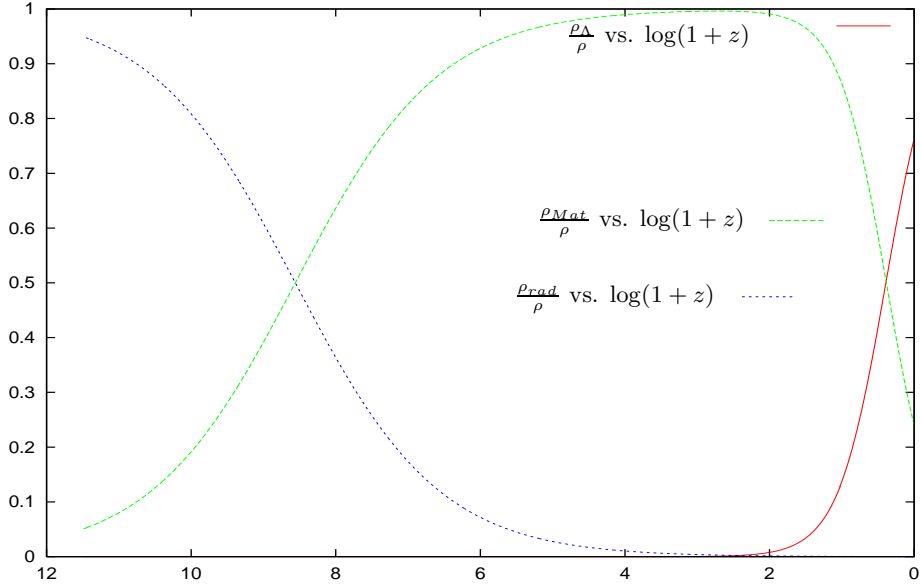}
\end{turn}
\caption{ $ \frac{\rho_{\Lambda}}{\rho}, \; \frac{\rho_{Matter}}{\rho} $ and 
$ \frac{\rho_{radiation}}{\rho} $ vs.  $ \log(1+z) $.} 
\label{ome}
\end{center}
\end{figure}

Main events in the universe after inflation are (see fig. \ref{ome}): 

\begin{itemize}
\item{Begining of the {\bf RD} era and end of inflation:  $ z \sim 10^{29} 
\; , \quad T_{reh} \sim 10^{16} $ GeV, $ t  \sim 10^{-36}$ sec.}

\item{Electro-Weak phase transition: $ z \sim 10^{15} \; , \quad
T_{\rm EW} \sim 100$ GeV, $ t\sim 10^{-11} $ sec.}

\item{QCD phase transition (confinement): $ z \sim 10^{12} \; , 
\quad T_{\rm QCD} \sim 170$ MeV, $ t \sim 10^{-5} $ sec.}

\item{Big bang nucleosynthesis (BBN):  $ z \sim  10^9 \; , \quad \ln(1+z)\sim 
21  \; , \quad  T \simeq 0.1$ MeV, $ t \sim 20 $ sec.}

\item{Radiation-Matter equality: $ z \simeq 3200  \; , \; \ln(1+z) 
\simeq 8 \; , \;  T \simeq 0.7$ eV,  $ t \sim 57000 $ yr.}

\item{CMB last scattering: $ z \simeq 1100 \; , \; \ln(1+z) \simeq 7  
\; , \;  T \simeq 0.25$ eV,  $ t \sim 370000 $ yr.}

\item{Matter-Dark Energy equality: $ z  \simeq 0.47  \; , \; \ln(1+z) 
\simeq 0.38 \; , \;  T \simeq 0.345$ meV $ t \sim 8.9 $ Gyr.}

\item{Today: $ z = 0 \; , \; \ln(1+z) = 0 \; , \;  T = 2.725$K $=0.2348 $ 
meV,  $ t \equiv t_0 = 13.72 $ Gyr.}
\end{itemize}

In fig. \ref{ome} we plot $ \rho_{\Lambda}/\rho, \; \rho_{Matter}/\rho $ 
and $ \rho_{radiation}/\rho $ as functions of $ \log(1+z) $ where 
$ \rho_{\Lambda} =\Lambda,
\; \rho_{Matter} = \Omega_M/a^3 $ and $ \rho_{radiation} = \Omega_r/a^4 $.
Notice that $ \rho_{\Lambda} + \rho_{Matter} +  \rho_{radiation} = \rho $.

\medskip

In summary, the Friedmann equation (\ref{fri}) can be written as
\be\label{frgen}
H^2(t) = H^2_0 \left[\Omega_{\Lambda} + \frac{\Omega_M}{a^3} 
+ \frac{\Omega_r}{a^4}\right] \; .
\ee
The temperature of the universe in the post-inflation radiation dominated era 
(reheating temperature $ T_r $) is bounded from below in order to explain the 
baryon asymmetry and the Big Bang Nucleosynthesis (BBN). This amounts
to a further constraint on the inflationary model. The BBN constraint
is the milder. If  the observed baryon asymmetry is
produced at the electroweak scale, the  constraint on the  reheating
temperature is $ \gtrsim 100 $ GeV, however the origin of the baryon
asymmetry may be at the GUT scale in which case the reheating temperature 
should be $  T_r > 10^9 $GeV \cite{kt}.

\subsection{The Horizon and Flatness problems in non-inflationary cosmology 
and their inflationary resolution.}\label{dos}

In this section we discuss the horizon and flatness
problems that arise in cosmology when there is not an
inflationary era before the RD and MD stages.

\subsubsection{The horizon problem}

The particle horizon is the size of the causally connected region at a 
given time $ t $. It is given by,
\be\label{dfhp}
d(t) = a(t) \; \int_{t_{min}}^t \frac{dt'}{a(t')} \; .
\ee
where $ t_{min} $ stands for the minimal time where the classical 
geometry (\ref{FRW}) applies. The comoving size of the horizon
\be\label{confo}
\eta \equiv \int_{t_{min}}^t \frac{dt'}{a(t')} \; ,
\ee
is just the conformal time. $ \eta $ is the maximum comoving
distance traveled by a photon since $ t_{min} $. Objects separated by
comoving distances larger than $ \eta $ were never in causal
contact \cite{libros}. 

For a decelerated matter or radiation dominated geometry, it follows that
\be\label{a}
a(t) =  a(t_s) \; \left(\frac{t}{t_s}\right)^b
\quad {\rm with} \quad b = \frac12 \quad {\rm for ~ RD} \quad
{\rm and} \quad b = \frac23 \quad {\rm for ~ MD} \; .
\ee
Here $ t_s $ stands for $ t_r $ or $ t_{eq} $, the starting times of the RD or MD eras, 
respectively.

Inserting eq.(\ref{a}) into eq.(\ref{dfhp}) yields,
\be
d(t) =\frac1{1-b}\left[ t - t_{min} \left(\frac{t}{t_{min}}\right)^b \right]
\ee
Therefore, since $ 0 < b < 1 $ we can take the $ t_{min} \to 0 $ limit with the result
\be\label{hpd}
d(t) = \frac{t}{1-b} = \frac{b}{1-b} \; \frac1{H(t)} = 
\frac1{1-b} \; \left[\frac{a(t)}{a(t_s)}\right]^\frac1{b} \; .
\ee
Notice that $ d(t) \to 0 $ for $ t  \to 0 $ implying that the regions in causal contact 
were smaller and smaller for $ t  \to 0 $. This goes against
an homogeneous and isotropic  universe at early times.

On the contrary, for late times  $ d(t) $ grows {\bf faster} in 
eq.(\ref{hpd}) than the scale factor since $ 1/b > 1 $. 

At the time of matter-radiation equality
$ t \simeq 81300 $ yr and $ z \simeq 3200 $. Therefore, $ d_0 $, the 
particle horizon today is dominated by the matter dominated era 
$ (b=2/3) $:
\be\label{dt}
d_0  = \frac2{H_0} = 2 \; \times 13.9 \; {\rm Gyr} = 8.55 \; {\rm Gpc}
\ee
ignoring the present accelerated expansion of the universe.

The photons seen today as the CMB decoupled from matter at 
$ z_d \simeq 1100 $, that is $ t_d \simeq 370000 $ yr. A sphere with the 
size $ d(t_d) $ of the particle horizon at decoupling is the last 
scattering surface of the photons. The expansion of the universe stretches 
this sphere and the angle subtended today by this  particle horizon is
\be\label{teta}
\theta_d = ( z_d + 1 ) \; \frac{d(t_d)}{d_0} 
= \frac1{\sqrt{z_d + 1}} \simeq 0.03 = 1.7^o \; ,
\ee
where we used eq.(\ref{hpd}) with $ b=2/3 $. Thus, the sky should split 
into 
\be\label{parches}
\frac{4 \; \pi}{\theta_d^2} = 4 \; \pi \; ( z_d + 1 ) \simeq 13800
\ee
patches which had never communicated before the CMB formed. Hence, one would expect a 
different
CMB temperature at each patch. The {\bf horizon problem} is that the CMB temperature
in all these patches is the {\bf same} up to fluctuations of order $ 0.01 \% $, 
as we know from COBE and WMAP.

\medskip

We have ignored in this subsection the presence of dark energy. Taking into
account the cosmological constant as in sec. \ref{edad} slightly changes 
the coefficients in eqs.(\ref{teta}) and (\ref{parches}):
$$
\theta_d = \frac{1.112}{\sqrt{z_d + 1}} \simeq 0.034 = 1.9^o \quad , \quad
\frac{4 \; \pi}{\theta_d^2} \simeq 11170 \; ,
$$
where we used eqs.(\ref{etaa}) and (\ref{horiz}). This analysis shows
that the presence (or absence) of dark energy is irrelevant to the
horizon problem since dark energy started to dominate the Universe 
expansion fairly recently at $ z \simeq 0.5 $.

\subsubsection{The flatness problem}

Present data \cite{WMAP5,pdg} supports the spatially flat FRW geometry  
eq.(\ref{FRW})
\be\label{omeg}
\Omega_0 \equiv \frac{\rho_0}{\rho_c} = 1.003 \pm 0.02 \; . 
\ee
where $ \Omega_0 \equiv \Omega({\rm today}),\;\rho_0 \equiv 
\rho({\rm today}) $ and  $ \rho_c $ is defined by eq.(\ref{rhocrit}). 
However, one can consider the more general homogeneous and isotropic FRW 
metric eq.(\ref{FRWk}) in which case the Friedmann equation (\ref{fri}) 
takes now the form
\be \label{frik}
 H^2(t) = \frac{\rho}{3 M^2_{Pl}} - \frac{k}{a^2}
\ee
Hence,
\be \label{ome1}
\Omega = \frac{\rho}{\rho_c} = 1 + \frac{k}{a^2 \; H^2}
\ee
For a decelerated universe eq.(\ref{a}) we obtain
$$
\Omega - 1 = \frac{k}{a^2(t_s) \; b^2} \; 
\left(\frac{t}{t_s}\right)^{2 - 2 \, b} \; .
$$
Since $ 2 - 2 \, b > 0 $ this quantity {\bf grows} with time. Hence, in 
order to explain the flatness of the universe today [see eq.(\ref{omeg})], 
the universe must have been {\bf even much flatter} in early times.

\subsubsection{The solution to the horizon problem in inflation}
\label{soluhor}

The solution of the horizon problem provided by inflation
can be understood within de Sitter inflation which is an 
approximation to slow-roll inflation.

The particle horizon during de Sitter inflation [see eq.(\ref{fa})] is 
given by
\be\label{dinfla}
d(t) = e^{H \; t} \int_{t_{min}}^t dt' \;  e^{-H \; t'} = 
\frac1{H}\left[e^{H(t- t_{min})}-1\right]\simeq
\frac1{H} \; \frac{a(t)}{a(t_{min})}
\ee
Here, the size of the causally connected regions grows {\bf exactly} as 
the scale factor while in
decelerated geometries the particle horizon grows {\bf faster} than the 
scale factor [see eq.(\ref{hpd})].
In addition, for $ t_{min} $ deep in the past, $ d(t) \to \infty $ since 
$ a(t)/a(t_{min}) \to \infty $ for fixed $ t $.
This is consistent with an homogeneous and isotropic  universe at early 
times since homogeneity and isotropy
can establish in a bigger and bigger causally connected region.

The horizon problem arises when $ d(t_d) \ll d_0 $ as shown by eq.(\ref{teta}).
Now, since the contribution to  $ d(t_d) $ from the inflationary epoch 
[see  eq.(\ref{dinfla})] can be {\bf very large} if $  a(t_{end}) \gg a(t_{min}) $, 
where $ t_{end} $ stands for the end of inflation, 
inflation {\bf can solve} the horizon problem. We derive now a lower bound for the ratio 
\be\label{N}
e^{N_{tot}} \equiv \frac{a(t_{end})}{a(t_{min})} \; .
\ee
$ N_{tot} $ is the total number of efolds during inflation.

In order to explain the smallness of the 
CMB anisotropy today, the particle horizon taken at the end of inflation 
and then red-shifted today
must be at least of the size of the particle horizon today 
[eq.(\ref{dt})]. That is, the visible universe today was contained inside
the horizon during inflation,
\be\label{cot1}
\frac1{H} \; \frac{a_0}{a(t_{min})} \simeq 
\frac1{H} \; \frac{a(t_{end})}{a(t_{min})} \; \left(\frac{a_{eq}}{a_r}\right)_{RD} \; 
\left(\frac{a_0}{a_{eq}}\right)_{MD} \geq \frac{3.362\ldots}{H_0}
\ee
where $ a_r $ and $ a_{eq} $ stand for the scale factor at the beginning and at the end 
of the RD era, respectively and we used the particle horizon today given by 
eq.(\ref{horiz}) which takes into account the present accelerated phase of the universe.
We assume for simplicity a sudden transition from the inflationary to the RD era,
namely $ a(t_{end})=a(t_r) $. In reality there must be an intermediate reheating
stage after inflation where radiation is abundantly created becoming the dominant 
component in the Universe and establishing the onset of the RD era.

We now insert in eq.(\ref{cot1}) the following relations obtained from eqs.(\ref{z}) 
and (\ref{fa}),
\be\label{aeq}
a_{eq} = \frac{a_0}{1 + z_{eq}} \quad ,  \quad \frac{a_{eq}}{a_r} 
= \sqrt{\frac{t_{eq}}{t_r}} 
= \sqrt{\frac{H_r}{H_{eq}}} \; .
\ee
Notice that the Hubble parameter at the end of inflation and the beginning 
of the radiation era $ H_r $ is smaller than the Hubble parameter at the 
beginning of inflation $ H $. As we show in sec. \ref{3C}, 
$ H_r \simeq H/\sqrt{N} $ where $ N \sim 60 $ is the number of efolds since
the cosmologically relevant modes exit the horizon till the end of 
inflation. For simplicity, we used in 
eq.(\ref{dinfla}) the Hubble parameter $ H $ at the beginning of inflation.
Replacing the above results, eqs.(\ref{cot1}) and (\ref{aeq}), 
into eq.(\ref{N}) yields
\be\label{cot2}
e^{N_{tot}} \geq \frac{3.362\ldots}{H_0} \; \sqrt{H_{eq}\; H} \; 
\frac{N^\frac14}{1 + z_{eq}}
\ee
Let us consider the Friedmann equation at the transition from RD to MD  
(at matter-radiation equality)
$$
H_{eq}^2 = \frac{2 \; \rho_m(t_{eq})}{3 M^2_{Pl}} \,.
$$
where $ \rho_m(t_{eq}) = \rho_r(t_{eq}) $ stand for the matter and 
radiation densities at matter-radiation equality. Since,
$$
\frac{\rho_m(t_{eq})}{\rho_m(t_0)} = \left(\frac{a_0}{a_{eq}}\right)^3  
\qquad {\rm and}  \qquad \rho_m(t_0) = \Omega_M \; \rho_c \; ,
$$
we find using also eq.(\ref{rhocrit})
\be\label{heq}
H_{eq} = \sqrt{2 \; \Omega_M} \; \left(\frac{a_0}{a_{eq}}\right)^\frac32 \;
H_0 = \sqrt{2 \; \Omega_M} \; (1 + z_{eq})^\frac32 \; H_0 \; .
\ee
Inserting eq.(\ref{heq}) into eq.(\ref{cot2}) yields
\be\label{ntot1}
e^{N_{tot}} \geq 3.362\ldots \; \sqrt{\frac{H}{H_0}} \; 
\left(\frac{2 \; \Omega_M \; N}{1 + z_{eq}}\right)^\frac14 \; .
\ee
and
\be\label{ar}
a_r = 100 \; \beta \; \left(\frac{2 \; \Omega_M \; N}{1 + z_{eq}}\right)^\frac14 \; 
\sqrt{\frac{H_0}{M_{Pl}}} \; ,
\ee
where
\be\label{calR}
\beta \equiv \sqrt{\frac{10^{-4} \; M_{Pl}}{H}} \; .
\ee
Introducing in eqs.(\ref{ntot1})-(\ref{calR}) the explicit values $ 1 + z_{eq} = 3200,
\; N \simeq 60 $ and those from table I yields,
\be\label{vinN}
N_{tot} \geq 64.8 - \log\beta \quad , \quad 
e^{N_{tot}} \geq \frac{1.39}{\beta} \; 10^{28} \quad , \quad
a_r \simeq 2.5 \; \beta \;  10^{-29} \; .
\ee
As we shall see below [see eq.(\ref{myH})], 
$ {\beta} \gtrsim 1 $ for generic slow-roll inflationary models 
reproducing the CMB data. 

We conclude that an inflationary stage before the RD era {\bf solves} the 
horizon problem provided inflation lasts  {\bf at least sixty-four
efolds}.

\subsubsection{The solution to the flatness problem in inflation}

The total density today $ \Omega_0 $ can be related to the total density at
the beginning of inflation following eq.(\ref{ome1}):
$$
\sqrt{|\Omega_0 - 1|} = \sqrt{|\Omega(t_{min}) - 1|} \; 
\frac{a(t_{min}) \; H(t_{min})}{a_0 \; H_0} \; .
$$
Now, eq.(\ref{cot1}) precisely ensures that 
\be\label{cota2}
\frac{a(t_{min}) \; H(t_{min})}{a_0 \; H_0} \leq 1 \; .
\ee
Therefore, having at least sixty-two efolds of inflation guarantees that 
$ \sqrt{|\Omega_0 - 1|} \leq \sqrt{|\Omega(t_{min}) - 1|} $.
There is no need to fine-tune $ \Omega(t_{min}) $ to unity as it was the 
case in absence of inflation. 
The observed value of $ \Omega_0 $ eq.(\ref{omeg}) can be explained
as the result of a value for $ \Omega(t_{min}) \sim \Omega_0 $ provided 
eq.(\ref{cota2}) is valid which is guaranteed by $ N_{tot} \gtrsim 64 $.

\subsubsection{The Entropy of the Universe}

The entropy of the universe today is dominated by the photons
(the CMB) and the neutrinos and takes the value
$$
S \sim  d_0^3 \; [s_{\gamma} + s_{\nu}]= 0.97 \times 10^{89}
$$
where $ d_0 $ is the particle horizon today [eq.(\ref{dt})], 
$ s_{\gamma} $ and $ s_{\nu} $ are the entropies per unit comoving volume 
of the photon and the neutrinos
$$
s_{\gamma} = \frac{2 \; \pi^2}{45} \; g_{\gamma} \; T_{\gamma}^3 \quad , \quad 
s_{\nu} = \frac{7 \; \pi^2}{180} \; g_{\nu} \; T_{\nu}^3 \; ,
$$
$ g_{\gamma} = 2 $ counts the photon polarizations, $ g_{\nu} = 2 $ counts the 
neutrino states, $ T_{\gamma} = 2.725 $ K and $ T_{\nu}= (4/11)^\frac13 \; T_{\gamma}$ 
are the CMB temperature and the neutrino temperature, respectively \cite{libros}.

An important problem in cosmology has been to explain such huge value for 
the entropy today \cite{kt,mass}. 

\medskip

Let us show that the entropy is constant from the microscopic evolution 
equation (\ref{conener}). Let us consider the energy $ E(t) $ inside a
comoving volume $ V_c $:
$$
E(t) = \rho(t) \; a^3(t) \;  V_c
$$
while the physical volume grows as $ V(t) =  a^3(t)  \;  V_c $.
Multiplying eq.(\ref{conener}) by $ V(t) $ yields,
$$
 V_c \; a^3(t) \; \left[\dot{\rho}+3 \; H(t) \; \left( \rho+p\right)  
\right] = \dot{E(t)} + p \;  \dot{V(t)} =0
$$
Therefore, according to the first principle of thermodynamics
$$
T \; dS = dE + p \; dV = 0
$$
and entropy is conserved.

Eq.(\ref{conener}) is valid both in classical and quantum
field theory as shown in \cite{chalo,chalo2}. Namely, the entropy
remains {\bf constant} according to the {\bf microscopic} evolution
equations. Entropy grows upon coarse-graining of degrees of
freedom when quantum decoherence happens as it is the case during inflation. 
Namely, inflation stretches the lengths by a enormous factor of at least
$ \sim e^{64} \sim 10^{28} $ making classical the
quantum description of   matter \cite{ps,chalo2}.

More precisely, the huge number of ultrarelativistic
particles created during the reheating phase between slow-roll 
quasi-de Sitter inflation and the RD era must be 
described by a {\bf density matrix} (probably out of thermal
equilibrium) and not by a pure quantum state by the
the end of reheating. Let us estimate the entropy during reheating.

The entropy by the end of reheating and the beginning of the RD era
is dominated by ultrarelativistic
particles within a horizon size patch. The horizon size by the end
of inflation follows from eqs.(\ref{dinfla}) and (\ref{N})  to be
$ d(t_{end}) \sim e^{N_{tot}} \; / H $. Approximating the expansion during 
reheating by the RD scale factor, the horizon size patch gets redshifted by
$ \sqrt{H/H_{reh}} $, where $ H_{reh} $ stands for the Hubble parameter by 
the end of reheating and the patch size results
\be\label{dreh}
d_{reh} \sim \sqrt{\frac{H}{H_{reh}}} \;  d(t_{end}) \sim 
\frac{e^{N_{tot}}}{\sqrt{H \; H_{reh}}} \; .
\ee
The reheating temperature $ T_{reh} $ is related to the Hubble parameter 
$ H_{reh} $ by the Friedmann equation:
\be\label{frier}
 H_{reh}^2 = \frac{\pi^2 \; g_{rh}}{90 \; M_{Pl}^2} \;  T_{reh}^4 \; ,
\ee
where $ g_{rh} $ is the number of ultrarelativistic degrees of freedom
by the end of reheating and we used $ \rho_r = (\pi^2 \; g_{rh}/30) \;  T_{reh}^4 $ 
for the energy density of the ultrarelativistic particles.
The entropy by the end of reheating is,
\be\label{sreh}
S_{rh}\sim \frac{2 \; \pi^2 \; g_{rh}}{45} \; T_{reh}^3 \; d_{reh}^3 \; .
\ee
Using eqs.(\ref{dreh})-(\ref{sreh}) and imposing that this value accounts 
for the entropy today $ \sim  10^{89} $ yields
$$
S_{rh}\sim 2^\frac32 \; \left(\frac{2 \; \pi^2 \; 
g_{rh}}{45}\right)^\frac14
\; e^{3 \; N_{tot}} \; \left(\frac{M_{Pl}}{H}\right)^\frac32 \geq  10^{89}
$$
That is,
\be\label{62}
N_{tot} \geq 62.4 - \frac12 \; \log \beta - \frac1{12} \; 
\log\frac{g_{rh}}{1000}
\; . 
\ee
We thus obtain a lower bound on $ N_{tot} $ similar to
eq.(\ref{vinN}) for the solution of the horizon problem.

\medskip

In summary a number of efolds $ N_{tot} \gtrsim 64 $ during inflation 
reproduces the value of the entropy of the universe today and solves both 
the horizon and flatness problems.

\subsubsection{The Age of the Universe}\label{edad}

Usually, the age of the universe is computed in cosmic time. 
At the beginning of the matter dominated 
era the Universe was only $ \sim 10^5 $ yrs old. Hence, to compute the age of 
the universe today we can restrict ourselves
to the matter and dark energy dominated eras. Therefore, neglecting radiation
the Friedmann equation (\ref{fri}) in cosmic time takes the form
$$
\frac1{a} \; \frac{da}{dt} = H_0 \; \sqrt{\Omega_{\Lambda} + 
\frac{1-\Omega_{\Lambda}}{a^3}} \; ,
$$
where  $ \Omega_{\Lambda} = \rho_{\Lambda}/\rho_c $ and then
\be\label{tcomo}
H_0 \; t = \int_0^a \frac{da}{a \; \sqrt{\Omega_{\Lambda} + \frac{1-\Omega_{\Lambda}}{a^3}}}=
\frac2{3 \sqrt{\Omega_{\Lambda}}} \; {\rm Arg \; Sinh}
\left[\sqrt{\frac{\Omega_{\Lambda}}{1-\Omega_{\Lambda}}} \; a^{\frac32} \right]
\ee
Therefore, the scale factor grows as
\be\label{acom}
a(t) = \left(\frac{\Omega_{\Lambda}}{1-\Omega_{\Lambda}}\right)^{\frac13}
\left[\sinh \left(\frac32 \;
\sqrt{\Omega_{\Lambda}} \;  H_0 \; t \right)\right]^{\frac23} \; .
\ee
We have today $ t = t_0 $ and $ a(t_0) = 1 $. We thus obtain from eq.(\ref{tcomo})
$$
t_0 = \frac2{3  \; H_0 \; \sqrt{\Omega_{\Lambda}}} \;
{\rm Arg \; Sinh} \left[\sqrt{\frac{\Omega_{\Lambda}}{1-\Omega_{\Lambda}}}\right]
$$
Here, $ \Omega_{\Lambda} = 0.726 $ \cite{WMAP5} yields for the age of the Universe in cosmic time
$$
t_0 = \frac{0.9887}{H_0} = \frac{9.667}{h} \,  {\rm Gyr} = 13.71 \, {\rm Gyr}
$$
where we used $ h = 0.705 $ \cite{WMAP5}. $ t_0 $ grows monotonically with $ \Omega_{\Lambda} $.
In particular, $ t_0 \buildrel{\Omega_{\Lambda} \to 0}\over= 2/(3 \;H_0) $.

From eq.(\ref{acom}) we notice that $ a(t) $ exhibits the matter dominated behaviour
$ \sim t^{\frac23} $ for early times $ t \ll t_0 $ and the de Sitter behaviour 
$ \sim e^{H_0 \; \Omega_{\Lambda} \; t} $ for late times $ t \gtrsim t_0 $.

For times $ t \lesssim 1 $ Gyr, $ z \lesssim 4 ,  \; \Omega_{\Lambda} $ can be neglected
in eq.(\ref{frgen}) and the cosmic time results
$$
t = \frac1{H_0} \int_0^a \frac{a \; da}{\sqrt{\Omega_r + \Omega_M \; a}} =
\frac{2 \; \sqrt{\Omega_r}}{3 \;  H_0 \; \Omega_M}
\left[\left(a-\frac{2 \; \Omega_r}{\Omega_M}\right)
\sqrt{1+ \frac{\Omega_M}{\Omega_r} \; a }+ \frac{2 \; \Omega_r}{\Omega_M}\right] \; .
$$

The Friedmann equation (\ref{fri}) in conformal time eq.(\ref{confo})
takes the form
$$
\frac1{a^2} \; \frac{da}{d\eta} = H_0 \; \sqrt{\Omega_{\Lambda} + \frac{1-\Omega_{\Lambda}}{a^3}} \; .
$$
Thus \cite{bri},
\be\label{etaa}
H_0 \; \eta  = \int_0^a \frac{da}{a^2 \; \sqrt{\Omega_{\Lambda} + \frac{1-\Omega_{\Lambda}}{a^3}}}=
\frac1{3^\frac14 \; \Omega_{\Lambda}^\frac16  \; \left(1-\Omega_{\Lambda}\right)^\frac13 } \;
F(\phi,q) \; .
\ee
Here, $ F(\phi,q) $ is the elliptic integral of first kind, 
\be\label{integeli}
\cos \phi = \frac{(1-\sqrt3)\Omega_{\Lambda}^\frac13 \; a + (1-\Omega_{\Lambda})^\frac13}
{(1+\sqrt3)\Omega_{\Lambda}^\frac13 \; a + (1-\Omega_{\Lambda})^\frac13} \quad {\rm and} \quad 
q = \sin \left(\frac{5 \, \pi}{12}\right) = \frac{1+\sqrt3}{2 \, \sqrt2} \; .
\ee
The age of the universe in conformal time $ \eta_0 $ follows by setting 
$ a = 1 $ in eqs.(\ref{etaa})-(\ref{integeli}),  
\be\label{horiz}
\eta_0 = \frac{3.3623\ldots}{H_0} = \frac{32.88\ldots}{h}  \,  {\rm Gyr} =
\frac{10.1\ldots}{h} \, {\rm Gpc} = 47.6 \, {\rm Gyr} = 14.3 \, {\rm Gpc} 
\; ,
\ee
where we used $ \Omega_{\Lambda} = 0.726 $ and  $ h = 0.705 $ \cite{WMAP5}.
Notice that $ \eta_0 $ coincides with the particle horizon today 
eq.(\ref{dfhp}).

From eq.(\ref{etaa}) we obtain the scale factor as function of the conformal time,
\be\label{jaco}
a(\eta) =  \left(\frac{1-\Omega_{\Lambda}}{\Omega_{\Lambda}}\right)^\frac13 \;
\frac{1-{\rm cn}(u,q)}{\sqrt3-1 + (1+\sqrt3) \; {\rm cn}(u,q)} \; 
\ee
where $ u \equiv 3^\frac14 \; \Omega_{\Lambda}^\frac16  \; \left(1-\Omega_{\Lambda}\right)^\frac13
\;  H_0 \; \eta $ and cn$(u,k)$ is the Jacobi cosinus function \cite{jel}.

Since the integral in eq.(\ref{etaa}) converges for $ a = \infty $, the conformal
time ranges in the finite interval:
$$
0 \leq  \eta \leq  \eta_1 \equiv 
\frac{2.8044\ldots}{\Omega_{\Lambda}^\frac16  \; \left(1-\Omega_{\Lambda}\right)^\frac13 } 
\; \frac1{H_0} = 4.554\ldots \; \frac1{H_0} \; .
$$
It follows from eqs.(\ref{etaa}) and (\ref{jaco}) that the scale factor behaves for 
early times as matter dominated,
$$
a(\eta) \buildrel{\eta \to 0}\over= \frac14 \; \left(1-\Omega_{\Lambda}\right) \; 
\left( H_0 \; \eta\right)^2 \to 0 \; ,
$$
while for times $ \eta $ approaching  $ \eta_1 $, $ a(\eta) $ exhibits a de Sitter
behaviour,
$$
a(\eta) \buildrel{\eta \to\eta_1 }\over= \frac1{\sqrt{\Omega_{\Lambda}} \;
H_0 \; (\eta_1-\eta)} \to \infty \; .
$$
We see that the comoving particle horizon keeps growing, asymptotically 
reaching the limiting size $ \eta_1 \simeq 63.2 \, {\rm Gyr}= 19.4  \, 
{\rm Gpc} $ for $ a(\eta) \to \infty $.

\subsection{Inflationary Dynamics in the Effective Theory of 
Inflation}\label{tres}

As discussed  in the previous section, inflation was originally proposed to
solve the flatness, horizon and entropy problems \cite{guthsato} thus
becoming an important paradigm in cosmology. At the same time, it
provides a natural mechanism for the generation of scalar density
fluctuations that seed large scale structure as well as that of  tensor perturbations
(primordial gravitational waves), thus explaining the
origin of the temperature anisotropies in the cosmic microwave
background (CMB). Inflation is the statement that
the cosmological scale factor $ a(t) $ has a
\emph{positive acceleration}, namely,   $ \ddot{a}(t)/a(t) > 0 $.
Hence, eq.(\ref{accel}) requires the equation of state $ p/\rho < -1/3 \, $. 

\begin{figure}[h]
\includegraphics[height=8cm,width=12cm]{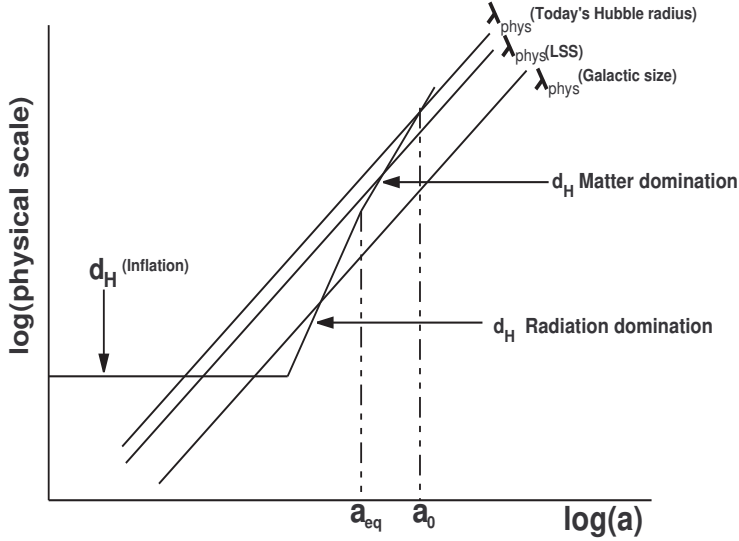}
\begin{center}
\caption{Logarithm of physical scales vs. logarithm of the scale
factor. The Hubble radius $ d_H $ is shown for the inflationary
(De~Sitter), radiation dominated and matter dominated stages. The
physical wavelengths  for today's Hubble radius $ d_H(today) $, and
a galactic scale $ \lambda_{gal} $ are shown.  
We see that modes that leave the horizon during inflation get back
during the RD and MD eras with wavelengths today between the galactic
sizes and today's Hubble radius.}
\label{inflation}
\end{center}
\end{figure}

Inflation gives rise to a remarkable phenomenon: physical
wavelengths grow \emph{faster} than the size of the Hubble radius
$$
d_H = \frac{a(t)}{\dot{a}(t)} = \frac1{H(t)} \; , 
$$
indeed,
\be
\frac{\dot{\lambda}_{phys}}{\lambda_{phys}} = \frac{\dot{a}}{a}=
H(t)= \frac{\dot{d}_H}{d_H}+ d_H \, \frac{\ddot{a}}{a} \; ,
\label{inflad} 
\ee 
\noindent Therefore, during inflation ($ \ddot{a} > 0 $), eq.(\ref{inflad}) states that 
 \emph{physical wavelengths become larger than the Hubble
radius}. Once a physical wavelength becomes larger than the Hubble
radius, it is causally disconnected from physical processes. The
inflationary era is followed by the radiation dominated and matter
dominated stages where the acceleration of the scale factor
becomes negative since $ p/\rho = 1/3 $ in a radiation dominated
era and $ p=0 $ in a matter dominated era [see eq.(\ref{accel})].
With a negative acceleration of the scale factor, the Hubble
radius grows {\bf faster} than the scale factor, and wavelengths that
were outside the Hubble radius, can now re-enter the Hubble radius. This is depicted
in fig. \ref{inflation}.
This is the main concept behind the inflationary paradigm for the
generation of the CMB temperature fluctuations as well as for providing
the seeds for large scale structure formation: quantum
fluctuations generated early in the inflationary stage exit the
Hubble radius during inflation, and eventually re-enter the horizon during the
matter dominated era. 

The basic mechanism for generation of
temperature anisotropies and primordial gravitational waves
through inflation is the following \cite{kt,libros,mass}: the
energy momentum tensor is split into the classical fluid component $
T^{\mu \nu}_{fluid} $ [eq.(\ref{fluido})] that drives the
classical FRW metric plus small quantum fluctuations $ \delta \,T^{\mu \nu} $, namely 
$$
T^{\mu \nu} = T^{\mu \nu}_{fluid}+ \delta \,T^{\mu \nu} \; .
$$ 
The quantum fluctuations of the matter fields induce quantum
fluctuations in the metric (space-time geometry) $ \delta \, G^{\mu \nu} =
\delta\,T^{\mu \nu}/M^2_{Pl} $. In the linearized approximation,
the different wavelengths of the perturbations evolve
independently. After a given wavelength exits the Hubble radius,
the corresponding perturbation becomes \emph{causally
disconnected} from microphysical processes. Perturbations that
re-enter the horizon late in the RD era and during the MD era, induce small
fluctuations in the space-time metric which generate fluctuations in
the matter distribution driving acoustic oscillations in the
photon-baryon fluid. At the last scattering surface, when photons
decouple from the plasma these oscillations are imprinted in the
power spectrum of the temperature anisotropies of the CMB and seed
the inhomogeneities which generate structure upon gravitational
collapse\cite{libros,hu}. 

The horizon problem, namely why the
temperature of the CMB is nearly homogeneous and isotropic (to one
part in $ 10^5 $) is solved by an inflationary epoch because the
wavelengths corresponding to the Hubble radius at the time of
recombination were \emph{inside} the Hubble radius hence in causal
contact during inflation (see sec. \ref{soluhor}). 
This mechanism is depicted in fig. \ref{inflation}. 

Slow-roll inflation leads to a gaussian and nearly scale
invariant spectrum of adiabatic scalar (curvature) and tensor
(gravitational waves) primordial fluctuations
These predictions of inflationary models
make the inflationary paradigm robust and have been spectacularly 
confirmed by the WMAP data for curvature (temperature) fluctuations 
\cite{WMAP1,WMAP3,WMAP5}. WMAP has also provided
perhaps the most striking validation of inflation as a mechanism
for generating \emph{superhorizon}  fluctuations, through the
measurement of the anticorrelation peak in the
temperature-polarization (TE) angular power spectrum at $ l \sim
150 $ corresponding to superhorizon
scales \cite{WMAP1,WMAP3}.

Let us now estimate the wavenumbers of the cosmologically relevant modes
that reenter the horizon today. They have a wavelength at most of the 
order of the horizon today [eq.(\ref{horiz})]
$$
\eta_0 = \frac{3.362\ldots}{H_0} \simeq \frac{10.1}{h}  \,  
{\rm Gpc} \simeq 14.3 \,  {\rm Gpc}
$$
Notice that the conformal time eq.(\ref{confo}) coincides at any time
with the comoving size of the horizon eq.(\ref{dfhp}).

Actually, a more precise calculation in eq.(\ref{kqu}) yields for the 
CMB quadrupole modes (the longest observed wavelengths) 
\cite{quadru2,quamc}
\be\label{kqu1}
 k_Q = \frac{3.342\ldots}{\eta_0-\eta_{LSS}} = 1.014 \times H_0 =
 0.238 \; {\rm Gpc}^{-1} = 1.52 \times 10^{-42} \; {\rm GeV} \; ,
\ee
where LSS stands for last scattering surface.
Notice that half the quadrupole wavelength today is very close to the 
horizon size today eq.(\ref{horiz}).

According to eqs.(\ref{ar})-(\ref{vinN}) the physical wavenumber $ k $
of a mode today and its physical wavenumber $ k^{init} $ at the 
beginning of inflation are related by the redshift factor
\be\label{ar1}
k^{init} = \frac{e^{N_{tot}}}{a_r} \; k = 2.49 \; 10^{56} \; 
\beta^{-1} \; e^{N_{tot}-64} \; k \; .
\ee
The total number of efolds during inflation must be $ N_{tot} \geq 64 $
according to eq.(\ref{vinN}). We choose $ N_{tot} = 64 $ as benchmark 
value since the detailed analysis of the fast-roll 
explanation of the quadrupole supression (see secs. \ref{ntot64} and 
\ref{fastroll}) favours  $ N_{tot} = 64 $ \cite{quamc}. 
Another {\bf hint} to choose $ N_{tot} 
\simeq 64 $ comes from the WMAP preferred  value $ n_s \simeq 0.96 $ 
(see sec. \ref{fastroll}). 

In particular, we find for the quadrupole mode from  eqs.(\ref{kqu1}) and 
(\ref{ar1})
\be\label{qini}
k_Q^{init} = 3.79 \; \beta^{-1} \; e^{N_{tot}-64} \;
10^{14} \; {\rm GeV} \; .
\ee
[Recall that $ 1 \, {\rm GeV} = 1.564 \times  10^{41} \; 
({\rm Gpc})^{-1} $].

CMB and LSS observations allow to detect the $k$-modes over four orders 
of magnitude, in a range going from $ \sim 1 $ Mpc to $ \sim 10^4 $ Mpc: 
$ k_Q < k < 10^4 \; k_Q $. At the beginning of inflation this corresponds 
to the physical wavenumbers
\be\label{rangok}
3.8 \;  10^{14} \; {\rm GeV} \;  {\beta}^{-1} \; e^{N_{tot}-64} < k^{init} <
 3.8 \; 10^{18} \; {\rm GeV} \; {\beta}^{-1} \; e^{N_{tot}-64} \; .
\ee
These energy values are to be compared with the inflaton mass scale 
[eq.(\ref{myH})] $ m \sim 10^{13} $ GeV.  We see that the cosmologically 
relevant modes are in a range going from $ \sim 30 $ inflaton masses $ m $
till the Planck scale for typical values $ N_{tot} \sim 64 $ and 
$ H \sim 10^{-4} \; M_{Pl} $.
These  Planck scale $k$-values are high energy modes within the effective theory of 
inflation since $ k \gg M $ for them.

\medskip

A $k$-mode crosses the horizon when $ k \; \eta \sim 1 , \; \eta $
being the comoving horizon size eq.(\ref{confo}). Actually,
it is convenient to redefine $ \eta $ after inflation as,
\be\label{confo2}
\eta \equiv \int_{t_{end}}^t \frac{dt'}{a(t')} = \int_{a_r}^a
\frac{da}{a^2 \; H(a)} \; ,
\ee
where $ a_r \sim 10^{-29} $  eq.(\ref{vinN}). Alternatively,
the horizon crossing condition for a $k$-mode can be taken
$ k = H(t) \; a(t) $ as depicted in fig. \ref{acus}.
Using eqs.(\ref{frgen}) and (\ref{confo2}) we find
\be\label{etaz}
\eta(z) = \frac{I(z)}{H_0} \quad {\rm where} \quad I(z) \equiv
\int^{\frac1{z+1}}_{a_r}\frac{da}{\sqrt{\Omega_r
+ \Omega_M \; a + \Omega_{\Lambda} \; a^4}}
\ee
and the wavevector $ k(z) $ that reenters the horizon at redshift 
$ z $ is given by
\be\label{klz}
 k(z) \sim \frac{H_0}{I(z)} \; .
\ee
We normalize $ k(z) $ such that $ k(z=0) = k_Q $ given by eq.(\ref{kqu1}) 
and we use that $ I(0) = 3.3623 $. This yields,
\be\label{kz}
k(z) = 3.409 \; \frac{H_0}{I(z)} \; .
\ee
For $ z \gtrsim 1 , \; \Omega_{\Lambda} $ can be neglected in the
integrand of  $ I(z) $ and we obtain
\be\label{Izanal}
 I(z) = \frac{2 \; \sqrt{\Omega_r}}{\Omega_M}\left[
\sqrt{1+\frac{\Omega_M}{\Omega_r \; (z+1)}}-1\right]=
0.0659 \; \left[\sqrt{1+\frac{3227}{z+1}}-1\right] \; .
\ee

\medskip

The corresponding $\ell$-multipole is given by the maximum of the 
spherical Bessel function \cite{libros} $ j_\ell(x) $ where 
\be\label{defx}
x \equiv k \; (\eta_0-\eta_{LSS}) \gg 1 \; ,
\ee
and
\be \label{etaLSS}
\eta_0-\eta_{LSS} = \frac1{H_0} \; \int^1_{\frac1{1+z_{LSS}}}
\frac{da}{\sqrt{\Omega_r + \Omega_M \; a + \Omega_{\Lambda} \; a^4}} \; , 
\ee 
is the comoving distance between today and the last scattering surface 
(LSS). We find using table I and $ z_{LSS} =1100 $,
\be\label{329}
\eta_0-\eta_{LSS} = \frac{3.296}{H_0}  \; . 
\ee 
Hence, $ k^{init} $ (at the beginning of inflation)
and $ x $ are related by
\be \label{rangk}
k^{init} = 1.135 \; 10^{14} \; {\rm GeV} \; e^{N_{tot}-64} \;
\beta^{-1} \; x  \; .
\ee
and we find for $ x(z) $ 
\be\label{xz}
x(z) = k(z) \; (\eta_0-\eta_{LSS}) = \frac{11.24}{I(z)} \; .
\ee
where we used eqs.(\ref{ar1}), (\ref{kz}), (\ref{defx}) and (\ref{329}).

For large $ \ell $ and $ x $ the maximum of $ j_\ell(x) $ is at 
$ \ell \sim x $ where Watson's formula applies \cite{jel}
\be\label{wats}
j_\ell(x) \buildrel{\ell , \; x \gg 1}\over= \sqrt{\frac{\pi}{2 \, x}} \; 
\left(\frac{3 \; y}{\ell+\frac12}\right)^\frac13 \; S(y) \; ,
\ee
where,
$$
y \equiv \frac{\left(x^2 - [l+1/2]^2\right)^{3/2}}{3 \; [l+1/2]^2} \quad 
{\rm and} \quad S(y) \equiv J_{\frac13}(y) +  J_{-\frac13}(y) \; .
$$
Here $ J_{\pm\frac13}(y) $ are Bessel functions.
The function $ y^{\frac13} \; S(y) $ has its maximum at 
$ y = 0.685948 \ldots $ and hence eq.(\ref{wats})
gives the maximum of $ j_\ell(x) $ at
\be \label{maxjl}
\ell+\frac12 = x \left[ 1 - \frac{0.808617\ldots}{x^{\frac23}} + 
{\cal O}\left(x^{-\frac43}\right)\right] \quad , \quad
x = \left(\ell+\frac12\right) \; \left[ 1 + 
\frac{0.808617\ldots}{\left(\ell+\frac12\right)^{\frac23}}
+ {\cal O}\left(\left[\ell+\frac12\right]^{-\frac43}\right)\right] \quad ,
\; x , \; \ell \gg 1 \; .
\ee
Therefore, $ \ell+\frac12 $ is {\bf smaller} than $ x $ at the maximum 
of  $ j_\ell(x) $ by an amount of the order $ x^{\frac13} $. 
In particular, at  $ \ell = 2 $ eq.(\ref{maxjl}) yields
$ x = 3.597\ldots $ for  the position of the maximum which is only 
$8\%$ larger than the exact result $ x=3.342\ldots $
Namely, the physical quadrupole ($ l = 2 $) wavemodes today $ k_Q $ are related 
to the particle horizon by \cite{libros}   
\be\label{kqu}
 k_Q  \; (\eta_0-\eta_{LSS}) = 3.342\ldots \; ,
\ee
where the spherical Bessel function 
$ j_2(k \; [\eta_0-\eta_{LSS}]) $ takes its maximum value, 
and $ \eta_0-\eta_{LSS} $ is given by eq.(\ref{329}).
Therefore, using the present value for $ H_0 $ \cite{WMAP5}
we obtain $  k_Q = 0.238  \; ({\rm Gpc})^{-1} $. 

The modes reentering during the MD and $\Lambda$D eras are in the range
$$
0 \leq z \leq 3200 \quad , \quad 2 \leq \ell \leq  392 \quad , \quad 
k_Q^{init} \leq k^{init} \leq 
4.44 \; 10^{16} \; {\rm Gev} \; e^{N_{tot}-64} \; {\beta}^{-1} \; . 
$$
where we used eqs.(\ref{Izanal}), (\ref{xz}) and (\ref{maxjl}).
For example, $ \ell(z=1100) = 161 $ and $ \ell(z = 1678) = 220 $ 
(first peak in fig. \ref{acus}) reenter in the MD era. We display in Table II the 
reentering redshift $ z $, the wavenumber $ k^{init} $ and the corresponding
CMB multipole $ \ell $. 

\begin{table}
\begin{tabular}{|c|c|c|}
\hline  $z$  &  $ k^{init} \; e^{64-N_{tot}} \; \beta $ &  $ \ell $    \\
\hline 0 & $ 3.7 \; 10^{14} \; {\rm Gev}  $ 
& $ 2 $ \\ \hline
$ 1100 $ & $ 1.9 \; 10^{16} \; {\rm Gev}  $ &
$ 161 $ \\ \hline
$ 1678 $ &  $ 2.6 \; 10^{16} \; {\rm Gev}  $ &
$ 220 $ \\ \hline
$ 3200 $ &  $ 4.5 \; 10^{16} \; {\rm Gev}  $ & $ 385 $ 
\\ \hline
\end{tabular} 
\caption{The reentering redshift $ z $, the initial wavenumber $ k $ 
and the corresponding CMB multipole $ \ell $.}
\end{table}

We have from eqs.(\ref{Izanal}) and (\ref{xz}) for $ x \gtrsim 3, \; \ell \gtrsim 2 $,
\bea\label{kdel}
&&x(z) = 5.35 \; \sqrt{\Omega_r} \; (z+1)\left[1+
\sqrt{1+\frac{\Omega_M}{\Omega_r \; (z+1)}}\right] =
0.0493 \; (z+1)\left[1+\sqrt{1+\frac{3227}{z+1}}\right] \; , \cr \cr
&& z+ 1 = \frac{10.14 \; x^2}{x + 79.55} \; .
\eea
For example, we get  $ \ell \simeq x \simeq 10^4 $ for $ z \simeq 10^5 $ 
which is well after BBN. For such modes $ k^{init}\simeq  {\beta}^{-1} 
\; 10^{18} \; e^{N_{tot}-64}$ GeV.

\noindent
We want to draw the attention on the fact that $ k^{init} < M_{Pl} $ for $ \ell 
< 2 \times 10^4 \; e^{N_{tot}-64} \; \beta^{-1} $ and $ z <  2 \times 10^5 \; 
e^{N_{tot}-64} \; \beta^{-1} $ as follows from eqs.(\ref{rangk}), (\ref{maxjl}) 
and (\ref{kdel}). Namely, the range eq.(\ref{rangok}) of CMB detectable modes 
{\bf does not contain} trans-planckian wavenumbers for the value $ N_{tot} \sim 64 $ 
derived in sec. \ref{ntot64}. Truly trans-Planckian modes $ k \gg M_{Pl} $ are 
unobservable through the CMB-LSS data. Information about trans-Planckian modes
could be obtained perhaps in the future through the 21cm H line 
\cite{21cm}. The CMB multipoles $ \ell < 200 $ exhibiting features 
\cite{WMAP1,WMAP3,WMAP5} are definitely {\bf sub-Planckian} if $ N_{tot} \sim 64 $, 
since they have $ k^{init} < 2 \; {\beta}^{-1} \; e^{N_{tot}-64} \; 10^{16}$ GeV 
according to eqs.(\ref{rangk}) and (\ref{maxjl}).

\subsubsection{Inflation and Inflaton field dynamics}

A simple implementation of the inflationary scenario is based on a single
scalar field, the \emph{inflaton} with a Lagrangian density 
\be
\mathcal{L} = a^3(t)\left[\frac{\dot{\varphi}^2}2 -
\frac{(\nabla\varphi)^2}{2a^2(t)}-V(\varphi) \right] \; ,
\ee 
where $ V(\varphi) $ is the inflaton potential. Since the universe
expands exponentially fast during inflation, gradient terms 
are  exponentially suppressed and can be neglected.
At the same time, the exponential stretching of spatial lengths
classicalize the physics and permits a classical treatment.
One can therefore consider an homogeneous and classical inflaton field 
$ \varphi(t) $ which obeys the evolution equation
\be\label{eqno} 
{\ddot \varphi} + 3 \, H(t) \; {\dot \varphi} + V'(\varphi) = 0 \; . 
\ee 
in the isotropic and homogeneous FRW metric eq.(\ref{FRW}) which is sourced by the 
inflaton. 

The energy density and the pressure for a spatially homogeneous inflaton 
are given by
\be\label{enerpres} 
\rho = \frac{\dot{\varphi}^2}2+ V(\varphi)
\quad , \quad p  =\frac{\dot{\varphi}^2}2-V(\varphi) \; . 
\ee
The scale factor $ a(t) $ obeys the Friedmann equation eq.(\ref{fri}) which here 
takes the form
\be\label{frinf}
H^2(t) = \frac1{3 M^2_{Pl}} \left[\frac12 \; \dot \varphi^2 + V(\varphi)\right] \; .
\ee
The time derivative of the Hubble parameter takes the form
\be
{\dot H}(t) = -\frac{\dot{\varphi}^2}{2 \;  M^2_{Pl}}
\ee
where we used eqs.(\ref{eqno}) and (\ref{frinf}).
This shows that $  H (t) $ {\bf decreases monotonically} with time.

The inflaton fields starts at $ t = 0 $ with some chosen values of $ \varphi $ 
and  $ \dot \varphi $ and evolves together with the scale factor according to
eqs.(\ref{eqno}) and (\ref{frinf}). The inflaton clearly rolls down the slope
of the potential going towards a local minimum of $ V(\varphi) $.
The basic constraint on the inflationary potential is
\be\label{vmin}
V(\varphi_{min})= V'(\varphi_{min})= 0 \; .
\ee
That is, the inflaton potential {\bf must vanish} at its minimum $ \varphi_{min} $
in order to have a finite number of efolds.
The inflaton evolves from its initial value (which is model dependent)
towards the minimum $ \varphi_{min} $. If $ V(\varphi_{min}) > 0 $, we see from 
eq.(\ref{frinf}) that inflation will be eternal. That is, a de Sitter phase will 
continue forever with the inflaton at the constant value $ \varphi_{min} $.

There are two main classes of inflaton potentials leading to two main  
classes of inflation.

\noindent
(a) In {\bf small field inflation} the minimum of the potential is at a 
non-zero value $ \varphi_{min} \neq 0 $ and the inflaton field starts near 
(or at) $ \varphi = 0 $ evolving towards $ \varphi =\varphi_{min} $.
These are typically discrete symmetry ($ \varphi \to - \varphi $) breaking 
potentials \cite{infnue}, 
\be\label{nueva}
V(\varphi) = \frac{\lambda}4 \left( \varphi^2- 
\frac{m^2}{\lambda} \right)^2 =
-\frac{m^2}2 \; \varphi^2 +\frac{\lambda}4  \; \varphi^4 + 
\frac{m^4}{4 \; \lambda} \quad , \quad {\rm new \; inflation} \; .
\ee
For historical reasons small field inflation is often called {\bf new 
inflation}.

(b) In {\bf large field inflation} the minimum of the potential is at 
$ \varphi_{min} = 0 $ and the inflaton field starts  at $ \varphi \gg M $ 
evolving towards  $ \varphi = 0 $.
These are typically unbroken symmetry potentials \cite{infcao}, 
\be\label{caotica}
V(\varphi)= +\frac{m^2}2  \; \varphi^2 +\frac{\lambda}4  \; \varphi^4 
\quad , \quad {\rm chaotic \; inflation} \; .
\ee
For historical reasons large field inflation is often called {\bf  chaotic inflation}.

\medskip

As we discussed in sec. \ref{uno}, inflation should last at least $ N_{tot} \gtrsim 64 $ 
efolds in order to solve the entropy, horizon and flatness problems. Inflation can produce 
such large number of efolds provided it lasts enough time. This can be achieved if the 
inflaton evolves slowly (slow-roll), namely $ \dot{\varphi}^2 \ll V(\varphi) $. 
This implies from eq.(\ref{enerpres}) that
$$
\rho = -p \simeq  V(\varphi) \simeq  {\rm constant},
$$
as the equation of state leading to a de Sitter universe. 
Eq.(\ref{frinf}) yields as scale factor 
\be\label{groso}
a(t) \simeq e^{H \; t} \quad,\quad H \simeq \sqrt{\frac{V(\varphi)}{3 \; M^2_{Pl}}} 
\ee
[see eq.(\ref{fa})]. However, eq.(\ref{groso}) is only an
approximation to the slow-roll inflationary dynamics [see the discussion in sec. 
\ref{3C} and eqs.(\ref{chislr}) and (\ref{asr})]. 

\medskip

While inflationary dynamics is typically studied in terms of a
\emph{classical} homogeneous inflaton field as explained above, such 
classical field must be understood as the expectation value of a 
\emph{quantum field} in an isotropic and homogeneous quantum state. In
ref.\cite{chalo,chalo2,tsu} the \emph{quantum dynamics} of inflation was
studied for inflaton potentials belonging to the two main classes small and large field
inflation discussed above.

The  initial quantum state was taken to be a gaussian wave
function(al) with vanishing or non-vanishing expectation value of
the field. This state evolves in time with the full inflationary
potential which features an unstable (spinodal) region for $ \varphi^2 <
m^2/(3 \; \lambda) $ where $ V''(\varphi) < 0 $ in the broken symmetric 
case eq.(\ref{nueva}). Just as in the case of Minkowski space time, there 
is a band of spinodally or
parametrically unstable wavevectors, in which the
amplitude of the quantum fluctuations grows exponentially fast
\cite{tranfas,chalo}. Because of the
cosmological expansion wave vectors are redshifted into the
unstable band and when the wavelength of the unstable modes
becomes larger than the Hubble radius these modes become
\emph{classical} with a large amplitude and a frozen phase. These
long wavelength modes assemble into a classical coherent and
homogeneous condensate, which obeys the equations of motion of the
classical inflaton\cite{chalo,chalo2,tsu}.  This phenomenon of
classicalization and the formation of a homogeneous condensate
takes place during the \emph{first} $ 5-10 $ efolds after the
beginning of the inflationary stage. The {\bf non perturbative}
quantum field theory treatment in refs.\cite{chalo,chalo2,tsu} shows that 
this rapid redshift and classicalization justifies the use of an 
homogeneous classical inflaton leading to the following robust
conclusions \cite{chalo,chalo2,tsu}:
\begin{itemize}
\item{ The quantum fluctuations of the inflaton are of two
different kinds: 

\noindent
(a) Large amplitude quantum inflaton fluctuations generated
at the beginning of inflation through spinodal instabilities or parametric
resonance depending on the inflationary scenario chosen. They have
at the beginning of inflation physical 
wavenumbers in the range of (see \cite{chalo,chalo2,tsu})
\be\label{modosa}
k \lesssim 10 \; m \; ,
\ee
and they become superhorizon a few efolds after the beginning of 
inflation.
The phase of these long-wavelength inflaton fluctuations freeze 
out and their
amplitude grows thereby effectively forming a homogeneous
\emph{classical} inflaton condensate. The study of more general initial
quantum states featuring highly excited distribution of quanta lead
to similar conclusions \cite{tsu}: during the first few efolds of
evolution the rapid redshift produces a classicalization of
long-wavelength  inflaton fluctuations and the emergence of a homogeneous
coherent  inflaton condensate obeying the \emph{classical equations of
motion} in terms of the inflaton potential. 

\noindent
(b) Cosmological scales relevant for the observations \emph{today} 
between $ \sim 1 $ Mpc and the horizon today had first crossed 
(exited) the Hubble radius inside a window of about $ 10 $ e-folds from
$ \sim 63 $ to $ \sim 53 $ efolds before the end of inflation \cite{kt}.
These correspond to small fluctuations of high physical wavenumbers
at the beginning of inflation in the
range given by eq.(\ref{rangok}). Since [eq.(\ref{myH})] 
$ m \sim 10^{13} $ GeV,  we see that the large amplitude modes 
eq.(\ref{modosa}) for typical values $ N_{tot} \sim 64 $ and 
$ H \sim 10^{-4} \; M_{Pl} $ are {\bf below} the wavenumbers $k$ of the 
cosmologically relevant modes eq.(\ref{rangok}).}
\item{During the rest of the inflationary stage the dynamics is described
by this classical homogeneous condensate that obeys the classical 
equations of motion with the inflaton potential. Thus, inflation even if
triggered by an initial quantum state or density matrix of the
quantum field, is effectively described in terms of a classical 
homogeneous scalar condensate. }
\end{itemize}

The body of results emerging from these studies provide a
justification for the description of inflationary dynamics in
terms of a \emph{classical} homogeneous scalar field. The conclusion
is that  after a few initial e-folds during which the unstable
wavevectors are redshifted well beyond the Hubble radius, all
what remains for the ensuing dynamics is a homogeneous classical
condensate, plus small quantum fluctuations corresponding to the wave 
$k$-modes.

These {\bf small} quantum fluctuations include scalar curvature and tensor
gravitational fluctuations. They must be treated
together with the inflaton fluctuations in the unified gauge invariant
approach we present in sec. \ref{gistf} \cite{hu}.
In the treatment of {\bf large} amplitude quantum inflaton fluctuations,
gravitational fluctuations can be safely neglected \cite{chalo}.

\medskip

Inflation based on a scalar inflaton field should be considered as
an {\bf effective theory}, namely,  not necessarily a fundamental
theory but as a low energy limit of a microscopic fundamental
theory. The inflaton may be a coarse-grained average of
fundamental scalar fields, or a composite (bound state) of fields
with higher spin, just as in superconductivity. Bosonic fields do
not need to be fundamental fields, for example they may emerge as
condensates of fermion-antifermion pairs $ < {\bar \Psi} \Psi> $
in a grand unified theory (GUT) in the cosmological background. In
order to describe the cosmological evolution it is enough to consider
the effective dynamics of such condensates.  The relation between
the low energy effective field theory of inflation and the
microscopic fundamental theory is akin to the relation between the
effective Ginsburg-Landau theory of superconductivity \cite{gl} and the
microscopic BCS theory, or like the relation of the $ O(4) $ sigma
model, an effective low energy theory of pions, photons and nucleons 
(as skyrmions), with quantum chromodynamics (QCD) \cite{quir}. The
guiding principle to construct the effective theory is to include
the appropriate symmetries \cite{quir}. Contrary to the sigma model
where the chiral symmetry strongly constraints the
model \cite{quir}, only general covariance can be imposed on the
inflaton model.

\medskip

In summary, the physics during inflation is characterized by:
\begin{itemize}
\item{Out of equilibrium matter field evolution in a rapidly
expanding space-time
dominated by the vacuum energy. The scale factor is quasi-de Sitter:
 $ a(t) \simeq e^{H \, t}.$}
\item{Extremely high energy density at the scale of 
$\lesssim 10^{16}$ GeV.}
\item{Explosive particle production at the beginning of inflation
due to spinodal or parametric {\bf instabilities} for new and chaotic
inflation, respectively \cite{chalo,chalo2}.}
\item{The enormous redshift as a consequence of a large number of e-folds 
($ \sim 64 $) classicalizes the dynamics: an  {\bf assembly} of
(superhorizon) fluctuations behave as the classical and homogeneous
inflaton field. The inflaton which is a long-wavelength
condensate slowly rolls down the potential hill towards its
minimum \cite{chalo2}.}
\item{Quantum non-linear phenomena eventually {\bf shut-off} the 
instabilities and {\bf stop} inflation \cite{chalo,chalo2,tsu}.}
\end{itemize}

As indicated above eq.(\ref{rangok}), the cosmologically relevant 
fluctuations have at the beginning of inflation physical
wavelengths in a range reaching the Planck scale 
$$ 
3.3  \; 10^{-32} \; e^{64-N_{tot}} \; \beta  \; {\rm cm} 
\lesssim \lambda^{init}
= 2 \, \pi/k^{init} \lesssim 3.3  \; 10^{-28} \; e^{64-N_{tot}} \; \beta 
\; {\rm cm} \; ,
$$
These fluctuations become macroscopic 
through the huge redshift during inflation and the subsequent expansion of
the universe with wavelengths today in the range 
$ 1 \, {\rm Mpc} \lesssim \lambda_{today} \lesssim 10^4 $ Mpc.
Namely, a total redshift of $ 10^{56} $. During this process these quantum
fluctuations classicalize just due to the huge stretching of the lengths. 
A field theoretical treatment shows that the quantum density matrix of the
inflaton becomes diagonal in the inflaton field representation as 
inflation ends \cite{chalo2,ps}. 

\subsubsection{Slow-roll, the Universal Form of the Inflaton Potential 
and the Energy Scale of Inflation} \label{potuniv}

The inflaton potential $ V(\varphi) $ must be a slowly varying
function of $ \varphi $ in order to permit a slow-roll solution for
the inflaton field $ \varphi(t) $ which guarantees a total number of 
efolds $ \sim 64 $ as discussed in secs. \ref{dos} and \ref{tres}.
Slow-roll inflation corresponds to a fairly flat potential and the
slow-roll approximation usually invokes a hierarchy of dimensionless
ratios in terms of the derivatives of the potential 
\cite{libros,hu,fluc,WMAP1}.
We recast here the slow-roll approximation as an expansion in $ 1/N $ 
where $ N \sim 60 $ is the number of efolds since the cosmologically 
relevant modes exit the horizon till the end of inflation \cite{1sN}.

We start by writing the inflaton potential in dimensionless variables as \cite{ciri}
\be\label{v1}
V(\varphi) = M^4 \; v\left(\frac{\varphi}{M_{Pl}}\right)  \; ,
\ee
where $ M $ is the energy scale of inflation and $ v(\psi) $ a dimensionless
function. The effective theory of inflation is consistent provided 
$  M \ll  M_{Pl} $ which is actually the case as shown by the WMAP data of the CMB
anisotropy amplitude [see below, eqs.(\ref{Mwmap}) and (\ref{valorM})].
The inflaton potential eq.(\ref{v1}) has the slow-roll property built-in
since a change of $ \varphi $ of the order $ \lesssim M $ leads to a small
change on $ V(\varphi) $ thanks to $  M \ll  M_{Pl} $.

In the slow-roll regime higher time derivatives can be neglected in the
evolution eqs.(\ref{eqno}) and (\ref{frinf}) with the result
\be\label{sr1}
3 \, H(t) \; {\dot \varphi} + V'(\varphi) = 0 \quad ,  \quad 
H^2(t) = \frac{V(\varphi)}{3 M^2_{Pl}} 
\ee
These first order equations can be solved in closed from as
\be \label{Nefo}
N[\varphi] = -\int_{\varphi/M_{Pl}}^{\varphi_{end}/M_{Pl}}  \;
v(\psi) \; \frac{d \psi}{dv} \; d\psi \;  \; .
\ee
where $ \psi = \varphi/M_{Pl} $ and $ N[\varphi] $ is the number 
of e-folds since the field $ \varphi $ exits the 
horizon till the end of inflation (where $ \varphi $ takes the value 
$ \varphi_{end} $). This is in fact the slow roll solution 
of the evolution equations eqs.(\ref{eqno}) and (\ref{frinf}) 
in terms of quadratures.

Eq.(\ref{Nefo}) indicates that $ N[\varphi] $ scales as $ \psi^2 $ and therefore
the field $ \varphi = \psi \; M_{Pl} $ is
of the order $ \sqrt{N} \sim \sqrt{60} $ for the cosmologically relevant modes.
Therefore, we propose as universal form for the inflaton potential \cite{1sN}
\be \label{V} 
V(\varphi) = N \; M^4 \; w(\chi)  \; ,
\ee  
\noindent where $ \chi $ is a dimensionless, slowly varying field 
\be\label{chifla} 
\chi = \frac{\varphi}{\sqrt{N} \;  M_{Pl}}  \; ,
\ee 
More precisely, we choose $ N \equiv N[\varphi] $ as the number of e-folds since a 
pivot mode $ k_0 $ exits the horizon till the end of inflation.
Eq.(\ref{V}) includes all well known slow-roll families of inflation models such as new inflation
\cite{infnue}, chaotic inflation \cite{infcao}, natural inflation \cite{natu}, etc.

\medskip

The dynamics of the rescaled field $ \chi $ exhibits the slow time 
evolution in terms of the \emph{stretched} dimensionless cosmic time 
variable, 
\be \label{tau} 
\tau =  \frac{t \; M^2}{M_{Pl} \; \sqrt{N}}  \quad , \quad  
{\cal H} \equiv \frac{H \; M_{Pl}}{\sqrt{N} \; M^2} = {\cal O}(1) \; .
\ee 
The rescaled variables $ \chi $ and $ \tau $ change slowly with time. 
A large change in the field amplitude $ \varphi $ results in a small 
change in the $ \chi $ amplitude, a change in $ \varphi \sim  M_{Pl} $ 
results in a $ \chi $ change $ \sim 1/\sqrt{N} $. The form of the 
potential, eq.(\ref{V}), the rescaled dimensionless inflaton field 
eq.(\ref{chifla}) and the time variable $ \tau $ make {\bf manifest} the 
slow-roll expansion as a consistent 
systematic expansion in powers of $ 1/N $ \cite{1sN}.  

We can choose $ |w''(0)| = 1 $ without loosing generality. Then, 
the inflaton mass scale around zero field is given by a see-saw formula
\be \label{m}
m^2  = \left| V''(\varphi=0) \right| = 
\frac{M^4}{M_{Pl}^2}  \quad ,  \quad m = \frac{M^2}{M_{Pl}} \; .
\ee
The Hubble parameter when the cosmologically relevant modes exit the 
horizon is given by
\be \label{hub}
H  = \sqrt{N} \; m \, {\cal H} \sim 7 \; m \; ,  
\ee
where we used that $  {\cal H} \sim 1 $. As a result, $ m\ll M $ and 
$ H \ll M_{Pl} $. The value of $ M $ is determined by the amplitude of the
CMB fluctuations within the effective theory of inflation. We obtain in 
sec. \ref{tres} [see eqs.(\ref{valorM}) and (\ref{myH})]: $ M \sim 0.70
\; 10^{16} $ GeV, $ m \sim 2.04 \; 10^{13} $ GeV and $ H \sim 10^{14} $ 
GeV for generic slow-roll potentials eq.(\ref{V}).

The energy density and the pressure [eq.(\ref{enerpres})] in terms of the 
dimensionless rescaled field $ \chi $ and the slow time variable $ \tau $ 
take the form,
\be\label{enepre2} 
\frac{\rho}{ N \; M^4} = \frac1{2\;N} \left(\frac{d\chi}{d \tau}\right)^2 
+ w(\chi) \quad ,\quad 
\frac{p}{ N \; M^4} = \frac1{2\;N} \left(\frac{d\chi}{d \tau}\right)^2 - 
w(\chi) \; . 
\ee
The equations of motion (\ref{eqno}) and (\ref{frinf}), in the same 
variables become
\bea \label{evol} 
&&  {\cal H}^2(\tau) = \frac{\rho}{ N \; M^4} =
\frac13\left[\frac1{2\;N} 
\left(\frac{d\chi}{d \tau}\right)^2 + w(\chi) \right] \quad , \cr \cr
&& \frac1{N} \;  \frac{d^2
\chi}{d \tau^2} + 3 \;  {\cal H} \; \frac{d\chi}{d \tau} + w'(\chi) = 0 
\quad .
\eea 
In addition, eqs.(\ref{evol}) imply for the derivative of the Hubble 
parameter
$$
\frac{d{\cal H}}{d \tau} = - \frac12 \; 
\left(\frac{d\chi}{d \tau}\right)^2 \; .
$$
Notice that,
$$
\frac{d}{d \; \tau} \ln a = N \;  {\cal H} \; .
$$
The slow-roll approximation follows by neglecting the
$ 1/N $ terms in eqs.(\ref{evol}). Both
$ w(\chi) $ and $ {\cal H}(\tau) $ are of order $ N^0 $ for large $ N $. 
Both equations make manifest the slow-roll expansion as an expansion in 
$ 1/N $.

Eq.(\ref{Nefo}) in terms of the field $ \chi $ takes the form
\be \label{Nchi}
-\int_{\chi_{exit}}^{\chi_{end}}  \; \frac{w(\chi)}{w'(\chi)} \; d\chi = 1 
\; .
\ee
This gives $ \chi= \chi_{exit} $ at horizon exit as a function of the couplings 
in the inflaton potential $ w(\chi) $.

Inflation ends after a finite number of efolds provided [see eq.(\ref{vmin})]
\be\label{condw}
w(\chi_{end}) = w'(\chi_{end}) = 0 \; .
\ee
So, this condition is enforced in the inflationary potentials.

\medskip

\noindent
For the quartic degree potentials $ V(\varphi) $ 
eqs.(\ref{nueva})-(\ref{caotica}), the corresponding dimensionless 
potentials $ w(\chi) $ take the form
\bea\label{wnue}
&& w(\chi) = \frac{y}{32} \left(\chi^2 - \frac8{y}\right)^{\! 2} 
= -\frac12 \; \chi^2 
+ \frac{y}{32} \; \chi^4 + \frac2{y}\quad , \quad {\rm BNI: \; binomial \; 
new \; inflation} \; , \\ \cr 
&& w(\chi) = \frac12 \; \chi^2 + \frac{y}{32} \; \chi^4 \quad , 
\quad {\rm BCI: \; binomial \; chaotic \; inflation} \; , \label{wchao}
\eea
where the coupling $ y $ is of {\bf order one} and 
$$
\lambda = \frac{y}{8 \; N} \left( \frac{M}{M_{Pl}}\right)^4 \ll 1 \quad 
{\rm since} \quad  M \ll  M_{Pl} \; .
$$
For a general potential $ V(\varphi) $ we can always eliminate the linear 
term by a shift in the field $ \varphi $ without loosing generality,
\be \label{serie}
V(\varphi) = V_0 \pm \frac12 \; m^2 \; \varphi^2 + 
\sum_{n=3}^{\infty} \frac{\lambda_n}{n} \; \varphi^n \; ,
\ee
and 
\be\label{seriew}
w(\chi) = w_0 \pm \frac12 \; \chi^2 + \sum_{n=3}^{\infty} 
\frac{G_n}{n} \; \chi^n \; ,
\ee
where the dimensionless coefficients $ G_n $ are of order one.
We find from eqs. (\ref{V}) and (\ref{chifla}),
\be\label{Gn}
 V_0 = N \; M^4 \;  w_0 \quad , \quad 
\lambda_n =  \frac{G_n \; M^4}{N^{\frac{n}2-1} \; M_{Pl}^n} \; ,
\ee
In particular, we get comparing with eqs.(\ref{nueva}), (\ref{caotica}),
(\ref{wnue}) and (\ref{wchao}),
\be\label{lambdaG4}
\lambda_3 = \frac{G_3}{\sqrt{N}}
\; \frac{M^4}{M_{Pl}^3}  \quad ,  \quad \lambda = \lambda_4  = 
\frac{G_4}{N} \left(\frac{M}{M_{Pl}}\right)^4 \quad , \quad G_4 = 
\frac{y}8 \; .
\ee
We find the dimensionful couplings $ \lambda_n $ suppressed by the nth 
power of $ M_{Pl} $
as well as by the factor $ N^{\frac{n}2-1} $. Notice that this suppression 
factors are natural and come from the ratio of the two relevant energy 
scales here: the Planck mass and the inflation scale $ M $.

\medskip

In new inflation with the potential of eq.(\ref{wnue}), the inflaton 
starts near the local maximum $ \chi = 0 $ and keeps rolling down the 
potential hill till it reaches the absolute minimum $ \chi = \sqrt{8/y} $. 
In chaotic inflation [eq.(\ref{wchao})] the inflaton starts at some 
significant value $ \chi \gtrsim 3 $ and rolls down the potential hill till
it reaches the absolute minimum $ \chi= 0 $. The initial values of $ \chi $
and $ \dot\chi $ must be chosen to have a total of $ \gtrsim 64 $ efolds of
inflation. In all cases $ \chi(0) $ and $ \dot\chi(0) $ turn to be of order
one.

\medskip

There are two {\it generic} inflationary regimes: slow-roll and fast-roll 
depending on whether
\bea
&& \frac1{2\;N} \left(\frac{d\chi}{d \tau}\right)^2 \ll w(\chi) \quad  : 
\quad {\rm  slow-roll \; regime} \cr \cr
&& \frac1{2\;N} \left(\frac{d\chi}{d \tau}\right)^2 \sim w(\chi) \quad  : 
\quad {\rm  fast-roll \; regime}
\eea
Both regimes show up in {\bf all} inflationary models in the class 
eq.(\ref{V}). A fast-roll stage emerges from generic initial 
conditions for the inflaton field. This fast-roll stage is generally  very
short and is followed by the slow-roll stage (see sec. \ref{3C}).
The slow-roll regime is an attractor in this dynamical system.

The possibility of using $ 1/N $ as an expansion to study
inflationary dynamics was advocated in
ref.\cite{mangano}. The analysis above confirms this early
suggestion and establishes the slow-roll expansion as a systematic
expansion in $ 1/N $  where $ N $ is
the number of e-folds since the cosmologically relevant
modes exited the horizon till the end of inflation \cite{1sN}.

\medskip

The generalization of the universal form eq.(\ref{V}) to inflationary
models with more than one field is straightforward
$$
V(\varphi_1,\varphi_2) = N \; M^4 \; 
w\left(\frac{\varphi_1}{\sqrt{N} \;  M_{Pl}},
\frac{\varphi_2}{\sqrt{N} \;  M_{Pl}}\right) \; .
$$
(See ref.\cite{infwmap} for hybrid inflation).

\medskip

Eq.(\ref{V}) for the inflaton potential resembles the
moduli potential coming from supersymmetry breaking,
\be\label{susy}
V_{susy}(\varphi) =  
m_{susy}^4 \; v\!\left(\frac{\varphi}{M_{Pl}}\right) \; ,
\ee
where $ m_{susy} $ stands for the supersymmetry breaking scale. 
Potentials with such form were used in the inflationary context in 
refs.\cite{natu,susy}. In our context, eq.(\ref{susy}) implies that 
$  m_{susy} \sim 10^{16}$ Gev. That is, the supersymmetry breaking scale
$ m_{susy} $ turns out to be at the GUT scale $ m_{susy} \sim M_{GUT} $.

It must be stressed that the validity of the inflaton potential 
eq.(\ref{V}) is independent of whether or not there is an underlying 
supersymmetry. In addition, the observational support on inflaton 
potentials like eq.(\ref{V}) can be taken as a first signal 
of the presence of supersymmetry in a cosmological context.
No experimental signals of supersymmetry are known so far despite
the enormous theoretical work done on supersymmetry since 1971.

\subsubsection{Inflationary Dynamics: Homogeneous Inflaton}\label{3C}

\begin{figure}[h]
\begin{center}
\begin{turn}{-90}
\psfrag{"lna.dat"}{$ \ln a(\tau) $ vs. $ \tau $}
\includegraphics[height=10cm,width=7cm]{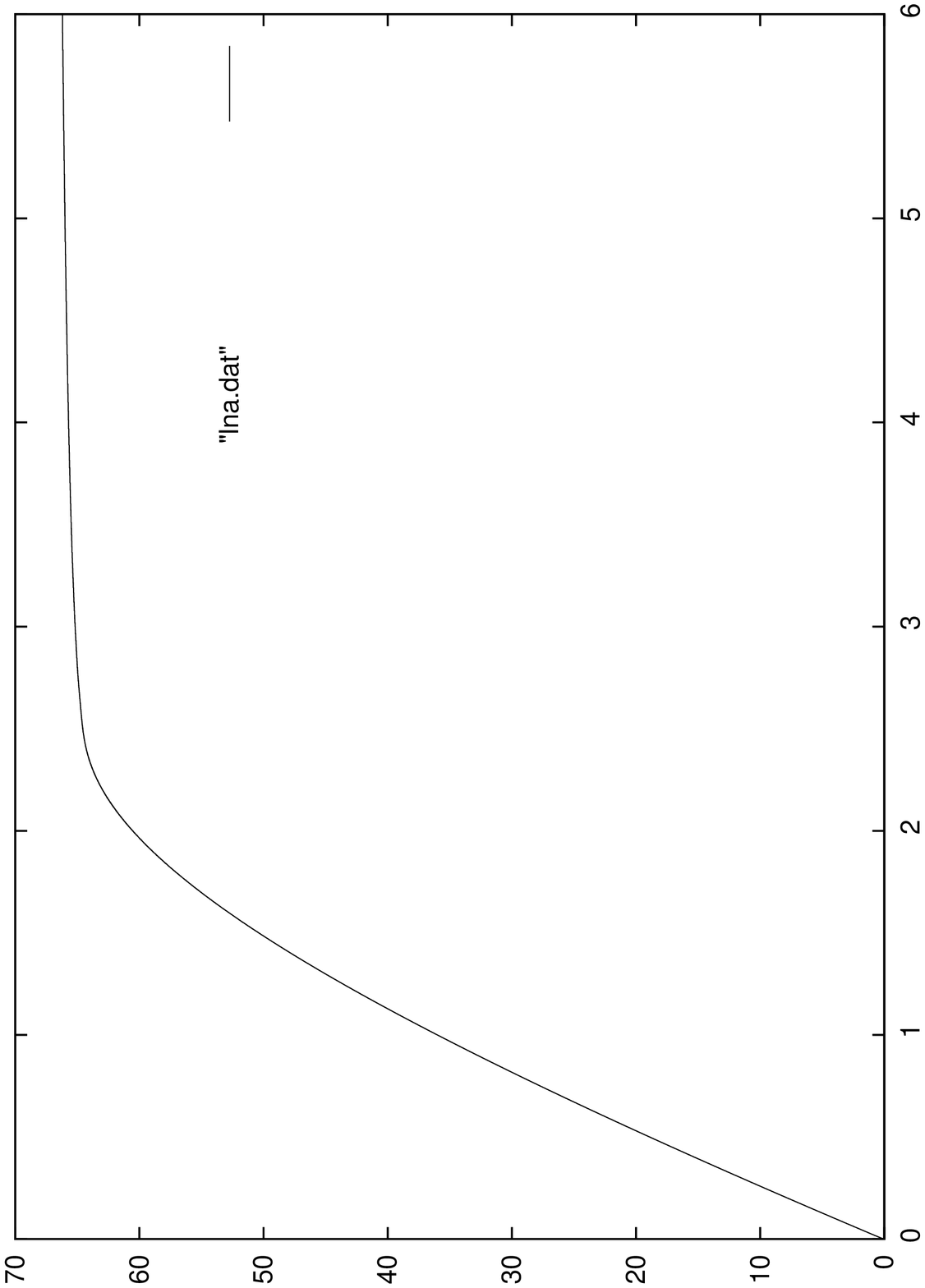}
\psfrag{"pe.dat"}{$ p/\rho $ vs. $ \tau $}
\includegraphics[height=10cm,width=7cm]{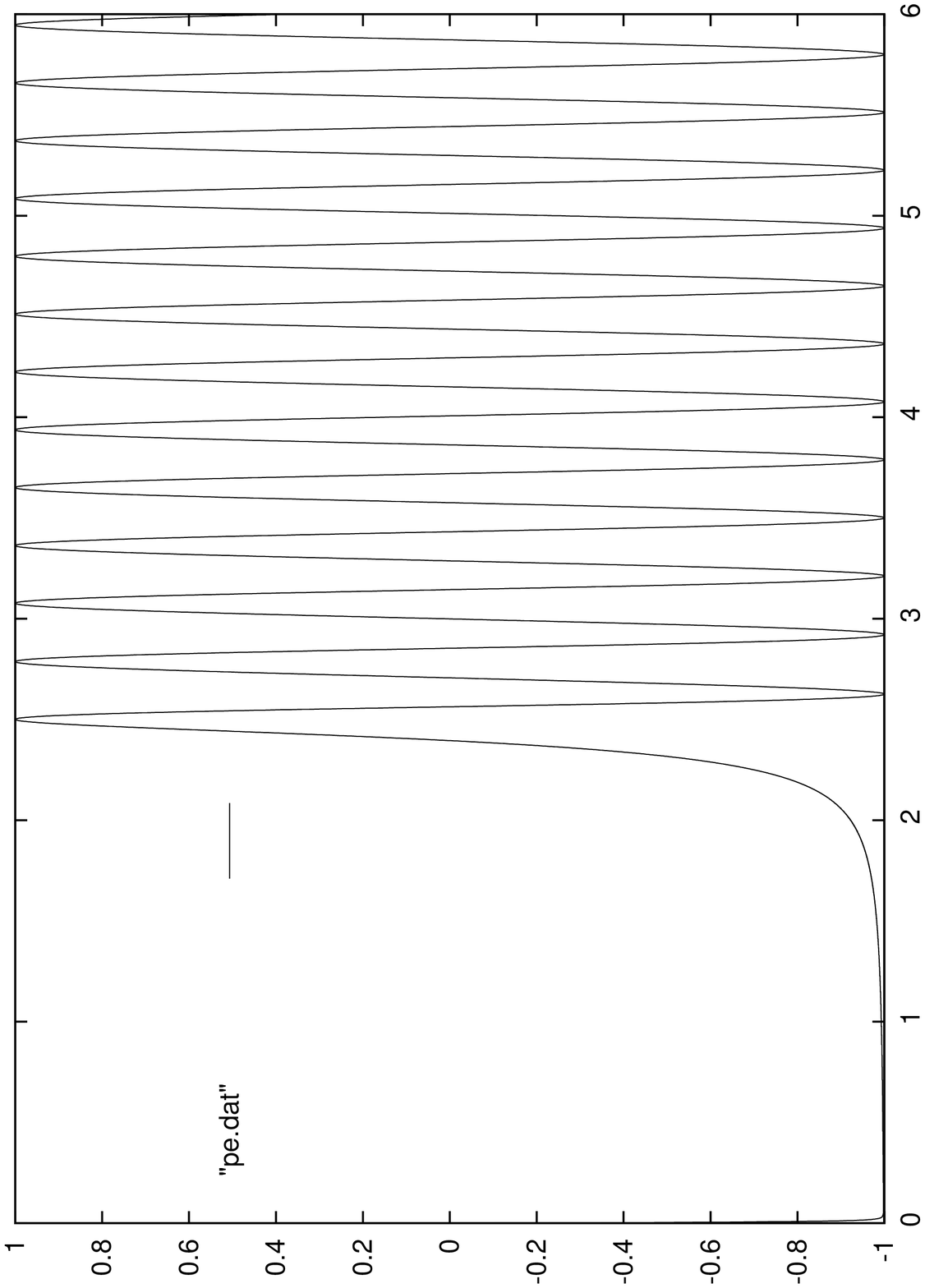}
\end{turn}
\caption{Upper panel: $ \ln a(\tau) $ which is the number of e-folds
as a function of the stretched cosmic time eq.(\ref{tau}). 
We see that $ a(\tau) $ grows exponentially with time
(quasi-de Sitter inflation) for $ \tau < \tau_{end} \simeq 2.39 $.
Lower panel: the equation of state  $  p / \rho $ vs. $ \tau $ 
[eq.(\ref{eqnofstate})]. We have $ p / \rho \simeq - 1 $ after the 
fast-roll stage and before the end of inflation. That is, for
$ 0.0247 \lesssim\tau <   \tau_{end} \simeq 2.39 . \;  p / \rho $ 
oscillates with zero average: $ <p / \rho> = 0 $ during the subsequent 
matter dominated era. Both figures are for the inflaton potential
eq.(\ref{wnue}) with $ y = 1.26 $ and $ N_{tot} = 64 $ efolds of 
inflation. $ p / \rho $ is plotted for  $ 0 <\tau < 0.0365 $ allowing 
to see the fast-roll stage in fig. \ref{fR1}.} 
\label{infla1}
\end{center}
\end{figure}

To leading order in the slow-roll approximation
(neglecting $1/N$ corrections) the inflaton equations
of motion (\ref{evol}) are solvable in terms of quadratures
\be\label{tausr}
\tau = - \int_{\chi(0)}^\chi d\chi' \; 
\frac{\sqrt{3 \; w(\chi')}}{ w'(\chi')}\; .
\ee
where we used that 
\be\label{hsr}
{\cal H}(\tau) = \sqrt{\frac{w(\chi)}3} + 
{\cal O}\left(\frac1{N}\right) \; ,
\ee
to leading order in $ 1/N $. We find for the new inflationary model 
(broken symmetric potential) eq.(\ref{wnue}) from eqs.(\ref{enerpres}), 
(\ref{tausr}) and (\ref{hsr}):
\bea\label{chislr}
&& \chi(\tau) = \chi(0) \; e^{\sqrt{\frac{y}6} \; \tau} 
+{\cal O}\left(\frac1{N} \right)
= \sqrt{\frac8{y}} \;  e^{-\sqrt{\frac{y}6} \; (\tau_{end}-\tau)} 
+{\cal O}\left(\frac1{N} \right)
\; , \\ \cr
&& {\cal H}(\tau) =\sqrt{\frac2{3 \; y}}
\left[ 1 -  e^{-\sqrt{\frac{2 \; y}3} \; (\tau_{end}-\tau)}
\right] +{\cal O}\left(\frac1{N} \right) \; , \cr \cr
&& \frac{p}{\rho} = -1 + \frac{y}{6 \; N} \; 
\frac1{\sinh^2\left[\sqrt{\frac{y}6}(\tau_{end}-\tau)\right]} 
+{\cal O}\left(\frac1{N^2} \right) \label{ecesta} \\ \cr
&& {\rm for} \; 0 \leq \tau \leq \tau_{end} = \sqrt{\frac3{2 \; y}} \; 
\ln \left[\frac8{\chi^2(0) \; y}\right] + {\cal O}\left(\frac1{\sqrt{N}} 
\right) \label{taufin} \; .
\eea
Inflation ends when the equation of state becomes $ p/\rho \sim 0 $. 
According to eq.(\ref{ecesta}), $ p/\rho $ vanishes when
$ \tau_{end} -\tau \sim {\cal O}\left(1/\sqrt{N} \right) $. Therefore,
expressions eqs.(\ref{chislr})-(\ref{ecesta}) are valid as long as
$$
0 \leq \tau \leq \tau_{end} -  {\cal O}\left(\frac1{\sqrt{N}} \right) 
\quad {\rm where} \quad   {\cal O}\left(\frac1{\sqrt{N}}\right)  > 0 \; .
$$
That is, eqs.(\ref{chislr}) hold while the inflaton is not very near the
minimum of the potential $ \chi_{end} = \sqrt{8/y} $.

We obtain for the scale factor $ a(t) $ integrating the Hubble parameter 
$ {\cal H}(\tau) $
\bea \label{asr}
&& \log \frac{a(t)}{a(0)} = \sqrt{\frac2{3 \; y}} \; N \; \tau -
\frac{N}8 \; \chi^2(0) 
\left[e^{\sqrt{\frac{2 \; y}3} \; \tau} - 1 \right] =\\ \cr
&&= \sqrt{\frac{2 \; N}{3 \; y}} \; m \; t - \frac18 \; 
\left(\frac{\varphi(0)}{M_{Pl}}\right)^2 \; 
\left[e^{\sqrt{\frac{2 \; y}{3 \; N}} \; m \; t} - 1 \right] \; ,
\nonumber
\eea
where we used eqs.(\ref{tau} ) and (\ref{chislr}). It must be noticed
that $ a(t) $ {\bf is not} a de Sitter scale factor, even in the large 
$ N $ limit at fixed $ \tau $.

\begin{figure}[h]
\begin{center}
\begin{turn}{-90}
\psfrag{"fi.dat"}{$ \chi(\tau) $ vs. $ \tau $}
\psfrag{"fip.dat"}{$ {\dot \chi}(\tau) $ vs. $ \tau $}
\includegraphics[height=12cm,width=8cm,keepaspectratio=true]{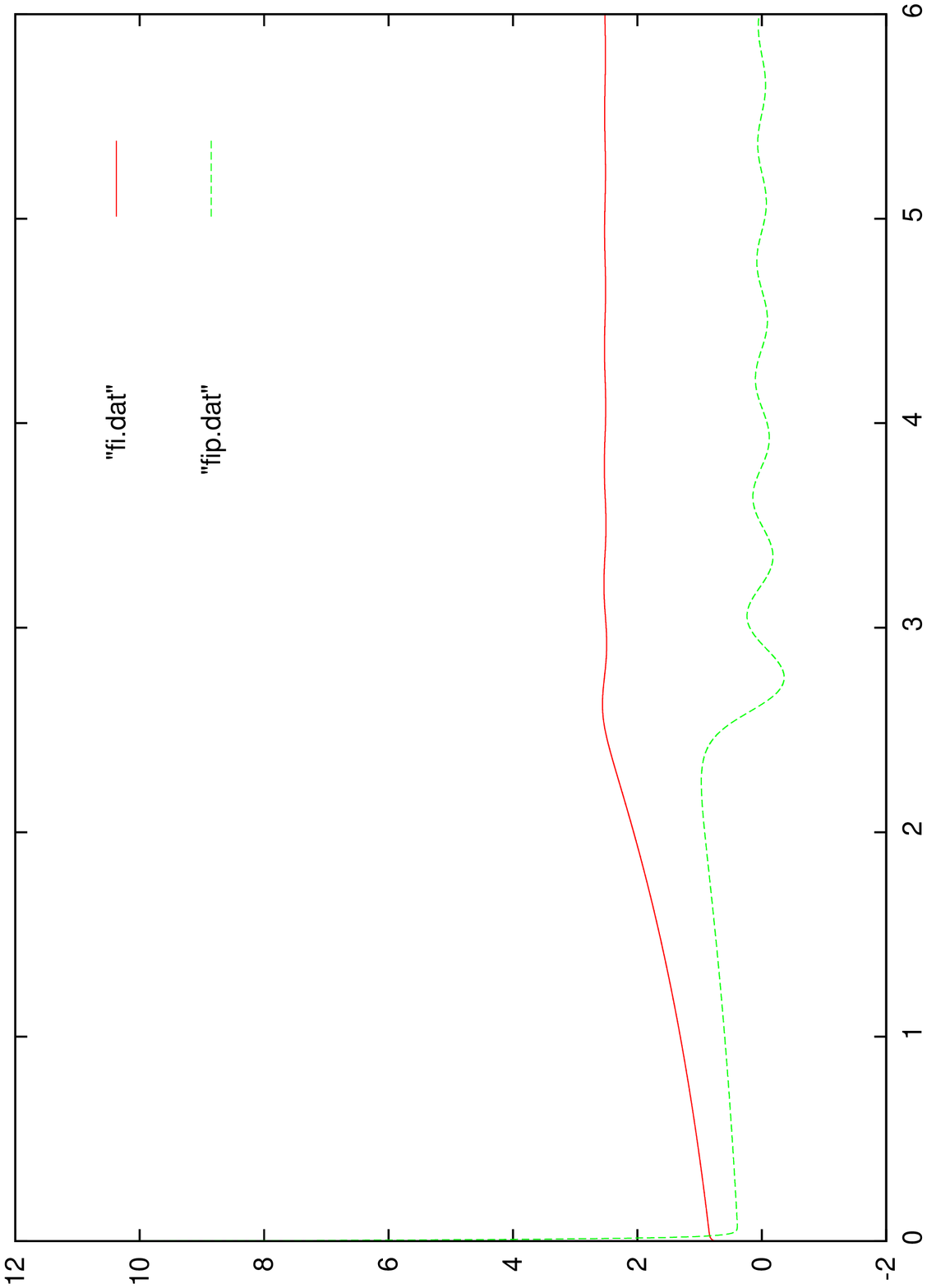}
\psfrag{"ush.dat"}{ $ 1/{\cal H}(\tau) $ vs. $ \tau $}
\includegraphics[height=12cm,width=8cm,keepaspectratio=true]{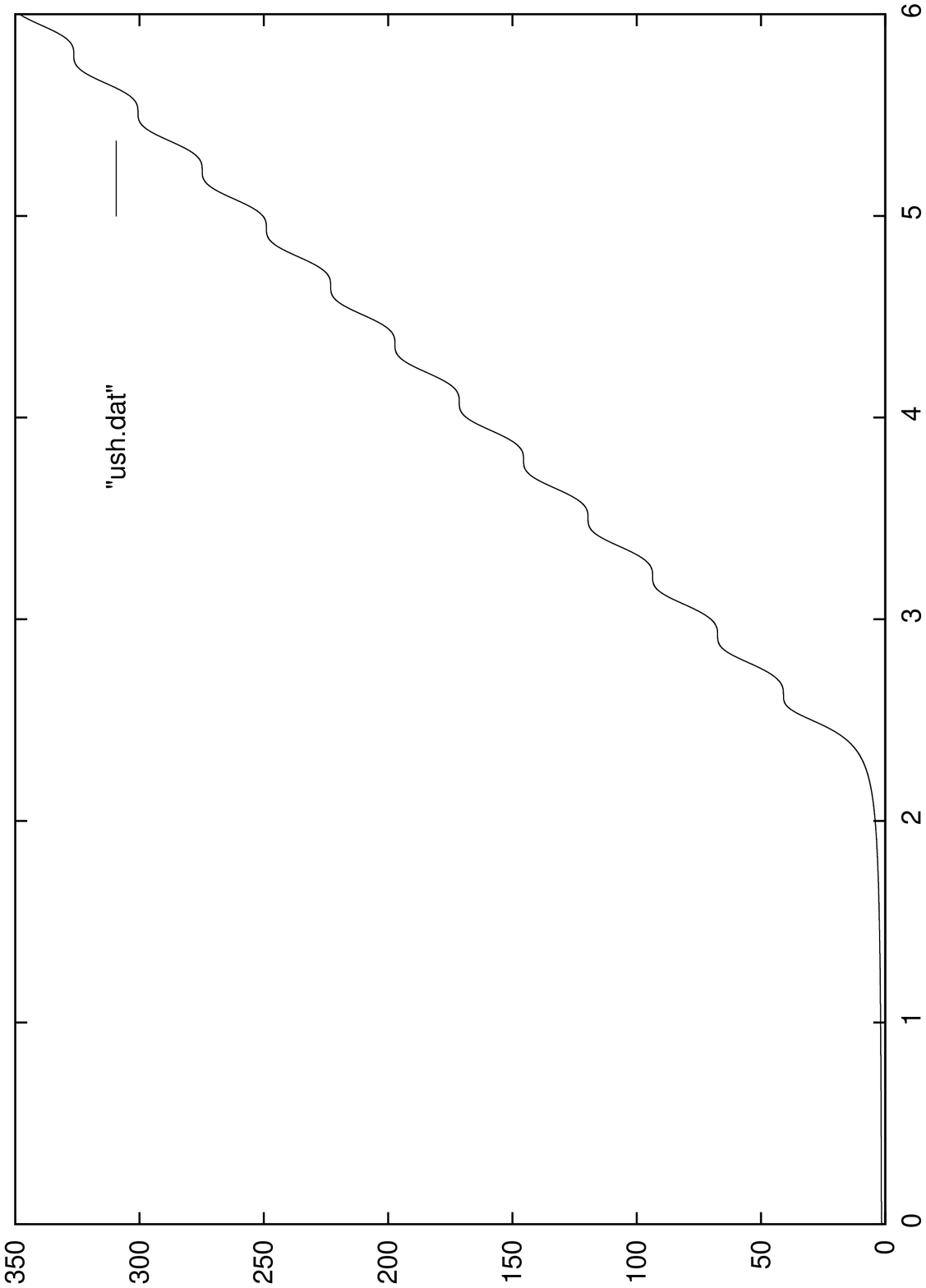}
\end{turn}
\caption{Upper panel: $ \chi(\tau) $ and  $ \dot\chi(\tau)$
as a function of the stretched  cosmic time $ \tau $ for $ \chi(0) = 
0.740 $ and initial kinetic energy equal to the initial potential energy 
eq.(\ref{cifr}) which implies $ {\dot \chi}(0) = 12.6 $. After a short 
fast-roll stage for $ \tau\lesssim 0.0247 $ 
the inflaton field slowly rolls toward its absolute minimum at 
$ \chi = \sqrt{8/y} \simeq 2.52\ldots, \; { \dot\chi } = 0 $. 
$ \chi(\tau) $ and  $ \dot\chi(\tau)$ are plotted for 
$ 0 <\tau < 0.034 $ allowing to see the fast-roll stage in fig. \ref{fR1}.
Lower panel: $ 1/{ \cal H} $ vs. $ \tau $.
$ 1/{ \cal H} $ grows slowly during inflation $ \tau < \tau_{end} \simeq 
2.39 $ and grows as $ 1/{\cal H} \simeq \frac32 \; N \; (\tau-\tau_{end}) $
in the subsequent matter dominated era.}
\label{infla2}
\end{center}
\end{figure}

\begin{figure}[h]
\begin{center}
\begin{turn}{-90}
\psfrag{"chi.dat"}{ $ {\cal H}(\tau) $ vs. $ \tau $}
\includegraphics[height=12cm,width=8cm,keepaspectratio=true]{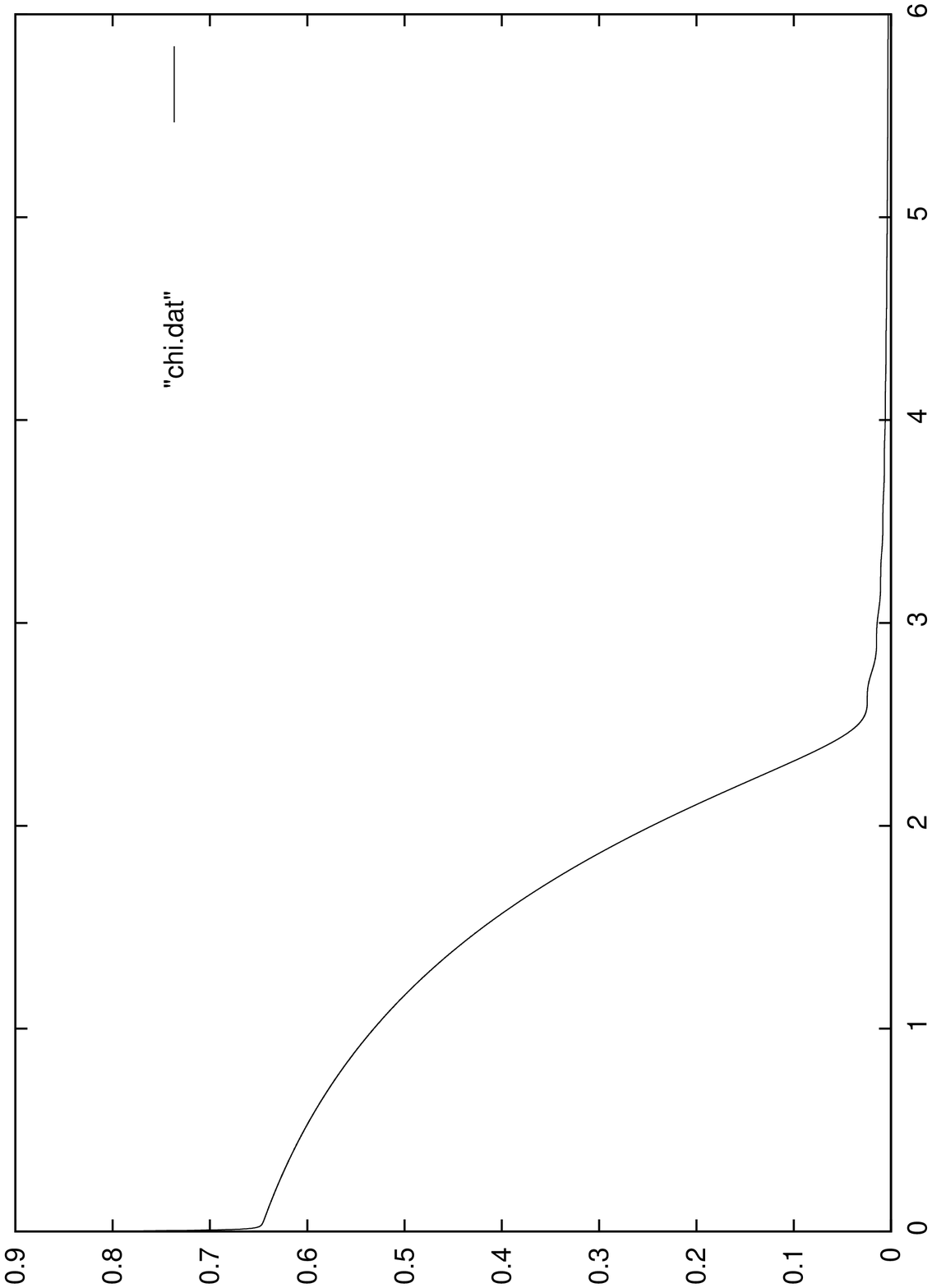}
\end{turn}
\caption{ $ {\cal H}(\tau) $ as a function of the stretched  cosmic time 
$ \tau $ for the same inflaton potential, initial conditions and coupling 
$ y $ as in figs. \ref{infla1} and \ref{infla2}. In fig. \ref{fR1}
$ {\cal H}(\tau) $ is plotted for $ 0 <\tau < 0.034 $ allowing to more 
clearly see the fast-roll stage.}
\label{hubZ}
\end{center}
\end{figure}

As a pedagogical example of inflationary dynamics we display the main
features for the new inflationary model (broken symmetric potential)
eq.(\ref{wnue}) for $ y = 1.26 $. This value of $ y $ is favoured by the 
MCMC analysis of the CMB and LSS data 
(see secs. \ref{mcmc} and \ref{fastroll}). 
We solved eqs.(\ref{evol}) numerically and plot the results 
in figs. \ref{infla1}, \ref{infla2} and \ref{hubZ}. 
We choose initial kinetic energy equal to the initial potential energy
of the inflaton (and hence zero initial pressure), 
\be\label{cifr}
\frac1{2 \, N} \; {\dot \chi}^2(0) = w(\chi(0)) 
\ee
(no slow-roll initially). We choose the initial value of the inflaton
such that the total number of efolds during inflation is 
$ N_{tot} \simeq 64 $ full-filling the constraints of sec. \ref{dos}.
This is achieved for $ \chi(0) = 0.740 , \;  {\dot \chi}(0) = 12.6 $. 
Qualitatively similar results are obtained for any values for 
$ \chi(0) $ and $ {\dot \chi}(0) $ of order one.

The inflaton reaches very soon, by $ \tau \simeq 0.0247 $ and 
$ \ln a = 1.0347 $, a slow-roll stage as displayed in 
figs. \ref{infla1}, \ref{infla2} and \ref{hubZ}.
This is a general property and implies that the slow-roll regime is
an {\bf attractor} for this dynamical system \cite{bgzk}.
We see a de Sitter-like expansion during the slow-roll stage 
$ 0.0247 \lesssim \tau \lesssim 2.39 $ during which
the Hubble parameter decreases slowly and monotonically. 
Both $ \chi(\tau) $ and $ {\dot \chi}(\tau) $ 
grow at the same pace during slow-roll according to eq.(\ref{chislr}) with
$$
{\dot \chi}(\tau) \simeq \sqrt{\frac{y}6} \;  \chi(\tau) \; .
$$
[Compare with fig. \ref{infla2} where $ \sqrt{y/6} = 0.4583 $].

Inflation ends at $ \tau = \tau_{end} \simeq 2.39 $ when the pressure 
becomes positive. This time corresponds to $ t_{end} \simeq 0.901 \; 10^{-36} $ sec
according to eqs.(\ref{tau}) and (\ref{myM}).
The slow-roll analytic expression eq.(\ref{taufin}) yields
$$
\tau_{end}^{slow-roll} = \sqrt{\frac3{2 \; y}} \; 
\ln \left[\frac8{\chi^2(0) \; y}\right] \simeq 2.67
$$
which exceeds by 12 \%, namely by 
$ {\cal O}\left(1/\sqrt{N} \right) \simeq 0.13 $ for $ N = 60 $, 
the exact numerical result $ \tau_{end} \simeq 2.39 $.

At the end of inflation the number of efolds is $ \ln a \simeq 
64 $, the inflaton is near its minimum $ \chi = \sqrt{8/y} \simeq 2.52 , 
\; \dot \chi $ starts to oscillate around zero and $ \cal{H}(\tau) $ 
begins a rapid decrease (see figs. \ref{infla1}, \ref{infla2} and 
\ref{hubZ}). At this time the inflaton field is no longer slowly coasting 
in the $ w''(\chi) < 0 $ region but rapidly approaching
its equilibrium minimum. When inflation ends the inflaton is at its minimum
up to corrections of order $ 1/\sqrt{N} $. Therefore, we see from
the Friedmann eq.(\ref{evol}) and eqs.(\ref{chislr}) that
\be\label{reduH}
\frac1{N} \left(\frac{d\chi}{d \tau}\right)^{\! 2}(\tau_{end}) = 
{\cal O}\left(\frac1{N}\right) \quad , \quad 
w(\chi(\tau_{end})) = {\cal O}\left(\frac1{N}\right) \quad {\rm therefore,} \quad 
{\cal H}(\tau_{end}) = {\cal O}\left(\frac1{\sqrt{N}} \right) \; ,
\ee
while $ {\cal H}(0) = {\cal O}(1) $. Namely, the Hubble parameter decreases by a factor
of the order $ \sqrt{N} \sim 8 $ during slow-roll inflation. We see in fig. \ref{hubZ}
that the exact $ \cal{H}(\tau) $ decreases by a factor six during slow-roll inflation,
confirming the slow-roll analytic estimate.

We can compute the total number of inflation efolds $ N_{tot} $ to leading 
order in slow-roll inserting the analytic formula for $ \tau_{end} $ 
eq.(\ref{taufin}) in eq.(\ref{asr}) with the result,
\be\label{ntotsr}
N_{tot} = \frac{N}{y} \left[\ln\left( \frac8{\chi^2(0) \; y}\right) - 1 + 
\frac18 \; y \; \chi^2(0) \right] + 
{\cal O}\left(\frac1{\sqrt{N}} \right) \; .
\ee
In the case considered, $ y =1.26 $ and $ \chi(0) = 0.740 $, 
eq.(\ref{ntotsr}) yields $ N_{tot} = 1.22 \; N $, about $ 14 \% $ in excess
of the exact result $ N_{tot} = 64 $. We find here again an error 
$ {\cal O}\left(1/\sqrt{N} \right) \simeq 0.13 $ for $ N = 60 $.

\medskip

In the upper panel of  fig. \ref{infla1} we depict $ \ln a(\tau) $ vs. 
$ \tau $, that is the number of e-folds as a function of time [by 
definition $ N_e(t)\equiv\ln a(\tau) $ ]. 

\medskip

We have carried out analogous numerical studies in scenarios of
chaotic inflation with  similar results: if the initial kinetic
energy of the inflaton is of the same order as the potential energy,
a \emph{fast-roll} stage is \emph{always} present. 

An initial state for the \emph{inflaton} (inflaton classical
dynamics) with approximate \emph{equipartition} between kinetic and
potential energies is a more \emph{general} initialization of
cosmological dynamics in the effective field theory than slow-roll
which requires that the inflaton kinetic energy is much smaller
than its potential energy. The most
\emph{generic} initialization of the inflaton dynamics in the
effective field theory leads to a \emph{fast-roll} stage followed by
slow-roll inflation \cite{quadru2}.

\medskip

The equations of motion (\ref{evol}) can be linearized around the absolute minimum
of the potential  $ \chi_{end} = \sqrt{8/y} $ with asymptotic solution
\bea\label{aproxmd}
&&\chi(\tau) \buildrel{\tau \gg 1}\over= \sqrt{\frac8{y}} +\frac2{\sqrt3 \; N \; \tau} \;
\cos\left[\sqrt{2 \; N}\left(\tau-\tau_{end}\right)\right]\left[1 + 
{\cal O}\left(\frac1{\tau}\right)\right]
\cr \cr
&& {\cal H}(\tau) \buildrel{\tau \gg 1}\over= \frac2{3 \; N \; \tau} \left[1 + 
{\cal O}\left(\frac1{\tau}\right)\right]
\; .
\eea
or, in dimensionful variables
\bea
&&\varphi(t) \buildrel{m \; t \gg 1}\over= \frac{m}{\sqrt{\lambda}} 
+\frac2{\sqrt3} \frac{M_{Pl}}{m \; t} \;
\cos\left[\sqrt2 \; m \; \left(t-t_{end}\right)\right]\left[1 + 
{\cal O}\left(\frac1{m \; t}\right)\right]
\cr \cr
&& H(t) \buildrel{m \; t \gg 1}\over= \frac2{3 \; t} \left[1 + 
{\cal O}\left(\frac1{m \; t}\right)\right]
\; .
\eea
Notice that this asymptotic behaviour is independent of the coupling $ y $ and is in
agreement with the numerical integration of eqs.(\ref{evol}) plotted in fig. \ref{infla2}.

The energy density and the pressure take in this regime the form
$$
\frac{\rho}{ N \; M^4} =  \frac4{3 \; N^2 \; \tau^2} \quad , \quad \frac{p}{\rho} =
- \sin\left[2 \; \sqrt{2 \; N}\left(\tau-\tau_{end}\right)\right] =
- \sin\left[2 \; \sqrt2 \; m \; \left(t-t_{end}\right)\right] \; ,
$$
where we have used eqs.(\ref{enepre2}) and (\ref{aproxmd}). 
[Compare with fig. \ref{infla1} where the oscillation frequency of $ p/\rho $ takes the value
$ 2 \; \sqrt{2 \; N} \simeq 21.91 $.]

The inflationary stage is here followed by a matter dominated stage as we see
from figs. \ref{infla1} and \ref{infla2}. This is always the case when one considers
an homogeneous inflaton field. In order to reach a radiation dominated era, inhomogeneous
modes, namely $k$-modes with nonzero $ k \gg m $, must be present and dominating the energy
of the universe. 

\medskip

As discussed in section \ref{dos} a successful inflationary 
scenario requires at least a total number of efolds $ \sim 64 $ to solve the horizon 
and entropy problems.

Wavelengths of cosmological relevance today exited the Hubble radius about 
$ N \sim 60 $ e-folds before the \emph{end} of inflation within a window of
width $ \Delta N \sim 10 $. This window corresponds to a small interval
$ \Delta \tau \lesssim 0.5 $. A scale $ k_0 $ which exits the horizon 
about 3 or 6 efolds after the beginning of inflation is chosen as pivot.
We see from figs. \ref{infla1} and \ref{infla2} that this corresponds to 
$ \tau \sim 0.076 $ or $ 0.15 $ and  $ \chi_{exit} \sim 0.9 $ in both cases.
We have from eq.(\ref{chifla}) in dimensionful variables,
\be\label{fiexit}
\varphi_{exit} \sim 7 \; M_{Pl} \; .
\ee
The value of the inflaton field at horizon exit is $ \chi_{exit} \sim 0.9 $
{\bf smaller} than one for new inflation and the coefficients $ G_n $ in 
the inflaton potential $ w(\chi) $ [eq.(\ref{seriew})] are of order one. 
Therefore, higher order terms in $ w(\chi) $
can be neglected. This is the basic reason why one can restrict 
to quartic potentials in new inflation (see discussion in sec. 
\ref{potfi2n}).

It is interesting to obtain the energy scale of the inflaton field
at symmetry breaking and horizon crossing. For the broken-symmetry new 
inflation potential eq.(\ref{wnue}), we have for the symmetry breaking 
scale $ \chi_{min} = \sqrt{8/y} $ which corresponds to the dimensionful 
inflaton field [see eq.(\ref{chifla})] \cite{fi6}:
$$
\varphi_{min} = \sqrt{N} \;  M_{Pl} \; \chi_{min} = 
\sqrt{\frac{8 \; N}{y}} \;  M_{Pl} \sim 20 \;  M_{Pl} \; .
$$
We want to draw attention to the fact that the energy density for such 
fields $ \varphi_{min} $ and $ \varphi_{exit} $ is of the order 
$ N \; M^4 \ll  M_{Pl}^4 $ due to eq.(\ref{V}). Therefore,
the effective theory of inflation {\bf does} apply for such fields 
$ \varphi_{exit} $ (see sec. \ref{validez}).

We consider translationally and rotationally invariant
cosmology where the only source of inhomogeneities are (small)
quantum fluctuations. Indeed, inhomogeneities cannot be excluded
at the beginning of inflation but the redshift of scales during
inflation by at least a factor $ e^{64} \sim 10^{28} $ effectively erases
all eventual initial inhomogeneities. As a consequence of this, all
structures and inhomogeneities in the present universe originated from the 
small quantum fluctuations during inflation (see sec. \ref{gistf}).

\subsubsection{Fixing the Total Number of Inflation e-folds from 
Fast-Roll and the CMB Quadrupole suppression}\label{ntot64}

We shall discuss in sec. \ref{fastroll} how an early fast-roll stage of 
inflation can explain the CMB quadrupole suppression. Such analysis 
shows that in order to explain the CMB quadrupole 
suppression, the quadrupole mode must exit the horizon by the end of 
fast-roll. 

The physical quadrupole ($ l = 2 $) wavemode today $ k_Q $ is given by
eq.(\ref{kqu1}) $  k_Q = 0.238  \; ({\rm Gpc})^{-1} $,
while at the begining of inflation (see Table VII),
\be\label{kQini}
k_Q^{init} = 11.5 \; m = 1.39 \times 10^{14} \; {\rm GeV} \; . 
\ee
The total number of inflation efolds $ N_{tot} $ equals $ N $ 
plus the number of efolds from the beginning of inflation till
the pivot scale $ k_0 $ exits the horizon. According to Table VII,
\be\label{NTn6}
N_{tot} \simeq N_{CosmoMC}  + 6.2 = N_{WMAP} + 3  \; ,
\ee
where $ N_{CosmoMC} $ and $ N_{WMAP} $ stand for the number of efolds
since the respective pivot scale $ k_0 $ exits the horizon till the end of 
inflation.

Using eqs.(\ref{kqu1}) and (\ref{kQini}) the redshift $ z_{init} $ at the 
begining of inflation takes the value,
$$
\frac1{a_{init}} = 1 + z_{init} \equiv \frac{k_Q^{init}}{k_Q}
= 0.915 \times 10^{56} \simeq e^{129} \; .
$$
Equating this relation with the redshift at the begining of inflation
eq.(\ref{ar1}) yields
\be\label{H57}
N_{tot} \simeq 63 + \ln{\beta}  \; . 
\ee
Therefore, the early fast-roll stage not only explains the CMB quadrupole 
supression but in addition  {\bf fixes the total} number of efolds during 
inflation. Combining this with the lower bound eq.(\ref{vinN}) on 
$ N_{tot} $ that solves the horizon problem yields,
$$
\ln{\beta} \geq 0.9 \quad , \quad \beta  \geq 2.46 \quad {\rm and} \quad
N_{tot} \geq 63.9 \; . 
$$
This bound for  $ \beta $ is very close to the 
new inflation best fit eq.(\ref{myM}) $ \beta \sim 2 $.
We conclude that the lower bound eq.(\ref{vinN}) should be practically
saturated and
\be\label{NTinfN}
N_{tot} \simeq 64 \quad , \quad \beta \simeq 2.5 \; .
\ee
These values of $ N_{tot} $ and  $ \beta $ a fortiori fufill also
the lower bound eq.(\ref{62}) from the entropy. Furthermore, our MCMC 
simulations give good fits for $ N \sim 60 $. $ N_{tot} \sim 64 $ 
corresponds to $ N_{CosmoMC} \sim 57 $ and $ N_{WMAP} \sim 61 $ according 
to eq.(\ref{NTn6}). These values are 
indistinguishable in practice from $ N \sim 60 $
in the slow-roll ($1/N$) expansion since $ 1/57 - 1/60 \simeq 0.0009 $
and $ 1/61 - 1/60 \simeq 0.0003 $.

Combining the lower bound for $ N_{tot} $  eq.(\ref{vinN}) with 
eqs.(\ref{calR}) and (\ref{H57}) yields a lower bound for $ \beta $ and 
consequently an upper bound for $ H $,
$$
\beta \geq 2.5 \quad , \quad H \leq 0.4 \times 10^{14} {\rm GeV} \; .
$$
This is compatible with the estimate eq.(\ref{myH}) and the best fit eq.(\ref{myM}) 
for $ H $.

The value of $ H_r \simeq H/\sqrt{N} $ [see eq.(\ref{reduH})] can be bounded 
from {\bf below} in order to explain the baryon asymmetry and the BBN. 
The bound depends on the mechanism responsible of these phenomena. 
For example, if the baryon asymmetry follows from the out of equilibrium 
decay of GUT bosons we have $ H_r > 10^9 $GeV \cite{kt}. 
This implies $ \beta < 180 $ and the  {\bf lower} bound from eq.(\ref{H57}),
\be\label{coti}
N_{tot} \simeq 63 + \ln{\beta} < 68.2 \; ,
\ee
which is respected by our best value $ N_{tot} \simeq 64 $
from eq.(\ref{NTinfN}).

In summary, combining the value of the observed CMB anisotropy amplitude
and the MCMC analysis of the CMB+LSS data including the early fast-roll
stage leading to the CMB quadrupole suppression {\bf fixes}
$ N_{tot} \simeq 64 $. 

The fact that several independent theoretical arguments and data 
(MCMC analysis of CMB+LSS data with BNI plus the fast-roll stage
in secs. \ref{mcmc:trino} and \ref{BNIFR},  simple sharp transition from 
inflation to RD which entails eq.(\ref{ar1}) and the $ a_r $ value 
eq.(\ref{vinN}) derived in sec. \ref{soluhor})
combine consistently yielding $ N_{tot} \simeq 64 $ 
is a {\bf major success} of the effective theory of inflation. 

\subsection{Gauge invariant Scalar and Tensor Fluctuations }\label{gistf}

The effective field theory of slow-roll inflation has two main
ingredients: the classical Friedmann equation in terms of a
\emph{classical} part of the energy momentum tensor described
by a homogeneous and isotropic condensate,  and a quantum part.
The latter features scalar fluctuations determined by a gauge
invariant combination of the inflaton field 
and metric fluctuations, and a tensor component, namely
gravitational waves. A consistency condition for this description  
is that the contributions from the fluctuations to the
energy momentum tensor be much smaller than those from the
homogeneous and isotropic condensate. The effective field theory
must include renormalization counterterms so that it is insensitive
to the possible ultraviolet singularities of the short wavelength
fluctuations. Different initial conditions on the mode functions of
the quantum fluctuations yield different values for their
contribution to the energy momentum tensor.

Criteria for acceptable initial conditions must include the
following: (i) back reaction effects from the quantum fluctuations
should not modify the inflationary dynamics described by the
inflaton, (ii) the ultraviolet counterterms that renormalize  the
energy momentum tensor should not depend on the particular choice of
initial conditions, namely different initial conditions \emph{should
not} introduce new ultraviolet divergences: a single renormalization
scheme, independent of initial conditions, should render the energy
momentum tensor UV finite. This set of criteria imply that the
ultraviolet allowed states have their large $ k $ behaviour
constrained up to the fourth order in $ 1/k $ \cite{motto2}. 

Notice that only gauge invariant fluctuations and the full gauge invariant
energy-momentum tensor of scalar and tensor fluctuations are physically 
meaningful.

\subsubsection{Scalar  Curvature Perturbations}\label{4A}

The gauge invariant curvature perturbation of the comoving
hypersurfaces is given in terms of the Newtonian (Bardeen) potential 
$ \psi $ and inflaton fluctuation  $ \delta \varphi $ by \cite{hu}
\be \label{curvature} 
\mathcal{R}= -\psi-\frac{H}{\dot{\varphi} } \; \delta \varphi  \; . 
\ee 
where $ \dot \varphi  $ stands for the derivative of the
inflaton field $ \varphi  $ with respect to the cosmic time $ t $.

In longitudinal gauge and in cosmic time the metric including
the fluctuation takes the form
$$
ds^2= (1 + 2 \; \phi) \; dt^2 - a^2(t) \; (1 - 2 \; \psi) \; d\vec{x}^2 
\quad ,
$$
where $ \phi $ is the other Newtonian or Bardeen potential. 
In single-field inflation one can impose
the condition $ \phi = \psi $ \cite{hu}.

It is convenient to introduce the gauge invariant potential \cite{hu}, 
\be \label{u}
u(\vx,t)=-z \;  \mathcal{R}(\vx,t) \; , 
\ee where 
\be \label{za} 
z \equiv  a(t) \; \frac{\dot{\varphi}(t)}{H(t)} \;  .
\ee 
The action for the gauge invariant curvature fluctuations field $ u(\vx,t)
$ follows from the Einstein-Hilbert action for the gravitational field 
plus the action for the inflaton in the cosmological space-time. The 
quadratic part of the action takes in conformal
time $ \eta $ [eq.(\ref{confo})] the form \cite{hu},
\be \label{accu}
S = \frac12 \; \int d\eta \; d^3 x \left[ (\partial_{\eta} u)^2 -  
(\nabla u)^2 + \frac{\partial^2_{\eta}z}{z} \; u^2 \right] \; .
\ee
The gauge invariant field $ u(\vx,t) $ expanded in terms of conformal time 
mode functions and creation and annihilation operators takes the form 
\cite{hu}
\be \label{curvau} 
u(\vx,\eta)=\int \frac{d^3k}{(2 \; \pi)^{\frac32}}\left[
\alpha_\mathcal{R}(\vk) \; S_\mathcal{R}(k;\eta) \; e^{i\vk\cdot\vx}  +
\alpha^\dagger_\mathcal{R}(\vk) \; S^*_\mathcal{R}(k;\eta)  \; e^{-i\vk\cdot\vx}\right] \; ,
\ee 
where the operators obey canonical commutation relations
\be \label{ccr}
\left[ \alpha_\mathcal{R}(\vk), \; \alpha^\dagger_\mathcal{R}(\vk') \right]
 = \delta(\vk-\vk') \; ,
\ee
the vacuum state is annihilated by the operators $ \alpha_\mathcal{R}(k)
$ and the mode functions obey the equation that follows from 
eq.(\ref{accu}) 
\be\label{fluces}
\Bigg[\frac{d^2}{d\eta^2}+k^2- \frac1{z} \frac{d^2z}{d\eta^2}
\Bigg]S_\mathcal{R}(k;\eta) =0 \,. 
\ee 
Canonical commutation relations for the field $ u(\vx,t) $ entail
that these solutions $ S_\mathcal{R}(k;\eta) $ are normalized through 
their Wronskian as
\be\label{wronskian}
W[S_\mathcal{R}(k;\eta),S^*_\mathcal{R}(k;\eta)]= S_\mathcal{R}(k;\eta) \; 
S^{*'}_\mathcal{R}(k;\eta) - S'_\mathcal{R}(k;\eta) \; 
S^*_\mathcal{R}(k;\eta) = i \; .
\ee
(here prime stands for derivative with respect to the conformal time).

Eq.(\ref{fluces}) is a Schr\"odinger-type differential equation in the 
variable $ \eta $. The potential felt by the fluctuations 
\be\label{WC}
W_\mathcal{R}(\eta) \equiv \frac1{z} \; \frac{d^2 z}{d \eta^2}
\ee
can be expressed in terms of the inflaton potential and its derivatives.
In order to achieve this, it is more convenient to pass to cosmic time, 
in terms of which, 
\be 
\frac{d^2 z}{d \eta^2} = a^2~(\ddot z~ +~ H \; \dot z) \; . 
\ee 
From eqs.(\ref{za}) and (\ref{WC}) and using the inflation equations of motion 
(\ref{eqno})-(\ref{frinf}), the potential $ W_\mathcal{R}(\eta) $ can be written 
as \cite{quadru2}
\be\label{W} 
W_\mathcal{R}(\eta)  = a^2(\eta) \;
H^2(\eta) \left[ 2 - 7 \, \epsilon_v + 2 \, \epsilon_v^2 - 
\frac{\sqrt{8 \; \epsilon_v} \; V'}{M_{Pl} \; H^2} 
- \eta_v (3 -  \epsilon_v) \right] \; , 
\ee 
where we take for the sign of the square root $ \sqrt{\epsilon_v} $ the sign of 
$ \dot{\varphi} $ and
\bea\label{srp}
&& \epsilon_v \equiv  \frac1{2 \; M_{Pl}^2} \; \frac{\dot{\varphi}^2}{H^2} \; , \\ \cr
&& \eta_v \equiv  M_{Pl}^2 \; \frac{V''(\varphi)}{V(\varphi)} \; .
\label{etadef}
\eea
$ \epsilon_v $ and $ \eta_v $ are the known slow-roll parameters \cite{hu}.
Notice that eqs.(\ref{W})-(\ref{etadef}) are {\bf exact} (no slow-roll approximation).

\medskip

In terms of the dimensionless variables eqs.(\ref{V})-(\ref{tau}) we obtain for the 
potential $ W_\mathcal{R}(\eta) $,
\be\label{Wb}
W_\mathcal{R}(\eta) =  a^2(\eta)
\;  {\cal H}^2(\eta) \; m^2 \; N \left[ 2 - 7 \; \epsilon_v + 2 \; \epsilon_v^2 - 
\sqrt{\frac{8 \; \epsilon_v }{N}}
\frac{w'}{{\cal H}^2}  - \eta_v (3 -  \epsilon_v)\right] \quad ,
\ee
while the slow-roll parameters take the form 
\be\label{slrsd}
\epsilon_v = \frac1{2 \; N} \; \frac1{{\cal H}^2} \; 
\left( \frac{d \chi}{d \tau}\right)^2 
\quad , \quad   \eta_v = \frac1{N}  \; \frac{w''(\chi)}{w(\chi)} \; .
\ee
We explicitly see that the slow-roll parameters are suppressed by powers of $ 1/N $.
This result is valid for {\bf all} models in the class defined by eq.(\ref{V}) 
regardless of the precise form of $ w(\chi) $. 

Another exact formula for $ \epsilon_v $ is
\be\label{epsH}
\epsilon_v = - \frac{\dot H}{H^2} = \frac1{N} \; \frac{d}{d\tau} \frac1{\cal H} \; ,
\ee
where we used eqs.(\ref{fri}), (\ref{accel}) and (\ref{srp}). 

Using eqs.(\ref{evol}) and (\ref{slrsd}), we obtain to first order in $ 1/N $,
\be\label{slrN}
\epsilon_v = \frac1{2 \; N} \; \left[\frac{w'(\chi)}{w(\chi)} \right]^2+
{\cal O}\left(\frac1{N^2}\right)
\ee
We can obtain the general expression of the conformal time integrating by parts in
 eq.(\ref{confo}),
$$
\eta = \int \frac{dt}{a(t)} = \int \frac{da}{a^2 \; H} = 
- \frac1{a \; H} -\int\frac{\dot H}{a \; H^2}
 \; dt =  - \frac1{a \; H} + \int\frac{\epsilon_v}{a^2 \; H} \; da 
$$
where we used eq.(\ref{epsH}).
In the slow-roll approximation we can consider $ \epsilon_v $ constant and pull out it of 
the integral as follows
$$
\int\frac{\epsilon_v}{a^2 \; H} \; da =
\epsilon_v \; \int \frac{da}{a^2 \; H} \left[1 + {\cal O}\left(\frac1{N}\right) \right]
=  \epsilon_v \; \eta  \left[1 + {\cal O}\left(\frac1{N}\right) \right]\; .
$$ 
We thus get the result
\be\label{adeta}
\eta = -\frac1{a \; H \; \left[1 -\epsilon_v +  
{\cal O}\left(\frac1{N^2}\right) \right] }
\quad {\rm or} \quad a(\eta) =  -\frac1{\eta \; H \; \left[1 -\epsilon_v +
{\cal O}\left(\frac1{N^2}\right) \right] } \; .
\ee
We can now write the potential  $ W_\mathcal{R}(\eta) $ 
during slow-roll and to leading order in $ 1/N $
\be \label{eqnz}
W_\mathcal{R}(\eta)= \frac2{\eta^2} \Bigg[1+ \frac32\; (3
\; \epsilon_v-\eta_v) \Bigg] = \frac{\nu^2_\mathcal{R}-
\frac14}{\eta^2} \quad , \quad \nu_\mathcal{R} = \frac32+ 3 \,
\epsilon_v -\eta_v +  {\cal O}\left(\frac1{N^2}\right) \; , 
\ee 
where we used eqs.(\ref{hsr}) and (\ref{Wb})-(\ref{adeta}).

\medskip

In the slow-roll regime we can consider $ \epsilon_v $ and $ \eta_v $ constants in time
in eq.(\ref{eqnz}). The general solution of eqs.(\ref{fluces})-(\ref{WC}) is then given by 
\be S_\mathcal{R}(k;\eta) = A_\mathcal{R}(k) \;
g_{\nu_\mathcal{R}}(\eta) + B_\mathcal{R}(k) \; g^*_{\nu_\mathcal{R}}(\eta) \; , 
\ee 
where the function $ g_\nu(\eta) $ is given by 
\be\label{gnu}
g_{\nu }(k;\eta)  =  \frac12 \; i^{\nu +\frac12} \;
\sqrt{-\pi \eta} \; H^{(1)}_{\nu }(-k \; \eta) \; , 
\ee
$ A_\mathcal{R}(k), \;  B_\mathcal{R}(k) $ are constants determined by the 
initial conditions and $ H^{(1)}_{\nu }(z) $ is a Hankel function.

For wavevectors deep inside the Hubble radius $ | k \, \eta | \gg 1
$ the mode functions have the asymptotic behavior
\be
g_{\nu}(k;\eta) \buildrel{\eta \to
-\infty}\over= \frac1{\sqrt{2 \, k}} \; e^{-i \; k \; \eta} \quad ,  \quad
g^*_{\nu}(k;\eta) \buildrel{\eta \to -\infty}\over=
\frac1{\sqrt{2k}} \; e^{ i \; k \; \eta} \; , \label{fnuasy}
\ee
and for $ \eta \to 0^- $, the mode functions behave as:
\be\label{geta0}
g_\nu(k;\eta)\buildrel{\eta \to 0^-}\over=
\frac{\Gamma(\nu)}{\sqrt{2 \, \pi \; k}} \; \left(\frac2{i \; k \;
\eta} \right)^{\nu - \frac12} \; .
\ee
The complex conjugate formula holds for $ g^*_{\nu}(k;\eta) $.

In particular, in the scale invariant case $ \nu=\frac32 $ which is
the leading order in the slow-roll expansion, the mode functions 
eqs.(\ref{gnu}) simplify to
\be\label{g32}
g_{\frac32}(k;\eta) =
\frac{e^{-i \; k \;\eta}}{\sqrt{2k}}\left[1-\frac{i}{k \;\eta}\right]\; .
\ee

\subsubsection{Tensor Perturbations}\label{sec:tensor}

Taking into account small fluctuations the FRW geometry eq.(\ref{FRW}) 
becomes in conformal time
$$
ds^2= a^2(\eta) \left( d\eta^2- d\vec{x}^2\right)+a^2(\eta) \; h_{\mu\nu} \; dx^\mu\; dx^\nu \; .
$$
where $  h_{\mu\nu} \ll 1 $ stands for the metric perturbation with $ h_{00} = h^i_i = 0 $. 
In the transverse traceless gauge $ h_{0i} = \partial^i h_{ij} =  0 $ and there are two 
independent polarization states. These are usually denoted by $ \lambda = \times $ and $ + $. 
The quantum modes associated with these two states are called gravitons. The action for the 
metric fluctuations follows by expanding the Einstein-Hilbert action to quadratic order in 
$ h_{\mu\nu} $ and it takes the form \cite{hu}
\be \label{accg}
S = \frac12 \; \left(\frac{M_{Pl}}2\right)^2 \int d\eta \; d^3 x \; a^2(\eta)
\; \partial_{\mu}  h^i_j \; \partial^{\mu}  h_i^j \; .
\ee
Tensor perturbations correspond to minimally coupled  massless fields with two physical 
polarizations. Tensor perturbations (gravitational waves) are gauge invariant.
The quantum fields (gravitons) are written as 
\be \label{tensh}
h^i_j(\vec x,\eta) = \frac2{a(\eta) \; M_{Pl}}
\sum_{\lambda=\times,+} \int \frac{d^3k}{(2 \; \pi)^\frac32}
 \; \epsilon^i_j(\lambda,\vec{k}) \left[e^{i \vec{k}\cdot \vec{x}} \; \alpha_{T,\lambda}(\vk)
\; \,S_T(k,\eta)+e^{-i \vec{k}\cdot \vec{x}} \; \alpha^\dagger_{T,\lambda}(\vk)
\; \,S^*_T(k,\eta) \right] \; ,
\ee
\noindent where $ \lambda $ labels the two standard
transverse and traceless polarizations $ \times $ and $ + $. The
operators $ \alpha_{T,\lambda}(\vk), \; \alpha^\dagger_{T,\lambda}(\vk') $
obey canonical commutation relations eq.(\ref{ccr}), 
and $ \epsilon_{ij}(\lambda,\vec{k}) $ are
the two independent symmetric and traceless-transverse tensors constructed from
the two independent polarization vectors transverse to $
\hat{\bf{k}} $, chosen to be real and normalized such that $
\epsilon^i_j(\lambda,\vec{k})\, \; \epsilon^j_k(\lambda',\vec{k})=\delta^i_k
\; \delta_{\lambda,\lambda'} $.

\medskip

It follows from eq.(\ref{accg}) that the mode functions $ S_T(k;\eta) $ obey the 
differential equation,
\be\label{Sten} 
S^{''}_{T}(k;\eta)+\left[k^2-
\frac{a''(\eta)}{a(\eta)}\right]S_{T}(k;\eta) = 0 \; . 
\ee
In the slow-roll regime we find from eq.(\ref{adeta}),
\be \label{eqnC}
\frac{a''(\eta)}{a(\eta)} = \frac{\nu^2_{T}-\frac14}{\eta^2} \quad , \quad
\nu_T = \frac32+\epsilon_v + {\cal O}\left(\frac1{N^2}\right) \; . 
\ee 
and the general solution of eq.(\ref{Sten}) is then given by 
\be 
S_T(k;\eta) = A_T(k) \;
g_{\nu_T}(\eta) + B_T(k) \; g^*_{\nu_T}(\eta) \; , 
\ee 
where the function $ g_\nu(\eta) $ is given by eq.(\ref{gnu})
and $ A_T(k), \;  B_T(k) $ are constants determined by the initial conditions.

\medskip

We see that both scalar curvature and tensor perturbations obey in conformal time
Schr\"odinger-type differential equations (\ref{fluces}) and (\ref{Sten}) with potentials: 
\be\label{WRT}
W_\mathcal{R}(\eta) = \frac1{z} \; \frac{d^2 z}{d \eta^2} \quad ,  \quad 
W_T(\eta) = \frac{a''(\eta)}{a(\eta)} \; ,
\ee
respectively.

\subsubsection{Initial conditions}

We treat both scalar and tensor perturbations on the same footing by
focusing on mode functions solutions of the general equation 
\be
\left[\frac{d^2}{d\eta^2}+k^2 - \frac{\nu^2 -\frac14}{\eta^2}
\right]S(k,\eta) = 0 \; . \label{geneq} 
\ee 
where $ \nu $ can be considered time independent during the slow-roll 
regime. For general initial conditions we write 
\be \label{genS} 
S (k;\eta) = A (k)\,g_{\nu }(k;\eta) + B (k) \, g^*_{\nu }(k;\eta) \; . 
\ee 
where the solution $ g_{\nu }(k;\eta) $ is given by  eq.(\ref{gnu}).
For the specific cases of scalar curvature or tensor perturbations, the 
mode functions  and coefficients $ A(k), \; B(k) $ will feature a subscript
index $ \mathcal{R} $ and $ T $, respectively.

The coefficients $ A(k), \; {B}(k) $  for the general solution
eq.(\ref{genS}) are determined by an initial condition on the mode
functions $ S(k;\eta) $ at a given initial conformal time
$ {\bar \eta} $, namely
\be
B (k) =  -i[g_{\nu }(k;{\bar \eta}) \; S' (k;{\bar \eta})-g'_{\nu
}(k;{\bar \eta}) \; S (k;{\bar \eta})] \quad , \quad
A (k ) =  -i[g^{*'}_{\nu }(k;{\bar \eta}) \; S (k;{\bar \eta})-g^*_{\nu
}(k;{\bar \eta}) \; S' (k;{\bar \eta})] \label{A} \; . 
\ee
The constancy of the Wronskian  eq.(\ref{wronskian}) and eq.(\ref{genS}) 
imply the constraint,
\be 
|A (k)|^2-| B (k)|^2 = 1 \label{constraint} \; . 
\ee 
The S-vacuum state $ |0\rangle_\mathcal{S} $ is annihilated by the
operators $ \alpha_{\vk} $ associated with the modes $ S (k;\eta) $.
A different choice of the coefficients $ A (k); \; B(k) $ determines
different choices of vacua, the Bunch-Davies vacuum corresponds to $
A(k)=1, \;  B(k)=0 $.   An  illuminating representation of these
coefficients can be gleaned by computing the expectation value of
the number operator in the Bunch-Davies vacuum. Consider the
expansion of the fluctuation field both in terms of Bunch-Davies
modes $ g_\nu(k;\eta) $ and in terms of the general modes $ S
(k;\eta) $, for example for the  scalar curvature $ \mathcal{R} $ (similarly for
tensor fields with a subscript $ T $ and corresponding normalization)
\begin{equation}\label{expa} 
u(\vk,\eta)= a_{\vk} \;  g_{\nu_\mathcal{R}}(k;\eta) + a^{\dagger}_{-\vk} \; 
g^*_{\nu_\mathcal{R}}(k;\eta) \equiv \alpha_\mathcal{R}(\vk)\,S_\mathcal{R}(k;\eta)+
\alpha^\dagger_\mathcal{R}(\vk)\,S^*_\mathcal{R}(k;\eta) \; ,
\end{equation}
\noindent the creation and annihilation operators are related by a
Bogoliubov transformation
\be
\alpha^\dagger_\mathcal{R}(\vk)=A_\mathcal{R}(k) \;  a^\dagger_{\vk} - B_\mathcal{R}(k) \;  
a_{-\vk} \quad , \quad \alpha_\mathcal{R}(\vk) = { A}^*_\mathcal{R}(k) \;  a_{\vk} -  
{B}^*_\mathcal{R}(k) \; a^\dagger_{-\vk}  \; .
\ee
The Bunch-Davis vacuum $ |0\rangle_{BD} $ is annihilated by $ a_{\vk} $,
hence we find the expectation value
\be
{}_{BD}\langle 0|\alpha^\dagger_\mathcal{R}(\vk) \; \alpha_\mathcal{R}(\vk)
|0\rangle_{BD}= | B_\mathcal{R}(k)|^2= N_\mathcal{R}(k)  \; .
\ee
Where $ N_\mathcal{R}(k) $ is interpreted as the number of S-vacuum particles
in the Bunch-Davis vacuum. In combination with the constraint
eq.(\ref{constraint}) the above result suggests the following
representation for the  coefficients $ A(k); \; B(k) $
\be
A_\mathcal{R}(k) = \sqrt{1+N_\mathcal{R}(k)} \; e^{i\theta_A(k)}~~;~~
B_\mathcal{R}(k)=\sqrt{N_\mathcal{R}(k)} \; e^{i\theta_{B}(k)} \label{bogonum} \; ,
\ee 
\noindent where $ N_\mathcal{R}(k), \; \theta_{A,{B}}(k) $ are real.
The only relevant phase is the difference 
\be \label{phasediff}
\theta_k=\theta_{B}(k)-\theta_A(k) \; . 
\ee 
Notice that we provide the initial conditions at a given conformal time $ {\bar \eta} $
which is obviously the same for all $k$-modes. This is the
consistent manner to define the initial value problem (or Cauchy
problem) for the fluctuations. This is different from what is often
done in the literature when an ad-hoc dependence on $ k $ is given
to $ {\bar \eta} $ \cite{otros}.

\subsubsection{The power spectrum of adiabatic scalar and tensor 
perturbations}

The power spectrum of curvature perturbations $ \mathcal{R} $ is given by the
expectation value $ <\mathcal{R}^2> $ in the state with general initial conditions.
$ <\mathcal{R}^2> $ follows from eq.(\ref{u}) and (\ref{curvau}) using the fact that the
operators  $ \alpha_\mathcal{R}(\vk) $ annihilate the state: $ \alpha_\mathcal{R}(\vk)|0> = 0 $.
We obtain,
\be\label{pot1}
< \mathcal{R}^2(\vx,\eta)> = \displaystyle
<\frac{u^2(\vx,\eta)}{z^2(\eta)}> =\int_0^{\infty} 
\frac{\displaystyle |S_\mathcal{R}(k;\eta)|^2}{\displaystyle z^2(\eta)} \; 
\frac{k^2 \, dk}{2 \, \pi^2} \; .
\ee
The power spectrum at time $ \eta $ is customary defined as
\be\label{pot2}
P_\mathcal{R}(k,\eta) \equiv \displaystyle \frac{k^3}{2 \; \pi^2} \; 
\frac{\displaystyle|S_\mathcal{R}(k;\eta)|^2}{\displaystyle z^2(\eta)} 
\ee
such that it is the power per unit logarithmic interval in $ k $
$$
<\mathcal{R}^2(\vx,\eta)> =\int_0^{\infty} \frac{dk}{k} \; P_\mathcal{R}(k,\eta) \; .
$$
Therefore, the scalar power by the end of inflation results,
\be\label{curvapot}
P_\mathcal{R}(k)  = {\displaystyle \lim_{\eta \rightarrow 0^-}} \; P_\mathcal{R}(k,\eta)
={\displaystyle \lim_{\eta \rightarrow 0^-}} \; \displaystyle \frac{k^3}{2 \; \pi^2} \; 
\frac{\displaystyle|S_\mathcal{R}(k;\eta)|^2}{\displaystyle z^2(\eta)} \; .
\ee 
This primordial power $ P_\mathcal{R}(k) $ is dimensionless and
is chosen to follow the usual conventions \cite{hu}.

From eq.(\ref{WC}) and (\ref{eqnz}) we see that in the slow-roll regime $ z(\eta) $
behaves  as 
\be\label{zeta2} 
z(\eta) = \frac{z_0}{ (-k_0 \; \eta)^{\nu_R-\frac12}} \; , 
\ee 
where $ z_0 $ is the value of $ z $ when the pivot scale $ k_0 $ exits the horizon,
that is at $ \eta = \eta_0 = -1/k_0 $. Combining this
result with the small $ \eta $ limit eq.(\ref{geta0}) we find from
eqs.(\ref{curvapot}) and (\ref{zeta2}), 
\be \label{powR}
P_{\mathcal{R}}(k) = P^{BD}_{\mathcal{R}}(k)\Big[1+ D_\mathcal{R}(k) \Big] \; ,
\ee 
where we introduced the transfer function for the initial
conditions of curvature perturbations: 
\bea \label{DofkR}
D_\mathcal{R}(k) &=& 2 \; | {B}_\mathcal{R}(k)|^2 -2 \;
\mathrm{Re}\left[A_\mathcal{R}(k) \;
B^*_\mathcal{R}(k)\,i^{2\nu_R-3}\right] = \cr \cr &=& 2 \;
N_\mathcal{R}(k)-2 \; \sqrt{N_\mathcal{R}(k)[1+N_\mathcal{R}(k)]} \;
\cos\left[ \theta_k - \pi \left(\nu_R - \frac32 \right) \right] 
\eea
and 
\be\label{potinBD} 
P^{BD}_\mathcal{R}(k) = \left( \frac{k}{2 \, k_0}\right)^{3 - 2 \; 
\nu_\mathcal{R}} \;
\frac{\Gamma^2(\nu_\mathcal{R})}{\pi^3} \; \left(\frac{k \; H}{a(t) 
\; {\dot \varphi}}\right)_{exit}^2  \; . 
\ee 
The index $ _{exit} $ refers to the time where the pivot scale $ k_0 $ 
exits the horizon. In terms of the slow-roll parameter $ \epsilon_v $, 
using eq.(\ref{srp}) and $ k = a_{exit} \; H $, expression
(\ref{potinBD}) takes the form 
\be  \label{potBD}
P^{BD}_\mathcal{R}(k) = \left( \frac{k}{2 \, k_0}\right)^{n_s - 1}
\; \frac{\Gamma^2(\nu_\mathcal{R})}{\pi^3} \; \frac{H^2}{2 \;
\epsilon_v  \; M_{Pl}^2 }\equiv |{\Delta}_{k\;ad}^{\mathcal{R}}|^2 \; 
\left(\frac{k}{k_0}\right)^{n_s - 1} \; , 
\ee 
where in the slow-roll regime the amplitude 
$ |{\Delta}_{k\;ad}^{\mathcal{R}}|^2 $ is given by 
\be \label{ampR} 
|{\Delta}_{k\;ad}^{\mathcal{R}}|^2= \frac1{8 \, \pi^2 \; \epsilon_v } \; 
\left(\frac{H}{M_{Pl}}\right)^2 \left\{ 1 +(3 \; 
\epsilon_v - \eta_v  )\left[\ln 4 +
\psi\left(\frac32\right)\right]+ {\cal O}\left(\frac1{N^2}\right)\right\} 
\; ,
\ee
and $ n_s $ stands for the spectral index
\be\label{ns}
 n_s - 1 = 3 - 2 \; \nu_\mathcal{R} = 2 \; \eta_v  - 6 \; \epsilon_v  \; ,
\ee
$ \psi(z) $ is the digamma
function and $ \psi\left(\frac32\right) = - 1.463510026.....\;$ .

To leading order in $ 1/N $ and in terms of the dimensionless variables 
eqs.(\ref{V})-(\ref{tau}), the amplitude 
$ |{\Delta}_{k\;ad}^{\mathcal{R}}|^2 $ eq.(\ref{ampR}) takes the form
\be \label{ampliI}
|{\Delta}_{k\;ad}^{\mathcal{R}}|^2  = \frac{N^2}{12 \, \pi^2} \;
\left(\frac{M}{M_{Pl}}\right)^4 \; \frac{w^3(\chi)}{{w'}^2(\chi)} \; .
\ee
where $ \chi \equiv \chi_{exit} $ stands for the inflaton field at 
horizon exit.

The power spectrum of tensor perturbations in the state with general
initial conditions follows from eq.(\ref{tensh}) 
in analogous way to the scalar power eq.(\ref{pot1})-(\ref{pot2}),
\be\label{potT}
P_T(k) \buildrel{\eta \to  0^-}\over=
\frac{4 \; k^3}{M_{Pl}^2 \; \pi^2} \;  \Big|\frac{S_T(k;\eta)}{a(\eta)} \Big|^2  =  
P^{BD}_{T}(k)\Big[1+ D_T(k) \Big] \; .
\ee 
The different prefactor in the scalar [eq.(\ref{curvapot})] and tensor [eq.(\ref{potT})] 
power spectra  comes
from the extra factor $ M_{Pl}^2/2 $ in front of the graviton action eq.(\ref{accg})
and the fact that the graviton has two independent polarizations.

We find in the Bunch-Davis initial state using eqs.(\ref{adeta}) and (\ref{geta0})
to leading order in $ 1/N $,
\be\label{ptbd}
P_T^{BD}(k) = \frac8{\pi^3} \;  \frac{H^2}{M_{Pl}^2} \; \Gamma^2(\nu_T) \; 
\left(\frac{k}2\right)^{3-2 \; \nu_T} =|{\Delta}_{k}^T|^2 \; 
\left(\frac{k}{k_0}\right)^{n_T}
\ee
where
\be\label{ampliT}
|{\Delta}_{k}^T|^2 =  \frac2{\pi^2}  \;  \frac{H^2}{M_{Pl}^2}
\left[1 + {\cal O}\left(\frac1{N}\right)\right] \quad {\rm and} \quad 
n_T = 3-2 \; \nu_T = - 2 \; \epsilon_v \; .
\ee
The transfer function $ D_T(k) $ in eq.(\ref{potT}) for the initial 
conditions of tensor perturbations is given by
\be  \label{Dofkh} 
D_T(k) = 2 \; | {B}_T(k)|^2 -2
\; \mathrm{Re}\left[A_T(k)B^*_T(k)\,i^{2\nu_T-3}\right] = 2 \;
N_T(k)-2 \;\sqrt{N_T(k)[1+N_T(k)]}\, \cos\left[ \theta_k - \pi
(\nu_T - \frac32 ) \right] 
\ee 
The ratio of tensor to scalar fluctuations thus follows from 
eqs.(\ref{ampR}) and (\ref{ampliT})
\be\label{defr}
r \equiv \frac{|{\Delta}_{k}^T|^2}{|{\Delta}_{k\;ad}^{\mathcal{R}}|^2} 
= 16 \; \epsilon_v +{\cal O}\left(\frac1{N^2}\right) \; .
\ee
We thus obtain for the spectral index $ n_s $ and the ratio $ r $
to leading order in $ 1/N $ using eqs.(\ref{srp}), (\ref{etadef}),
(\ref{ns}) and (\ref{slrN}),
\bea \label{indi}
&&n_s - 1 = -\frac3{N} \; \left[\frac{w'(\chi)}{w(\chi)} \right]^2
+  \frac2{N}  \; \frac{w''(\chi)}{w(\chi)} +
{\cal O}\left(\frac1{N^2}\right)\quad , \cr \cr 
&&r = \frac8{N} \; \left[\frac{w'(\chi)}{w(\chi)} \right]^2 +
{\cal O}\left(\frac1{N^2}\right)\quad .
\eea
In eqs.(\ref{indi}) the value of $ \chi $ is evaluated at
horizon exiting setting $ N[\chi] = N = 60 $.

Since $ n_s $ and $ r $ are computed when the mode $ k = k_0 $
exits the horizon they are both $k$-dependent. However, their 
$k$-dependence is subleading in $ 1/N $. It can be shown that the 
running of  $ n_s $ is of the order 
$ 1/N^2 $ \cite{hu,1sN},
\be\label{run}
\frac{d n_s}{d \ln k}= - \frac2{N^2} \; \frac{w'(\chi) \;
w'''(\chi)}{w^2(\chi)} - \frac6{N^2} \; \frac{[w'(\chi)]^4}{w^4(\chi)}
+ \frac8{N^2} \; \frac{[w'(\chi)]^2 \; w''(\chi)}{w^3(\chi)}\quad .
\ee
Therefore, the running is negligible $ \sim 2 \times 10^{-4} $ for 
$ N \sim 60 $.

The higher order slow-roll parameters $ \xi_v $ and $ \sigma_v $ 
\cite{hu} turn to be of the order $ N^{-2} $ and $ N^{-3} $, 
respectively:
\bea\label{xisig}
&&\xi_v = M^4_{Pl} \; 
\frac{V'(\varphi) \; V^{'''}(\varphi)}{V^2(\varphi)} = 
\frac1{N^2} \; \frac{w'(\chi) \; w'''(\chi)}{w^2(\chi)}
\quad , \\ \cr 
&&\sigma_v = M^6_{Pl}\; 
\frac{\left[V^{'}(\varphi)\right]^2\,V^{(IV)}(\varphi)}{V^3(\varphi)}
= \frac1{N^3} \; \frac{[w'(\chi)]^2 \; 
w''''(\chi)}{w^3(\chi)} \; .\label{sigmv}
\eea
Notice that the concavity $ w''(\chi) $ of the inflaton potential at 
horizon exit can be expressed in terms of the observables $ n_s $ and 
$ r $ from eqs.(\ref{indi}) as
\be\label{conca}
n_s - 1 + \frac38 \; r = \frac2{N}  \; \frac{w''(\chi)}{w(\chi)} \; .
\ee
Hence, knowing the observed values of $ n_s $ and $ r $ determines
whether the inflaton potential is convex or concave at horizon exit.
A negative value of $ w''(\chi) $ will rule out binomial chaotic inflation
eq.(\ref{wchao}).

\subsubsection{The energy scale of inflation and the quasi-scale
invariance during inflation.}

Since, $ w(\chi) $ and  $ w'(\chi) $ are of order one, 
we find from eq.(\ref{ampliI})
\be\label{Mwmap}
\left(\frac{M}{M_{Pl}}\right)^2 \sim \frac{2
\, \sqrt3 \, \pi}{N} \; |{\Delta}_{k\;ad}^{\mathcal{R}}| \simeq  0.897
\times 10^{-5} \; .
\ee
where we used $ N \simeq 60 $, set $ k = k_0 $ with  
$ k_0 = 0.002 \; ({\rm Mpc})^{-1} $ the WMAP pivot scale and ref.\cite{WMAP5}
\be\label{aniso}
|{\Delta}_{k\;ad}^{\mathcal{R}}| = (4.94 \pm 0.1)\times 10^{-5} \; .
\ee
This fixes the scale of inflation to be
\be\label{valorM}
M \sim 2.99 \times 10^{-3} \; M_{Pl} \sim 0.73
\times 10^{16}\,\textrm{GeV} \; .
\ee
This value {\em pinpoints the scale of the potential} during inflation 
to be at the GUT scale suggesting a deep connection between inflation 
and the physics at the GUT scale in cosmological space-time.

As a consequence we get for the inflaton mass and
the Hubble parameter during inflation from eq.(\ref{m})-(\ref{hub}),
\be\label{myH}
m = \frac{M^2}{M_{Pl}} \sim 2.18 \times 10^{13} \, \textrm{GeV} 
\quad , \quad H \sim 10^{14} \, \textrm{GeV} 
\ee
Notice that these values for the inflation scale $ M $ and the inflaton 
mass are {\bf model independent} within the class of models 
eq.(\ref{V}). In addition, we see that 
$$
m \simeq 0.003 \; M \; .
$$
Namely, the inflaton is a {\bf very light} field in this context.
We can therefore expect infrared and scale invariant phenomena here.

Since $ M/M_{Pl} \sim 3 \times 10^{-3} $ [eq.(\ref{valorM})],
we {\bf naturally} find from eq.(\ref{lambdaG4}) the order of
magnitude of the cubic and quartic couplings, 
\be  \label{natu} 
\lambda_3 \sim 10^{-6} \; m \quad ,  \quad \lambda=\lambda_4 
\sim 10^{-12} \quad . 
\ee 
These relations are a {\bf natural} consequence
of the validity of the effective field theory and of slow-roll and
relieve the {\bf fine tuning problem}.  We emphasize that the
`see-saw-like' form of the couplings is a consequence of the form of the 
potential eq.(\ref{V}) and is valid for all inflationary models within
the class defined by eq.(\ref{V}).
This analysis reveals that small couplings are
naturally explained in terms of powers of the ratio between the
inflationary and Planck scales \emph{and} integer powers of 
$ 1/\sqrt{N} $ \cite{1sN}.

Eqs.(\ref{valorM})-(\ref{natu}) apply in order of magnitude
to all inflationary potentials within the universal slow-roll class 
eq.(\ref{V}). We provide in eq.(\ref{myM}) the values for 
$ M , \; m $ and the coupling valid for the inflaton potential giving 
the best MCMC fit to the present CMB and LSS data.
These values are consistent with the estimates 
eqs.(\ref{valorM})-(\ref{natu}), as it must be.

\medskip

We see that $ |n_s -1| $ as well as  the ratio $ r $ turn out to be of 
order $ 1/N $ for generic inflationary models. The case $ n_s = 1 $ and 
$ r = 0 $ corresponds to the Harrison-Zeldovich spectrum which is 
exactly scale invariant [see eq.(\ref{potBD})].
This nearly scale invariance is a natural property of
inflation which is described by a quasi-de Sitter space-time geometry.
This can be understood intuitively as follows: 
the geometry of the universe is scale invariant during de Sitter stage 
since the metric takes in conformal time the form 
\be\label{desiconf}
ds^2 = \frac1{(H \; \eta)^2}\left[ (d \eta)^2 - (d \vec x)^2 \right] \; .
\ee
Therefore, the primordial power generated is scale invariant except
for the fact that inflation is not eternal and lasts for $N$ efolds.
Hence, the  primordial spectrum is scale invariant up to $ 1/N $ 
corrections. The Harrison-Zeldovich (HZ) values $ n_s = 1, \; r = 0 $
correspond to a critical point as discussed in ref.\cite{1sN}.
This gaussian fixed point is {\bf not} the inflationary model that 
reproduces the data but the inflation model hovers around it 
in the renormalization group sense with an almost scale invariant 
spectrum of scalar fluctuations during the slow-roll stage.

The HZ point $ n_s = 1, \; r = 0 $ follows in the $ N \to \infty $
limit, that is, for eternal inflation the scale invariant de Sitter metric
eq.(\ref{desiconf}). 
The HZ point is ruled out by CMB+LSS data. It must be noticed that
the HZ point is theoretically excluded in new inflation as shown in
figs. \ref{nsrb}, \ref{ene} and \ref{nsr}. Chaotic inflation admits the
HZ point but in a highly unphysical and unrealistic singular limit in
which the inflaton potential identically vanishes as shown in sec. \ref{HZe},
eqs.(\ref{HZ1}) and (\ref{HZ2}).

\medskip

The fact that $ r \sim 1/N $ [eq.(\ref{indi})] shows that
the tensor fluctuations are suppressed by a factor $ N \sim 60 $ 
compared with the curvature scalar fluctuations.
This suppression can be explained as follows: 
the scalar curvature fluctuations are quantum 
fluctuations around the classical inflaton while the tensor 
fluctuations are just quantum zero-point fluctuations.
The inflaton is present through the quantity $ z(t) =a(t) \; \dot{\varphi}(t)/H(t) $ 
[eq.(\ref{za})] both in the scalar curvature fluctuations equation (\ref{fluces}) 
and in the scalar power spectrum eqs.(\ref{curvapot}) and (\ref{potinBD}),
while solely the scalar factor $ a(t) $ appears in the tensor fluctuations 
eq.(\ref{Sten}) and in the tensor power spectrum eq.(\ref{potT}).
Since the matter distribution is homogeneous and isotropic, 
it can only source scalar fluctuations in the geometry.

\medskip

The observation of a nonzero $ r $ will have {\bf twofold} relevance.
First, it would be the {\bf first} detection of (linearized)
gravitational waves as predicted by Einstein's General Relativity.
Second, $ r > 0 $ indicates the presence of gravitons,
namely, {\bf quantized} gravitational waves at tree level.

\medskip

Neutrino oscillations and neutrino masses $ m_{\nu} $ 
are currently explained in the see-saw mechanism as follows \cite{ita},
\be\label{neum}
\Delta m_{\nu} \sim \frac{M^2_{Fermi}}{M_R} 
\ee
where $ M_{Fermi} \sim 250$ GeV is the Fermi mass scale, 
$ M_R \gg  M_{Fermi} $ is a large energy scale and $ \Delta m_{\nu} $ 
is the difference between the neutrino masses for different flavors. 
The observed values for $ \Delta m_{\nu} \sim 0.009 - 0.05 $ eV 
naturally call for a mass scale $ M \sim 10^{15-16}$ GeV close to 
the GUT scale\cite{ita}. 

We see thus, that the energy scale $ \sim 10^{16}$ GeV
appears in fundamental physics in at least three independent ways:
grand unification scale, inflation scale and the scale $ M_R $
in the neutrino mass formula.
\section{Theoretical predictions, MCMC data analysis, early fast-roll stage and
CMB quadrupole suppression.}
\subsection{Ginsburg-Landau polynomial realizations of the Inflaton 
Potential}

In the Ginsburg-Landau spirit the potential is a polynomial in the field
starting by a constant term \cite{gl}. Linear terms can always be 
eliminated by a constant shift of the inflaton field. The quadratic term 
can have a positive or a negative sign. In the first case 
the symmetry $ \varphi \to - \varphi $ is unbroken (unless the potential 
contains terms odd in $ \varphi $), in the latter case  the symmetry 
$ \varphi \to - \varphi $ is spontaneously broken since the minimum of the potential 
is at $ \varphi \neq 0 $. 

Inflaton potentials with $ w''(0) > 0 $ lead to chaotic (large field) 
inflation while inflaton potentials with $ w''(0) < 0 $ lead to new (small
field) inflation.

The inflaton potential must be bounded from below, therefore the next potential beyond
the quadratic potential is the quartic one with a positive quartic coefficient.

The request of renormalizability restricts the degree of the inflaton potential to four.
However, since the theory of inflation is an effective theory,  potentials of degrees
higher than four are acceptable.

In the context of an effective theory or Ginsburg-Landau model it is highly 
unnatural to set $ m=0 $ \cite{gl}. This corresponds to be exactly at the critical point 
of the model where the mass vanishes, that is,  in the statistical mechanical context
the correlation length is infinite.
In fact, the WMAP result unfavouring the $ m = 0 $ 
choice (purely $ \varphi^4 $ potential, see \cite{WMAP1,WMAP3,WMAP5} and section \ref{mcmc}) 
supports this purely theoretical argument against the $ \varphi^4 $ monomial potential.
We want to stress that excluding the quadratic mass term in the potential 
$ V(\varphi) $ implies to fine tune to zero the mass term which is 
only justified at isolated points (critical points in the statistical mechanical sense). 
Therefore, from a physical  point of view, the pure quartic potential
is a weird choice implying to fine tune to zero the coefficient of the 
mass term. In other words, one would be considering a field with 
self-interaction but lacking of the mass term.

\medskip 

Dropping the cubic term implies that $ \varphi \to -\varphi $ is a symmetry of the 
inflaton potential. As stated in \cite{ciri},
we do not see reasons based on fundamental physics to choose a zero or a 
nonzero cubic term.
Only the phenomenology, that is the fit to CMB data, can decide for the moment 
on the value of the cubic term. The MCMC analysis of the WMAP plus LSS data 
shows that the cubic term can be ignored for new inflation (see section \ref{mcmc} and
\cite{mcmc}). 

\medskip 

A model with only one field is clearly unrealistic since the inflaton would 
then describe a stable and ultra-heavy ($ \sim 10^{13}$GeV) particle. It is
necessary to couple the inflaton with lighter particles, in which case the inflaton
can decay into them. 
There are many available scenarios for inflation. Most of them add
other fields coupled to the inflaton. This variety of inflationary
scenarios may seem confusing since several of them are compatible with the
observational data \cite{WMAP1,WMAP3,WMAP5}. Indeed, future observations should
constraint the models more tightly excluding some families of them. 
The hybrid inflationary \cite{hibri} models are amongst those strongly disfavoured by
the WMAP data since they give $ n_s > 1 $ in most of their parameter space 
contrary to the WMAP results \cite{WMAP3,WMAP5}.
The regions of parameter space where hybrid inflation yields $ n_s < 1 $
are equivalently covered by one-field chaotic inflation \cite{infwmap}.

The variety of inflationary models shows the {\bf power} of the
inflationary paradigm. Whatever the correct microscopic model for
the early universe would be, it should include inflation with the generic 
features we know today. Many inflatons can be considered (multi-field inflation),
but such family of models introduce extra features as non-adiabatic (isocurvature)
density fluctuations, which in turn become strongly constrained by the WMAP 
data \cite{WMAP1,WMAP3,WMAP5}.

\medskip 

Our approach is different to the inflationary flow equations \cite{flujo}
where the inflaton potential {\bf changes} (that is,  {\bf the model changes}) 
along the flow. We work with a {\bf given} potential within the Ginsburg-Landau 
(GL) framework, that is the trinomial potential.
We investigate the physics of this potential in the parameter space driven by 
the data through the Monte Carlo Markov Chains.
In our work, $ n_s $ and $ r $ are computed analytically to order $ 1/N $.

\medskip 

Following the spirit of the Ginsburg-Landau theory of phase transitions \cite{gl},
the simplest choice for the inflaton potential is a quartic trinomial for the inflaton 
potential
\cite{ciri,1sN}:
\be \label{wxi}
w(\chi)= w_0 \pm \frac12 \; \chi^2 + \frac{h}3 \; \sqrt{\frac{y}2} \; \chi^3 +
\frac{y}{32} \; \chi^4 \; .
\ee
where the coefficients $ w_0, \; h $ and $ y $ are dimensionless and of order 
one and the signs $ \pm $ correspond to large and small field inflation, 
respectively (chaotic and new inflation, respectively).

We find from eqs.(\ref{seriew}) and (\ref{wxi})  that
$$
2 \; h = \frac{G_3}{\sqrt{G_4}} \; ,
$$
and since  the strength of the couplings are usually
as $ G_4 \sim G_3^2 $, we conclude that the parameter $ h $ measures the 
{\bf asymmetry } of the inflaton potential.

Inserting eq.(\ref{wxi}) in eq.(\ref{V}) yields,
\be\label{VI}
V(\varphi)= V_0 \pm \frac{m^2}2 \; \varphi^2 +  \frac{ m
\; g }3 \; \varphi^3 + \frac{\lambda}{4}\; \varphi^4 \; .
\ee
where the mass $ m^2 > 0 $ and the couplings $ \; g $ and $ \lambda $ are given
by the following see-saw-like relations, 
\be 
m = \frac{M^2}{M_{Pl}} \qquad ,  \qquad g = h \; \sqrt{\frac{y}{2 \; N}} 
\; \left(\frac{M}{M_{Pl}}\right)^2  \qquad ,  \qquad \lambda  = 
\frac{y}{8 \; N} \left( \frac{M}{M_{Pl}}\right)^4 \label{acoi} 
\qquad ,  \qquad  V_0 = N \; M^4 \; w_0 \; . 
\ee 
Notice that $ y \sim {\cal O}(1) \sim h $ guarantee that $ g  \sim 
{\cal O}(10^{-6}) $ and $ \lambda  \sim {\cal O}(10^{-12}) $ without
any fine tuning as stressed before in eq.(\ref{natu}) \cite{1sN}. 
Namely, the smallness
of the couplings  directly follow from the form of the inflaton potential
eq.(\ref{V}) and the amplitude of the scalar fluctuations that fixes $ M \ll M_{Pl} $ in
eqs.(\ref{Mwmap}) and (\ref{valorM}).

We now study in secs. \ref{binocao}-\ref{binocom} binomial inflaton potentials. That is, 
we start by setting for simplicity the cubic coupling to zero in eq.(\ref{wxi}).
Trinomial inflaton potentials are investigated in secs. \ref{tci} and \ref{tni}.

\subsubsection{Binomial inflaton potentials for chaotic inflation}\label{binocao}

Let us consider chaotic inflation with the binomial potential eq.(\ref{wchao})
\be\label{binoc}
w(\chi) = \frac12 \; \chi^2 + \frac{y}{32} \; \chi^4 \quad , \quad 
{\rm chaotic \; inflation} \; ,
\ee
Chaotic inflation is obtained by choosing the initial field $ \chi $ in 
the interval $ (0,+\infty) $. The inflaton  $ \chi $ slowly rolls down the slope
of the potential from its initial value till the absolute minimum of the 
potential at the origin.

It is convenient to define the field variable $z$ by:
\be\label{defz}
z \equiv \frac{y}8 \; \chi^2 \; .
\ee
In terms of $ z $ the chaotic trinomial potential takes the form
$$
w(\chi) = \frac{2 \; z}{y} \; \left(2 +  z \right) \; .
$$
When $ z \lesssim 1 $ we are in the quadratic regime where $ w(\chi) $ is approximated
by the $ \chi^2 $ term. For $ z \gtrsim 1 $ we go to the non-linear regime in $ z $
and both terms in $ w(\chi) $ are of the same order of magnitude.

Eq.(\ref{Nchi}) defines the field $ \chi $ or equivalently $ z $, 
in terms of the coupling $ y $. We insert there eq.(\ref{binoc}) for $ w(\chi) $ and obtain
\be\label{cbino}
y = z + \log\left(1 + z \right) 
\ee
We see that $ z $ results to be a monotonically increasing function of the coupling $ y $
for $ 0 < y, \; z < +\infty $ and $ 2 < \chi < 2 \; \sqrt2 $. Recall that $ \chi $ and 
$ z $ correspond to the time of horizon exit.

We obtain from eqs.(\ref{ampliI}), (\ref{indi}), (\ref{run}),  
and (\ref{binoc}) the spectral index, its running, the ratio $ r $ and 
the amplitude of adiabatic perturbations for binomial chaotic inflation,
\bea\label{nsbicao}
n_s&=&1- \frac{y}{N \; z}\frac{3 \; z^2 + 5 \; z +4}{(z+2)^2} \; , \\ \cr
\frac{d n_s}{d \ln k} &=& \frac{y^2}{z^2 \; N^2}\left[-3 \; 
\frac{z \; (z+1)}{(z+2)^2}
- 24 \; \frac{(z+1)^4}{(z+2)^4}+ 8 \; \frac{(3 \; z + 1) 
(z+1)^2}{(z+2)^3}\right]\; , \\ \cr
r&=& \frac{16 \; y}{N \; z}\frac{(z+1)^2}{(z+2)^2} \quad , \quad
\label{ampbicao}
|{\Delta}_{k\;ad}^{\mathcal{R}}|^2  = \frac{N^2}{12 \, \pi^2} \; 
\left(\frac{M}{M_{Pl}}\right)^4 \; \frac{z^2 \; \left(2+z \right)^3}{y^2 
\;  \left(1 + z \right)^2} \; .
\eea
In chaotic binomial inflation, the limit $ z \to 0^+ $ implies weak 
coupling $ y \to 0^+ $, with
$$
y \buildrel{z \to 0^+}\over= 2  \, z +{\cal O}(z^\frac32) \quad , \quad
\chi \buildrel{z \to 0^+}\over= 2 + {\cal O}(z) \; .
$$
That is the quadratic potential and we find,
\bea\label{ycerocao}
&&n_s\buildrel{y \to 0^+}\over=1- \frac2{N} \quad , \quad
\frac{d n_s}{d \ln k} \buildrel{y \to 0}\over= -\frac2{N^2} \cr \cr
&& r \buildrel{y \to 0^+}\over=\frac8{N}\quad , \quad
|{\Delta}_{k\;ad}^{\mathcal{R}}|^2  \buildrel{y \to 0^+}\over= 
\frac{N^2}{6 \, \pi^2} \; 
\left(\frac{M}{M_{Pl}}\right)^4 \; .\label{cuad}
\eea
The results in the $ y \to 0^+ $ coincide with those for the purely 
quadratic monomial potential $ \frac12 \; \chi^2 $.

In the limit $ z \to +\infty $ which implies $ y\to +\infty  $ (strong 
coupling), we have
\bea
&&y \buildrel{z \to +\infty}\over=  z + \log z +{\cal O}(1) \quad , \quad
\chi \buildrel{z \to +\infty}\over= 2 \; \sqrt2 + 
{\cal O}\left(\frac{\log z}{z}\right) \; , \cr \cr
&&n_s \buildrel{y \to +\infty}\over=1- \frac3{N}\left[1 + 
\frac{\log z}{z} 
+{\cal O}\left(\frac1{z}\right)\right] \; ,
\cr \cr
&&r \buildrel{y \to +\infty}\over=\frac{16}{N}\left[1 + 
\frac{\log z}{z} +{\cal O}\left(\frac1{z}\right)\right]\; . \cr\cr
&&|{\Delta}_{k\;ad}^{\mathcal{R}}|^2  \buildrel{y \to +\infty}\over= 
\frac{N^2}{12 \, \pi^2} \; 
\left(\frac{M}{M_{Pl}}\right)^4 \frac{z}{\left[1 + \frac{\log z}{z} +
{\cal O}\left(\frac1{z}\right)\right]^2} \; .
\eea
Hence, $ \chi, \; n_s $ and $ r $ in the limit $ y\to +\infty  $
coincide with those of the purely quartic monomial potential $ \chi^4 $:
\be \label{cuar}
\chi = 2 \, \sqrt2 \quad , \quad
n_s = 1 - \frac3{N} \quad , \quad \frac{d n_s}{d \ln k} = -\frac3{N^2} 
\quad , \quad r = \frac{16}{N} \; .
\ee
We have for $ N = 60 $,
$$
n_s = 0.95  \quad ,  \quad \frac{d n_s}{d \ln k} = -0.00083 
\quad ,  \quad r = 0.27 \; .
$$
This value for $ r $ is in tension with the WMAP+SDSS data which indicate
$ r < 0.22 \; (95\% CL)$. Therefore, the purely quartic monomial potential
is excluded at more than 95\% CL \cite{WMAP1,WMAP3,WMAP5}.

\subsubsection{Binomial inflaton potentials for new inflation}\label{binonue}

Let us now consider new inflation with the binomial potential eq.(\ref{wnue}),
\be\label{binon}
w(\chi) = \frac{y}{32} \left(\chi^2 - \frac8{y}\right)^{\! 2} = -\frac12 \; \chi^2 
+ \frac{y}{32} \; \chi^4 + \frac2{y}\quad , \quad {\rm new \; inflation} \; .
\ee
New inflation is obtained by choosing the initial field $ \chi $ in 
the interval $ (0,+\sqrt{8/y}) $. The inflaton  $ \chi $ slowly rolls down the slope
of the potential from its initial value till the absolute minimum of the 
potential $ \sqrt{8/y} $.  

In terms of $ z $ the new inflation trinomial potential takes the form
$$
w(\chi) = \frac2{y} \; \left(1 - z \right)^2 \; .
$$
Inserting eq.(\ref{binon}) into eq.(\ref{Nchi})
for the number of efolds, we find the field 
$ \chi $ at horizon crossing related to the coupling $ y $,
\be\label{nbino}
y = z -1 - \log z  
\ee
where $ 0 \leq z \leq 1 $ since $ 0 \leq \chi \leq \sqrt{8/y} $ and 
eq.(\ref{defz}). $ z $ turns to be a monotonically decreasing function of 
$ y : \; z $ decreases from $ z = 1 $ till $ z = 0 $ when
$ y $ increases from $ y = 0 $ till $ y = \infty $.
When  $ z \to 1 , \; y $ vanishes quadratically as,
$$
y \buildrel{z \to 1}\over= \frac12 \; (1-z)^2 \; .
$$
It is interesting to study the concavity of the potential
eq.(\ref{binon}) for the inflaton field at horizon crossing.
We find from eq.(\ref{binon})
$$
w''(\chi) = \frac38 \; y \; \chi^2 - 1 = 3 \; z  - 1 \; .
$$
We see that $ w''(\chi) $ vanishes at $ z = \frac13 $, that is at
$ y = \ln 3 - 2/3 = 0.431946\ldots $. [This is usually called
the spinodal point]. Therefore,
\be\label{sigwseg}
 w''(\chi) > 0 \quad {\rm for}  \quad y < 0.431946\ldots
\quad {\rm and} \quad  w''(\chi) < 0 \quad {\rm for}  \quad y > 
0.431946\ldots \; .
\ee
Notice that for chaotic inflation eq.(\ref{binoc}) we always have 
$ w''(\chi) > 0 $. The negative concavity case  $ w''(\chi) < 0 $ for
$ y > 0.431946\ldots $ is {\bf specific} to new inflation 
eq.(\ref{binon}). Moreover $ w''(\chi) $ can be expressed as a linear
combination of the observables $ n_s $ and $ r $ as shown in 
eq.(\ref{conca}).

\medskip

\begin{figure}[h]
\begin{center} 
\begin{turn}{-90}
\psfrag{"nsN.dat"}{New binomial inflation, $ n_s $ vs. $ y $}
\psfrag{"nsC.dat"}{Chaotic binomial inflation, $ n_s $ vs. $ y $}
\includegraphics[height=10cm,width=7cm]{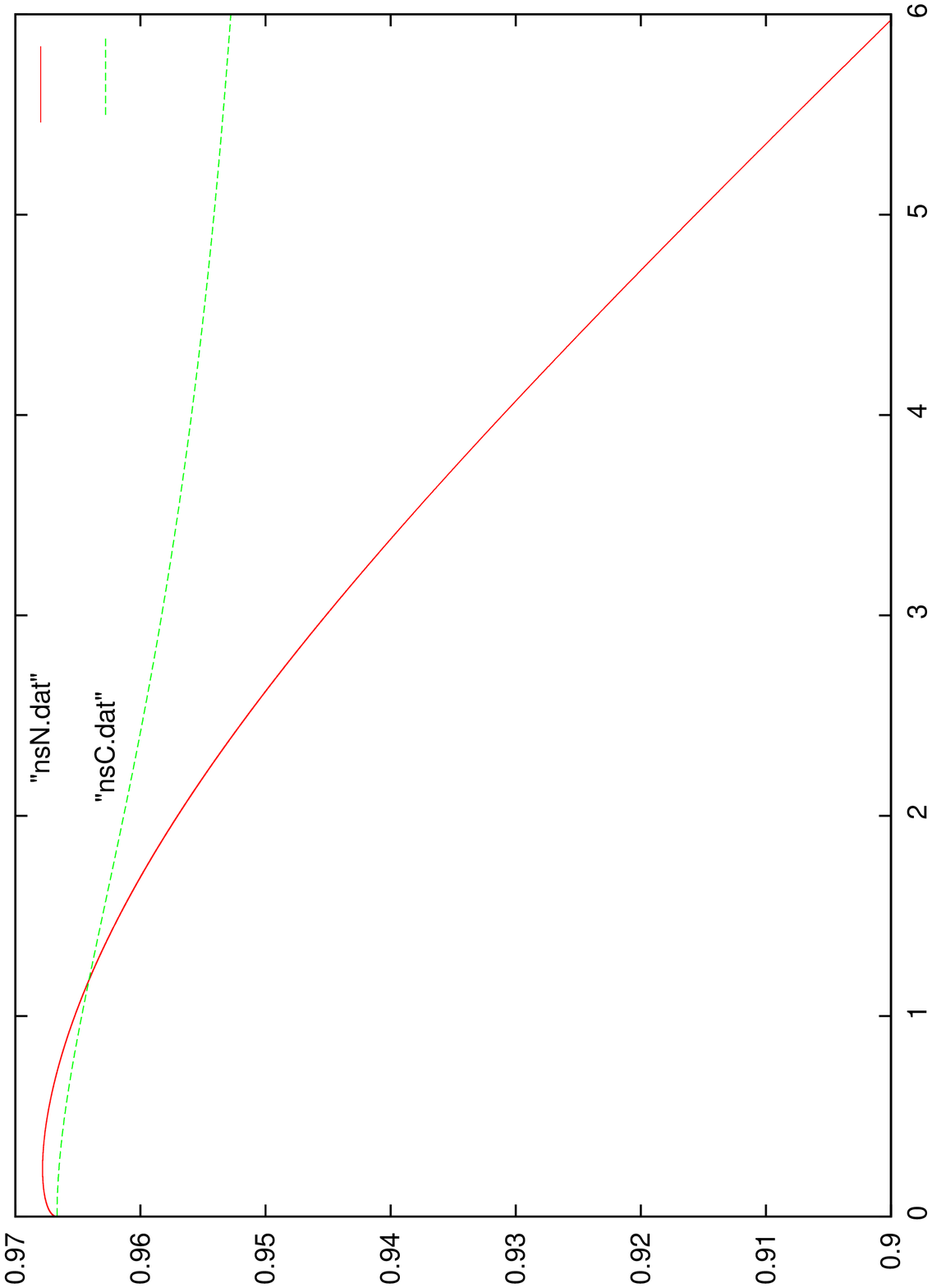}
\psfrag{"rN.dat"}{New binomial inflation, $ r $ vs. $ y $}
\psfrag{"rC.dat"}{Chaotic binomial inflation, $ r $ vs. $ y $}
\includegraphics[height=10cm,width=7cm]{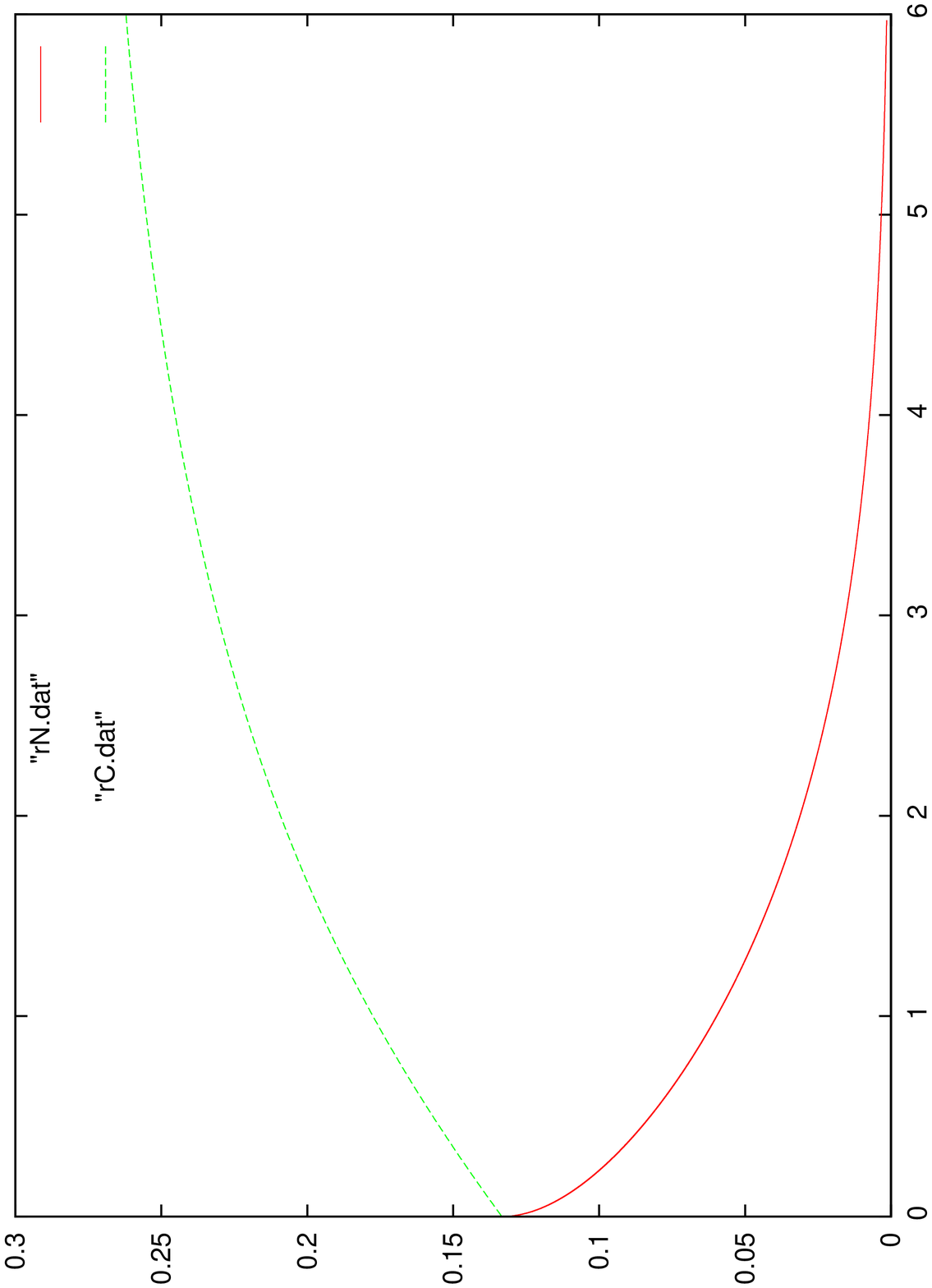}
\end{turn}
\caption{Binomial Inflation. Upper panel: $ n_s $ vs. the quartic 
coupling $ y $ for new binomial and chaotic binomial inflation. There 
are maximum values for $ n_s $: $  n_s \leq 0.9678 \ldots $ and  
$ n_s \leq  1- 2/N = 0.9666\ldots $, respectively.
Lower panel: $ r $ vs. the quartic coupling $ y $ for new binomial and 
chaotic binomial inflation.
The range of $ r $ in new binomial inflation is exactly {\bf below}
its range in chaotic binomial inflation. The ranges of $ n_s $ have an 
overlap.} 
\label{nsyryb}
\end{center}
\end{figure}

We obtain from eqs.(\ref{ampliI}), (\ref{indi}), (\ref{run}) and 
(\ref{nbino}) the spectral index, its running, the ratio $ r $ and the 
amplitude of adiabatic perturbations for binomial new inflation,
\bea\label{nsbinew}
n_s &=& 1 - \frac{y}{N} \; \frac{3 \; z + 1}{(1-z)^2}   \quad , \quad
\frac{d n_s}{d \ln k} = \frac{y^2 \; z}{N^2 \; (1-z)^4}\left(24 \; z^2 
- 35 \; z +3\right) \quad , \cr \cr
r &=& \frac{16 \; y}{N}\; \frac{z}{(1-z)^2} \label{rbinew}  \quad  , \quad
|{\Delta}_{k\;ad}^{\mathcal{R}}|^2  = \frac{N^2}{12 \, \pi^2} \; 
\left(\frac{M}{M_{Pl}}\right)^4 \; \frac{(1-z)^4}{y^2 \; z}\quad . 
\eea
In new binomial inflation the limit $ z \to 1^- $ implies weak coupling
$ y \to 0^+ $, that is, the potential is quadratic around the absolute
minimum $ \chi = \sqrt{8/y} $ and we find,
\bea\label{cobino}
&& n_s \buildrel{y \to 0}\over= 1 - \frac2{N} 
\quad , \quad  r \buildrel{y \to 0}\over= \frac8{N} \quad , \quad
\chi  \buildrel{y \to 0}\over= \sqrt{\frac8{y}} \to \infty \; , \\ \cr
&& \frac{d n_s}{d \ln k} \buildrel{y \to 0}\over= 
-\frac2{N^2} \quad , \quad 
|{\Delta}_{k\;ad}^{\mathcal{R}}|^2  \buildrel{y \to 0}\over= 
\frac{N^2}{3 \, \pi^2} \; 
\left(\frac{M}{M_{Pl}}\right)^4 \; \; ,\label{ampbin}
\eea
which coincide with $ n_s , \; d n_s/d \ln k $ and $ r $ 
for the monomial quadratic potential in chaotic inflation 
eq.(\ref{ycerocao}). There is an extra factor two in 
$ |{\Delta}_{k\;ad}^{\mathcal{R}}|^2 $ for new inflation
and $ y \to 0^+ $ due to the fact that $ w''(\sqrt{8/y}) = 2 $ at 
the absolute minimun of the potential eq.(\ref{binon}) while 
$ w''(0) = 1 $ at the absolute minimum of the potential eq.(\ref{binoc}).

\medskip

In the limit $ z \to 0^+ $ which implies $ y\to +\infty $ (strong 
coupling), we have
$$ 
z \buildrel{y \to +\infty}\over= e^{-y-1} \to 0^+ 
$$
and
\bea\label{binonewyG}
&&n_s \; \; \buildrel{y \gg 1}\over=1 - \frac{y}{N}
\quad , \quad r \; \;  \buildrel{y \gg 1}\over= \frac{16 \; y}{N} \; 
e^{-y-1}\quad ,\quad
\chi  \; \; \buildrel{y \gg 1}\over= \sqrt{\frac8{y}}\; e^{-(y+1)/2} 
\to 0 \; , \cr \cr
&& \frac{d n_s}{d \ln k} \; \; \buildrel{y \gg 1}\over= -\frac{3 \; 
y^2 \; e^{- (y+1)}}{N^2 \;} 
\quad , \quad |{\Delta}_{k\;ad}^{\mathcal{R}}|^2  \; \; 
\buildrel{y \gg 1}\over=  \; \; \
\frac{N^2}{12 \, \pi^2} \; \left(\frac{M}{M_{Pl}}\right)^4 \; 
\frac{e^{y+1}}{y^2} \quad .
\eea
Notice that the slow-roll approximation is no longer valid when the 
coefficient of $ 1/N $ becomes much larger than unity. Hence, the results 
in eq.(\ref{binonewyG}) are valid for $ y \lesssim N $. We see that in 
this strong coupling regime $ r $ becomes very small and $ n_s $ becomes well 
below unity. The WMAP+LSS results exclude $ n_s \lesssim 0.9 $ 
\cite{WMAP1,WMAP3,WMAP5}. Therefore, this strong coupling limit $ y \gg 1 $ 
is ruled out. 

\subsubsection{Contrasting the results of new and chaotic binomial 
inflation.}\label{binocom}

We plot in figs. \ref{nsyryb}  $ n_s $ and  $ r $ vs. the coupling $ y $, 
respectively.
The curves are restricted to the region $ 0 < r < 0.3 $ and $ n_s > 0.9 $
compatible with the WMAP data \cite{WMAP3,WMAP5}.

\begin{figure}[h]
\begin{center}
\begin{turn}{-90}
\psfrag{"nsrN.dat"}{New binomial inflation, $ r $ vs. $ n_s $}
\psfrag{"nsrC.dat"}{Chaotic binomial inflation, $ r $ vs. $ n_s $}
\includegraphics[height=12cm,width=8cm,keepaspectratio=true]{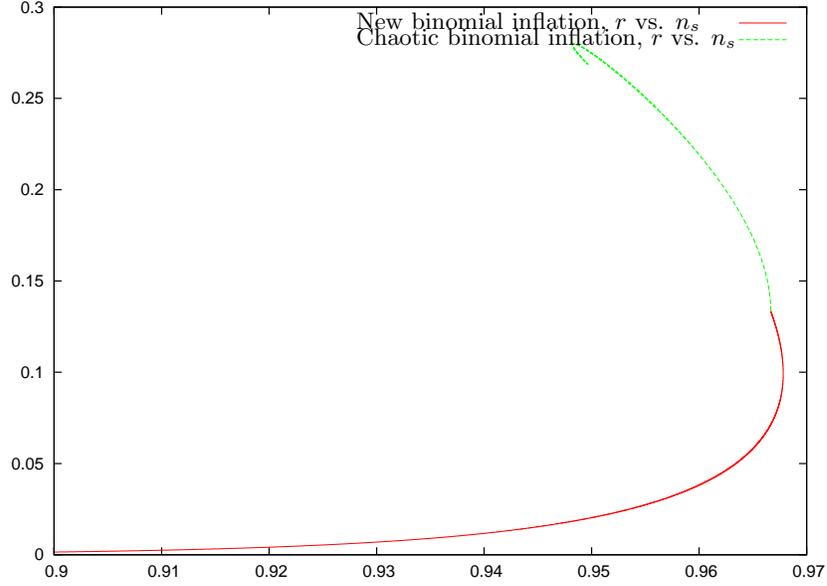}
\end{turn}
\caption{ $ r $ vs. $ n_s $ for new binomial and chaotic binomial inflation. 
We set here $ N = 60 $ and restrict the plot to the region $ n_s > 0.9 $. $ r $ 
in new binomial inflation is exactly 
{\bf below} its range in chaotic binomial inflation.} 
\label{nsrb}
\end{center}
\end{figure}

We find from sec. \ref{binocao} that in binomial chaotic inflation $ n_s $
and $ r $ cover the following range of values for $ N = 60 $:
\bea\label{rangoc}
&&0.94821\ldots = 1 - \frac{3.1072\ldots}{N}
\leq n_s \leq 1- \frac2{N} = 0.9666\ldots \quad ,  \cr \cr
&& 0.1333 \ldots= \frac8{N} \leq 
r \leq \frac{16.802\ldots}{N} = 0.2800\ldots 
\quad {\rm binomial ~ chaotic ~ inflation} \; .
\eea
We find from sec. \ref{binonue} that in binomial new inflation the 
following range of values for $ n_s $ and $ r $ are covered:
\be\label{rangoN}
 n_s \leq 1- \frac{1.93051\ldots}{N} = 0.96782 \ldots
\quad ,  \quad 0 < r \leq \frac8{N}=0.1333\ldots 
\quad {\rm binomial ~ new ~ inflation} \; .
\ee
The maximum value for $ n_s $ corresponds to $ z = 0.4582 \ldots, \;
y = 0.2387 \ldots , \; r = 0.1192\ldots $.
We see that the range of $ r $ in new binomial inflation is exactly 
{\bf below} its range in chaotic binomial inflation [see eqs.(\ref{rangoc})
and (\ref{rangoN})] while the ranges of $ n_s $ have an overlap.

It must be noticed that there are maximum values for $ n_s $ {\bf both} in 
new and chaotic binomial inflation: 

\noindent
$ n_s \leq 0.9678 \ldots $ and  $  n_s \leq 0.97 $ in new and chaotic 
binomial inflation, respectively. An observed  value of  $ n_s $ above any 
of these bounds would rule out the corresponding model.

\begin{figure}[h]
\begin{center}
\begin{turn}{-90}
\psfrag{"grn.dat"}{$ g(z)^{1/4} $ vs. $ r $ for New Binomial Inflation}
\psfrag{"grc.dat"}{$ g(z)^{1/4} $ vs. $ r $ for Chaotic Binomial Inflation}
\includegraphics[height=12cm,width=8cm,keepaspectratio=true]{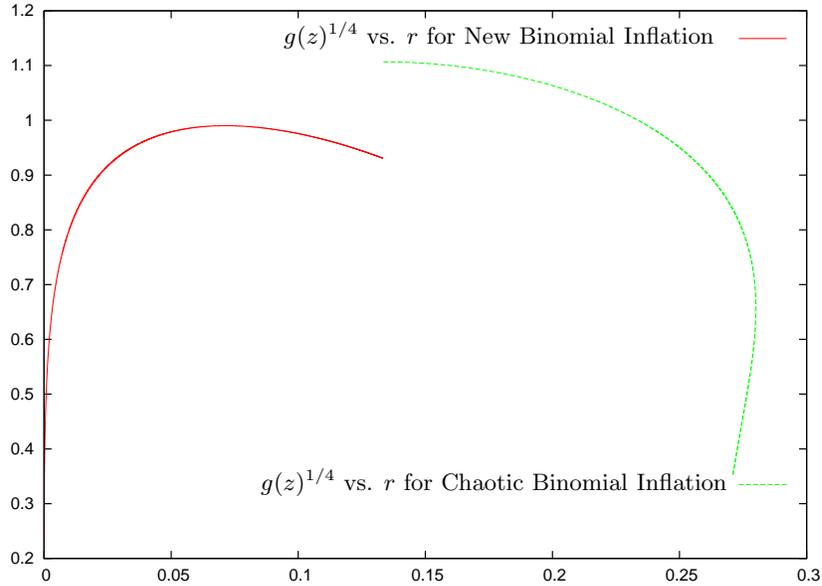}
\end{turn}
\caption{ $ g^\frac14(z) $ vs. $ r $ for new binomial and chaotic binomial inflation.
We consider the domain $ 0 < r < 0.3 , \; n_s > 0.9 $ allowed by the WMAP data.
The endpoints of the new and chaotic inflation curves at $ r = 8/N = 0.1333\ldots $
correspond to $ (3/4)^\frac14 = 0.930604\ldots $ and 
$ (3/2)^\frac14=1.106682\ldots $, respectively.
The endpoints do not coincide because the second derivatives of the binomial potential at 
the minimum differ by a factor two in new and chaotic inflation [see explanation below 
eq.(\ref{ampbin})].}
\label{amp}
\end{center}
\end{figure}

\medskip

We plot in fig. \ref{nsrb} $ r $ vs. $ n_s $ for new binomial and chaotic binomial inflation.
Notice that the two curves join at the quadratic monomial point 
$ n_s = 1- 2/N = 0.966\ldots, \; r = 8/N = 0.133\ldots $.

Notice that in new inflation, $ n_s > 0.92 $ {\bf necessarily implies} $ r > 0.042 $ 
(see fig. \ref{nsrb}). That is, within
new binomial inflation, a lower bound for $ n_s $ as provided by WMAP data implies a lower
bound for $ r $. We will see this property generalized to trinomial new inflation 
in sec. \ref{tni} \cite{infwmap,mcmc}.

\medskip

The observed values of $ n_s $ and  $ r $ therefore determine the value of the {\bf couplings}
(for the binomial potential only one coupling $ y $). Once $ y $ is known,
the {\bf energy scale} of inflation $ M $ is determined by the observed value of the scalar
fluctuations amplitude combined with eq.(\ref{ampliI}). We find
\bea\label{amplg}
&&\frac{M}{M_{Pl}} = \sqrt{\frac{2 \; \pi \; {\Delta}_{k\;ad}^{\mathcal{R}}}{N}} \; g^\frac14(z)
= 2.27 \; 10^{-3}  \times  g^\frac14(z) \; , \cr \cr
&& M = 0.554 \; 10^{16} \; {\rm GeV} \times g^\frac14(z) \; ,
\eea
where we set $ N = 60 $, used the WMAP value for $ {\Delta}_{k\;ad}^{\mathcal{R}} $ 
eq.(\ref{aniso}) and we called:
\bea
&& g(z) \equiv \frac{3 \; y^2 \; (z+1)^2}{z^2 \; (z+2)^3} \quad {\rm binomial ~ chaotic ~ 
inflation} \quad , \cr \cr
&&  g(z) \equiv \frac{3 \; z \; y^2}{(1-z)^4} \quad {\rm binomial ~ new ~ inflation} \quad .
\eea
Notice that for $ 0 \leq y \leq \infty $,
$$
\frac32 \geq g(z) \geq 0 \quad {\rm binomial ~ chaotic ~ inflation}\quad ,\quad 
0.96099\ldots \geq g(z) \geq 0 \quad {\rm binomial ~ new ~ inflation} \quad .
$$
We plot in fig. \ref{amp} $ g^\frac14(z) $ vs. $ r $ for new binomial and 
chaotic binomial inflation. Notice that $ g^\frac14(z) = {\cal O}(1) $ in 
the domain of values  $ 0 < r < 0.3 $ and $ n_s > 0.9 $
compatible with the WMAP data \cite{WMAP3,WMAP5}. We thus see from 
eq.(\ref{amplg}) and fig. \ref{amp} that the value of the inflation 
{\bf energy scale} is in the GUT range for all values of $ n_s $ and  
$ r $ allowed by the observations.

\noindent
We obtain from eq.(\ref{amplg}) at the coupling value $ y = 1.26 $ that 
best fit the WMAP5+SDSS+SN data (new inflation, see Table VI),
\be \label{myM}
M = 0.543 \times 10^{16} \;  {\rm GeV} \quad ,  \quad
m = 1.21 \times 10^{13} {\rm GeV ~~~ and} \quad H \simeq 6 \times  10^{13}
 {\rm ~GeV ~~~ for}  \quad   y = 1.26  \; . 
\ee
Notice that these values {\bf agree} with the generic estimates 
eq.(\ref{valorM})-(\ref{myH}).

\subsection{Trinomial Chaotic Inflation: 
 Spectral index, amplitude ratio, running index and 
limiting cases}\label{tci}

We consider now the trinomial potential with unbroken symmetry 
investigated in ref.\cite{ciri,mcmc,hbkp}:
\be\label{VC}
V(\varphi)=  \frac{m^2}2 \; \varphi^2 + \frac{ m \; g }3 \; \varphi^3 + 
\frac{\lambda}{4}\; \varphi^4 \; ,
\ee
where $ m^2 > 0 $ and $ g $ and $ \lambda $ are dimensionless couplings.

The corresponding dimensionless potential $ w(\chi) $ eqs.(\ref{V})-(\ref{chifla} )
takes the form
\be\label{trinoC}
w(\chi) = \frac12 \; \chi^2 + \frac{h}3 \; \sqrt{\frac{y}2} \; \chi^3 +
\frac{y}{32} \; \chi^4 \; ,
\ee
where the quartic coupling $ y $ is dimensionless as well as 
the asymmetry parameter $ h $. The couplings in eq.(\ref{VC}) and 
eq.(\ref{trinoC}) are related by eq.(\ref{acoi}).

Chaotic inflation is obtained by choosing the initial field $ \chi $ in 
the interval $ (0,+\infty) $. The inflaton  $ \chi $ slowly rolls down the slope
of the potential from its initial value till the absolute minimum of the 
potential at the origin.

Computing the number of efolds from eq.(\ref{Nchi}), we find the field 
$ \chi $ at horizon crossing related to the couplings $ y $ and $ h $.
Without loss of generality, we choose $ h < 0 $ and 
work with positive fields $ \chi $. 

The potential eq.(\ref{trinoC}) has extrema at $ \chi = 0 $ and $ \chi_{\pm} $ given by,
\be\label{minD}
\chi_{\pm} = \sqrt{\frac8{y}}\left[ -h \pm i \; \Delta \right] 
\quad , \quad \Delta \equiv \sqrt{1-h^2} \; .
\ee
That is,  for $ |h| < 1 , \; w(\chi) $ has 
only one real extremum at $ \chi = 0 $ while for  $ |h| \geq 1 , \; w(\chi) $ has
three real extrema. There is always a minimum
at  $ \chi = 0 $ since $ w''(0) = 1 $.  At the non-zero extrema we have
$$
w''(\chi_{\pm}) = -2 \; \Delta \; \left(\Delta \pm i \; h \right) \; .
$$
We have for  $ |h|>1 $,
$$
\chi_{\pm} = \sqrt{\frac8{y}}\left[ -h \pm \sqrt{h^2-1} \right] \; , \quad {\rm and}
\quad w''(\chi_{\pm}) =2 \,  \sqrt{h^2-1}\left(\sqrt{h^2-1} \mp h \right) \; .
$$
Hence, for any $ h < -1 $, we have $ w''(\chi_{+}) > 0 $ while $ w''(\chi_{-}) < 0 $.
Notice that $ \chi_{\pm} > 0 $  for $ h < -1 $.

Therefore, we have chaotic inflation for positive field $ \chi $ in the regime $ |h| < 1 $
using the inflaton potential eq.(\ref{trinoC}).

We can also have  chaotic inflation with the potential  eq.(\ref{trinoC})
for negative field if $  h > 3/\sqrt8 $, but for $  |h| > 3/\sqrt8 $
the absolute minimum is no more at  $ \chi = 0 $ but at $ \chi_{+} $.
Since $ w(\chi_{+}) < 0 $ for $  |h| > 3/\sqrt8 $ we have to subtract
in this case the value  $ w(\chi_{+}) $ from  $ w(\chi) $ 
in order to enforce eq.(\ref{condw}).

We consider in subsections \ref{prih}, \ref{potcha} and \ref{HZe} the regime $ -1 < h \leq 0 ,
\; \chi \geq 0 $. The case $ h < -1 $ is analyzed in subsection \ref{hmemu}.

\begin{figure}
\includegraphics[height=8cm,width=12cm,keepaspectratio=true]{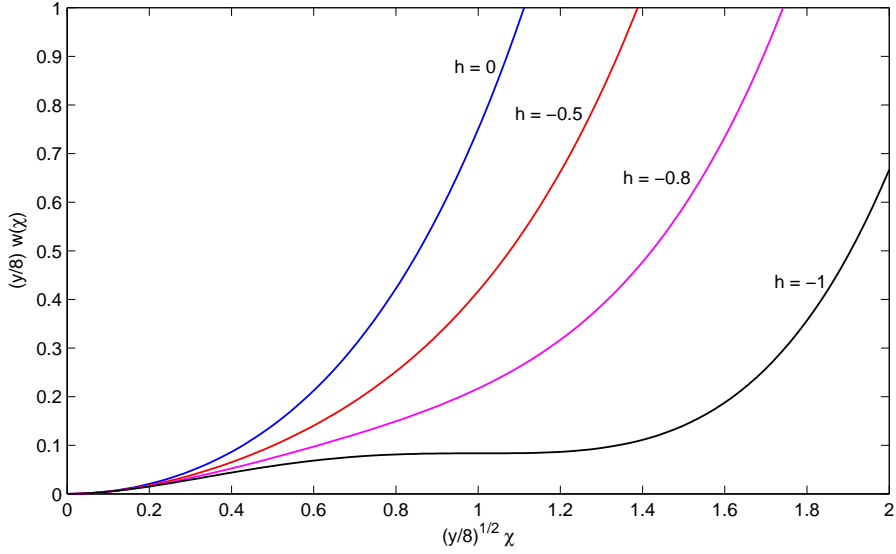}
\caption{Trinomial Chaotic Inflation. We plot here the chaotic inflation trinomial
potential  $ \frac{y}8 \; w(\chi) $ [eq.(\ref{trinoC}) with positive quadratic term] 
vs. the field variable $ \sqrt{z}=\sqrt{y/8} \; \chi $ for different values of the 
asymmetry parameter $ h $, namely, $ h =0, \; -0.5, \; -0.8 $ and $ -1 $. 
Notice the inflection point at $ \sqrt{z}= 1 $ when $ h=-1 $.}
\label{vv}
\end{figure}

\subsubsection{The small asymmetry regime: $ -1 < h \leq 0 $.}\label{prih}

In chaotic inflation the inflaton field slowly rolls down the slope
of the potential from its initial value till the absolute minimum of the 
potential at $ \chi = 0 $.

\medskip

It is convenient to use the field variable $ z $ defined by eq.(\ref{defz}).
$ z = y \; \chi^2/8 $. In terms of $ z $ the chaotic trinomial potential takes the form
$$
w(\chi) = \frac4{y} \; z \; \left(1+\frac43 \, h \; \sqrt{z} + \frac12 \, z \right) \; .
$$
When $ z \lesssim 1 $ we are in the quadratic regime where $ w(\chi) $ is approximated
by the $ \chi^2 $ term. For $ z \gtrsim 1 $ we go to the non-linear regime in $ z $
and all three terms in $ w(\chi) $ are of the same order of magnitude.

\medskip

In fig. \ref{vv}, we plot $ y \; w(\chi)/8 $ as a function of $ \sqrt{z} $ for several
values of $ h \geq - 1 $. We see that the potential becomes flatter as $ h $ decreases.
For $ h = - 1 $, both $ w'(\chi) $ {\bf and } $ w'(\chi) $ vanish at
$ \sqrt{z} = 1 $. The case $ h = - 1 $ is singular since the inflaton gets stuck
an infinite amount of time at the point $ \sqrt{z} = 1 , \; \chi = \sqrt{8/y} $.

\medskip

By inserting eq.(\ref{trinoC}) for $ w(\chi) $ into  eq.(\ref{Nchi}) for $ N[\chi] $
and setting $ N[\chi] = N $ we obtain the field $ \chi $ or equivalently $ z $, 
at horizon exit, in terms of the coupling $ y $ and the asymmetry parameter $ h $,
\be\label{ctrino}
y = z + \frac43 \; h \; \sqrt{z} + \left(1 - \frac43 \; h^2\right) \;
\log\left(1+2 \, h \; \sqrt{z} + z \right) 
 -\frac{4\,h}{3 \, \Delta} \left(\frac52 - 2 \, h^2\right)
\left[\arctan\left(\frac{h+\sqrt{z}}{\Delta}\right)
-\arctan \left(\frac{h}{\Delta}\right)\right] 
\ee
This defines the field $ z $ as a monotonically increasing function of the coupling $ y $
for $ 0 < y, \; z < +\infty $. Recall that $ \chi $ and $ z $ corresponds
to the time of horizon exit.

\medskip

We obtain from eqs.(\ref{ampliI}), (\ref{indi}), (\ref{run}), and 
(\ref{trinoC}) the spectral index, its running, the ratio $ r $ and the 
amplitude of adiabatic perturbations,
\bea\label{nscao}
n_s&=&1- \frac{y}{2 \, N \; z}\left[\frac{ 3 \; \left(1+2 \, h \; 
\sqrt{z} + z \right)^2}{
\left(1+\frac43 \, h \; \sqrt{z} + \frac12 \, z \right)^2}-
\frac{1+4 \, h \; \sqrt{z} + 3 \, z}{1+\frac43 \, h \; \sqrt{z} + 
\frac12 \, z}\right] \; , \\ \cr\cr
\frac{d n_s}{d \ln k} &=& \frac{y^2}{2 \,  z^2 \; N^2}\left[ - 
\frac{\left(1+2 \, h \; \sqrt{z} + z \right)\left(h \; \sqrt{z} + 
\frac32 \, z \right)}
{\left(1+\frac43 \, h \; \sqrt{z} + \frac12 \, z \right)^2} - 
\frac{3 \; \left(1+2 \, h \; \sqrt{z} + z \right)^4}{
\left(1+\frac43 \, h \; \sqrt{z} + \frac12 \, z \right)^4} + 
\right. \cr \cr\cr
&+& \left.2 \; \frac{\left(1+2 \, h \; \sqrt{z} + z \right)^2
\left(1+4 \, h \; \sqrt{z} + 3 \, z\right)}{\left(1+\frac43 \, h \; 
\sqrt{z} + 
\frac12 \, z \right)^3} \right]\; , \label{runcao}\\ \cr\cr
r &=& \frac{4 \, y}{N \; z}\frac{\left(1+2 \, h \; \sqrt{z} + 
z \right)^2}{
\left(1+\frac43 \, h \; \sqrt{z} + \frac12 \, z \right)^2} \; ,
\label{rcao} \\ \cr\cr
\label{ampcao}
|{\Delta}_{k\;ad}^{\mathcal{R}}|^2  &=& \frac{2 \, N^2}{3 \, \pi^2} \; 
\left(\frac{M}{M_{Pl}}\right)^4 \;
\frac{z^2 \; \left(1+\frac43 \, h \; \sqrt{z} + 
\frac12 \, z \right)^3}{y^2 \; \left(1+2 \, h \; \sqrt{z} + 
z \right)^2} \; .
\eea
In chaotic inflation, the limit $ z \to 0^+ $ implies
$ y \to 0^+ $ (shallow limit), we have in this limit:
\bea
&&y \buildrel{z \to 0^+}\over= 2  \, z +{\cal O}(z^\frac32) \quad , \quad
\chi \buildrel{y \to 0^+}\over= 2 + {\cal O}(y) \quad , \quad
n_s\buildrel{y \to 0^+}\over=1- \frac2{N}\; ,  \cr \cr
&& r \buildrel{y \to 0^+}\over=\frac8{N}\quad , \quad
|{\Delta}_{k\;ad}^{\mathcal{R}}|^2  \buildrel{y \to 0^+}\over= \frac{N^2}{6 \, \pi^2} \; 
\left(\frac{M}{M_{Pl}}\right)^4 \; .\label{cuadh}
\eea
The results in the $ y \to 0^+ $ limit are {\bf independent} of the asymmetry $ h $
and coincide with those for the  purely quadratic monomial potential $ \chi^2/2 $.

In the limit $ z \to +\infty $ which implies $ y\to +\infty  $ (strong coupling), we have
for fixed $ h > - 1 $,
\bea
&&y \buildrel{z \to +\infty}\over=  z + \frac43 \; h \; \sqrt{z} + 
\left(1 - \frac43 \; h^2\right) \; \log z +{\cal O}(1) \quad , \quad
\chi \buildrel{y \to +\infty}\over= 2 \; \sqrt2 + 
{\cal O}\left(\frac1{\sqrt{z}}\right)
\; , \cr \cr
&&n_s \buildrel{y \to +\infty}\over=1- \frac3{N}\left[1 + 
\frac43 \; \frac{h}{\sqrt{z}} + 
\left(1 - \frac43 \; h^2\right) \; \frac{\log z}{z} +
{\cal O}\left(\frac1{z}\right)\right] \; , \cr \cr
&&r \buildrel{y \to +\infty}\over=\frac{16}{N}\left[1 + 
\frac43 \; \frac{h}{\sqrt{z}} + 
\left(1 - \frac43 \; h^2\right) \; \frac{\log z}{z} +
{\cal O}\left(\frac1{z}\right)\right]\; . \cr\cr
&&|{\Delta}_{k\;ad}^{\mathcal{R}}|^2  \buildrel{y \to +\infty}\over= 
\frac{N^2}{12 \, \pi^2} \; \left(\frac{M}{M_{Pl}}\right)^4 
\frac{z}{\left[1 + \frac43 \; \frac{h}{\sqrt{z}} + 
\left(1 - \frac43 \; h^2\right) \; \frac{\log z}{z} +
{\cal O}\left(\frac1{z}\right)\right]^2} \; .
\eea
For $ h = 0 , \; n_s $ and $ r $ in the limit $ y\to +\infty  $
coincide with those of the purely quartic monomial potential 
$ \chi^4 $ given by eq.(\ref{cuar}).

\subsubsection{The flat potential limit $ h \to -1^{+} $} \label{potcha}

We consider chaotic inflation in the regime $ -1 < h \leq 0 , \; 
\chi \geq 0 $.

At $ h = -1 $ the potential eq.(\ref{trinoC}) exhibits an inflexion point 
at $ \chi_0 \equiv \sqrt{8/y} $. Namely, $ w'(\chi_0) =  
w''(\chi_0) = 0 $ while $ w(\chi_0) = 2/(3 \; y ) > 0 $. That is, this 
happens at $ z_0 = y \; \chi_0^2/ 8 = 1 $.

Therefore, for $ h > - 1 $ but very close to $ h = -1 $, the field 
evolution strongly slows down near the point $ \chi = \chi_0 $. This 
strong slow down shows up in the calculation of observables when the 
field  $ \chi $ at horizon crossing
is $ \chi > \chi_0 $, namely, for $ z > z_0 = 1 $. For $ \chi < \chi_0 $, 
that is $ z < 1 $, the slow down of the field evolution will only appear 
if $ z \simeq 1 $. Therefore, the limit $ h = -1 $ is singular since the 
inflaton field gets trapped at the point $ z = 1 $.

\medskip

Let us derive the $ h \to -1^{+} $ limit of $ y , \; n_s $ and $ r $ from
eqs.(\ref{ctrino})-(\ref{ampcao}) in the regimes $ z < 1 $ and $ z > 1 $, 
respectively.

\medskip

When $ h \to -1^{+} $ we see from eq.(\ref{minD}) that $ \Delta \to 0 $ 
and the arguments of
the two arctan  in eq.(\ref{ctrino}) diverge. Hence, the arctan tends to
$ +\pi/2 $ or $ -\pi/2 $. Depending on whether $ z < 1 $ or $ z > 1 $, 
the $ \pi/2 $'s terms cancel out or add, respectively. The special case 
$ z = 1 $ is investigated in the next subsection \ref{HZe}.

\medskip

In the  case  $ z < 1 $ we get:
\bea\label{yL}
&&y = z - \frac43  \; \sqrt{z} - \frac23 \; \log\left(1-\sqrt{z} \right) 
+\frac23\frac{\sqrt{z}}{1-\sqrt{z}} \quad , \quad h \to -1^{+}\quad, \quad z<1 \; ,
\\ \cr
n_s&=&1- \frac{y}{2 \, N \; z}\left[ 3 \; \frac{\left(1-\sqrt{z}\right)^4}{
\left(1-\frac43  \; \sqrt{z} + \frac12 \, z \right)^2}-
\frac{\left(1-\sqrt{z} \right) \left(1-3 \, \sqrt{z} \right)}{1-\frac43  \; \sqrt{z} + 
\frac12 \, z}\right] \quad , \quad h \to -1^{+}\quad, \quad z<1 \; , \\ \cr\cr
r &=& \frac{4 \, y}{N \; z}\frac{\left(1-\sqrt{z}\right)^4}{
\left(1-\frac43  \; \sqrt{z} + \frac12 \, z \right)^2} \quad , 
\quad h \to -1^{+}\quad, \quad z<1 \; ,
\label{rcaoL} \cr\cr
|{\Delta}_{k\;ad}^{
\mathcal{R}}|^2  &=& \frac{2 \, N^2}{3 \, \pi^2} \; 
\left(\frac{M}{M_{Pl}}\right)^4 \;
\frac{z^2 \; \left(1-\frac43 \; \sqrt{z} + \frac12 \, z \right)^3}{y^2 \; \left(1-\sqrt{z}\right)^4}
\; .
\eea
In particular, in the regime $ z \to 1^{-} $ we find,
\bea\label{nsrlim}
&&n_s\buildrel{z \to 1^{-}}\over= 1 - \frac4{N} \quad , \quad
r \buildrel{z \to 1^{-}}\over= \frac{96}{N} \; \left(1-\sqrt{z}\right)^3 \to 0 
\quad , \quad h \to -1^{+} \; , \cr\cr
&& y \buildrel{z \to 1^{-}}\over= \frac23\frac1{1-\sqrt{z}} \to \infty
\quad , \quad 
\chi \buildrel{z \to 1^{-}}\over= 4 \; \sqrt3 \; \sqrt{1-\sqrt{z}} \to 0 
\quad , \quad 
|{\Delta}_{k\;ad}^{\mathcal{R}}|^2 \buildrel{z \to 1^{-}}\over=\left(\frac{N}{8 \, \pi} \;
\frac{M^2}{M_{Pl}^2} \; y \right)^2\label{amplim}
\; .
\eea
That is, in the limit $  h \to -1^{+} , \; z \to 1^{-} $
the ratio $ r $ becomes {\bf very small} while the
spectral index takes the value $ n_s = 0.92 $. The ratio $ r $ tends
to zero in the regime $ z \to 1^{-} , \;  \chi \to \chi_0 = \sqrt{8/y} $
because  $ w'(\chi_0) = 0 $ and $ r $ is proportional to
$ {w'}^2(\chi) $ according to eq.(\ref{indi}).

The $ z \to 1^{-} $ regime for $  h \to -1^{+} $ is a strong coupling
limit since $ y \to +\infty $ as shown by eq.(\ref{amplim}). 
In addition, eq.(\ref{amplim}) shows that for large
$ y $ one must keep the product $ y \; M^2 $ fixed
since it is determined by the amplitude of the adiabatic perturbations.
We see from eq.(\ref{amplim}) that $ {\tilde M} \equiv \sqrt{y} \; M $ 
becomes the energy scale of inflation in the $ y \to \infty $ limit:
from eq.(\ref{valorM}), $ {\tilde M} \sim 10^{16}$GeV according to the 
observed value of $ |{\Delta}_{k\;ad}^{\mathcal{R}}|/N $ displayed in 
eq.(\ref{Mwmap}), while $ M $ and $ m $ vanish in the  $ y \to \infty $ 
limit
$$
M = \frac{\tilde M}{\sqrt{y}} \buildrel{ y \to \infty }\over= 0 \quad ,  
\quad m = \frac{M^2}{M_{Pl}} = \frac{ {\tilde M}^2}{y \; M_{Pl}} 
\buildrel{ y \to \infty }\over= 0 \quad .
$$
In fig. \ref{ene} we display $ r $ vs. $ n_s $ for fixed values of the 
asymmetry parameter $ h $ and the coupling $ y $ varying along the curves.
The red curves correspond to chaotic inflation. 
In the bottom of fig. \ref{ene} we can see the curve for $ h = -0.999 $.
The limiting curve for $ h = -1 $ (not drawn) will reach the point 
$ n_s = 0.92 $ and the bottom line $ r = 0 $ as described by 
eq.(\ref{nsrlim}).

In the case $ z > 1 $, we get from eqs.(\ref{ctrino})-(\ref{ampcao}),
\bea\label{yL2}
y &\buildrel{h \to -1^{+}}\over=& \frac{\pi}3 \; \sqrt{\frac2{h+1}} + 
{\cal O}\left([h+1]^0 \right) \to +\infty  \quad, \quad 
\chi \buildrel{h \to -1^{+}}\over= 2 \; \sqrt{\frac{3 \; z}{\pi}} \; 
\left[ 2 \; (h+1) \right]^\frac14 \to 0  \quad , \quad z > 1 \; ,   \cr\cr
n_s &\buildrel{h \to -1^{+}}\over=& 1- \frac{\pi}{3 \, \sqrt2 \, N \; z}\frac1{\sqrt{h+1}}
\left[ 3 \; \frac{\left(1-\sqrt{z}\right)^4}{
\left(1-\frac43  \; \sqrt{z} + \frac12 \, z \right)^2}-
\frac{\left(1-\sqrt{z} \right) \left(1-3 \, \sqrt{z} \right)}{1-\frac43  \; \sqrt{z} + 
\frac12 \, z}\right] + {\cal O}\left([h+1]^0 \right) \quad, \quad z > 1 \; , \cr \cr\cr
r  &\buildrel{h \to -1^{+}}\over=& \frac{8 \, \pi}{3 \, \sqrt2\; N \; z}\frac1{\sqrt{h+1}}
\frac{\left(1-\sqrt{z}\right)^4}{\left(1-\frac43  \; \sqrt{z} + \frac12 \, z \right)^2}  
+ {\cal O}\left([h+1]^0 \right)\quad, \quad z > 1 \; .
\label{rcaoL2} 
\eea
In this strong coupling regime 
the indices become {\bf very large} and hence in contradiction with the data.
In addition, the slow-roll expansion cannot be trusted when the coefficients
of $ 1/N $ become large compared with unit. In conclusion, for $ h \to -1^{+} $,
the case of large field $ z > 1 $ is excluded by the data.

\subsubsection{The singular limit $ z = 1 $ and then $ h \to -1^{+} $ 
yields the Harrison-Zeldovich spectrum}\label{HZe}

We study in this section the case $ z = 1 $ in trinomial chaotic inflation.
We obtain from eq.(\ref{ctrino})
\be\label{yz1}
y \buildrel{z=1}\over= 1 + \frac43 \; h  + \left(1 - \frac43 \; h^2\right) \;
\log\left[2 \, (1+h)\right] -\frac{4\,h}{3 \, \Delta} \left(\frac52 - 2 \, h^2\right)
\left[\arctan\left(\frac{h+1}{\Delta}\right)
-\arctan \left(\frac{h}{\Delta}\right)\right]  \; .
\ee
For  $ z = 1 $, we find from eqs.(\ref{nscao})-(\ref{ampcao}) and (\ref{yz1}) in the limit
$ h \to -1^{+} $,
\bea\label{HZ1}
&& y \buildrel{z=1, \; h \to -1^{+}}\over= \frac{\pi}3 \; \sqrt{\frac1{2(h+1)}}
- \frac13 \; \log(h+1) +  {\cal O}\left([h+1]^0\right) \; , \cr \cr
&& n_s \buildrel{z=1, \; h \to -1^{+}}\over= 1 +
\frac{2 \; \pi}{N} \; \sqrt{2 \, (h+1)} \left[1+ {\cal O}\left(h+1\right)\right] \; ,\cr \cr
&& r \buildrel{z=1, \; h \to -1^{+}}\over=\frac{16 \; \pi}{N} \; \sqrt2 \; (h+1)^{\frac32}
\left[1+ {\cal O}\left(h+1\right)\right] \; ,\cr \cr
&&|{\Delta}_{k\;ad}^{\mathcal{R}}|^2 \buildrel{z=1, \; h \to -1^{+}}\over=
\frac{N^2}{36} \; \left[\frac{M}{\pi \; M_{Pl} \;  (h+1)^{\frac14}}\right]^4 \; .
\eea
Therefore, we reach the Harrison-Zeldovich spectrum $ n_s = 1, \; r = 0 $ as a 
limiting value. This is a strong coupling regime $ y \to \infty $ where in addition
we must keep the ratio $ M/(h+1)^{\frac14} $ fixed for $ h \to -1^{+} $
since  it is determined by the amplitude of the adiabatic perturbations.
That is, we must keep
$$
{\bar M} \equiv  \frac{M}{(h+1)^{\frac14}} \quad {\rm fixed~~ as~~} h \to -1^{+} \; ,
$$
while $ M $ as well as $ m $ go to zero. Actually, the {\bf whole} potential $ V(\varphi) $
eq.(\ref{VC}) {\bf vanishes} in this limit since,
\bea\label{HZ2}
&&m = \frac{M^2}{M_{Pl}} \quad \buildrel{z=1, \; h \to -1^{+}}\over= \quad 
\frac{{\bar M}^2}{M_{Pl}} \; \sqrt{h+1} \to 0 \; , \cr \cr
&& g = h \; \sqrt{\frac{y}{2 \; N}} \; \left(\frac{M}{M_{Pl}}\right)^2  \quad
\buildrel{z=1, \; h \to -1^{+}}\over= \quad - \sqrt{\frac{\pi}{6 \; N}} \; 
\left(\frac{\bar M}{M_{Pl}}\right)^2 \; \left( \frac{1+h}2 \right)^{\frac14} \to 0 \; ,
\cr \cr
&& \lambda  = \frac{y}{8 \; N} \left( \frac{M}{M_{Pl}}\right)^4 
\quad \buildrel{z=1, \; h \to -1^{+}}\over= \quad 
\sqrt{\frac{\pi}{24 \; N}} \; \left(\frac{\bar M}{M_{Pl}}\right)^4 \; \sqrt{\frac{1+h}2}\to 0
\eea
The inflaton field is therefore a {\bf massless} free field at $ z=1 $ and $ h \to -1^{+} $.
This explains why the corresponding spectrum is the scale invariant Harrison-Zeldovich one.
This is clearly a singular limit since one cannot obtain any inflation from an
identically zero potential. Namely, taking the flat limit $ h \to -1^{+} $ in the
spectrum computed for $ z=1 , \; h > -1 $ with a fixed  number of efolds $ N $, yields
the scale invariant Harrison-Zeldovich spectrum.

Notice that here we keep the number of efolds $ N $ {\em fixed} which makes
the potential to vanish since otherwise the field will be stuck at the 
point  $ z=1 $ when $ h = -1 $ leading to eternal inflation ($ N = \infty $).

In summary, this shows {\bf theoretically} that the Harrison-Zeldovich spectrum 
which is a strong coupling limit $ y \to \infty $, is
{\it highly unplausible and unrealistic} since it appears only in the singular limit 
$ z=1 , \; h \to -1^{+} $ where the inflaton potential identically vanishes. 
Recall that one can also get a Harrison-Zeldovich spectrum letting formally
the number of efolds $ N $ to infinity in eqs.(\ref{indi}), that is letting inflation
to last eternally.

Moreover, the Harrison-Zeldovich spectrum $ n_s = 1, \; r = 0 $ is excluded by the
data \cite{WMAP5}. 

\subsubsection{The high asymmetry $ h < - 1 $ regime.}\label{hmemu}

In order to fulfill the finite number of efolds condition eq.(\ref{condw}) for $ h < -1 $
we have to add a constant term to the chaotic inflationary potential eq.(\ref{trinoC}).
We therefore consider as inflaton potential in the  $ h < -1 $ regime,
\be\label{trino2}
w(\chi) = \frac12 \; \chi^2 + \frac{h}3 \; \sqrt{\frac{y}2} \; \chi^3 +
\frac{y}{32} \; \chi^4 + \frac2{y} \; G(h)  \; ,
\ee
The absolute minimum of this potential is at
\be
\chi_{+} = \sqrt{\frac8{y}}\left[ -h + D \right] \quad , \quad  D \equiv \sqrt{h^2-1}
\; , \quad h < - 1  \; .
\ee
and we have
\be\label{gdeh}
G(h) \equiv \frac83 \, h^4 - 4 \, h^2 + 1 + \frac83 \, |h| \, D^3  \quad , \quad
w''(\chi_{+}) =2 \,  \sqrt{h^2-1}\left(\sqrt{h^2-1} + |h| \right) > 0 \; .
\ee
That is, the inflaton mass squared in units of $ m^2 $ takes the value
$$ 
2 \; D \; ( D + |h| ) \; .
$$
Here, the inflaton field rolls down  the slope
of the potential from its initial value larger than $ \chi_{+} $ till the absolute minimum of 
the potential at $ \chi = \chi_{+} $.

By inserting eq.(\ref{trino2}) for $ w(\chi) $ into  eq.(\ref{Nchi}) for $ N[\chi] $
and setting $ N[\chi] \equiv  N $ we obtain the field $ \chi $ or equivalently 
$ z  = y \; \chi^2 / 8 $ , 
at horizon exit, in terms of the coupling $ y $ and the asymmetry parameter $ h $:
\bea\label{2trino}
&&y = z + \frac43 \; h \; \sqrt{z} + 1 +\frac23 \; h \; (D-h) 
+ 2 \, G(h) \; \log\frac{\sqrt{z}}{D-h} +
\frac{16}3 \;  h \; (h^2-1) \; (D-h) \; \log\left(\frac{\sqrt{z}+h+D}{2 \, D}\right) \; .
\eea
This defines the field $ z $ as a monotonically increasing function of the coupling $ y $
for 
$$ 
0 < y  < +\infty \quad , \quad  z_+ = (D - h)^2 < z < +\infty \; .
$$
Recall that $ \chi $ and $ z $ corresponds to the time of horizon exit. 

We obtain from eqs.(\ref{ampliI}), (\ref{indi}) and (\ref{trino2}) the 
spectral index,  the ratio $ r $ and the amplitude of adiabatic 
perturbations,
\bea\label{nscao2}
&& n_s=1 - \frac{y}{N}  \frac1{(\sqrt{z}+ h- D)^2}
\left[ \frac{6 \, z \; (\sqrt{z}+ h + D)^2}{\left[z +2 \, (D + \frac{h}3 ) \; 
\sqrt{z} - \frac23 \; h \; (D-h)-1\right]^2}
- \frac{1 + 4 \, h \; \sqrt{z} + 3 \, z}{z +2 \, (D + \frac{h}3 ) \; \sqrt{z} - 
\frac23 \; h \; (D-h)-1}
\right]  \; , \cr \cr
&& r =   \frac{16 \; y}{N} \; \frac{z}{(\sqrt{z}+h - D)^2} \; 
\frac{(\sqrt{z}+ h + D)^2}{\left[z +2 \, (D + \frac{h}3 ) \; \sqrt{z} - 
\frac23 \; h \; (D-h)-1\right]^2} \; , \\ \cr
&&|{\Delta}_{k\;ad}^{\mathcal{R}}|^2  = \frac{N^2}{12 \, \pi^2} \; 
\left(\frac{M}{M_{Pl}}\right)^4 \;
\frac{\left[G(h) + 2 \, z+ \frac83  \, h  \,  z^{3/2} + 
z^2\right] \; (\sqrt{z}+h - D)^2 \; \left[z +2 \, (D + \frac{h}3 ) \; \sqrt{z} -
\frac23 \; h \; (D-h)-1\right]^2}{y^2 \; z \; (\sqrt{z}+ h + D)^2}   \; . \nonumber
\eea

When  $ \sqrt{z} \to \sqrt{z_+} , \; y $ vanishes quadratically and $ \chi $ tends
to infinity as 
$$
y \buildrel{z \to z_+}\over= 2 \; \left(\sqrt{z} - \sqrt{z_+}\right)^2 + {\cal O} 
\left(\left[\sqrt{z} - \sqrt{z_+}\right]^3\right) \quad , \quad 
\chi \buildrel{z \to z_+}\over= \frac{2 \; \sqrt{z_+}}{\sqrt{z} - \sqrt{z_+}} 
\to \infty \; .
$$
In this limit the spectral index, the ratio $ r $ and the amplitude of adiabatic 
perturbations become, 
$$
 n_s \buildrel{y \to 0}\over= 1 - \frac2{N} \quad , \quad r \buildrel{y \to 0}\over= 
\frac8{N} \quad , \quad 
|{\Delta}_{k\;ad}^{\mathcal{R}}|^2 \buildrel{y \to 0}\over= \frac{N^2}{6 \, \pi^2} \; 
\left(\frac{M}{M_{Pl}}\right)^4 \; .
$$
These results are {\bf independent} of the asymmetry $ h $
and coincide with those for the  purely quadratic monomial potential $ \frac12 \; \chi^2 $.

We see here that
\be\label{regir}
\frac8{N} < r < \frac{16}{N} \quad , \quad 1 - \frac3{N} < n_s < 1 - \frac2{N} 
\quad {\rm for}\quad  0 < y < \infty \; .
\ee
Namely, the regime $ h < - 1 $ of chaotic inflation covers values of $ r $ {\bf larger} than 
the weak coupling limiting value $ r = 8/N $ eq.(\ref{cuadh}) and {\bf smaller} than the 
$ y \to \infty $ pure quartic potential value $ r = 16/N $ eq.(\ref{cuar}).

\subsection{Trinomial New Inflation: Spectral index, amplitude ratio,
  running index and limiting cases}\label{tni}
We consider here new inflation described by the trinomial potential with broken symmetry 
investigated in ref.\cite{ciri,infwmap,mcmc}
\be\label{VN}
V(\varphi)= V_0 - \frac{m^2}2 \; \varphi^2 + \frac{ m \; g }3 \; \varphi^3 + 
\frac{\lambda}{4}\; \varphi^4 \; ,
\ee
where $ m^2 > 0 $ and $ g $ and $ \lambda $ are dimensionless couplings.
The corresponding dimensionless potential $ w(\chi) $ takes the form
\be\label{trino}
w(\chi) = -\frac12 \; \chi^2 + \frac{h}3 \; \sqrt{\frac{y}2} \; \chi^3 +
\frac{y}{32} \; \chi^4 + \frac2{y} \; F(h)  \; ,
\ee
where the quartic coupling $ y $ is dimensionless as well as 
the asymmetry parameter $ h $. The couplings in eq.(\ref{VN}) and 
eq.(\ref{trino}) are related by,
\be 
g = h \; \sqrt{\frac{y}{2 \; N}} \; \left( \frac{M}{M_{Pl}}\right)^2  
\qquad ,  \qquad 
\lambda  = \frac{y}{8 \; N} \; \left( \frac{M}{M_{Pl}}\right)^4 
\label{aco} \; ,
\ee 
and the constant $ w_0 $ [see eq.(\ref{wxi})] is related to $ V_0 $ in eq.(\ref{VN}) by
$$
 w_0 \equiv \frac2{y} \; F(h) = \frac{V_0}{N \; M^4 } \; .
$$
The constant $ F(h) $ ensures that $ w(\chi_+) =  w'(\chi_+) = 0 $ 
at the absolute minimum $ \chi = \chi_+ = (\Delta + |h|) \; \sqrt{8/y} $ 
of the potential $ w(\chi) $ according to eq.(\ref{condw}). 
Thus, inflation does not run eternally. $ F(h) $ is given by
$$
F(h) \equiv \frac83 \, h^4 + 4 \, h^2 + 1 + \frac83 \, |h| \, \Delta^3 
\quad , \quad \Delta \equiv \sqrt{h^2 + 1} \; .
$$
The parameter $ h $ reflects how asymmetric is the potential.
Notice that  $ w(\chi) $ is invariant under the changes
$ \chi \to - \chi , \;  h \to - h $. Hence, we can restrict
ourselves to a given sign for $ h $. Without loss of 
generality, we choose $ h < 0 $ and 
work with positive fields $ \chi $ as in sec. \ref{tci}..

We have near the absolute minimum $ \chi = \chi_+ $,
\be \label{cudr}
w(\chi) \buildrel{\chi \to \chi_+}\over= \Delta ( \Delta + |h| ) \; (\chi -\chi_+)^2 +
{\cal O}\left(\sqrt{y}[\chi - \chi_+]^3 \right)  \; ,
\ee
That is, the inflaton mass squared in units of $ m^2 $ takes the value
$$ 
2 \; \Delta ( \Delta + |h| ) \; .
$$
Notice that the  inflaton mass squared takes the analogous value for chaotic inflation in the
$ h <-1 $ regime changing $ D \Longrightarrow \Delta $
while $ F(h) $ differs from $ G(h) $ given by eq.(\ref{gdeh}) only by the
sign of the $ 4 \; h^2 $ term.

Recall that $ y \sim {\cal O}(1) \sim h $ guarantees that $ g  \sim 
{\cal O}(10^{-6}) $ and $ \lambda  \sim {\cal O}(10^{-12}) $ {\it without}
any fine tuning as stressed in sec. \ref{potuniv}  \cite{1sN}. 

\begin{figure}[ht]
\begin{turn}{-90}
\includegraphics[height=12cm,width=8cm]{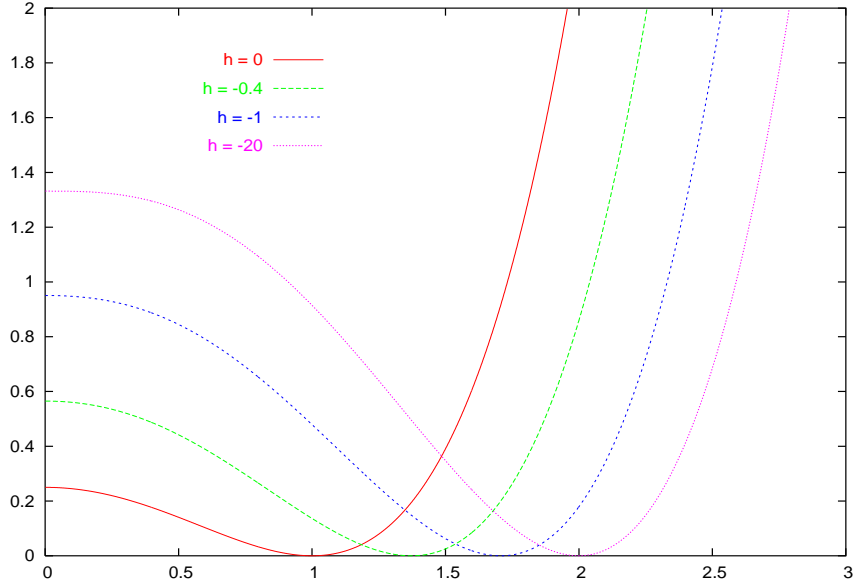}
\end{turn}
\caption{Trinomial New Inflation. The new inflation trinomial potential 
$ \frac{y \; w(\chi)}{8 \; (h^2 + 1)^2} $ [eq.(\ref{trino})] vs. the 
field variable $ \frac{\sqrt{y} \; \chi}{\sqrt8 \; \sqrt{h^2 + 1}} $ 
for $ \chi \geq 0 $ and different values of the asymmetry parameter 
$ h = 0, -0.4, -1, -20 $. We have normalized here the field variable and 
the potential by $h$-dependent factors in order to have a smooth 
$ |h| \gg 1 $ limit. }
\label{wh}
\end{figure}

In fig. \ref{wh}, we plot 
$$ 
\frac{y}{8 \; (h^2 + 1)^2}  \; w(\chi) 
\quad {\rm as ~ a ~ function ~ of} \quad  
\frac{\sqrt{y}}8  \; \frac{\chi}{\sqrt{h^2 + 1}}
$$ 
for several values of $ h < 0 $. We see that the position of the minimum of the potential 
$$ 
\frac{\sqrt{y} \; \chi_+}{\sqrt8 \; \sqrt{h^2 + 1}} = 1 + \frac{|h|}{\sqrt{h^2 + 1}} 
$$ 
grows as $ |h| $ grows. Similarly, the maximum of the potential at the origin
$$ 
\frac{y \; w(0)}{8 \; (h^2 + 1)^2} = \frac{F(h)}{4 \; (h^2 + 1)^2} 
$$ 
grows with  $ |h| $.

New inflation is obtained by choosing the initial field $ \chi $ in 
the interval $ (0,\chi_+) $. The inflaton $ \chi $ slowly rolls down the 
slope of the potential from its initial value till the absolute minimum 
of the potential $ \chi_+ $.

Computing the number of efolds from eq.(\ref{Nchi}), we find the field 
$ \chi $ at horizon crossing related to the coupling $ y $ and the 
asymmetry parameter $ h $. We obtain by inserting eq.(\ref{trino}) for 
$ w(\chi) $ into  eq.(\ref{Nchi}) and setting $ N[\chi] = N $,
\bea\label{ntrino}
&& y  = z - 2 \; h^2 -1 - 2  \; |h|  \; \Delta + \frac43 \; 
|h|  \; \left( |h| + \Delta - \sqrt{z} \right) + \cr \cr
&&+\frac{16}3 \; |h| \;  (\Delta + |h| ) \; \Delta^2  \; 
\log\left[\frac12 \left(1 + \frac{\sqrt{z} - |h|}{\Delta}\right)\right] - 
2 \, F(h) \, \log\left[\sqrt{z} \; (\Delta - |h|)\right] 
\quad , \quad z \equiv \frac{y}8 \; \chi^2 \; .
\eea
$ z $ turns to be a monotonically decreasing function of $ y $.
$ z $ decreases from  
\be\label{zmas}
z = z _+ = (\Delta + |h|)^2
\ee
till $ z = 0 $ when $ y $ increases from $ y = 0 $ till $ y = \infty $
while $ \chi $ decreases from $ \chi = \infty $ till $ \chi = 0 $.
When  $ \sqrt{z} \to \sqrt{z_+} , \; y $ vanishes quadratically as,
$$
y \buildrel{z \to z_+}\over= 2 \; \left(\sqrt{z} - \sqrt{z_+}\right)^2 + 
{\cal O} \left(\left[\sqrt{z} - \sqrt{z_+}\right]^3\right) \quad , \quad
\chi \buildrel{z \to z_+}\over= \frac{2 \; \sqrt{z_+}}{\sqrt{z} - 
\sqrt{z_+}} \to \infty \; .
$$
We obtain in analogous way from eqs.(\ref{ampliI}), (\ref{indi}), 
(\ref{run}) and (\ref{trino}) the spectral index, its running, the ratio 
$ r $ and the amplitude of adiabatic perturbations,
\bea\label{nstrino}
&& n_s=1 - 6 \,  \frac{y}{N} \, \frac{z \; (z + 2  \, h  \, 
\sqrt{z} -1)^2}{\left[F(h)  -2 \, z+ \frac83  \, h  \,  z^{3/2} +
z^2\right]^2} +  \frac{y}{N} \, \frac{ 3 \, z+ 4 \, h  \, \sqrt{z} -1}{
F(h)- 2 \, z+ \frac83  \, h  \,  z^{3/2} + z^2} \quad , \\ \cr\cr
\label{rtrino}
&&\frac{d n_s}{d \ln k}= - \frac2{N^2} \; \sqrt{z} \; y^2 \; 
\frac{(z + 2  \, h  \, \sqrt{z} -1)(h + \frac32 \; \sqrt{z})}{
\left[F(h)  -2 \, z+ \frac83  \, h  \,  z^{3/2} +z^2 \right]^2} 
- \frac{24}{N^2} \;  y^2 \; z^2 \; \frac{(z + 2  \, h  \, \sqrt{z} -1)^4}{
\left[F(h)  -2 \, z+ \frac83  \, h  \,  z^{3/2} +z^2 \right]^4} \cr\cr
&& + \frac8{N^2} \;  y^2 \; z \; \frac{ (3 \, z+ 4 \, h  \, \sqrt{z} -1)
(z + 2  \, h  \, \sqrt{z} -1)^2}{\left[F(h)  -2 \, z+ \frac83  \, h  \,  
z^{3/2} +z^2 \right]^3} \quad , \label{runnew} \\ \cr \cr
&& r = 16 \,   \frac{y}{N} \, \frac{z \; (z + 2  \, h  \, \sqrt{z} -1
)^2}{\left[F(h)  -2 \, z+ \frac83  \, h  \,  z^{3/2} +z^2
\right]^2}  \quad  , \label{rnue} \\ \cr\cr
&&|{\Delta}_{k\;ad}^{\mathcal{R}}|^2  = \frac{N^2}{12 \, \pi^2} \; 
\left(\frac{M}{M_{Pl}}\right)^4 \;
\frac{\left[F(h)- 2 \, z+ \frac83  \, h  \,  z^{3/2} + 
z^2\right]^3}{y^2 \; z \; (z + 2  \, h  \, \sqrt{z} -1)^2} \; .
\label{dtrino}
\eea

\subsubsection{The weak coupling limit $ y \to 0 $}

From eq.(\ref{ntrino}) we see that in the shallow limit 
$ y \to 0 , \; z $ tends to $ z_+ = (\Delta + |h|)^2 $.
$ y(z) $ has its minimum $ y = 0 $ at $ z =  z_+ $. We find from 
eqs.(\ref{ntrino})-(\ref{dtrino}), 
\bea\label{cotrino}
&& \chi \buildrel{y \to 0}\over= 2 \; \sqrt2 \; (\Delta + |h|) \; y^{-\frac12}
\to \infty \quad , \quad 
n_s \buildrel{y \to 0}\over= 1 - \frac2{N} \simeq 0.967 \quad , \quad  
r \buildrel{y \to 0}\over= \frac8{N}  \simeq 0.133 , \\ \cr
&& \frac{d n_s}{d \ln k} \buildrel{y \to 0}\over= 
-\frac2{N^2}\simeq -0.000556 \quad , \quad 
|{\Delta}_{k\;ad}^{\mathcal{R}}|^2  \buildrel{y \to 0}\over= 
\frac{N^2}{3 \, \pi^2} \; \left(\frac{M}{M_{Pl}}\right)^4 \; 
\Delta(\Delta+|h|) \; ,
\eea
which coincide with $ n_s , \; d n_s/d \ln k $ and $ r $ 
for the monomial quadratic potential. That is, the  $ y \to 0 $ limit
is $h$-independent except for $|{\Delta}_{k\;ad}^{\mathcal{R}}|$.
For fixed $ h $ and $ y \to 0 $ the inflaton potential eq.(\ref{trino}) 
becomes purely quadratic as we see from eq.(\ref{cudr}):
\be \label{maslim}
w(\chi) \buildrel{y \to 0}\over= \Delta ( \Delta + |h| ) \; (\chi -\chi_+)^2 +
{\cal O}(\sqrt{y} )  \; .
\ee
Notice that the amplitude of scalar adiabatic fluctuations eq.(\ref{cotrino})
turns out to be proportional to the square mass of the inflaton in this 
regime. We read this mass squared from eq.(\ref{maslim}): $ 2 \; \Delta ( \Delta + |h| ) $
in units of $ m^2 $. 
The shift of the inflaton field by $ \chi_+ $ has no observable consequences.
For $ h = 0 $ we recover the results of the monomial quadratic potential  
eq.(\ref{ycerocao}).

\subsubsection{The strong coupling limit $ y \to \infty $}

In the steep limit $ y \to \infty, \; z $ tends to zero for new inflation.
We find from eq.(\ref{ntrino})
\be\label{trikG}
y \buildrel{z \to 0}\over=- F(h) \; \log z -q(h) -1 + {\cal O}(\sqrt{z}) 
\to \infty \quad  ,\quad \chi \buildrel{y \to \infty}\over= 
\sqrt{\frac8{y}} \; \; e^{-\frac{y+1+q(h)}{2 \; F(h)}} \to 0 \; , 
\ee
where
$$
q(h) \equiv 2 \,  F(h) \log\left(\Delta-|h|\right) -
\frac23 \; \left( h^2 + |h| \; \Delta \right) 
\left\{ 8 \, \Delta^2 \, 
\log\left[\frac12 \left(1 - \frac{|h|}{\Delta}\right)\right] - 1 \right\}
\; ,
$$
$ q(h) $ is a monotonically increasing function of the asymmetry 
$ |h| : \; 0 \leq q(h) < \infty $ for $ 0 < |h| < \infty $.

Then, eqs.(\ref{nstrino})-(\ref{rtrino}) yield,
\bea\label{nsrtrikG}
&&n_s \; \; \buildrel{y \gg 1}\over=1 - \frac{y}{N \; F(h)}
\qquad , \qquad r \; \; \ \buildrel{y \gg 1}\over= 
\frac{16 \; y}{N \; F^2(h)} \; e^{- \frac{y+1+q(h)}{F(h)}}\quad ,
\cr \cr
&& \frac{d n_s}{d \ln k} \; \; \buildrel{y \gg 1}\over= -\frac{2 \; y^2 \; |h|}{N^2 \;
 F^2(h)} \, e^{- \frac{y+1+q(h)}{2 \; F(h)}}\quad , \cr \cr
&& |{\Delta}_{k\;ad}^{\mathcal{R}}|^2  \; \; \ \buildrel{y \gg 1}\over=  \; \; \
\frac{N^2}{12 \, \pi^2} \; \left(\frac{M}{M_{Pl}}\right)^4 \; 
\frac{F^3(h)}{y^2} \; e^{\frac{y+1+q(h)}{F(h)}} \quad .
\eea
In the case $ h = 0 $  we recover from the steep limit 
eqs.(\ref{trikG})-(\ref{nsrtrikG}) the results for new inflation 
in the large $ y $ regime: we have $ F(0) = 1 $ and $ q(0) = 0 $ and eq.(\ref{nsrtrikG}) 
becomes,
\bea\label{nsrtrikGh0}
&&n_s \buildrel{y \gg 1, h \to 0}\over=1 - \frac{y}{N}
\quad , \quad r \buildrel{y \gg 1, h \to 0}\over= 
\frac{16 \; y}{N} \, e^{-y-1}\quad ,
\\ \cr
&& \frac{d n_s}{d \ln k}\buildrel{y \gg 1, h \to 0}\over= 
-\frac{2 \; y^2 \; |h|}{N^2} \, e^{-y-1} \quad , \quad
|{\Delta}_{k\;ad}^{\mathcal{R}}|^2  \buildrel{y \gg 1, h \to 0}\over= 
\frac{N^2}{12 \, \pi^2} \; \left(\frac{M}{M_{Pl}}\right)^4 \; 
\frac{e^{y+1}}{y^2} \;  \quad .
\eea
Only the behaviour of the running is non-uniform in $ h $ for
$ y \to \infty $ as we see comparing with eq.(\ref{binonewyG}).

Notice that the slow-roll approximation is valid only when the 
coefficient of $ 1/N $ is smaller than $ N $, and so for $ y \lesssim N $.
For $ y \to \infty , \; n_s $ is well below one and $ r $ is very small.
This strong coupling limit is excluded since the WMAP+LSS data rule out
$ n_s \lesssim 0.9 $. 

\subsubsection{The extremely asymmetric limit $ |h| \to \infty $}

Eqs.(\ref{ntrino})-(\ref{dtrino}) have a finite limit for $ |h| \to \infty 
$ with $ y $ and $ z $ scaling as $ h^2 $. Define,
$$
Z \equiv \frac{z}{h^2} \quad , \quad Y \equiv \frac{y}{h^2} \; .
$$
We have $ 0 \leq Z \leq 4 $ for $ +\infty \geq Y \geq 0 $. 
Then, we find for $ |h| \to \infty $ 
from eqs.(\ref{ntrino})-(\ref{dtrino}) keeping $ Z $ and $ Y $ fixed,
\bea \label{hgra}
&& Y = Z - \frac43 \; \sqrt{Z} - 4 - \frac43 \; \log\frac{Z}4 +
\frac{16}{3 \; \sqrt{Z}}\; , \cr \cr
&& n_s = 1 - 6 \; \frac{Y}{N} \; \frac{Z^2 \; (\sqrt{Z} - 2)^2}{
[\frac{16}3 - \frac83 \; Z^{\frac32} + Z^2 ]^2 } + \frac{Y}{N} \;
\frac{3 \; Z - 4 \;\sqrt{Z} }{\frac{16}3 - \frac83 \; Z^{\frac32} + Z^2 }
\; , \\ \cr
&&\frac{d n_s}{d \ln k}= - \frac2{N^2} \;  Y^2 \; Z \; 
\frac{(\sqrt{Z} -2)(\frac32 \; \sqrt{Z}-1)}{
\left[\frac{16}3  - \frac83 \,  Z^{3/2} + Z^2 \right]^2} \cr\cr
&&- \frac{24}{N^2} \;  Y^2 \; Z^4 \; \frac{(\sqrt{Z} -2)^4}{
\left[\frac{16}3  - \frac83 \,  Z^{3/2} + Z^2 \right]^4} 
+ \frac8{N^2} \;  Y^2 \; Z^{\frac52} \; 
\frac{(3 \,\sqrt{Z} - 4)(\sqrt{Z} -2)^2}{
\left[\frac{16}3  - \frac83 \,  Z^{3/2} + Z^2 \right]^3}
\quad , \label{runh} \\ \cr \cr
&& r = 16 \; \frac{Y}{N} \; \frac{Z^2 \; (\sqrt{Z} - 2)^2}{
[\frac{16}3 - \frac83 \; Z^{\frac32} + Z^2 ]^2 } \quad , \quad 
|{\Delta}_{k\;ad}^{\mathcal{R}}|^2  = \frac{N^2 \; h^2}{12 \, \pi^2} \; 
\left(\frac{M}{M_{Pl}}\right)^4 \;
\frac{[\frac{16}3  - \frac83 \; Z^{\frac32} + Z^2]^2}{Y^2 \; Z^2  
\; (\sqrt{Z} - 2)^2} \; . \label{amplihg}
\eea
In the  $ |h| \to \infty $ limit, the inflaton potential takes the form
$$
W(\chi) \equiv \lim_{|h| \to \infty} \frac{w(\chi)}{h^2} =
\frac{32}{3 \; Y} - \frac13 \; \sqrt{\frac{Y}2} \; \chi^3 
+ \frac{Y}{32} \; \chi^4 \; .
$$
This is an asymmetric potential without quadratic term.
Notice that the cubic coupling has dimension of a mass in eq.(\ref{VN})
and hence this is {\bf not} a massless potential contrary to the quartic 
monomial $ \chi^4 $. 

In addition, eq.(\ref{amplihg}) shows that for large 
$ |h| $ one must keep the product $ |h| \; M^2 $ fixed
since this is determined by the amplitude of the adiabatic perturbations.
We see from eq.(\ref{amplihg}) that  in the $ |h| \to \infty $ limit, 
$ {\tilde M} \equiv \sqrt{|h|} \; M $ becomes the energy scale of 
inflation. $ {\tilde M} \sim 10^{16}$GeV  [eq.(\ref{valorM})] according to
the observed value of $ |{\Delta}_{k\;ad}^{\mathcal{R}}|/N $ displayed in 
eq.(\ref{Mwmap}), while $ M $ and $ m $ vanish as $ |h| \to \infty $,
$$
M = \frac{\tilde M}{\sqrt{|h|}} \buildrel{ |h| \to \infty }\over= 0 
\quad ,  \quad
m = \frac{M^2}{M_{Pl}} = \frac{ {\tilde M}^2}{|h| \; M_{Pl}} 
\buildrel{ |h| \to \infty}\over= 0 \quad .
$$
The MCMC analysis of the CMB+LSS data (see sec. \ref{mcmc:trino}) excludes 
large $ |h| $ in new inflation as shown in fig. \ref{z1hnsr}.

\subsubsection{Regions of $n_s$ and $r$ covered by New Inflation and by 
Chaotic Inflation.}\label{regions}

It follows from eqs.(\ref{rtrino}), (\ref{cotrino}) and (\ref{nsrtrikG}) 
that new inflation for $ h \leq 0 $ covers the narrow {\bf banana-shaped} 
sector between the black lines in the $(n_s,r)$ plane depicted in fig. 
\ref{ene}. We have in this region:
\be\label{cotsup}
0 < r < \frac8{N}  \quad , \quad n_s < 1 - \frac{1.9236 \ldots}{N} 
\quad {\rm for}\quad  \infty > y > 0 \; .
\ee
Chaotic inflation in the $ h < - 1 $ region covers the even narrower
{\bf complementary} strip eq.(\ref{regir}) for $ 8/N < r < 16/N $.
The zero coupling point $ y \to 0 \; , \; r = 8/N
\; , \; n_s = 1 -2/N $ being the border between the two regimes.

In summary, we see from fig. \ref{ene} that 
the regions of the trinomial new inflation and chaotic inflation 
are {\bf complementary} in the $(n_s,r)$ plane. New inflation describes the region of the 
$(n_s,r)$ plane between the two black lines while chaotic inflation describes the whole plane 
to the right of the rightmost black line.

\medskip

As we show in sec. \ref{mcmc} below, chaotic inflation for $ -1 < h \leq 0 $ 
covers a wide region depicted by 
the red lines in the  $(n_s,r)$ plane (see fig. \ref{ene}). However, as 
shown by fig. \ref{pdfzh}, this wide area is only a small corner of the 
field $ z $ - asymmetry $ h $ plane.
Since new inflation covers the banana-shaped region between the black 
curves, we see from fig. \ref{ene} that the most probable values of $ r $ are
{\bf definitely non-zero} within trinomial new inflation. Precise lower 
bounds for $ r $ are derived from MCMC in sec. \ref{mcmc} below.

\subsection{The Monte Carlo Markov Chain Method of Data Analysis}\label{mcmc}

Given a sufficiently rich set of empirical data and a theoretical model
with several free parameters that should describe these data,
one faces the problem of how to efficiently determine reliable bounds for
those free parameters. More precisely, one needs to reconstruct the
experimental probability distribution for the actual values of those
parameters by means of (some form of) statistical inference. 

The Monte Carlo Markov Chain (MCMC) method is a very efficient stochastic
numerical method to carry out such a reconstruction, which could be quite
difficult when the number of free parameters is not very small and the
probability distribution to be reconstructed has a complicated profile.

In the case under consideration, the theoretical model is the standard
cosmological model as presented in the previous
sections, while the set of empirical data is that listed in the
introduction [sec. \ref{uno}]. Among these, the CMB and LSS observations 
play a distinguished r\^ole for their richness and precision.

A key step in any inference approach is the determination of the
probability that a certain model, given certain values of its free
parameters, will predict specific observational data. This requires some
type of assumptions on the experimental uncertainties. Namely, after having
assessed as much as possible the systematic errors, one needs to adopt a
model for the statistical errors. This is especially important in the case
of CMB anisotropies which are seeded by primordial classicalized quantum
fluctuations which are intrinsically stochastic. In
sec.~\ref{gistf} these have been described as free--field
gaussian fluctuations, as a direct consequence of the linearized approach.
Primordial non-gaussianities and/or non-gaussianities induced by the
evolution from decoupling till now are of course allowed in principle but
not testable by present experimental accuracy, furthermore their
discussion and treatment is beyond the scope of this review article.

The primordial fluctuations, even if assumed gaussian, are nonetheless
quite different from statistical experimental errors, which are also
usually assumed to be a normally-distributed noise. In fact, the latter 
may in principle be reduced (or better, the statistical estimator may be 
improved) by increasing the number of observations. On the contrary, 
there is only one realization for primordial fluctuations which corresponds to
our observable universe. This is the source of the so-called cosmic 
variance.

To be specific, let us consider an ideal experiment capable of providing a
complete and noiseless sky map of the CMB temperature anisotropies. Any
such map is a function $ t=t(\bds n) $ yielding the temperature 
fluctuation in the direction $ \bds n $ with respect to the full sky mean 
value. According to the above assumption of gaussianity, these maps form a
normal ensemble with zero average and rotational-invariant covariance 
$ \avg{t(\bds n)t(\bds n')} $, that is 
\begin{equation*}
 \avg{t(\bds n)} = 0 \quad , \quad
  \avg{t(\bds n) \, t(\bds n')}=C(\bds n \cdot \bds n') \; .  
\end{equation*}
In a spherical harmonic decomposition we can then write (notice that the
monopole term $\ell=0$ vanishes by definition)
\begin{equation*}
  t(\bds n) = \sum_{\ell=1}^\infty\sum_{m=-\ell}^{m=\ell}a_{\ell m} 
  \, Y_{\ell m} (\bds n) \; ,
\end{equation*}
with
\begin{equation*}
 \avg{a_{\ell m}} = \avg{a^\ast_{\ell m}} = 0 \quad , \quad
  \avg{a_{\ell m} \, a^\ast_{\ell' m'}} = \delta_{\ell \ell'} \, \delta_{mm'} \; C_{\ell}
\end{equation*}
and
\begin{equation*}
C(\bds n \cdot \bds n') = \frac1{4\pi}\sum_{\ell} (2\,\ell+1)\, C_{\ell}\,
   P_l(\bds n \cdot \bds n') \; .
\end{equation*}
Thus, the multipoles $ C_{\ell} $ forming the so-called angular power
spectrum, fully characterize the statistical properties of any map 
$ t(\bds n) $. In turns out that the $ C_{\ell} $ are computable from the 
primordial gaussian fluctuations through their cosmological evolution from 
decoupling till the present age. Hence, they depend on the collection of 
cosmological parameters $ \lambda=\{\lambda_1,\lambda_2,\ldots\lambda_n\} 
$ , that is the free parameters in the cosmological model of choice. 

We considered in this example the best measured Temperature--Temperature 
(TT) correlation.  The same analysis can be repeated also for the other 
correlations, such as the Temperature--E\_modes (TE), the 
E\_modes--E\_modes (EE) and the 
B\_modes--B\_modes (BB) correlation multipoles (E\_modes and B\_modes are
special modes of the CMB polarization).

Apart from experimental errors, here and now in our universe we only
observe one specific map $ {\bar t}(\bds n) $, so that we can only try and
infer the value of the parameters $ \lambda $. Bayesian inference is based
on the notion of conditional probability\footnote{In abstract terms, if 
$ A $ and $ B $ are two possible events and with $ A\cap B $ we denote the
joint event, then
  \begin{equation*}
    \mathrm{Pr}(A\cap B) = \mathrm{Pr}(B\cap A) = 
\mathrm{Pr}(A|B) \; \mathrm{Pr}(B) = \mathrm{Pr}(B|A) \; \mathrm{Pr}(A)
  \end{equation*}
defines $ \mathrm{Pr}(A|B) $ and $ \mathrm{Pr}(B|A) $. Thus, since by 
definition $ \mathrm{Pr}(B)=\sum_A\mathrm{Pr}(A\cap B) $ if $ \{ A \} $
is a complete set of mutually exclusive events, we obtain Bayes' formula
\begin{equation*}
\mathrm{Pr}(A|B) = \frac{L(A|B)}{\sum_{A'}L(A'|B)}
\end{equation*} 
where $ L(A|B) \equiv \mathrm{Pr}(B|A)\,\mathrm{Pr}(A) $ is 
the {\em likelihood} of $ A $ given the observation of $ B $.}.  
The likelihood $ L $ of $ \lambda $, given the experimental data $ 
{\bar t}(\bds n) $, is related to the probability
of $ {\bar t}(\bds n) $ given  certain value of the parameters $ 
\lambda $ in the following way:
\begin{equation}\label{eq:bayesan}
  L(\lambda\,|\,{\bar t}\,) \propto  \mathrm{Pr}({\bar t}\,|\,
  C(\lambda))\, p(\lambda)
\end{equation}
where
\begin{equation}\label{eq:gaussian}
    \mathrm{Pr}(t\,|\,C) = [\mathrm{Det}(2\pi\,C)]^{-1/2}
    \exp\left[-\frac12 \int d^2n \int d^2n' \;t(\bds n)\,
      (C^{-1})(\bds n \cdot \bds n')\,t(\bds n')\right]
\end{equation}
is the Gaussian distribution for all possible maps for a given covariance
$ C(\bds n \cdot \bds n') $ and $ p(\lambda) $ is the prior probability on
the parameters, which collects any a priori bias on their actual values. 
In terms of the spherical amplitudes $ {\bar a}_{\ell m} $ corresponding 
to the observed map $ {\bar t}(\bds n) $ and of the multipoles 
$ C_{\ell} $ we have
\begin{equation*}
  \mathrm{Pr}({\bar t}\,|\,C) = \exp\left\{-\frac12 \sum_{\ell m}
    \left[\frac{|{\bar a}_{\ell m}|^2}{C_{\ell}} + \log(2\pi\,C_{\ell})
    \right] \right\}= \prod_{\ell}\left[(2\pi\,C_{\ell})^{-1/2}\exp\left(
      -\frac{{\bar C}_{\ell}}{2\,C_{\ell}} \right)\right]^{(2\ell+1)}
\end{equation*}
where
\begin{equation*}
{\bar C}_{\ell}=\frac1{2\,\ell+1}
\sum_{m=-\ell}^{m=\ell}|{\bar a}_{\ell m}|^2 
\end{equation*}
is the most likely value of $ C_{\ell} $, i. e. the value at which
$ \mathrm{Pr}({\bar t}\,|\,C) $ takes its maximum value as a function of 
the $ C_{\ell} $. The $ {\bar C}_{\ell} $ are the so-called pseudo-$ C_{\ell} $.
Notice that in this ideal example $ \mathrm{Pr}({\bar
t}\,|\,C) $ assigns independent weights to the multipoles $ C_{\ell} $; 
that is, the $ C_{\ell} $'s are correlated only by their dependence on the
cosmological parameters.

Finally, properly normalizing $ L(\lambda\,|\,{\bar t}) $ with respect to
$ \lambda $ yields the so--called posterior probability, which is just 
the sought experimental probability for the value $ \lambda $. 

The MCMC method reconstructs the profile of $L(\lambda\,|\,{\bar t})$
through the ergodic properties of sequences
$ \{\lambda^{(0)},\lambda^{(1)},\ldots,\lambda^{(k)},\ldots\} $ 
(the chains),
that start from points extracted with the prior probability and evolve
through a suitable acceptance/rejection one-step algorithm such as the
Metropolis rule:
\begin{equation*}
  W(\lambda^{(k+1)},\lambda^{(k)}) = g(\lambda^{(k+1)},\lambda^{(k)})
  \;{\rm min}\,\left\{1\,,\; \frac{L(\lambda^{(k+1)}\,|\,{\bar t}\,)\,
      g(\lambda^{(k+1)},\lambda^{(k)})}{L(\lambda^{(k)}\,|\,{\bar t}\,)\,
      g(\lambda^{(k)},\lambda^{(k+1)})} \right\}
\end{equation*}
Here $ W(\lambda,\lambda') $ is the actual one-step transition 
probability, while $ g(\lambda,\lambda') $ is an implementation-dependent 
proposal probability. The ergodic (for good proposal probabilities) and 
detailed balance properties of this $ W(\lambda,\lambda') $ guarantee that
very long chains will contain $ \lambda $ values distributed according to 
the posterior probability. In this case, the specific choice of $ 
g(\lambda,\lambda') $ will only affect the convergence rate of the 
process.

Of course, in a true experiment, such as WMAP, the observed sky map 
$ {\bar t}(\bds n) $ suffers from several errors and approximations, such 
as those introduced by the experimental noise, by the necessary map 
discretization (or {\em pixelization}) due to the finite resolving power 
or by the effects of removing various types of contamination sources in 
the sky. This complicates considerably any reliable calculation of that 
part of the likelihood (the Gaussian $ \mathrm{Pr}({\bar t}\,|\,C) $ in 
the ideal example above) which depends on $ \lambda $ only through the 
$ C_{\ell} $ multipoles. Moreover, the experimental likelihood weakly 
correlates the $ C_{\ell} $'s. At any rate no substantial change is 
implied in the MCMC application, since the experimental likelihood of the 
$ C_{\ell} $ has no relation at all with their dependence on the 
cosmological parameters.  In practice, a MCMC determination of the 
posterior probability for $ \lambda $ uses a likelihood code for the 
$ C_{\ell} $ multipoles provided by the experiment itself, such as WMAP, 
and some numerical program such as CMBFAST
or CAMB to compute the theoretical $ C_{\ell}(\lambda) $ within a given
cosmological model as a function of $ \lambda $. 
The publicly available open--source CosmoMC program
\cite{lewis} is a very useful and widely used package that integrates all
the necessary aspects of MCMC data analysis for cosmology.
 
\subsubsection{CMB, LSS and Inflation within a MCMC 
analysis.}\label{anamcmc}

Let us describe now the results of an accurate MCMC analysis of CMB and 
LSS data with the CosmoMC programme in which we have included the 
Trinomial New Inflation (TNI) and Trinomial Chaotic Inflation (TCI) 
models \cite{mcmc}.

The experimental data were those of CMB and LSS. In ref.~\cite{mcmc} the 
CMB data were the three years WMAP data, which contain also polarization 
maps and provide the dominating contribution, and also the small scale 
data (ACBAR, CBI2, BOOMERANG03). For LSS the SDSS data (release 4) were 
considered. In this MCMC analysis we neither marginalized over the the 
Sunayev-Zel'dovich amplitude nor included non-linear effects in the 
evolution of the matter spectrum. In any case, the relative corrections are 
not significant \cite{WMAP3}, especially in the present context. While 
ref.\cite{mcmc} was in the publication stage the five years WMAP data were
released \cite{WMAP5}. Also a new comprehensive compilation
of supernovae observations was published \cite{SN}.  We shall verify below 
that these new data do not change in any relevant way the overall picture 
drawn in ref.~\cite{mcmc}.

\begin{figure}
\includegraphics{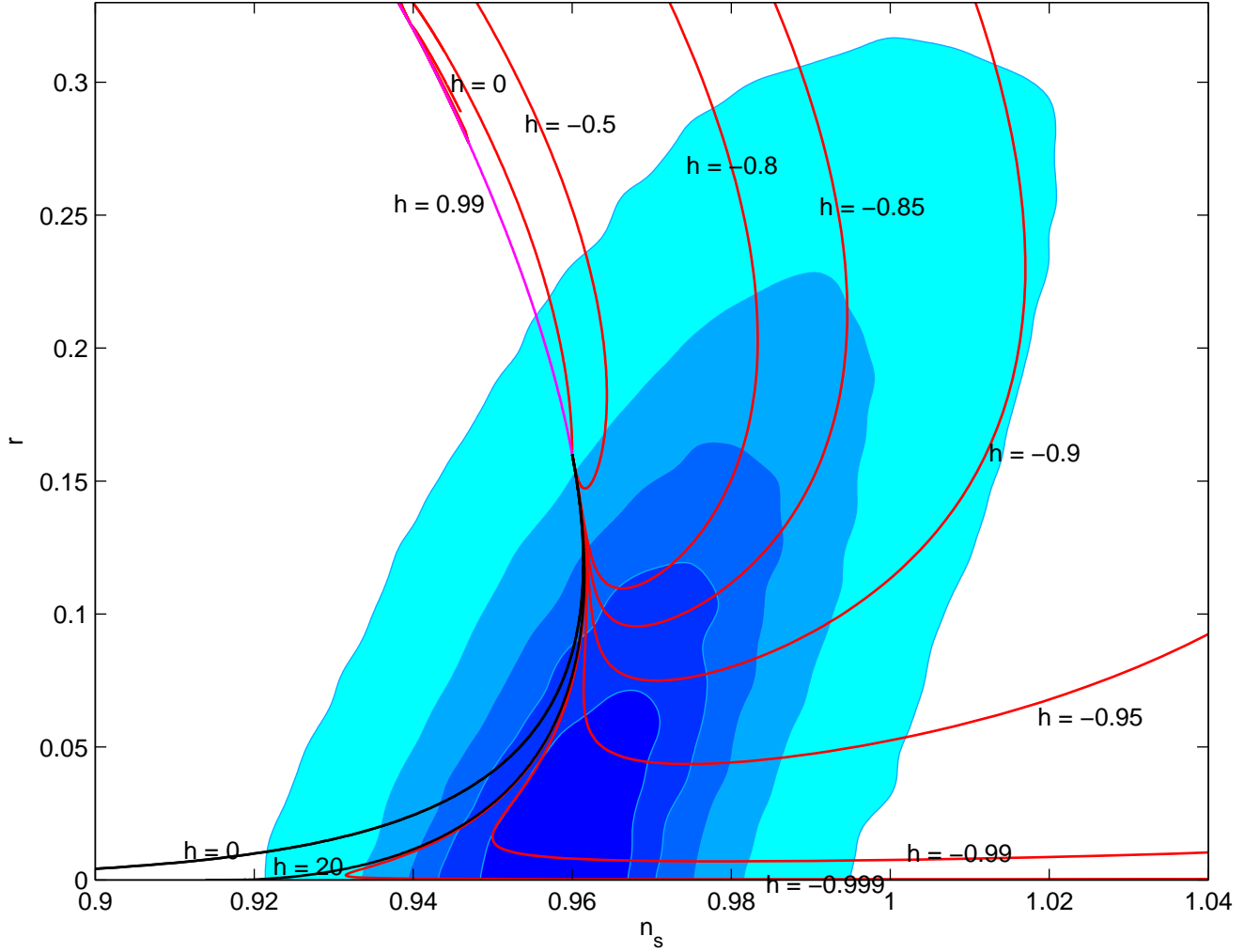}
\caption{Trinomial Inflation. We plot $ r $ vs. $ n_s $ for fixed values of
the asymmetry parameter $ h $ and the field $ z $ varying
along the curves. The red curves are those of chaotic inflation with
$ h\le 0 $ (only the short magenta curve has positive $ h $), while the black
curves are for new inflation. The color--filled areas correspond to $ 12\%, \; 
27\%, \;  45\%, \;  68 \% $ and $ 95 \% $ confidence levels according to the WMAP3 and
Sloan data. Red and black curves are drawn for $ N = 50 $. A slightly better agreement
with the data arises for $ N = 60 $.
The color of the areas goes from the darker to the lighter for increasing CL.
New inflation only covers a narrow area between the black lines while
chaotic inflation covers a much wider area but, as shown by fig. \ref{pdfzh},
this wide area is only a small corner of the field $ z $ - asymmetry $ h $ plane.
Since new inflation covers the banana-shaped region between the black curves, we see from
this figure that the most probable values of $ r $ are {\bf definitely non-zero} within
trinomial new inflation. Precise
lower bounds for $ r $ are derived from MCMC in eq.(\ref{cotinfr}).}
\label{ene}
\end{figure}

As discussed above, in a MCMC analysis one must also choose the prior
probability for the cosmological parameters. This implies restricting the 
search to a subset of the full parameter space by imposing so--called hard 
constraints, whose number and type depend on previously acquired 
information both experimental and theoretical. Since the central 
cosmological parameters in this analysis are the spectral index $ n_s $ of 
the adiabatic fluctuations and the ratio $ r $ of tensor to scalar 
fluctuations, we impose as hard constraint that $ n_s $ and $ r $ are 
restricted to the theoretical regions in the $ (n_s, \; r) $ plane 
described by TNI and TCI, respectively, as discussed in
subsection \ref{regions}. Our analysis differs in this {\bf crucial} 
aspect from previous MCMC studies involving the WMAP3 data \cite{otros}. 
As natural within slow-roll inflation, we also include the inflationary 
consistency relation $ n_T = -r/8 $ on the tensor spectral index. This 
constraint is in any case practically negligible.

Altogether we allow seven cosmological parameters to vary in our MCMC runs:
the baryonic matter fraction $ \omega_b $, the dark matter fraction $ 
\omega_c $, the optical depth $ \tau $, the ratio of the (approximate) 
sound horizon to the angular diameter distance $ \theta $, the primordial 
superhorizon power in the curvature perturbation at $ 0.05~{\rm Mpc}^{-1} 
, \; A_s $, the scalar spectral index $ n_s $ and the tensor-scalar ratio 
$ r $. The last three of these parameters characterize the primordial power
spectrum, while the first four affect the formation, evolution and 
propagation of the CMB after the reentering of superhorizon fluctuations.  
The inflationary cosmological models thus defined can be briefly identified
as $\Lambda$CDM+TNI model and $\Lambda$CDM+TCI model, according to the 
TNI potential eq.~\eqref{VN} and the TCI potential eq.~\eqref{VC}, 
respectively.

For comparison, we report also the results of a MCMC study within the 
standard $\Lambda$CDM model augmented by the tensor-scalar ratio $ r $ 
(the $\Lambda$CDM+$r $ model). That is, in this case we treated $ n_s $ and
$ r $ as unconstrained Monte Carlo parameters, using standard priors. The 
analysis with the $\Lambda$CDM+$r $ model is indeed by now quite standard 
and good priors are available already in CosmoMC.

\medskip

In the case of TNI, since the characteristic banana--shaped
allowed region in the $(n_s,r)$ plane is quite narrow and non--trivial, it
is convenient to use as MC parameters the two independent variables $ z $
and $ h $ of the trinomial inflationary setup. That is, $ n_s $ and $ r $
are parametrized in terms of $ z $ and $ h $ by the the analytic
expressions at order $ 1/N $, eqs.(\ref{nstrino}) and (\ref{rnue}). To be
more precise, rather than $ z $ we use the appropriate normalized variable
\begin{equation} \label{eq:z1}
  z_1 = 1 - \frac{z}{z_+} = 1 - \frac{z}{\big(\sqrt{h^2 + 1} + |h|\big)^2} 
\end{equation}
We recall that $ z $ contains the field at horizon crossing and the
coupling $ y $.  $ z_+ $ stands for $ z $ at the absolute minimum of the
potential.  The variable $ z_1 $ grows monotonically from $0$ to $1$ as the
coupling $y$ grows from $0$ to $\infty$ [see eq.~(\ref{ntrino}) and 
(\ref{zmas})].

Concerning priors, we keep the same, standard ones, of the $\Lambda$CDM+$r$
model for the first five parameters ($ \omega_b, \; \omega_c, \; \tau, \; 
\theta $ and $ A_s $), while we consider all the possibilities for $ z_1 $ 
and $ h $, that is $ 0<z_1<1 , \; 0<|h|<\infty $. Moreover, in order to 
correctly compare the results with those of the $\Lambda$CDM+$r$ model,
in which a flat prior distribution on $ r $ is indeed assumed,
we reweight the statistics to convert to a flat prior distribution on $ r $
(this is feasible since the relation between $ r $ and $ z_1 $ is monotonic
for any $ h $).

\medskip 

In the case of the $\Lambda$CDM+TCI model, it is more convenient to keep 
$ n_s $ and $ r $ as MC parameters and impose as hard priors that they lay 
in the region described by TCI (see fig.~\ref{ene}). This is because this 
region covers the major part of the probability support of $ n_s $ and 
$ r $ in the $\Lambda$CDM+$r$ model and the parametrization 
eqs.(\ref{nscao})-(\ref{rcao}) in terms of the coupling
parameters $ z $ and $ h $ becomes quite singular in the limit $ h \to -1
$. This is indeed the limit which allows to cover the region of highest
likelihood.

The distributions for the field variable $ z $ and the asymmetry parameter 
$ h $ are then recovered from the $ (n_s, \; r) $ distribution by a 
numerical change of variables, starting from an uniform two dimensional 
grid in the $z-h$ plane and using eqs.(\ref{nscao}) and (\ref{rcao}). This 
requires a rather accurate determination of the $ (n_s, \; r) $ 
distribution, which is obtained by running very long parallel chains (with 
a total number of samples close to five million for the results presented 
below). Quite naturally in this approach for TCI, the most likely values of
the cosmological parameters and the corresponding maximum of
the likelihood coincide to those of the $ \Lambda$CDM+$r $ model.

\medskip 

In all our MCMC runs we kept fixed the number of efolds $ N $ since the 
horizon exit of the pivot scale $ k_0 $ till the end of inflation. The 
reason is that the main physics that determines the value of $ N $ is 
{\bf not} contained in the available data but involves the reheating era. 
Therefore, although technically possible, it is not reliable to fit $ N $ 
with the CMB+LSS data solely within a pure, near scale-invariant slow 
roll scenario. Anyway, the precise value of $ N $ is certainly near $ N =60
$ \cite{kt,libros,mass}.
 
In this section we report the results of ref.\cite{mcmc} where the value 
$ N=50 $ was chosen as a reference baseline value for numerical analysis. 
Anyway, from eqs.(\ref{indi}) and (\ref{resca}) we see that both 
$ n_s - 1 $ and $ r $ scale as $ 1/N $.

\begin{figure}[ht]
\includegraphics[width=14cm]{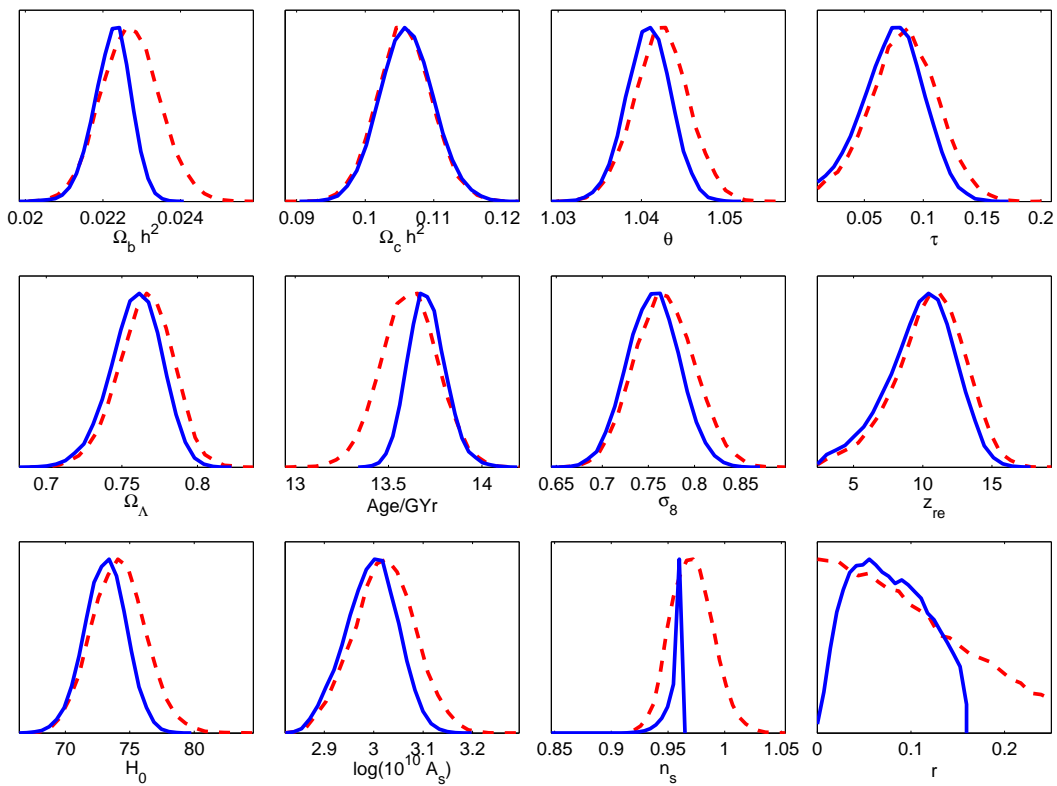}
\caption{Comparison of the marginalized probability distributions 
(normalized to have maximum equal to one) of the most relevant cosmological
parameters (both primary and derived) between the model $\Lambda$CDM+TNI 
(Trinomial New Inflation, solid blue curves) and the $\Lambda$CDM+$r $ 
model (dashed red curves). Here $ N=50 $ and WMAP3, small-scale CMB and 
SDSS data were used. }
\label{new_vs_std}
\end{figure}

\begin{figure}[ht]
\includegraphics[width=14cm]{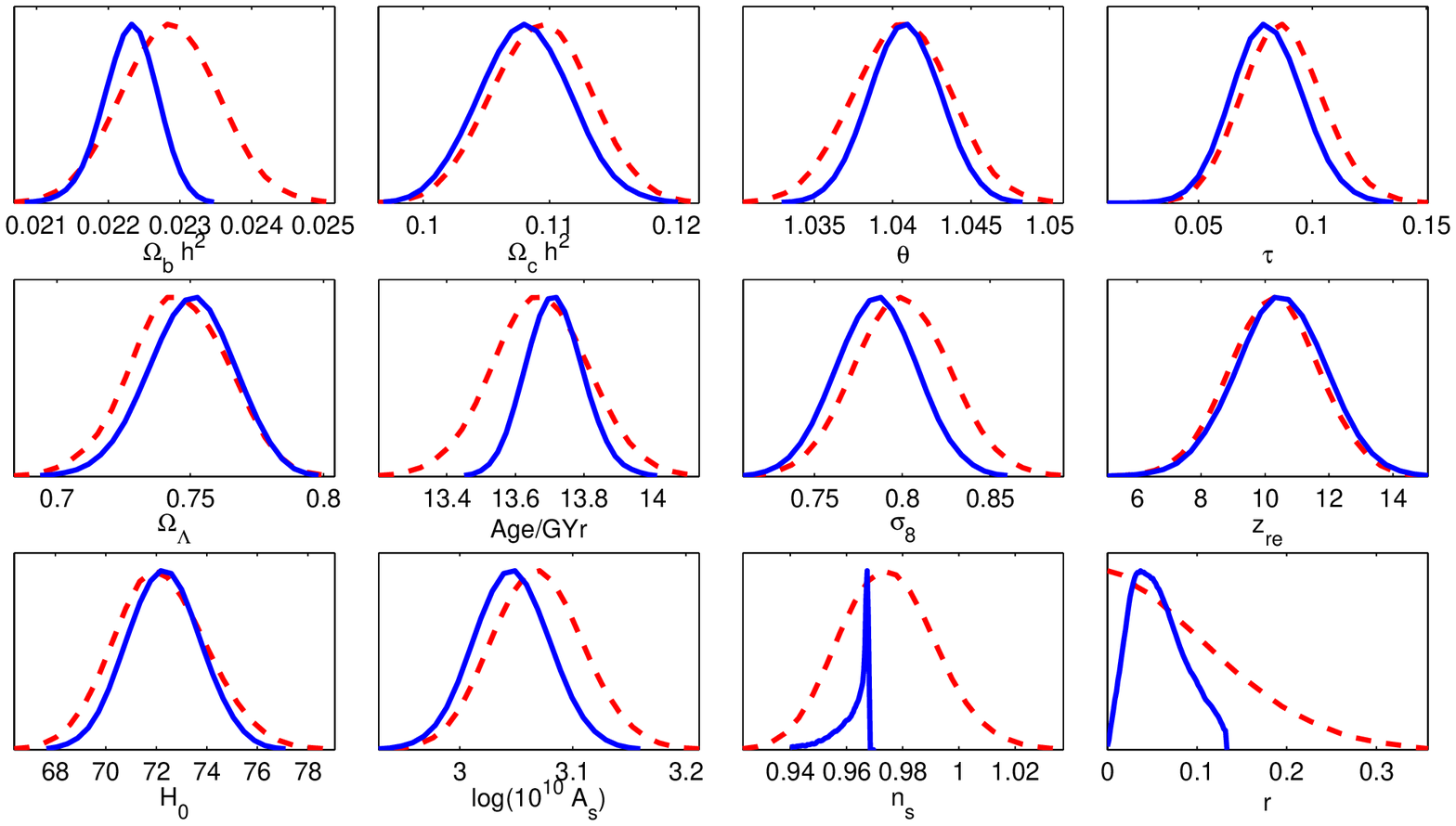}
\caption{Comparison of the marginalized probability distributions 
(normalized to have maximum equal to one) of the most relevant cosmological
parameters (both primary and derived) between $\Lambda$CDM+TNI (solid blue 
curves) and the $\Lambda$CDM+$r $ model (dashed red curves). Here $ N=60 $ 
and WMAP5, SN and SDSS data were used.}
\label{new_vs_std5}
\end{figure}

Therefore, decreasing or increasing $ N $ produces a scale transformation in
the $(n_s-1,r)$ plane, thus displacing the black and red curves in fig. 
\ref{ene} towards up and left or towards down and right. This produces, 
however, small quantitative changes in our bounds for $ r $ as well as in 
the most probable values for $ r $ and $ n_s $. We explicitly verified this
statement by performing also a MCMC analysis using the last WMAP5 release 
and setting $ N=60 $. The corresponding results are reported here for the 
first time. In summary, varying $ N $ from 50 to 60 has only minor, 
practically irrelevant effects in the MCMC fits we present
in this review. This is partially due to the fact that in the theoretical
formulas for $ n_s $ and $ r $, eqs.(\ref{nscao}), (\ref{cao}), 
(\ref{nstrino}) and (\ref{rtrino}), a change on $ N $ can be partially
compensated by a change on $ y $. The detailed analysis with variable $ N $
is at any rate beyond the scope of the present review.

Another {\bf hint} to increase $ N $ above 50 comes from the WMAP5 data 
that gives a slightly larger $ n_s $ value and using the theoretical upper 
limit for $ n_s $ eq.(\ref{cotsup}), which gives $ n_s < 0.9679 \ldots $ 
for $ N = 60 $. This bound on $ n_s $ is compatible with the $ n_s $ value from 
WMAP5+BAO+SN and no running \cite{WMAP5}.

In addition, the early fast-roll stage explanation of the quadrupole supression 
(see sec. \ref{fastroll} and refs. \cite{quadru1,quadru2,quamc})
allows to set an absolute wavelength scale for the primordial power
which fixes the total number of efolds of inflation so we 
checked the consistency of our assumptions about $ N $.

We have not introduced the running of the spectral index $ dn_s / d\ln k $ 
in our MCMC fits since the running [eq.(\ref{run})] must be very small of 
the order $ {\cal O}(N^{-2}) \sim 0.001 $ in slow-roll and for generic 
potentials \cite{1sN}. Indeed we found that adding $ dn_s / d\ln k $, as 
given by eq.~(\ref{runcao}) or eq.~(\ref{runnew}), to the MCMC analysis 
yields insignificant changes on the fit of $ n_s $ and $ r $.
On the contrary, when the running is introduced as a free parameter, then 
the fit of $ n_s $ and $ r $ gets worse and values for the running much 
larger than $ {\cal O}(N^{-2}) \sim 0.001 $ follow 
\cite{WMAP3,WMAP5,otros}. We 
think that the present data are {\bf not} yet precise enough to allow a 
determination of $ dn_s/ d\ln k $. That is, adding further parameters to 
the fit (like the running) does not improve the fit and does not teach 
anything new.

\medskip 

We provide in sec. \ref{BNIMCMC} the MCMC analysis of the Binomial New 
Inflation model (BNI) described in sec. \ref{binonue} \cite{quamc}, using 
the most recent CMB data (WMAP5 \cite{WMAP5} and ACBAR08 \cite{acbar08}) 
and setting $ N=60 $.  The results obtained are consistent with and 
complementary to the  analysis of the $\Lambda$CDM+TNI model
presented in the next subsection.

\subsubsection{MCMC results for Trinomial New Inflation.}\label{mcmc:trino}

\begin{figure}[ht]
\includegraphics[width=12cm]{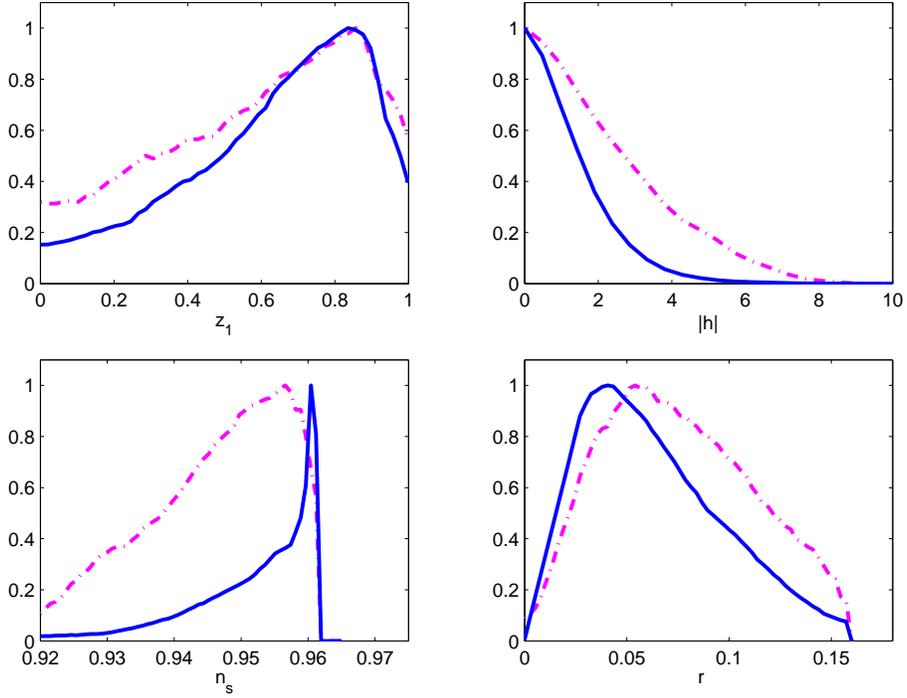}
\caption{$\Lambda$CDM+TNI for $ N=50 $ with WMAP3, small-scale CMB and
SDSS data. Upper panels: probability distributions (solid blue curves) 
and mean likelihoods (dot-dashed magenta curves), all normalized to have 
maximum equal to one, for the values of the normalized coupling at horizon 
exit $ z_1 $ and of the modulus $ |h| $ of the asymmetry of the potential. 
Lower panels: probabilities and mean likelihoods for the values of $ n_s $ 
and $ r $. Notice that $ n_s < 0.9615\ldots $ and $ r \le 0.16 $
since these are the theoretical upper bounds for $ n_s $ and $ r $ in 
trinomial new inflation with $ N=50 $. There appears a {\bf lower bound} 
for the tensor--scalar ratio: $ r > 0.016 \; (95\% \; {\rm CL}) ,\; 
r > 0.049 \; (68\% \; {\rm CL}) $. }
\label{z1hnsr}
\end{figure}

\begin{figure}[ht]
\includegraphics[width=12cm]{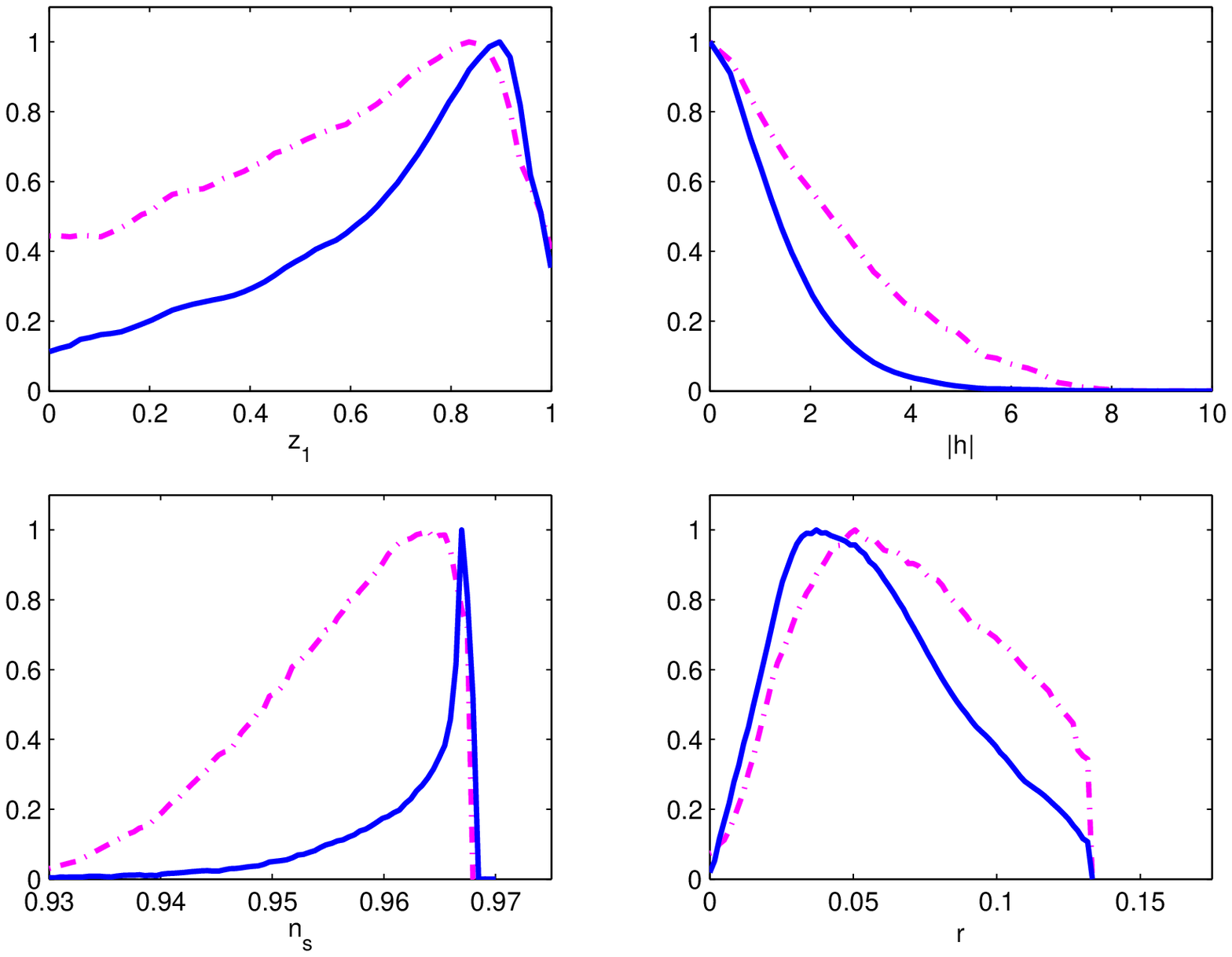}
\caption{$\Lambda$CDM+TNI for $ N=60 $ and with WMAP5, SN and SDSS data.
Upper panels: probability distributions (solid blue curves) and mean
likelihoods (dot-dashed magenta curves), all normalized to have maximum 
equal to one, for the values of the normalized coupling at horizon exit 
$ z_1 $ and of the modulus $ |h| $ of the asymmetry of the potential. Lower
panels: probabilities and mean likelihoods for the values of $ n_s $ and 
$ r $. Notice that here $ n_s < 0.9678\ldots $ and $ r \le 0.13 $ since
these are the theoretical upper bounds for $ n_s $ and $ r $ in trinomial 
new inflation with $ N=60 $. There appears a  a {\bf lower bound} for the 
tensor--scalar ratio: $ r > 0.017 \; (95\% \; {\rm CL}) ,\; r > 0.046 \; 
(68\% \; {\rm CL}) $. }
\label{z1hnsr5}
\end{figure}

\begin{figure}[ht]
\includegraphics[width=12cm]{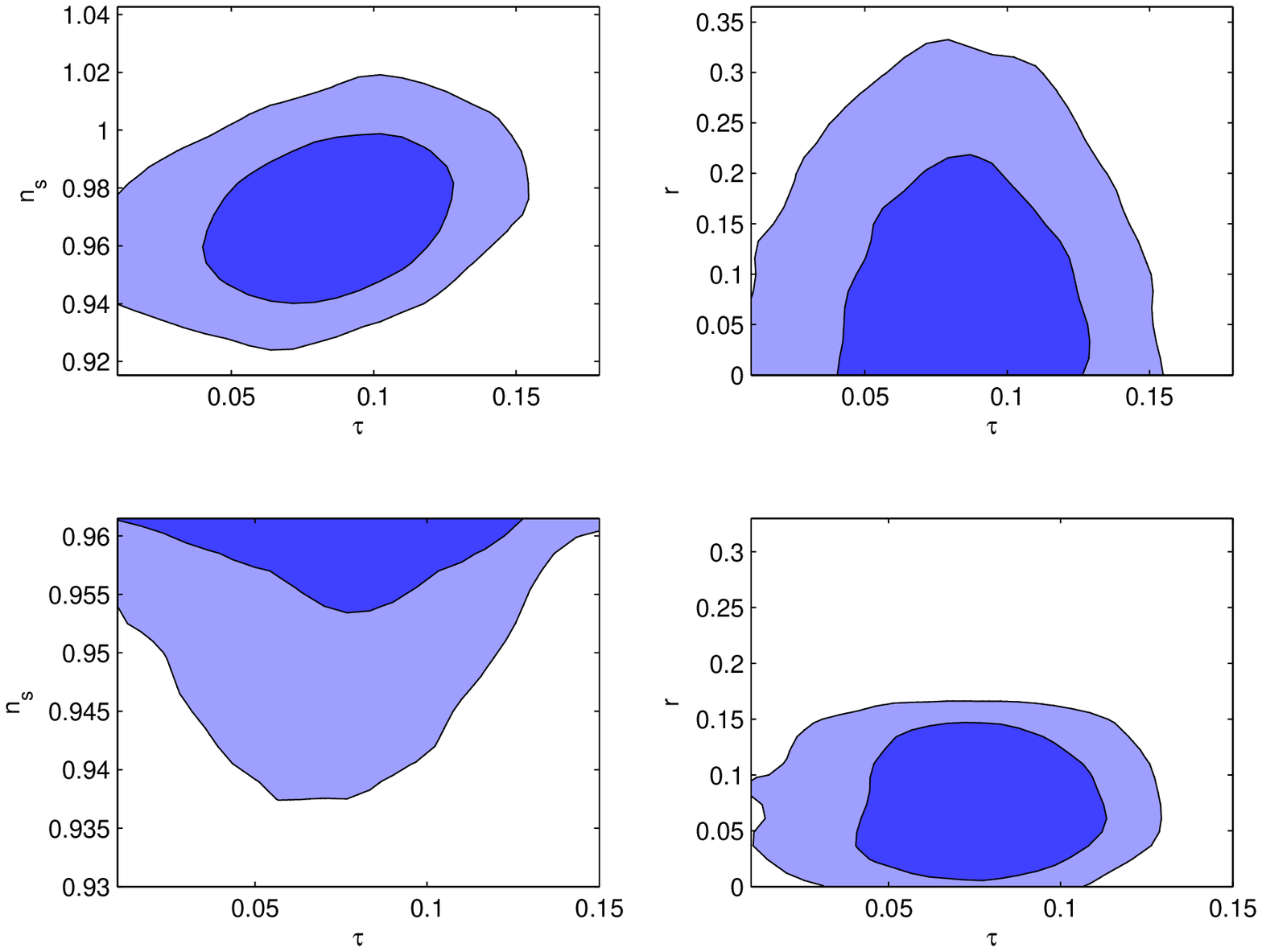}
\caption{Upper panels: 95\% and 68\% contour plots of joint probability 
$ (\tau,n_s) $ distribution (left) and $ (\tau,r) $ distribution
(right) in the $\Lambda$CDM+$r$ model. 
Lower panels: the two joint distributions in the $\Lambda$CDM+TNI model. 
Recall that in this case there is the  theoretical bound eq.(\ref{cotsup}) 
which gives $ n_s < 0.9615\ldots $ for $ N=50 $ and $ n_s < 0.9679\ldots $
for $ N=60 $.}
\label{taunsr}
\end{figure}

Our MCMC results for the $\Lambda$CDM+TNI model are summarized in
figs.~\ref{new_vs_std}--\ref{taunsr}.

In fig.~\ref{new_vs_std} we plot the marginalized probability distributions
(normalized to have maximum equal to one) of the most relevant cosmological
parameters, which are the primary ones allowed to vary independently in the
MCMC runs plus some derived ones. The solid blue curves refer to the runs 
with the hard priors specific to TNI, with the statistics reweighted to 
correspond to a flat prior distribution on $r$. The dashed red curves are 
those of the $\Lambda$CDM+$r $ model. As should have been expected from 
fig.~\ref{ene}, the really significant changes are restricted solely to 
$ n_s $ and $ r $. To provide further evidence of this, we compare in fig. 
\ref{taunsr} the joint probability distributions $ (\tau,n_s) $ and $ 
(\tau,r) $ of the $\Lambda$CDM+TNI model with those of the 
$\Lambda$CDM+$r$ model. We recall that the optical depth parameter 
$ \tau $ is strongly correlated with $ n_s $.

\medskip

For the sake of comparison, we provide in fig.~\ref{new_vs_std5} the same
probability distributions obtained with MCMC runs where $ N=60 $ and WMAP5
data are used together with SDSS data and the most recent SN compiled
observations. As anticipated earlier and as will be confirmed later on, no
really significant changes can be appreciated apart from a slightly tighter
determination for  $ r $.

\medskip

In the lower panels of figs.~\ref{z1hnsr} and~\ref{z1hnsr5} we provide an
enlarged version of the marginalized probability distributions for 
$ n_s $ and $ r $, together with their mean likelihoods, for both sets 
of data WMAP3+small-scale CMB+SDSS and WMAP5+SN+SDSS. In the upper 
panels we plot the marginalized probabilities and the mean likelihoods for
the normalized coupling at horizon exit $ z_1 $ and for the asymmetry 
$ h $, that are the two free parameters of the trinomial potential of new 
inflation. In this case the prior probability is flat over the full allowed
range of $ z_1 $. Again no significant differences can be observed, apart 
from those expected from the change from $ N=50 $ to $ N=60 $ which affects
the theoretical bounds on $ n_s $ and $ r $.

In figs.~\ref{z1hnsr} and~\ref{z1hnsr5} one can observe a significant 
difference between probability distributions and mean likelihood 
distributions for $ n_s $ and $ r $.
Probability and mean likelihood depend quite differently on the shapes and
parametrizations of the allowed $ (n_s, r) $ regions; if observational data
were well concentrated within the allowed $ (n_s, r) $ regions, both types
of distributions would have a gaussian-like shape over much narrower 
intervals and the difference would be much smaller. With the current data, 
the joint probability distributions over the parameters of the trinomial 
potential are very far from gaussian, as can be appreciated from 
fig.~\ref{pdfzh}.

The theoretical constraints narrow the allowed region of parameters in a 
very nontrivial way. Otherwise, the parameters could cover a much wider 
region. Hence, the theoretically constrained distributions can hardly be 
gaussian and in fact they are not. In the case of trinomial new inflation, 
the narrow banana--shaped region depicted in fig. \ref{ene} is responsible 
for the marked difference between marginalized probability distribution and
mean likelihood for $ n_s $, as shown in fig. \ref{z1hnsr}. Besides, the 
banana--shape of trinomial new inflation in the $ n_s-r $ plane, produces a
spike in the left lower panel of fig. \ref{z1hnsr}. The sharp cut on the 
right is due to the theoretical upper bound on $ n_s $ given by eq. 
(\ref{cotsup}), the sharp rise on the left of the
maximum is due to the marginalization over $ r $.

\medskip

Concerning most likely values ({\em i.e.} best fits), our results are 
summarized in table III.
\begin{table}
  \centering
  \begin{tabular}{|c||c|c||c|c|}
    & \multicolumn{2}{c||}{WMAP3 + small scale CMB+SDSS} &
      \multicolumn{2}{c|}{WMAP5+SN+SDSS}       \\ \hline
    parameter &$~\Lambda$CDM+$r~$ & $\Lambda$CDM+TNI & 
     $~~\Lambda$CDM+$r~~$ &  $~~\Lambda$CDM+TNI~~   \\ \hline             
     $100\Omega_bh^2$  & 2.224  &  2.219  &  2.250   &  2.237  \\ \hline
     $\Omega_ch^2$     & 0.107 &  0.106 &  0.107   &  0.108  \\ \hline
     $\theta$          & 1.043 & 1.041 &  1.038   &  1.041  \\ \hline
     $100\tau$         & 8.313 & 8.468  &  8.824   &  7.885  \\ \hline
     $H_0$             & 73.3 & 72.85 &  72.20    &  72.24  \\ \hline
     $\sigma_8$        & 0.773 & 0.766 &  0.784   &  0.786  \\ \hline
     $z_{\rm re}$        & 10.72 & 10.92 &  9.69   &  10.38  \\ \hline
     $\log[10^{10} A_s]$& 3.022 & 3.019 &  3.060    &  3.045  \\ \hline
     $n_s$             & 0.960 & 0.956 &  0.962   &  0.963  \\ \hline
     $r$          & 0.009 & 0.055  &  0.009    &  0.043  \\ \hline \hline
     $\Delta\chi^2$    & 0 & +0.15  &  0 &  +0.08\\ \hline
  \end{tabular}        
  \caption{Comparison of best fits between $~\Lambda$CDM+$r~$ and
    $\Lambda$CDM+TNI with the two data sets indicated on top. Also the 
    negligible variation of $\chi^2\equiv -2\log(likelihood)$ is reported}
\end{table}
The increase in $ \chi^2 $ from $\Lambda$CDM+$r$ to $\Lambda$CDM+TNI is 
roughly twice the standard deviation of $\chi^2$ over ten parallel chains 
with a given model and is therefore not significant.

Most importantly, as expected from fig.~\ref{ene} and quite evident from
figs.~\ref{new_vs_std} and ~\ref{new_vs_std5}, we find a {\bf lower bound} 
on $ r $ for $\Lambda$CDM+TNI:
\begin{equation}\label{cotinfr}
\begin{split}
  &r > 0.016 \quad (95\% \;   {\rm CL}) \quad ,\quad r > 0.049 \quad 
  (68\% \;  {\rm CL}) \quad ({\rm WMAP3 + smallscale~CMB+SDSS }) \\
  & r > 0.017 \quad (95\% \;   {\rm CL}) \quad ,\quad r > 0.046 \quad 
  (68\% \;  {\rm CL}) \quad ({\rm WMAP5+SN+SDSS } )\; .
\end{split}
\end{equation}

\medskip 

Let us now consider the MCMC results for the two parameters of the TNI potential,
$ z_1$ and $h$ (see the upper panels of figs.~\ref{z1hnsr} and~\ref{z1hnsr5} for
the the marginalized probabilities and the mean likelihoods obtained with flat
priors on $ z_1 $ and $ h $ themselves). The asymmetry $ h $ turns out to have very
little relevance, since its distribution is highly peaked near zero and we find
that
\begin{equation*}
  |h| < 2.5 \quad {\rm with} \; 95\% \;  {\rm CL} \;.
\end{equation*}
In fact, we could set it to a value of order 1 without a real loss of
generality. Then the symmetric choice $h=0$ would be the most natural, 
reducing
the $\Lambda$CDM+TNI model to the $\Lambda$CDM+BNI treated in more detail 
in sec. \ref{BNIMCMC} where also the MCMC results for the quadrupole 
supression are presented.

The $ z_1$ distribution exhibits a rather broad peak over most of its 
allowed values (recall that $0<z_1<1$ by construction); still the peak is 
sufficiently well defined to sensibly consider the best fits
\begin{equation*}
z_1 \simeq \begin{cases} 0.886 \quad & ({\rm WMAP3+small scale CMB+SDSS}) 
\\           0.867 \quad & ({\rm WMAP5+SN+SDSS}) \;.
             \end{cases}
\end{equation*} 
In particular for both dataset we find
\begin{equation*}
  z_1 < 0.95 \quad {\rm with} \; 95\% \;  {\rm CL} \;.
\end{equation*}
thus excluding the $ y \gg 1 $  strong coupling region in which 
$ z_1\to 1 $. This result can be read off directly from 
eq.(\ref{nsrtrikGh0}) and fig.~\ref{ene}, since 
$ y \gg 1 $ implies a too small spectral index $ n_s $. In the
opposite limit $ z_1\to 0 $ (that is, where one approaches the quadratic 
monomial potential at $ y = 0 $) the likelihood decreases because one gets
a too large value for $ r $ with respect to the experimental data. 

At $ h=0 $ one can easily convert the results for $ z_1 $ into results for
the coupling $ y $ of the TNI dimensionless potential eq.\eqref{trino}; we 
find the best fits
\be\label{fity}
  y \simeq \begin{cases} 1.28 \quad & ({\rm WMAP3+small scale CMB+SDSS}) \\
                         1.15 \quad & ({\rm WMAP5+SN+SDSS}) \;.
           \end{cases}
\ee
and the bound (valid for both datasets)
\begin{equation*}
  y < 2.93 \quad {\rm with} \; 95\% \;  {\rm CL} 
\end{equation*}
after conversion to a flat prior distribution over $y$.

\medskip 

In conclusion, the most likely trinomial potentials for new inflation are
almost symmetric (i. e. $ h = 0 $) and have moderate nonlinearity with the 
quartic coupling $ y $ of order $1$ [eq.(\ref{fity}]. Thus, we can take as
inflaton potential the binomial eq.(\ref{binon}).
The $ \chi \to - \chi $ symmetry is here spontaneously broken since the
absolute minimum of the potential is at $ \chi \neq 0 $.

\begin{figure}[ht]
\includegraphics[width=14cm]{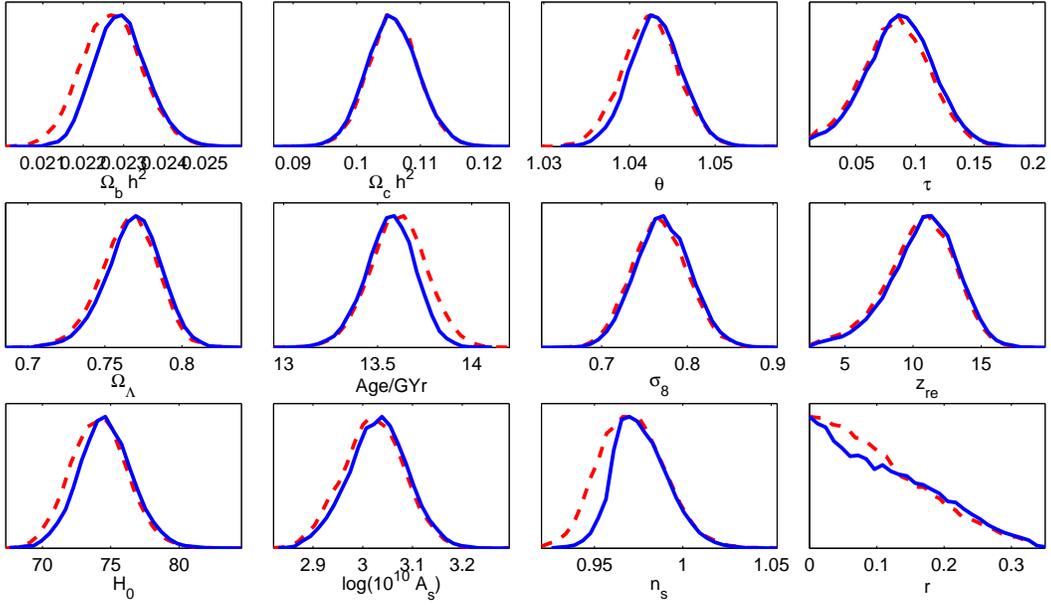}
\caption{Comparison of the marginalized probability distributions
(normalized to have maximum equal to one) of the most relevant cosmological
parameters (both primary and derived) between $\Lambda$CDM+TCI (Trinomial 
Chaotic Inflation, solid blue curves) and the $ \Lambda$CDM+$r $ model 
(dashed red curves).}
\label{chao_vs_std}
\end{figure}

\begin{figure}[ht]
\includegraphics[height=8.cm,width=14.cm]{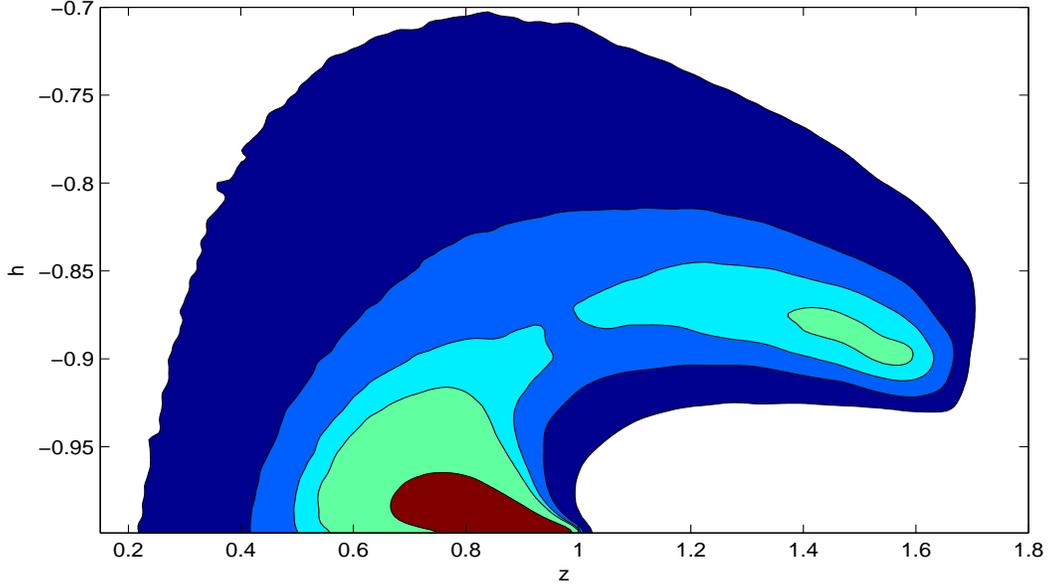}
\caption{Trinomial Chaotic Inflation.  12\%, 27\%, 45\%, 68\% and 95\%
confidence levels of the probability distribution in the field $ z $ - 
asymmetry $ h $ plane.
The color of the areas goes from brown to dark blue for increasing CL.
We see a strong preference of the data for a very asymmetric potential with 
$ h < -0.95 $
and  a significant nonlinearity $ 0.7 < z < 1 $ in chaotic inflation.}
\label{pdfzh}
\end{figure}

\begin{figure}[ht]
\includegraphics[width=14cm]{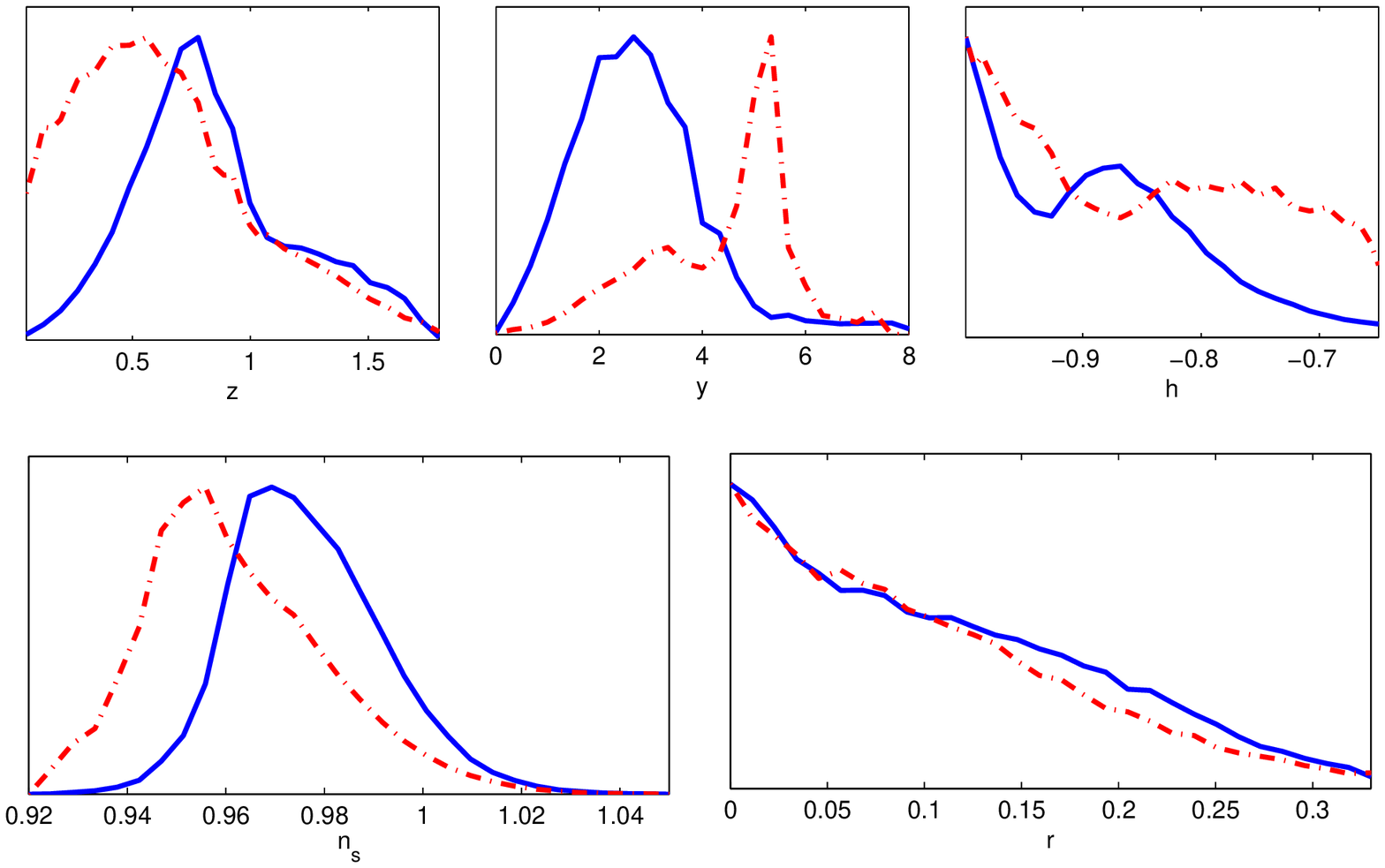}
\caption{Trinomial Chaotic Inflation. Upper panels: probability 
distributions (solid blue curves) and mean likelihoods (dot-dashed red 
curves), all normalized to have maximum equal to one, for the values of 
the normalized coupling at horizon exit $ z $, of the quartic coupling 
$ y $ and of the asymmetry $ h $ of the potential. Lower panels: 
probabilities and mean likelihoods for the values of $ n_s $ and $ r $. 
The data request a strongly asymmetric potential in chaotic inflation. 
That is, a strong breakdown of the $ \chi \to - \chi $ symmetry.}
\label{zyhnsr}
\end{figure}

\subsubsection{MCMC results for Chaotic Trinomial Inflation.}

Our results for the trinomial potential of chaotic inflation, 
eq.(\ref{trinoC}),
are summarized in figs.~\ref{chao_vs_std}, \ref{pdfzh} and \ref{zyhnsr}. 
In this case we report only MCMC results with the datasets WMAP3+small 
scale CMB+SDSS.
The arguments we shall present below are actually not sensitive to the 
small changes in the likelihoods over the $n_s-r$ plane induced by the 
inclusion of more recent experimental data.

In fig.~\ref{chao_vs_std} we plot the marginalized probability
distributions (normalized to have maximum equal to one) of the usual
cosmological parameters as in fig.~\ref{new_vs_std}. Again, solid blue
curves refer to the runs with the hard priors of chaotic inflation, while
the dashed red curves are those of the $\Lambda$CDM+$r$ model. As should
have been expected from fig.~\ref{ene}, there are no really significant
changes in any parameter.

In the upper panels of fig.~\ref{zyhnsr} we plot the marginalized
probabilities and the mean likelihoods for the parameters of the trinomial
potential of chaotic inflation, the normalized coupling at horizon exit
$z$, the quartic coupling $y$ and the asymmetry $h$. These are numerically
calculated from those of $n_s$ and $r$ (which are also reported in the
lower panels of fig.~\ref{zyhnsr}), by means of eqs.(\ref{nscao}), (\ref{rcao}) 
and (\ref{2trino}). The strong non--linearities of these
relations are responsible for the peculiar shapes of the probabilities and
mean likelihoods, and their marked relative differences, of the variables
in the upper panels.

Finally, in fig. \ref{pdfzh} we depict the confidence levels at $ 12\%, \;
27\%,\; 45\%,\; 68 \%$ and $ 95 \% $ in the $ (z , \; h) $ plane. The
chaotic symmetric trinomial potential $ h = 0 $ is almost certainly {\bf
  ruled out} since $ h < -0.7 $ at 95 \% confidence level.

We see that the maximum probability is for strong asymmetry $ h \lesssim
-0.9 $ and significant nonlinearity $ 0.5 \lesssim z \lesssim 1 $ and $ 2.5
\lesssim y \lesssim 5$. That is, {\em all three} terms in the trinomial
potential $ w(\chi) $ {\em do contribute to the same order}.

Notice that the range $ 0.5 < z < 1 $ corresponds for $ h \to -1^+ $ to the
region where the quartic coupling is quite significant: $ 4.207\ldots < y <
+\infty $ according to eq.(\ref{yL}).

Moreover, for these values of $ y $ and $ z $, we have 
$ \chi_{exit} \simeq 1.3,
\; \varphi_{exit} \sim 10 \; M_{Pl} $ according to eq.(\ref{defz}). 
The coefficients $ G_n $ in the general inflaton potential
$ w(\chi) $ [eq.(\ref{seriew})] are of order unit and $ \chi_{exit} 
\simeq 1.3 $ is {\bf larger} than unit for chaotic inflation. Therefore, 
higher order terms in $ w(\chi) $ may affect the couplings obtained with 
quartic potentials in chaotic inflation.

The probability is maximum on a highly special and narrow corner in the 
$(z,h)$ parameter space. This suggests: 
\begin{itemize}
\item[(i)] the data force both the asymmetry as well as the coupling to be 
{\bf large}.
\item[(ii)] the fact that $ z \sim 1 $ implies large $ y $ means that we
are in the nonlinear regime in $ z $, that is $ \chi_{exit} \simeq 1.3 $ 
larger than unit. This suggests that higher order terms in $ \chi $ may be 
added to $ w(\chi) $ and will be relevant.

\item[(iii)] If the preferred values are near the boundary of the parameter
  space, it could be that the true potential is {\bf beyond} that boundary.
  The region of parameter space in which trinomial chaotic inflation yields
  $ r \ll 1 $ is very narrow and highly non-generic and corresponds to an
  inflaton potential contrary to the Landau-Ginsburg spirit  \cite{gl} 
since adding
  higher degree terms in the field to the trinomial potential can produce
  relevant changes. This means that the fit for chaotic inflation can be
  unstable. The fit for new inflation is stable since for new inflation
the maximum probability happens for a
  moderate nonlinearity and therefore will not be much affected by higher
  degree terms. Finally, the $ r \to 0 $ regime is obtained near the
  singular point $ z = 1, \; h = - 1 $ where the inflaton potential
  vanishes for chaotic inflation. 
\item[(iv)]The MCMC runs appear to go towards the $ h = - 1 $ limiting
  chaotic potential exhibiting an inflexion point (see fig. \ref{vv}) which
  is at the boundary of the space of parameters. This may indicate that the
  true potential is {\bf not} within the class of chaotic potentials. Since
  the runs go towards the maximal value of the asymmetry parameter $ h $
  where the inflaton potential exhibits an inflexion point, this supports
the idea that the true potential {\bf must} definitely break the $ \chi
\to - \chi $ symmetry, as new inflaton potentials do spontaneously.  This
{\bf favours again} new inflation since the best fit to the trinomial new
inflation potential corresponds to small or zero asymmetry parameter $ h
$. The spontaneous breakdown of the $ \chi \to - \chi $ symmetry seems 
sufficient to obtain the best fit to the data without the presence of any
explicit symmetry breaking term $ \frac{h}3 \; \sqrt{\frac{y}2} \; \chi^3 $
[see eq.(\ref{trino})].
\end{itemize}

\subsection{Higher degree terms in inflaton potentials}\label{potfi2n}

In this section we study the dependence of the observables 
($ n_s, \; r $ and $ dn_s/d\ln k $) on the degree of the inflaton 
potential ($ 2 \, n $) for new and chaotic inflation and confront them 
to the WMAP data \cite{fi6}.
This study shows in general that fourth degree potentials ($ n=2 $)
provide the best fit to the data \cite{fi6}. We find that new inflation 
fits the data on an appreciable wider range of the parameters while
chaotic inflation does this in a much narrow range. Therefore,
amongst the families of inflationary models studied, new inflation
emerges again as a leading contender in comparison with chaotic inflation.
This analysis confirms the statement that within the
framework of effective field theories with polynomial potentials,
new inflation is a preferred model reproducing the present data
\cite{ciri,infwmap,mcmc}.

The main results presented in this section are \cite{fi6}:

\begin{itemize}
\item{The region in inflaton field space which is consistent with the
marginalized WMAP3 data can be explored in an expansion in $n_s-1+2/N$. }
 \item{ We find that the point $ n_s=1 -2/N= 0.96, \; r=8/N=0.16 $ which is in the region
allowed by the WMAP3 analysis belongs both to new inflation models
as a limiting point and to the simple chaotic inflation monomial, $ m^2 \, \varphi^2/2 $.
This point describes a
region in field and parameter space that separates small fields from
large fields,  and is a degeneracy point for the family of
models describing both chaotic and new inflation. }
\item{ For all members $ n = 2, \; 3, \; 4, ...$ of the new
inflation family, the small field region yields $ r<0.16 $ while the
large field region yields $ r>0.16 $. All members of the new inflation family
predict a small but negative running:
$$
-4 \; (n+1)\times 10^{-4} \leq  dn_s/d\ln k   \leq -2\times 10^{-4}  \; .
$$
This new inflation family features  a {\it large window} of
consistency with the WMAP and LSS data for $ n = 2 $ that narrows
for growing $ n $. If forthcoming data on tensor modes pinpoints the
tensor to scalar ratio to be $ r<0.1 $, we predict that the {\it
symmetry breaking scale} for these models is $ \varphi_{min} \sim
20\,M_{Pl} $ and that the scale of the field at which modes of
cosmological relevance today cross the Hubble radius is $ \varphi_{exit}
\sim 6 \; M_{Pl} $.}
\item{Chaotic inflationary models all yield a tensor to scalar
ratio $ r \geq 0.16 $,  where the minimum value $ r=0.16 $
corresponds to small amplitude of the inflaton  and coincides with
the value obtained from the monomial $ m^2\varphi^2/2 $.  The combined
marginalized data from WMAP3 yields a very small window
of field amplitude, around $|\varphi_{exit}|\sim 15~M_{Pl}$ within which
chaotic models are allowed by the data. These regions become
progressively smaller for larger $n$.  Some small regions in field
space feature peaks in the running of the scalar index but in most of 
the region consistent with the WMAP3 data the running is again negligible 
($ \sim 10^{-3} $). If future observations determine a tensor to scalar ratio $
r<0.16 $, this by itself will  {\bf rule out} a large family of
chaotic inflationary models.}
\end{itemize}

The precise value of $ N $ is certainly near $ N=60 $ as discussed in 
sec. \ref{ntot64}. We will take in this section the value $ N=50 $ as a reference 
baseline value for numerical analysis, but from the explicit expressions
given below, it becomes a simple rescaling to obtain results for arbitrary values 
of $ N $ [see eq.(\ref{resca}) below]. That is, we use the leading value
in  the $ 1/N $ expansion eqs.(\ref{ns}), (\ref{defr}) and (\ref{run})
to obtain their values for arbitrary $ N $, namely 
\be\label{resca} 
r[N] = r[50]  \; \frac{50}{N} \quad , \quad  
n_s[N] =  n_s[50]+\left(1- n_s[50]\right) \; \frac{N-50}{N} \quad , \quad
\frac{dn_s}{d\ln k }[N] = \frac{dn_s}{d\ln k }[50] \; \left(\frac{50}{N}\right)^2
\; . 
\ee
The combination of WMAP and SDSS (LRG) data yields \cite{WMAP3} 
\bea\label{wmapvals}
&& n_s = 0.958\pm 0.016~~~~(\mathrm{assuming~} r=0 \mathrm{~with ~no~ running})\\
&& r < 0.28 ~ (95\% \, CL) ~~\mathrm{no ~running}\quad , \quad
 r < 0.67 ~ (95\% \, CL) ~~\mathrm{with ~ running} \; .
\eea
The running must be very small and of the order $ \mathcal{O}(1/N^2)\sim 10^{-3} $
according to eq.(\ref{run}). Therefore, we can
safely consider $ dn_s/d\ln k =0 $ in our analysis. Figure 14 in the
first reference under \cite{WMAP3} and figure \ref{ene} here show that the
preferred value of $ n_s $ slowly grows with the preferred value of
$ r $ for $ r>0 $. We find approximately that 
\be 
\frac{\Delta n_s}{\Delta r} \simeq 0.12 \label{ley} 
\ee 
Therefore, for $ r\sim 0.1 $ the central value of $ n_s $ shifts from 
$ n_s = 0.958 ~(r=0) $ to $ n_s = 0.97~ (r=0.1) $ as can be readily gleaned 
from both quoted figures.

As a simple example that provides a guide post for comparison let us consider first 
the monomial potential 
\be 
V(\varphi) = \frac{\lambda}{2 \, n}  \, \varphi^{2
\, n}\,. \label{quadpot } 
\ee 
The case $ n=1 $ yields a satisfactory fit to the WMAP data 
\cite{WMAP1,WMAP3,WMAP5}. For these potentials it follows that, 
\be 
w(\chi) = \frac{\chi^{2 \, n}}{2 \, n}~~;~~ M^4 =
\lambda\,N^{n-1}\,M^{2 \, n}_{Pl}\, \, \label{potchi} 
\ee
Inflation ends at  $ \chi_{end} =0 $, and the value of the
dimensionless field $ \chi $ at $ \; N  $ efolds before the end
of inflation is 
\be \label{chi50} 
|\chi| = 2 \; \sqrt{n} \; . 
\ee
These results lead to  
\be \label{nsmono}
n_s-1 = -\frac{n+1}{N} \quad , \quad
r  = \frac{8 \, n}{N} \quad , \quad \frac{dn_s}{d\ln k }   =
 -\frac{n+1}{N^2}\; .
\ee
Taking $ N=50 $ as a baseline, these yield 
\be \label{indicesmono}
n_s-1 = -2 \, (n+1)\times 10^{-2}~\Big(\frac{50}{N}\Big)  \quad , \quad r = 0.16
\, n ~\Big(\frac{50}{N}\Big) \quad , \quad
\frac{dn_s}{d\ln k }=-4 \, (n+1)\times 10^{-4}~\Big(\frac{50}{N}\Big)^2 \; . 
\ee

\subsubsection{Family of models}

We study now the CMB observables $ n_s, \; r, \; dn_s/d\ln k $ for families of
new inflation and chaotic models determined by the following
inflationary potentials:
\bea
V(\varphi) & = & V_0 -\frac12 \;
m^2 \; \varphi^2 + \frac{\lambda}{2 \, n}\; \varphi^{2 \, n} \quad , \quad {\rm
broken ~ symmetry} \label{rota}
\\ \cr V(\varphi) & = & \frac12 \;  m^2 \; \varphi^2 +
\frac{\lambda}{2 \, n} \; \varphi^{2 \, n} \quad , \quad {\rm unbroken ~
symmetry} \; . \label{cao}
\eea
Upon introducing the rescaled field $ \chi $ given by eq.(\ref{chifla}), 
the family of rescaled potentials is
\bea
&& w(\chi) = w_0-\frac12 \; \chi^2 + \frac{g}{2 \, n} \;
\chi^{2 \, n}\quad , \quad {\rm broken ~ symmetry} \label{neww}  \\ \cr
&& w(\chi) = \frac12 \; \chi^2 + \frac{g}{2 \, n} \; \chi^{2 \, n} \quad
, \quad {\rm unbroken ~ symmetry} \; \, .\label{newcao1}
 \eea

$ w_0 $ and $ g $ are dimensionless and related to $ V_0 $ and $
\lambda $ by [compare with eq.(\ref{Gn})],
\be
V_0 = w_0 \;  N \;  M^4  \quad ,
\quad \lambda = \frac{M^4 \; g}{M^{2 \, n}_{Pl} \; N^{n-1}}   \; .
\ee
New inflation models described by the dimensionless potential
given by eq.(\ref{neww}) feature a minimum at $ \chi_{min} $ which 
must obey the conditions eq.(\ref{condw}). These conditions yield, 
\be \label{coupdef}
g = \chi^{2-2 \, n}_{min} \quad , \quad w_0 = \frac{\chi^2_{min}}{2 \, n}
\; \left(n-1\right)\; ,
\ee
$ \chi_{min} $ determines the scale of symmetry breaking $ \varphi_{min} $ 
of the inflaton potential upon the rescaling eq.(\ref{chifla}), namely
$ \varphi_{min} = \sqrt{N} \; M_{Pl} \;  \chi_{min} $.
It is convenient to introduce the dimensionless variable
\be
\label{defx2} x = \frac{\chi}{\chi_{min}}
\ee
Thus, the minimum of the potential is at $ x = 1 $ and the family of
inflation models eq.(\ref{neww})-(\ref{newcao1}) take the form
\bea
&&w(\chi) = \frac{\chi^2_{min}}{2 \, n}  \;   \left[ n \; (1-x^2) +
x^{2 \, n}-1\right] \quad , \quad {\rm broken ~ symmetry} \; ,
\label{newchi}\\
&&   w(\chi) = \frac{\chi^2_{min}}{2 \, n}  \left[n \;
x^2+x^{2 \, n}\right] \quad , \quad {\rm unbroken ~ symmetry} \; .
\label{caochi}
\eea
In terms of the variable $ x $, the small and
large field regions for the potential eq.(\ref{newchi}) correspond to $
x<1 $ and $ x>1 $, respectively. 

\subsubsection{Broken Symmetry models.}

Inflation ends when the inflaton field arrives to the minimum of the
potential. For the new inflation family of models eq.(\ref{newchi})
inflation ends for  
\be 
\chi_{end} = \chi_{min} \; . 
\ee 
In terms of the dimensionless variable $ x $, the condition 
eq.(\ref{Nchi}) becomes
\be 
\frac{2 \, n}{\chi^2_{min}} = I_n(X) \label{chi0I} \quad , \quad
X=\frac{\chi_{exit}}{\chi_{min}} \; , 
\ee
where $ \chi_{exit} $ is the inflaton field at horizon exit and 
\be \label{IXn}
I_n(X) \equiv \int_{X}^1  \frac{dx}{x} \; \frac{n \; (1-x^2) + x^{2 \, n}
- 1 }{1-  x^{2 \, n-2}}= \int_{X}^1 \frac{n-\sum_{m=0}^{n-1}
x^{2m}}{\sum_{m=0}^{n-2} x^{2m}} \; \frac{dx}{x} \; . 
\ee 
This integral can be computed in closed form as a finite sum of
elementary functions \cite{bri}.

For a fixed given value of $ X , \; \chi_{min} $ and
therefore the dimensionless coupling $ g $ are determined by the
equation (\ref{chi0I}). Once we obtain this value, the CMB
observables eqs.(\ref{ns}), (\ref{defr}) and (\ref{run})
are obtained by evaluating
the derivatives of $ w(\chi) $ at the value $ \chi_{exit} = \chi_{min}
\; X $ with the corresponding value of the coupling $ g $. Thus, a
study of the range of possible values for $ n_s, \; r, \; dn_s/d\ln
k $ is carried out by exploring the relationship between these
spectral indices as a function of $ X $. 
For this study we choose the baseline value $ N=50 $ from which the
indices can be obtained for arbitrary value of $ N $ by the relation
(\ref{resca}).

While the dependence of $ \chi_{min} $ and $ g $ upon the variable $ X $ 
must in general be studied numerically, their behavior in the
relevant limits, $ X \rightarrow (0, \; 1) $ for small field inflation and 
$ X >> 1 $ for large field inflation can be derived from 
eqs.(\ref{chi0I})-(\ref{IXn}).

For small field inflation and $ X \rightarrow 0 $, the lower
limit of the integral dominates leading to
\be
\chi^2_{min} \buildrel{X
\to 0}\over= \frac{2 \, n}{n-1} \; \frac1{\log\frac1{X}} \quad ,
\quad g \buildrel{X \to 0}\over= \left( \frac{n-1}{2 \, n} \;
\log\frac1{X}\right)^{n-1} \; , \label{smallX}
\ee
thus, as $ X \rightarrow 0 $ these are \emph{strongly coupled} models. 
This result has a clear and simple interpretation: for $ N =50 $ to be
the number of efolds between $ x=X $ and $ x=1 $ the coupling $ g $
must be large and the potential must be steep, otherwise there would
be many more efolds in such interval.

For small field inflation and $ X \rightarrow 1^{-} $ the integral 
$ I_n(X) $ obviously vanishes and
\be\label{Xten1}
\chi^2_{min} \buildrel{X \to  1^{-}}\over=\left( \frac2{1-X}\right)^2
\left[ 1 + \frac{2\, n - 1}9 \; (X-1) + {\cal O}(X-1)^2 \right]
\quad , \quad g \buildrel{X \to  1^{-}}\over= 
\left[\frac12 \; (1-X)\right]^{2\,(n-1)}
\rightarrow 0\; ,
\ee
thus, as $ X \rightarrow 1^{-} $, these are a \emph{weakly coupled family 
of models}.

\medskip

For large field inflation and $ X \gg 1 $, the integral $ I_n(X) $ is 
dominated by the term with the highest power, namely 
$ x^{2 \, n} $, leading to the behavior
\be\label{Xgran}
\chi^2_{min} \buildrel{X \gg 1}\over= \frac{4 \, n}{X^2}
\quad , \quad g \buildrel{X \gg 1}\over= 
\left( \frac{X^2}{4 \, n} \right)^{n-1} \; ,
\ee
which leads to a strongly coupled regime.

\medskip

Interesting and relevant information can be extracted by focusing 
on the region $ X \sim 1 $ which as  
discussed above corresponds to a weakly coupled family for broken
symmetry potentials. This is the region near the \emph{minimum} of
the potential and the integral $ I_n(X) $ can be evaluated simply by
expanding $ w(\chi) $ and its derivative near the minimum. To
leading order in $ (X-1) $ the condition eq.(\ref{chi0I}) leads to
eq.(\ref{Xten1}) and 
\be \label{quadcond} 
(\chi_c-\chi_{min})^2 = 4 
\quad {\rm or} \quad |\chi_c-\chi_{min}| = 2 \; . 
\ee 
This is precisely
eq.(\ref{chi50}) for $ n = 1 $ upon the shift $ \chi_c \to
\chi_c-\chi_{min} $. Namely, eq.(\ref{quadcond}) is the condition
eq.(\ref{chi50}) for the quadratic monomial potential with minimum
at $ \chi = \chi_{min} $ instead of $ \chi = 0 $ as in
eq.(\ref{potchi}). This is clearly a consequence of the fact that
near the minimum $ X = 1 $ the potential is quadratic, therefore for
$ X \sim 1 $ the quadratic monomial is an excellent approximation to
the family of higher degree potentials and more so because $ g \sim
0 $. For $ X\sim 1 $   we find to leading order in $ (X-1) $ the
values: 
\be \label{quadvals} 
n_s = 0.96+0.04~\Big(\frac{N-50}{N}\Big) \quad ,
\quad r = 0.16~\Big(\frac{50}{N}\Big) \quad , \quad
\frac{dn_s}{d\ln k } = - 0.0008~\Big(\frac{50}{N}\Big)^2 \; .
\ee 
The fact that the potential eq.(\ref{newchi})
is quadratic around the minimum $ X = 1 $ explains why we have in
this limit identical results for new inflation with the potential
eq.(\ref{newchi}) and chaotic inflation with the monomial
potential $ m^2 \; \varphi^2/2 $.

The values eq.(\ref{quadvals}) of $ r, \; n_s $ for $ N
\sim 50 $ yield a good fit to the available CMB data.

\medskip

\begin{figure}[h!]
\begin{center}
\includegraphics[height=3.5in,width=3.5in,keepaspectratio=true]{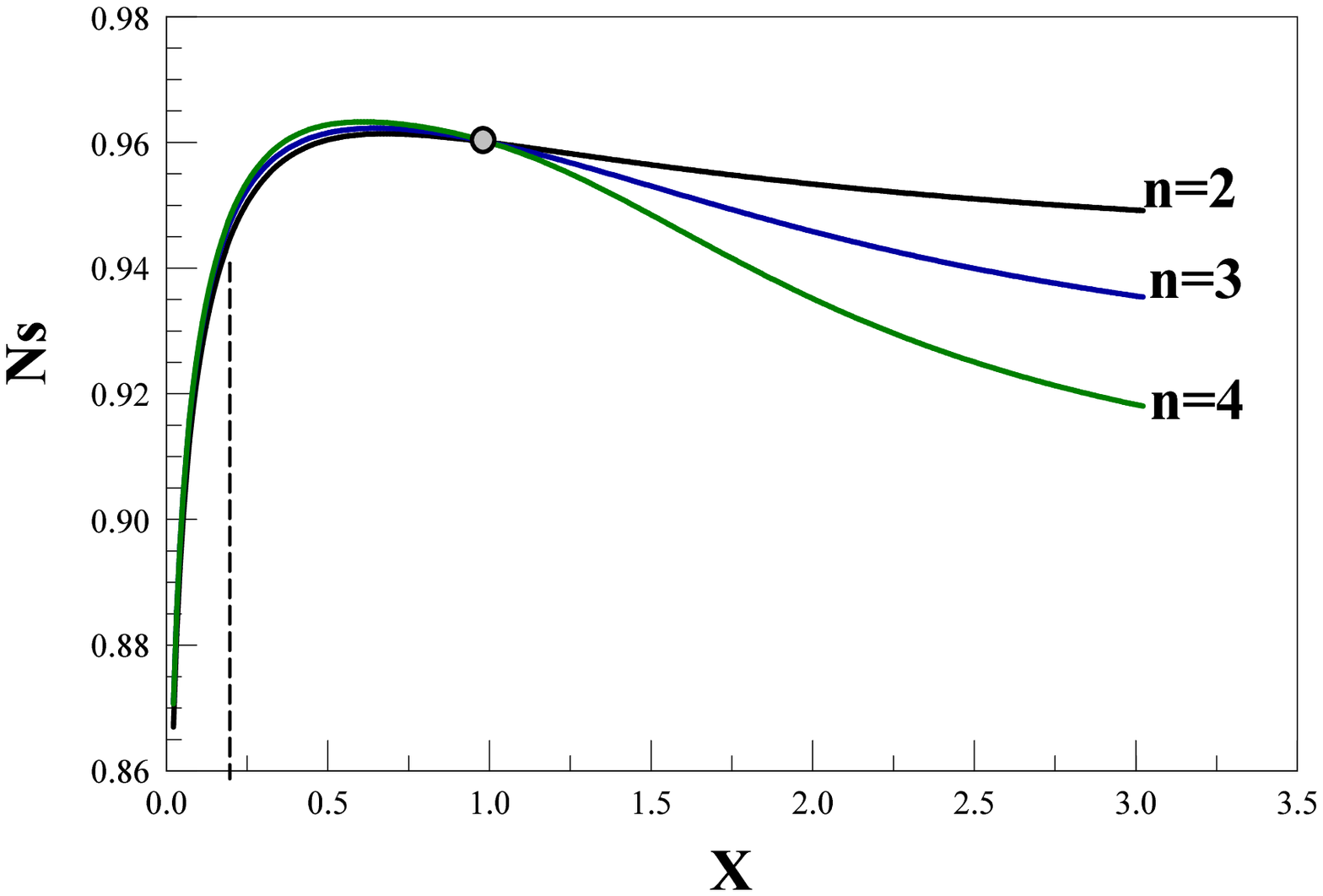} 
\includegraphics[height=3.5 in,width=3.5 in,keepaspectratio=true]{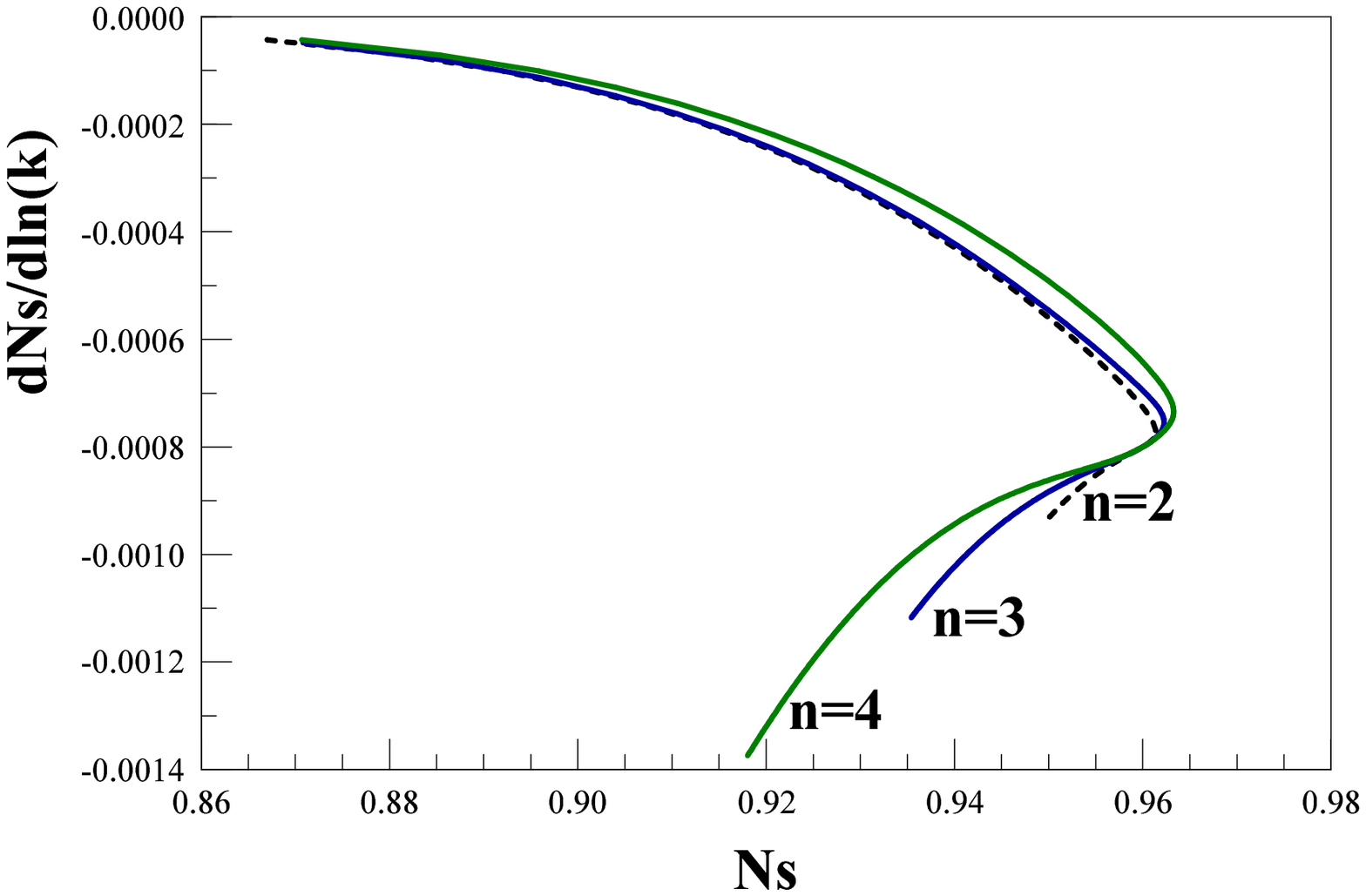}
\caption{Left panel: Scalar spectral index $ n_s $ for the degrees of the
potential $ n=2, \, 3, \, 4 $ for new inflation with $ N=50 $.  The
vertical line delimits the smallest value of $n_s$ (for
$ r=0 $) \cite{WMAP3,WMAP5}. The grey dot at $ n_s=0.96, \; X=1 $
corresponds to the value for the monomial potential $ n = 1 , \; m^2
\; \phi^2/2 $. Notice that the small field behavior is $ n $
independent. For arbitrary $ N $ the result follows directly 
from the $ N=50 $ value by using eq.(\ref{resca}).
Right panel: Running of the scalar index $ dn_s/d\ln k $ vs. $ n_s $ for 
degrees of the potential $ n=2, \, 3, \, 4 $ respectively for new i
nflation with $ N=50 $. The values for arbitrary $ N $ follow directly 
from the $ N=50 $ value by using eq.(\ref{resca}).} \label{fig:NsNI}
\end{center}
\end{figure}

\begin{figure}[h!]
 \begin{center}
 \includegraphics[height=3.5 in,width=3.5 in,keepaspectratio=true]{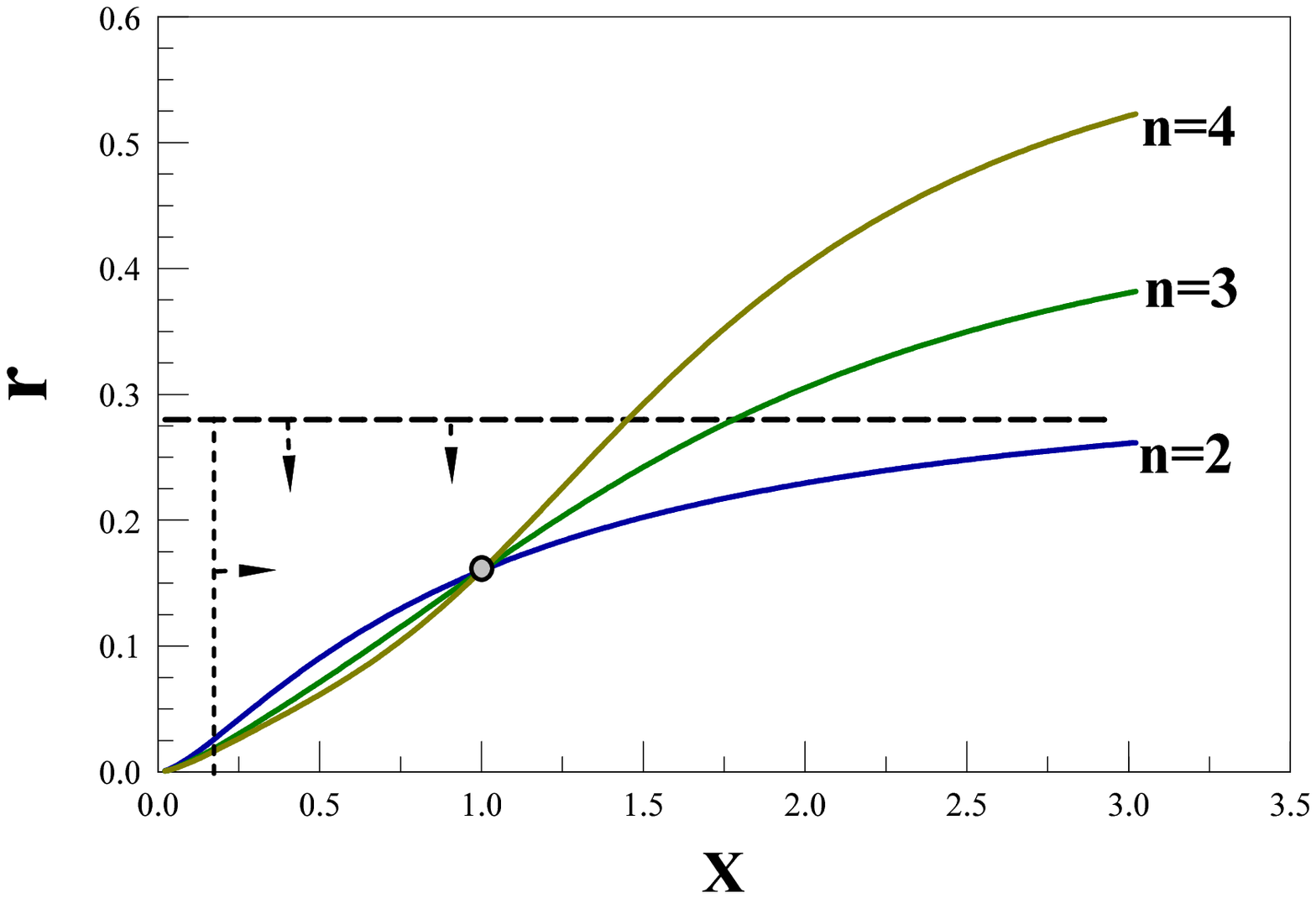}
\includegraphics[height=3.5 in,width=3.5 in,keepaspectratio=true]{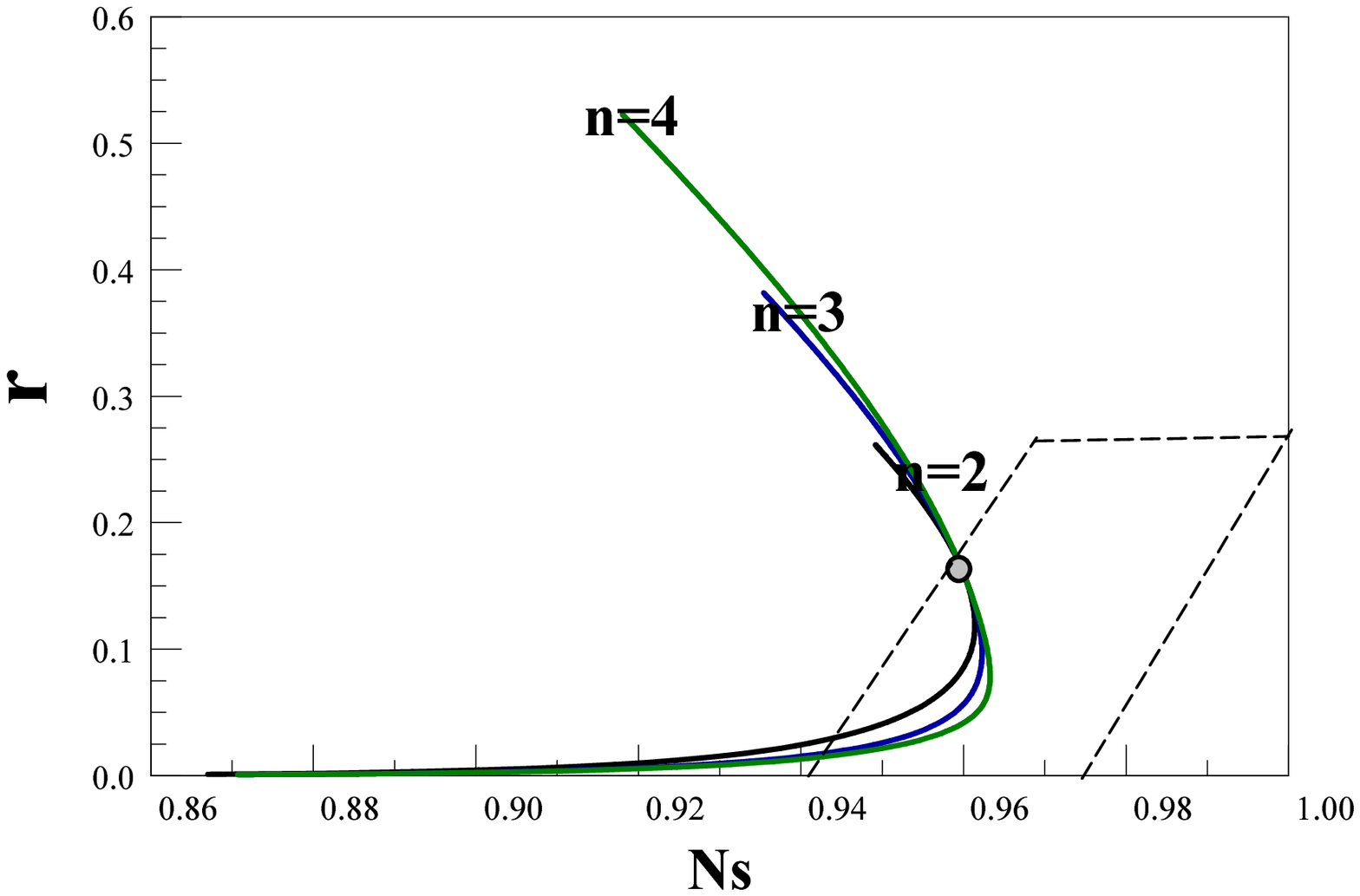} 
 \caption{Left panel: Tensor to scalar ratio $ r $ vs. $ X $
  for the degrees of the potential $ n=2, \; 3, \; 4 $ for
new inflation with $ N=50 $.
The horizontal dashed line corresponds to the upper limit 
$ r=0.28~(95\% CL) $ from WMAP3 without running. The vertical dashed line
determines the minimum value of $ X , \; X \sim 0.2 $, consistent
with the WMAP limits for $ n_s $ as in fig. \ref{fig:NsNI}.
The grey dot at $ X=1, \; r=0.16 $ corresponds to the value for the
monomial potential $ m^2 \; \phi^2/2 $. The small field limit is
nearly independent of $ n $. For arbitrary $ N $ the result follows 
directly from the $ N=50 $ value by using eq.(\ref{resca}).
Right panel: Tensor to scalar ratio $ r $ vs. $ n_s $
for degrees of the potential $ n=2, \, 3, \, 4 $ respectively for new 
inflation with $ N=50 $. $ r $ turns to be a {\bf double-valued} function
of  $ n_s $ exhibiting a maximum value for $ n_s $.
The values inside the box between the dashed lines
  correspond to the WMAP3 marginalized region of the
$(n_s,r)$ plane with $ (95\% CL): \; r < 0.28, \;  0.942+0.12 ~ r
\leq n_s \leq 0.974+0.12 ~ r $, see eq.(\ref{ley}). The grey dot
corresponds to the values for the monomial potential $ m^2\varphi^2/2 $
and the value $ X=1: \; r = 0.16, \; n_s = 0.96 $.}
 \label{fig:rni} 
 \end{center}
 \end{figure}

The value $ X \sim 0.2 $ determines the {\it minimum value}
of $ X $ for which $ n_s $ is consistent with the WMAP data for $ r=0 $ 
(see figs. \ref{fig:NsNI} and \ref{fig:rni}). For large values of 
$ X , \; n_s $ approaches asymptotically the values for the monomial 
potentials $ \varphi^{2 \, n} $ given by eq.(\ref{indicesmono}). 
For the larger degrees $ n $, the asymptotic behavior of  $ n_s $ and 
$ r $ settles at larger values of $ X $, which is a consequence of the 
larger region in which the coupling is small for larger  degrees $ n $. 

\medskip

Unlike the case of a pure monomial potential 
$ \lambda \, \varphi^{2 \, n} $ with $ n\geq 2 $, there is a {\bf large} 
region of field space within which the new inflation models given by 
eq.(\ref{rota}) are {\bf consistent} with the bounds from marginalized 
WMAP3 data and the combined WMAP3 + LSS data \cite{WMAP3}.

\medskip

Fig. \ref{fig:rni} displays $ r $ vs $ n_s $ for the values $
n= \, 2, \, 3, \, 4 $ in new inflation and indicate the trend with $ n $. 
While $ r $ is a monotonically increasing function of $ X, \; n_s $ 
features a \emph{maximum} as a function of $ X $, hence $ r $ becomes a 
{\bf double-valued} function of $ n_s $. The grey
dot at $ r=0.16, \; n_s =0.96 $ corresponds to the monomial
potential $ m^2 \; \varphi^2/2 $ for $ N=50 $. Values below the grey
dot along the curve in fig. \ref{fig:rni} correspond to small
fields $ X<1 $ while values above it correspond to large fields $
X>1 $. We see that
\emph{large fields systematically lead to larger values of}  $ r $.
Models that  fit the WMAP data to $ 95\%~CL $ are within the tilted
box in fig. \ref{fig:rni}.  The tilt accounts for the growth of the 
preferred value of $ n_s $ with $ r $ \cite{WMAP3} according to 
eq.(\ref{ley}).  

\medskip

Fig. \ref{fig:NsNI} displays the running of the scalar index
vs. $ n_s $ for the different members of the family of new inflation
showing clearly that running is all but negligible in the entire range of 
values consistent with the WMAP data. This was expected since the running 
in slow-roll is of the order $ \sim \frac1{N^2} \simeq 4 \times 10^{-4} $
[see eq.(\ref{run})] \cite{1sN}.

We note that $ dn_s/d\ln k $ is a
 monotonically \emph{decreasing} function of $ X $ approaching
 asymptotically the values for the monomials $ \varphi^{2 \, n} $ given by
 eq.(\ref{indicesmono}). 

 \subsubsection{Field reconstruction for new inflation}

 The above analysis suggests to study the \emph{inverse problem},
 namely, for a given member of the family labeled by $ n $, we may ask
 what is the value  $ \varphi_{exit} $ of the field at Hubble crossing and 
what is the scale $ \varphi_{0} $ of symmetry breaking of the potential 
which are consistent with the CMB+LSS data. This is tantamount to the 
program of reconstruction
 of the inflaton potential advocated in ref.\cite{hu}  and is
 achieved as follows: eq.(\ref{chi0I}) yields
 $ \chi_{min}=\chi_{min}[X] $ from which we obtain $ \chi_{exit} = 
\chi_{min} \; X $. These results are then input into the
 expression for $ n_s $ by evaluating the potential $ w(\chi) $ and its
derivatives at the value of
 $ \chi_{exit} $. This yields $ n_s=n_s[\chi_{exit}] $ which is then
 inverted to obtain $ \chi_{exit}= \chi_{exit}[n_s] $ and thus $ \varphi_{exit} $.

In the region $ X \sim 1 $ corresponding to the weakly coupled case, this
 reconstruction program can be carried out as a systematic series in
\be \label{deltaX}
\Delta_X \equiv  X -1 = \frac{\chi_{exit}}{\chi_{min}}-1   \; ,
\ee
by expanding the inflationary potential and its derivatives
 in a power series in $ x $ around $ x=1 $ in the integrand of $ I_n(X) $ [eq.
 (\ref{IXn})]. For $ X=1 $ the value of the scalar index $ n_s $ is
 determined by the simple monomial $ m^2 \; \varphi^2/2 $ which from eq.
 (\ref{nsmono}) for $ n=1 $ is given by $ n_s-1 = -2/N $.
Therefore, in terms of $ n_s $, the actual expansion parameter is $ n_s-1+2/N $.

We obtain $ n_s $ to first order in $ \Delta_X $ from
eqs.(\ref{ns}), (\ref{newchi}), (\ref{Xten1})  and
(\ref{deltaX}) with the result,
$$
 n_s-1 = -\frac2{N}\left[ 1 + \frac{2 \, n - 1}{18} \; \Delta_X + {\cal O}(\Delta_X^2)\right]
$$
then, by inverting this equation we find:
\be
\Delta_X(n_s,n) = X-1 = -\frac{9
 N}{2 \, n-1}\left( n_s-1+\frac2{N}\right) +
\mathcal{O}\left(\left[ n_s-1+\frac2{N}\right]^2\right) \; , \label{delns}
 \ee
and from eqs.(\ref{Xten1}) and (\ref{delns}) we find,
\be
\chi_{min}(n_s,n)  =
 \frac{2 \; (2 \, n-1)}{9 \, N \left| n_s-1+\frac2{N}\right|}\left[1-
 \frac{N}2 \left( n_s-1+\frac2{N}\right)  \right] +
\mathcal{O}\left(\left[ n_s-1+\frac2{N}\right]\right)
\label{chi0s}
\ee
The leading order ($ \propto 1/\Delta_X $) of this result for $ \chi_{min}(n_s,n) $
can be simply cast as eq.(\ref{quadcond}):
this is recognized as the condition to have $ 50 $ efolds for the 
quadratic monomial centered in the broken symmetry minimum [see 
discussion below eq.(\ref{quadcond})].

\medskip

Finally, the value of the (dimensionless) field $ \chi_{exit} $ at Hubble
crossing is determined from $ \chi_{exit}(n_s,n) = \chi_{min} \; [1+\Delta_X(n_s,n)] $ 
from which we obtain
 \be
\chi_{exit} = \frac{2 \; (2 \, n-1)}{9 \, N \left|
n_s-1+\frac2{N}\right|}\left[ 1- \frac{(2 \, n+17) \; N}{2 \;
(2 \, n-1)}\left( n_s-1+\frac2{N}\right)
 \right] + \mathcal{O}\left(\left[ n_s-1+\frac2{N}\right]\right) \; .
 \label{chi50s}
\ee
The coupling constant $ g $ can be also expressed in terms of $ n_s $
in this regime with the result,
$$
g = \left[ \frac{9 \, N \left| n_s-1+\frac2{N}\right|}{2 \; 
(2 \, n-1)}\right]^{2 \, n - 2}\to 0 \; ,
$$
which exhibits the weak coupling character of this limit.

\medskip

This analysis shows that the region in field space that
corresponds to the region in $ n_s $ that \emph{best fits the WMAP
data} can be systematically reconstructed in an expansion in
$ n_s-1+2/N $. This is yet another bonus of the $1/N$ expansion.
Although the above analysis can be   carried out to an arbitrary
order in $ n_s-1+2/N $, it is more convenient to perform a numerical
study of the region   outside from  $ X\sim 1 $ to find the values
of $ \chi_{exit} $ and $ \chi_{min} $ as a function of $ n_s $  
for fixed values of $ n, \; N $. 

\medskip

Finally, the values for the dimensionful  field $ \varphi $ are given
by $ \varphi_{exit} = \sqrt{N} \; M_{Pl} \; \chi_{exit} , \; \varphi_{min} =
\sqrt{N} \; M_{Pl} \; \chi_{min} $. For the  range of CMB parameters 
$ r < 0.1 $ and $ n_s \leq 0.96 $, the typical value of the 
{\it symmetry breaking scale} is $ \varphi_{min} \sim 20 \; M_{Pl} $ and 
the value of the inflaton field at
which cosmologically relevant wavelengths crossed the Hubble radius
during new inflation is $ \varphi_{exit} \sim 6 \; M_{Pl} $ with a weak
dependence on $ n $. For $ 0.1 < r < 0.16 $ we have $ |\varphi_{exit} -
\varphi_{min}| \sim  15 \; M_{Pl} $.

\medskip

We obtain for the coupling $ g $ in the $ X \to 0 $ limit which is a strong
coupling regime [see eq.(\ref{smallX})] where $ n_s \ll 1 $,
$$
g = \left[ \frac{N}4 \; \left( 1 -  \frac1{n} \right) (1 - n_s) \right]^{n-1} \; .
$$
Finally, we have the $ X \to \infty $ limit which is also a strong coupling
limit [see eq.(\ref{Xgran})] where $ n_s \to 1 - (n+1)/N $ and we find,
\bea
&&\chi^2_{min} = (4 \, n)^2 \left[\frac{N(1-n_s) 
-(n + 1)}{4 \, n (n-1) + 3 \, \left(1
+\frac1{n-2} \right)}\right]^{\frac1{n-1}} \to 0 \cr \cr
&& g = \frac{4 \, n (n-1) + 3 \, \left(1 +\frac1{n-2} \right)}{(4 \, n)^{n-1} \;
\left[N(1-n_s) - (n + 1)\right]} \to \infty \; .
\eea

\subsubsection{Chaotic inflation models.}

We now turn to the study of the family of chaotic inflationary
potentials given by eq.(\ref{caochi}).  Taking that the end of
inflation corresponds to $ x=0 $, the condition eq.(\ref{Nchi}) now
becomes 
\be 
\frac{2 \, n}{\chi^2_{min}} = J_n(X)  = \int^{X}_0 \frac{n
+x^{2 \, n-2}}{1+x^{2 \, n-2}}~ x~{dx}\,.\label{chis} 
\ee 
Again, this integral can be computed in closed form 
as a sum of elementary functions \cite{bri}. For general values of $ X $
the integral will be studied numerically, but the small $ X $ region
can be studied by expanding the integrand in powers of $ x^{2 \,
n-2} $, with the result 
\be
1=\frac{\chi^2_{min}\,X^2}{4}\left[1-\frac{n-1}{n^2} \; X^{2 \, n-2}+
\mathcal{O}(X^{4 \, n-4}) \right] \; . \label{chi0cao} 
\ee 
For small $ X $ and recalling that $ X = \chi_{exit}/\chi_{min} $ this relation yields
\be \label{xi50c} 
|\chi_{exit}| =2\left[1+\frac{n-1}{2 \, n^2} \; X^{2 \,
n-2}+\mathcal{O}(X^{4 \, n-4}) \right] 
\ee 
which is again, at dominant order the relation for the quadratic monomial potential
eq.(\ref{chi50}) for $ n = 1 $. This must be the case because the
small field limit is dominated by the quadratic term in the
potential. For small fields,  $ \chi_{min} \approx 2/X $ and the
coupling $ g $ vanishes as, 
\be 
g(X) \buildrel{X \to 0}\over=\left(\frac{X}2\right)^{2 \, n-2} \; . 
\ee 
The dependence of $ n_s $ and $ r $ in the full range of
$ X $ for several representative values of $ n $ is studied numerically.
In the small $ X $ regime, we obtain from eqs.(\ref{ns}), (\ref{defr}),
(\ref{caochi}) and (\ref{chi0cao}) the expressions,
\bea
&& n_s - 1 = -\frac2{N}\left[1- \frac{(2 \, n-1)(n-1)(n-2)}{2 \; n^2} \;
X^{2 \, n-2}+\mathcal{O}\left(X^{4 \, n-4}\right)\right] \label{evsX} \\
&& r =  \frac8{N}\left[1+ \frac{(2 \, n-1)(n-1)}{n^2} \; 
X^{2 \, n-2}+\mathcal{O}\left(X^{4 \, n-4}\right)\right]
\label{etavsX} 
\eea 
As $ X \rightarrow 0 , \; n_s $ and $ r $ tend to the result from the quadratic 
monomial potential, namely $  n_s - 1= -2/N , \; r = 8/N $ as must be the case 
because the quadratic term dominates the potential for $ X\ll 1 $.

For $ X\rightarrow 0 $ and $ N = 50 , \; n_s \rightarrow 0.96 $ and $ r \rightarrow 0.16 $ 
which are the values from the quadratic monomial potential $ m^2 \; \varphi^2/2 $.

For $ X\gg 1 $, the values of $ n_s, \; r $ 
for the monomial potentials $ \varphi^{2 \, n} $ are attained asymptotically,
namely, (for $ N = 50 $): $ n_s-1 =   -2 \, (n+1)\times 10^{-2}  , \; r = 0.16 \, n $.

\medskip

The range in which the chaotic family provides a
good fit to the WMAP data is \emph{very much smaller} than for new inflation.
In chaotic inflation
{\it only} for $ n=2 $ the range of $ n_s $ is
allowed by the WMAP data in a fairly extensive range of values of
$ X $, whereas for $ n=3, \; 4 $ (and certainly larger), there is a
\emph{relatively small} window in field space for $ X<1 $ which satisfies the
data for $ n_s $ and $ r $ simultaneously.

The tensor to scalar ratio $ r $ in chaotic inflationary models is
{\it larger} than $ 0.16 $ for all values of $ X $, approaching asymptotically
for large $ X $ the value $ r = 0.16 \, n $  associated to the monomial potentials $ \varphi^{2 \, n} $.

While the running  $ dn_s/d\ln k $ is again negligible, it is {\it strikingly
 different} from the new inflation case. \emph{Again} this study, in
 combination with those for $ n_s $ and $ r $ as functions of $ X $
 distinctly shows that {\bf only} $ n=2 $ in chaotic inflation is compatible with the
 bounds from the WMAP data, while for $ n=3,4 $ only a {\it small window}
 for $ X<1 $ is allowed by the data. 
$ dn_s/d\ln k $ takes negative as well as positive values for chaotic inflation,
in contrast with new inflation where  $ dn_s/d\ln k $ is always  $ < 0 $.

\begin{figure}[h]
 \begin{center}
 \includegraphics[height=3 in,width=3 in,keepaspectratio=true]{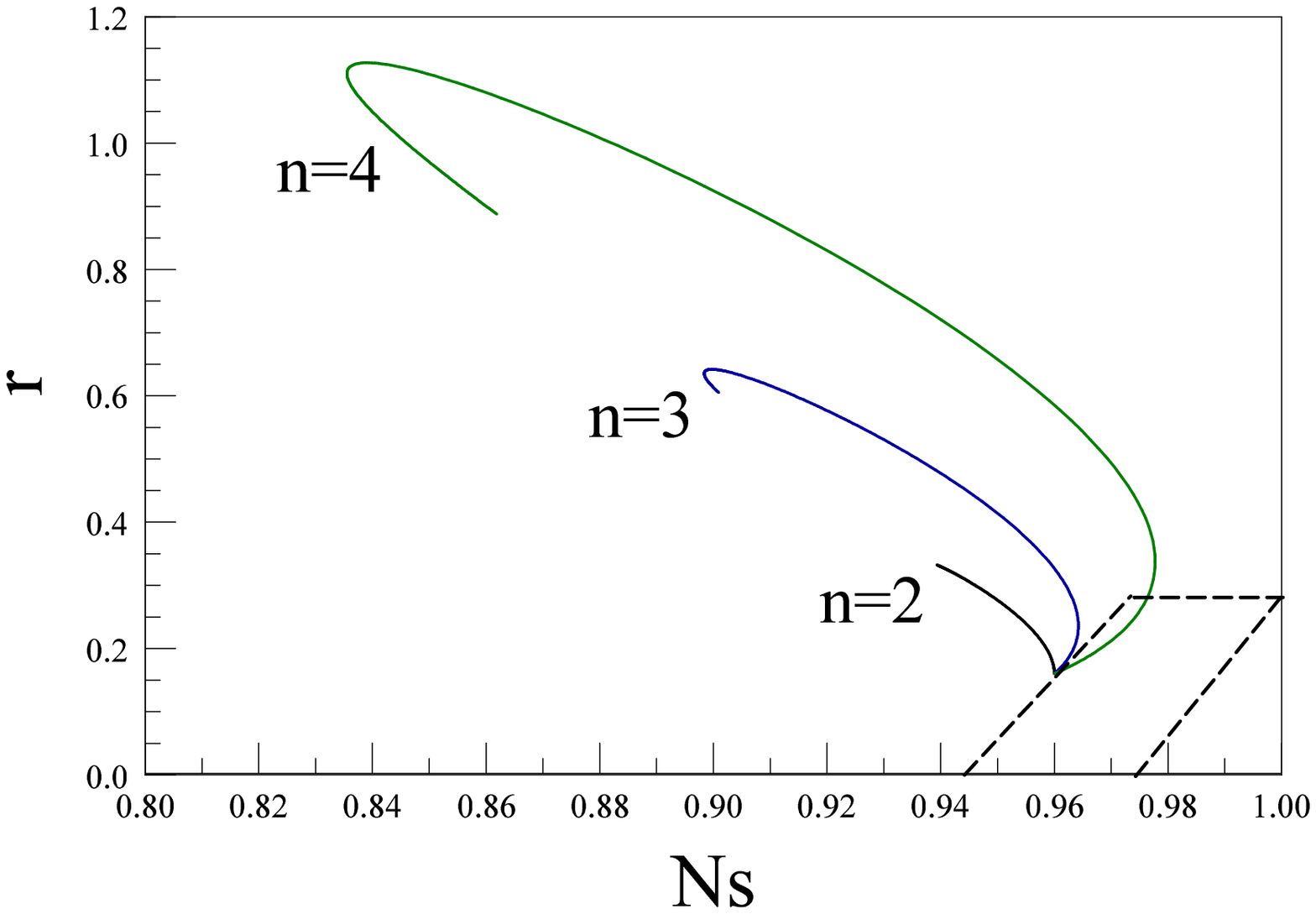}
\includegraphics[height=3.5 in,width=3.5 in,keepaspectratio=true]{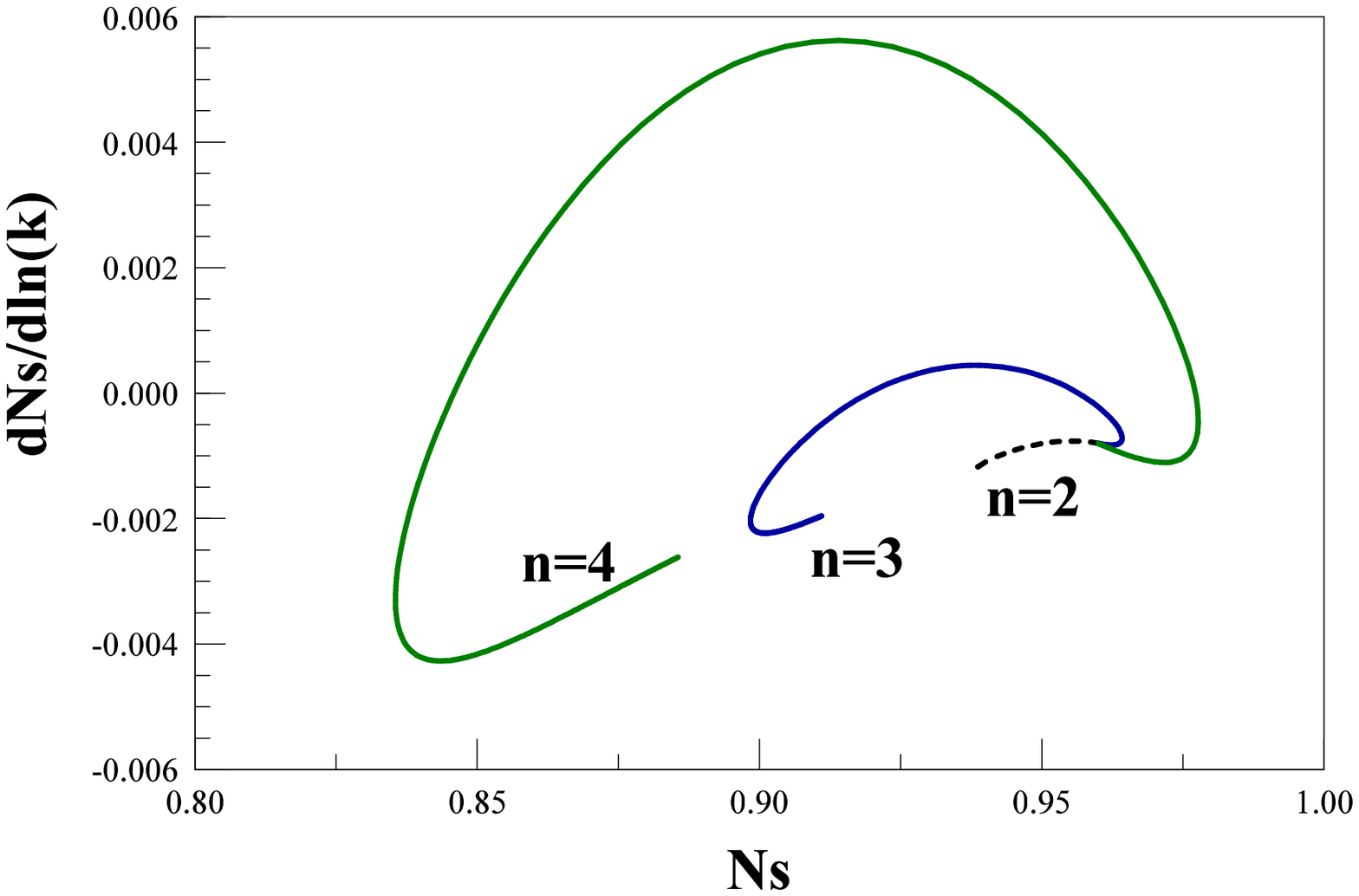}
 \caption{Left panel: Tensor to scalar ratio $ r $ vs. $ n_s $
  for degrees of the potential  $ n=2, \; 3, \; 4 $ respectively
for chaotic inflation with $ N=50 $.
The range of $ 95\%~CL $ as determined by WMAP3 \cite{WMAP3} is within
  the tilted box delimited by :
$ r < 0.28, \;  0.942+0.12 \, r \leq n_s \leq 0.974+0.12 \, r $, see
eq.(\ref{ley}). Right panel: Running of the scalar index $ dn_s/d\ln k $ 
vs. $ n_s $  for degrees of the potential
$ n=2, \, 3, \, 4 $ respectively for chaotic inflation with $ N=50 $. }
 \label{fig:rvsncao}
 \end{center}
 \end{figure}

The fact that the combined bounds on $ n_s, \; r $ and $ dn_s/d\ln k
$ from the WMAP3 data \cite{WMAP3} provide much more stringent
constraints on chaotic models is best captured by displaying $ r $
as a function of $ n_s $ in fig. \ref{fig:rvsncao}. The region
allowed by the WMAP data lies within the tilted box delimited by the
vertical and horizontal dashed lines that represent the $ 95\% CL $
band \cite{WMAP3}.

 A complementary assessment of the
 allowed region for this family of effective field theories is
 shown in fig. \ref{fig:rvsncao} which distinctly shows that {\bf only}
 the $ n=2 $ case of chaotic inflation is allowed by the WMAP3 data.

\subsubsection{Field reconstruction for chaotic inflation }

The reconstruction program proceeds in the same manner as in the
case of new inflation: the first step is to obtain $ \chi_{min}(X) $
from eq.(\ref{chis}). Then $ \epsilon_v $ and $ \eta_v $ are
obtained as a function of $ X $ which yields $ n_s(X) $. Inverting
this relation we find $ X=X(n_s) $ and finally $ \chi_{exit}(n_s) =
\chi_{min} \; X(n_s) $. While this program must be carried out
numerically, we can gain important insight by focusing on the small
$ X $ region and using eq.(\ref{chi0cao}).

From eqs.(\ref{ns}), (\ref{evsX}) and (\ref{etavsX}) we find 
\be
n_s-1+\frac2{N} = X^{2 \, n-2} \; \frac{(2 \,
n-1)(n-1)(n-2)}{n^2\,N}+\mathcal{O}\Big(X^{4n-4}\Big)\label{expacao}
\ee 
As $ X \rightarrow 0 $ it follows that $ n_s \rightarrow 1-2/N
$ which is the value for the scalar index for the quadratic monomial
potential $ m^2\varphi^2/2 $. However, for $n>2$ this limit is
approached from above, namely for $ n>2 $ it follows that
$ n_s>1-2/N $. The small $ X $ region corresponds to small departures
of $ n_s $ from the value determined by the quadratic monomial $
1-2/N $ but always larger than this value for $ n>2 $. In the small
field limit we reconstruct the value of $ \chi_{exit} $ in an expansion in
$ n_s-1+2/N $. The leading order in this expansion is obtained by
combining eqs.(\ref{xi50c}) and (\ref{expacao}), we obtain 
\be
\label{chifif} |\chi_{exit}|
=2\Bigg[1+\frac{\big(n_s-1+\frac2{N}\big)N}{2 \, (2 \, n-1) \,
(n-2)}\Bigg]+ {\cal O}\left( \left[n_s-1+2/N \right]^2\right) 
\ee
Obviously, this leading order term is singular at $ n=2 $, this is a
consequence of the result eq.(\ref{expacao}) which entails that for
$ n=2 $ the expansion must be pursued to higher order, up to $
X^{4n-4} $.

We find from eqs.(\ref{ns}), (\ref{evsX}) and (\ref{etavsX}) for $
n = 2 $, 
\be \label{neq2} 
n_s-1+\frac2{N} = - \frac{17}{24 \,
N} \; X^4 + \mathcal{O}\Big(X^6\Big) \quad , \quad n=2 \; , 
\ee
  therefore, 
\be\label{chin2} 
|\chi_{exit}| =2\Bigg[1+\sqrt{\frac{3
\, N}{136}\left(1 -\frac2{N}-n_s \right)}\,\Bigg] + {\cal
O}\left(n_s-1+\frac2{N} \right)\quad , \quad n=2 \; . 
\ee 
We see that the derivative of $ \chi_{exit} $ with respect to $ n_s $ is
singular for $ n=2 $ at $ n_s=1-2/N $. We note that for
$ n=2 $ there is a sign change with respect to the cases $ n>2 $ and
$ n_s-1+2/N \leq 0 $ as determined by eq.(\ref{neq2}).

It is clear that there is a small window in field space within which
chaotic models provide a good fit to the WMAP3 data, for $ N=50 $ we
find:
\bea \label{regiochi50} 
&& n =2 ~:   ~~0.95 \lesssim n_s \leq 0.960
~,~2.0 \leq |\chi_{exit}| \lesssim 2.25  \nonumber \\&& n =3~: ~~0.96
\leq n_s  \lesssim 0.965 ~,~2.0 \leq |\chi_{exit}| \lesssim 2.15 \nonumber \\
&& n =4~:   ~~0.96 \leq n_s \leq 0.975 ~,~2.0 \leq |\chi_{exit}| \lesssim
2.10  \; .
\eea 
Restoring the dimensions via eq.(\ref{chifla}) these
values translate into a narrow region of width $ \Delta_X \varphi \lesssim
1.5 \; M_{Pl} $ around the scale $ |\varphi_{exit}| \sim 15~M_{Pl} $.

Therefore, the joint analysis for $ n_s, \; r, \; dn_s/d\ln k $
distinctly reveals that: (i) chaotic models favor {\it larger}
values of $ r $ thus, larger tensor amplitudes, and (ii) chaotic models feature
{\it smaller} regions in field space consistent with the CMB and
large scale structure data.  {\it Only} the case $ n=2 $ features a
larger region of consistency with the combined WMAP3 data.

\subsubsection{Conclusions}

 We perform a systematic study of
\emph{families} of single field new and chaotic inflation slow-roll
models characterized by effective field theories with potentials of
the form \cite{fi6}
\bea
V(\varphi) &=&  V_0 -\frac12 \; m^2 \; \varphi^2 +
\frac{\lambda}{2 \, n}\; \varphi^{2n} \quad , \quad {\rm broken ~ symmetry}
\label{nuev}\\
V(\varphi) & = & \frac12 \;  m^2 \; \varphi^2 +
\frac{\lambda}{2 \, n} \; \varphi^{2 \, n} \quad , \quad {\rm unbroken ~
symmetry}\label{chau} \; .
\eea
Unlike the approach followed in \cite{WMAP1,WMAP3} based on the
inflationary flow equations \cite{flujo}, or more recent
studies which focused on specific inflationary
models \cite{salama}, or on statistical analysis of models \cite{flujo},
we implement an expansion in $ 1/N $ where $ N \sim
50 $ is the number of efolds before the end of inflation when
wavelengths of cosmological relevance today cross the Hubble radius
during inflation. We provide  in ref.\cite{fi6} an analysis of the dependence of CMB
observables ($ n_s, \; r $ and $ dn_s/d\ln k $) with the degree $ n $ of the potential and
establish the region in field space within which these families
provide a good agreement with the WMAP3 data combined with large
scale surveys.

\medskip

For new inflation models with potentials eq.(\ref{nuev}) there are two distinct
regions corresponding to values of the inflaton field  smaller
(small field) or larger (large field) than the symmetry breaking
scale. For this family  we find a {\bf wide range} in the $ (n_s,r) $ 
plane in which the different members $ n=2, \; 3, \; 4...$ are
allowed by the data both for small and large fields with negligible
running of the scalar index \cite{fi6}:
$$
-4 \; (n+1)\times 10^{-4} \leq  dn_s/d\ln k   \leq -2\times 10^{-4}  \; .
$$
For $ N=50 $ the values $ n_s=0.96, \; r=0.16 $ which are those
determined by the simple monomial potential $ m^2 \; \varphi^2/2 $
determine a divide and a degeneracy point in the field and parameter
space. Small field regions yield $ r< 0.16 $ while large field
regions correspond to $ r>0.16 $.

The $ 1/N $ expansion also provides a powerful tool to
implement a {\it reconstruction program} that allows to extract the
value of the field $ N $ efolds before the end of inflation, and in
the case of new inflationary models, the symmetry breaking scale.

We find that the region of field space favored by the WMAP3 data can
be explored in a systematic expansion in $ n_s-1+2/N $ \cite{fi6}. An analytic
and numerical study of this region lead us to conclude that if
forthcoming data on tensor modes favors $ r<0.16 $ then
new inflation {\bf is favored}, and we {\bf predict} for $ r < 0.1 $ that\
(i) the  {\bf symmetry breaking scale} is
$$
\varphi_{min} \sim 20~M_{Pl} \; ,
$$
and (ii) the value of the field when cosmologically relevant
wavelengths cross the Hubble radius is $ |\varphi_{exit}|\sim 6 \; M_{Pl} $ .

\medskip

The family of chaotic inflationary models characterized by the
potentials eq.(\ref{chau}) feature tensor to scalar ratios $ r \geq
0.16 $ (for $ N=50 $), with the minimum, $ r=0.16 $ obtained in
the limit of small inflaton amplitude and corresponds to the
monomial potential $ m^2 \; \varphi^2/2 $ which is again a degeneracy
point for this family of models.

The combined marginalized data from WMAP3 \cite{WMAP3} yields a very
small window within which chaotic models are allowed by the data,
the largest region of overlap with the $ (r,n_s) $ WMAP3 data
corresponds to $ n=2 $ and the width of the region decreases with
larger $ n $ \cite{fi6}. The typical scale of the field at Hubble crossing for
these models is $ |\varphi_{exit}|\sim 15~M_{Pl} $ (for $ N = 50 $). Some
small regions in field space consistent with the WMAP3 data feature
peaks in the running of the scalar index but in the region
consistent with the WMAP3 data in chaotic inflation the running is
again negligible. If future observations determine a tensor to
scalar ratio $ r<0.16 $, such bound will, all  by itself, {\bf rule
out} the  large family of chaotic inflationary models of the form
(\ref{chau}) for any $ n $ \cite{fi6}.

\subsection{The initial conditions for the scalar and tensor 
quantum fluctuations.}\label{supcua}

Scalar curvature and tensor (gravitational wave) quantum
fluctuations generated during the inflationary stage determine the
power spectrum of the anisotropies in the cosmic microwave
background (CMB) providing the seeds for large scale structure (LSS)
formation. Curvature and tensor fluctuations obey the wave equation  eq.(\ref{fluces})
and (\ref{Sten}), respectively, 
and the choice of a particular solution entails a choice of initial
conditions \cite{kt,libros,mass}. The power spectra of these
fluctuations depend in general on the initial conditions that define
the particular solutions. These   are usually chosen as
Bunch-Davies \cite{BD} initial conditions, which select positive
frequency modes asymptotically with respect to conformal time. The
quantum states in the Fock representation associated with these
initial conditions are known as Bunch-Davies states, the vacuum
state being invariant under the maximal symmetry group $ O(4,1) $ of
de Sitter space-time.

Alternative initial
conditions were also considered \cite{diferen}.  The requirement that
the energy momentum tensor be renormalizable constrains the UV
asymptotic behaviour of the Bogoliubov coefficients that encode
different initial conditions \cite{motto2}. The availability of high
precision cosmological data motivated a substantial effort to study
the effect of different initial conditions upon the angular power
spectrum of CMB anisotropies, focusing primarily in the high-$ \ell $
region near the acoustic peaks \cite{otros}. However, the exhaustive
analysis of the WMAP data \cite{WMAP3,WMAP5}
render much less statistical significance to possible effects on
small angular scales from alternative initial conditions
(see also sec. \ref{transPyR}).

\subsubsection{Initial conditions and the energy momentum tensor of
scalar and tensor perturbations.}\label{sec:inicon}

The effective field theory of slow-roll inflation has two main
ingredients: the classical Friedmann equations in terms of a
\emph{classical} part of the energy momentum tensor described by a
homogeneous and isotropic condensate,  and a quantum part. The
latter   features   scalar fluctuations determined by a gauge
invariant combination of the scalar field (inflaton) and metric
fluctuations, a tensor component, gravitational waves plus
contributions of further quantum fields (scalar, spinors, etc.)
A consistency condition for this description is that the contributions
from the fluctuations to the energy momentum tensor be much smaller
than those from the homogeneous and isotropic condensate. The
effective field theory must include renormalization counterterms so
that it is insensitive to the ultraviolet singularities of
the short wavelength fluctuations. Different initial conditions on
the mode functions of the quantum fluctuations yield different
values for their contribution to the energy momentum tensor.
Different initial conditions on the mode functions of the quantum
fluctuations yield different values for the energy momentum tensor.

Criteria  for acceptable initial conditions must include the
following: i) back reaction effects from the quantum fluctuations
should not modify the inflationary dynamics described by the
inflaton, ii) the ultraviolet counterterms that renormalize  the
energy momentum tensor should not depend on the particular choice of
initial conditions, namely different initial conditions \emph{should
not} introduce new ultraviolet divergences: a single renormalization
scheme, independent of initial conditions, should render the energy
momentum tensor UV finite. This set of criteria imply that the
ultraviolet allowed states have their large $ k $ behaviour
constrained up to the fourth order in $ 1/k $ \cite{motto2}. In ref.
\cite{motto2} only the energy momentum tensor of \emph{inflaton}
fluctuations was considered. However, the fluctuations of the
inflaton field are \emph{not} gauge invariant, and in order to
establish a set of criteria for UV allowed initial states in a gauge
invariant manner we studied the full gauge invariant energy
momentum of scalar and tensor fluctuations \cite{effpot,quant}.

\subsubsection{Scalar perturbations}\label{sec:scalar}

The gauge invariant energy momentum tensor for quadratic scalar
metric fluctuations has been obtained in ref.\cite{abramo,quant,quadru1} where the
reader is referred to for details. Its form simplifies in
longitudinal gauge, and in cosmic time it is given  by \cite{abramo,quant,quadru1}
\be\label{Too}
\langle T_{00} \rangle =  M^2_{Pl} \Bigg[12 \; H  \; \langle
\psi \dot{\psi} \rangle - 3 \; \langle (\dot{\psi})^2 \rangle +
\frac{9}{a^2(t)} \; \langle (\nabla \psi)^2\rangle \Bigg] 
+ \frac12 \; \langle (\dot{ \phi})^2\rangle + \frac{\langle
(\nabla  \phi)^2\rangle}{2\,a^2(t)} + \frac{V''(\Phi )}2 \; \langle
\phi^2 \rangle + 2 \; V'(\Phi ) \; \langle \psi \, \phi \rangle \; , 
\ee
\noindent where $ \Phi (t) $ stands for the
zero mode of the inflaton field, $ \phi(t,{\vec x}) $ for the
inflaton fluctuations around $ \Phi (t) , \; \psi(t,{\vec x}) $ is
the longitudinal gauge form of the Bardeen potential and the dots
stand for derivatives with respect to cosmic time. During inflation
the Newtonian potential and the Bardeen potential are the same in
the longitudinal gauge\cite{hu} and this property has
been used in the above expression.

In longitudinal gauge, the equations of motion in cosmic time for
the Fourier modes are\cite{hu} 
\bea \label{phieq}
&&\ddot{\psi}_{\vec k}+ \left(H -2 \; \frac{\ddot{\Phi}}{\dot{\Phi}
}\right)\dot{\psi}_{\vec k}+ \left[2 \; \left(\dot{H } -2 \; H \;
\frac{\ddot{\Phi} }{\dot{\Phi} }\right)+
\frac{k^2}{a^2(t)}\right]{\psi}_{\vec k}=0  \; , \cr \cr &&
\label{delfieqn} \ddot{ \phi}_{\vec k}+3 \; H \; \dot{ \phi}_{\vec
k}+\left[V''[\Phi ]+\frac{k^2}{a^2(t)} \right] \phi_{\vec k}+2 \;
V'[\Phi ] \; \psi_{\vec k}- 4 \; \dot{\Phi}  \; \dot{\psi}_{\vec
k}=0  \; \; , 
\eea 
\noindent with the constraint equation 
\be
\label{vin} \dot{\psi}_{\vec k}+H \; \psi_{\vec k}=
\frac1{2M^2_{Pl}} \;  \phi_{\vec k} \; \dot{\Phi}   \; . 
\ee 
Initial
conditions on the mode functions of the quantum fluctuations
correspond to an initial value problem at a fixed time hypersurface.
For modes  of cosmological relevance this time slice at which the
initial conditions are established is such that these modes are
\emph{subhorizon}.  Therefore, we must focus on the contribution to
the energy momentum tensor from subhorizon fluctuations, and in
particular in the large momentum region to assess the criteria for
UV allowed states.

For subhorizon modes with wavevectors $ k \gg a(t) \;  H $, the
solutions of the equation (\ref{phieq}) are \cite{hu} 
\be \label{apro} 
\psi_{\vec k}(t) \approx e^{\pm i k \eta} \Rightarrow
\dot{\psi}_{\vec k}(t) \sim \frac{i \, k}{a(t)} \; {\psi}_{\vec
k}(t) \; . 
\ee 
For $ k \gg a(t) \; H $ the constraint equation
(\ref{vin}) entails that \cite{abramo,quant,quadru1} 
\be \label{conss} 
\psi_{\vec k}(t) \approx \frac{i \, a(t)}{2\,M^2_{Pl}\,k} \; \dot{\Phi}   
\; \phi_{\vec k} \; . 
\ee 
In slow-roll, 
\be \label{fidotSR} \dot{\Phi}
= - \frac{V'(\Phi )}{3 \; H } \left[ 1 + {\cal
O}\left(\frac1{N}\right) \right] = - H \; M_{Pl} \;  \sqrt{2 \;
\epsilon_v} \left[ 1 + {\cal O}\left(\frac1{N}\right)\right] \; ,
\ee 
where the slow-roll parameters $ \epsilon_v, \; \eta_v $ are of
the order $ 1/N $ \cite{1sN} and given by eqs.(\ref{slrsd}). 
Therefore, for subhorizon modes, 
\be 
\psi_{\vec k}(t) \approx -i \; \sqrt{2 \;
\epsilon_v} \; \left(\frac{H \;  a(t) }{k} \right)
\frac{\phi_{\vk}}{2 \; M_{Pl}} \; . 
\ee 
These identities, valid in the limit $ k \gg a(t) \; H $ allow to obtain 
an estimate for the different contributions to $ T_{00} $. 
The first line of eq.(\ref{Too}), namely the contribution from the 
Newtonian potential mode with comoving wavevector $ k $ is 
\be 
\langle T^{(\psi)}_{00}\rangle \approx 6 \, \epsilon_v \, H^2 \langle (
\phi_{\vk} )^2 \rangle \label{Tphi} \; . 
\ee 
The first three terms in the second line of eq.(\ref{Too}) 
(the quadratic contribution from the scalar field fluctuations) is 
\be \label{Tvarphi} 
\langle T^{\phi}_{00} \rangle \approx 
\left(\frac{k}{a(t)}\right)^2 \langle
( \phi_{\vk})^2 \rangle \; , 
\ee 
and the cross term is: 
\be
\label{Tphivarphi} V'(\Phi) \;  \langle \psi_k   \; \phi_k \rangle
\approx \epsilon_v \, H^2 \,  \left(\frac{a(t) \;
H}{k}\right)\langle ( \phi_{\vk})^2 \rangle \; , 
\ee 
Therefore, in slow-roll, $ \epsilon_v, \; \eta_v \ll 1 $ and for 
subhorizon modes $ k \gtrsim a(t) \; H $, the leading contribution to 
the energy
momentum tensor for the scalar fluctuations is given by the
contribution from the inflaton fluctuations, namely 
\be 
\langle T_{00} \rangle \simeq \frac12 \; \langle (\dot{ \phi})^2\rangle +
\frac{\langle (\nabla \phi)^2\rangle}{2\,a^2(t)} + \frac{V''(\Phi
)}2 \; \langle \phi^2\rangle \,. \label{Toolead} 
\ee 
Furthermore, in terms of the slow-roll parameter 
$ \eta_v , \; V''= 3 \; \eta_v \;
H^2 $ and for subhorizon wavevectors with $ k\gg a(t) \; H $ the
last term in eq.(\ref{Toolead}) is subdominant and will be
neglected. Hence, the contribution to the energy momentum tensor
\emph{from subhorizon fluctuations during the slow-roll stage} is
determined by the subhorizon quantum fluctuations of the inflaton
and given by 
\be \label{T00sub} 
\langle T_{00} \rangle   \simeq
\frac12 \; \langle (\dot{ \phi})^2\rangle + \frac{\langle (\nabla
\phi)^2\rangle}{2\,a^2(t)} \; . 
\ee 
This analysis allows us to connect with the the results in 
ref.\cite{motto2} for inflaton fluctuations.

The inflaton fluctuation obeys the equation of motion 
\be \label{eqnofmot} 
\ddot{ \phi}_{\vk} +3 \; H \;
\dot{\phi}_{\vk}+\left[3 \,H^2 \,\eta_v +\frac{k^2}{a^2(t)} \right]
\phi_{\vk}= 0 \; . 
\ee 
In what follows it is convenient to pass to conformal time 
eq.(\ref{confo}) in terms of which, the FRW metric takes
the form eq.(\ref{metcon}) and the scalar factor is given in slow-roll
by eq.(\ref{adeta}).

In conformal time $ \eta $ the solution of eq.(\ref{eqnofmot}) is
given by 
\be \label{expalfa} 
\phi_{\vec{k}}(\eta) =
\frac1{a(\eta)}\left[\alpha_{\vk} \; S_\phi(k,\eta)+
\alpha^{\dagger}_{-\vk} \; S_\phi^*(k,\eta)\right] \; ,
\ee 
where the mode functions $ S_\phi(k,\eta) $ are
solutions of the wave equation
\be \label{phieqn}
\left[\frac{d^2}{d\eta^2}+k^2 +M^2 \; a^2(\eta)-
\frac{a''(\eta)}{a(\eta)} \right]S_\phi(k,\eta) = 0  \; , 
\ee 
here, 
\be \label{mass2} 
M^2 = V''(\Phi ) = 3 \;  H^2 \;
\eta_v \; . 
\ee  
and prime stands for derivative with respect to the
conformal time. Using eqs.(\ref{slrsd}) and (\ref{adeta}), 
this equation simplifies during slow-roll to 
\be\label{fluqeq} 
\left[\frac{d^2}{d\eta^2}+k^2 - \frac{\nu^2_\phi -\frac14}{\eta^2}
\right]S_\phi(k,\eta) = 0  \; , 
\ee 
where,
\be \label{Seqn2} 
\nu_\phi =
\frac32+\epsilon_v -\eta_v +  {\cal O}\left(\frac1{N^2}\right)\; .
\ee 
The scalar fluctuations $ {\phi}_{\vk} $ therefore obey wave equations
similar to the scalar curvature and tensor fluctuations
as follows comparing eqs.(\ref{eqnz}), (\ref{eqnC}) and (\ref{fluqeq}). 
These Hankel equations only differ on the value of the corresponding
index $ \nu $.

The operators $ \alpha_{\vk}, \; \alpha^{\dagger}_{\vk} $ in eq.
(\ref{expalfa}) obey the usual canonical commutation relations eq.(\ref{ccr}).

\subsubsection{Tensor perturbations}

The expectation value of the energy-momentum pseudo-tensor of tensor
perturbations in a quantum state has been obtained in ref.\cite{abramo,quant,quadru1} 
(see also ref.\cite{giova2}) and is given by
\be \label{tensorT00} 
\langle T^{(T)}_{00} \rangle = M^2_{Pl}
\Bigg\{ H  \; \langle
  \dot{h}_{kl} h_{kl} \rangle + \frac18
\left[ \langle \dot{h}_{kl} \; \dot{h}_{kl}\rangle +
\frac1{a^2(t)}\langle
  \nabla h_{kl} \;  \nabla h_{kl}  \rangle \right] \Bigg\} \; ,
\ee 
where the dot stands for derivative with respect to cosmic time.
Tensor perturbations (gravitational waves) are gauge invariant
and were analyzed in sec. \ref{sec:tensor}.

To leading order in slow-roll the mode functions for gravitational waves 
eqs.(\ref{Sten})-(\ref{eqnC}) obey the same equations of motion as for scalar fields but 
with vanishing mass, namely setting $ \eta_v =0 $ in eqs.(\ref{fluqeq})-(\ref{Seqn2}).

\subsubsection{The Transfer Function of Initial Conditions and its 
Asymptotic Behaviour}

For gauge invariant scalar perturbations, the analysis leading to
eq.(\ref{T00sub}) indicates that in order to study the energy
momentum tensor for general initial conditions it is enough to
consider the leading order in the slow-roll expansion. Consistently
with neglecting the contributions from the Newtonian potential as
well as the term proportional to $ V''[\Phi] $ for the inflaton
fluctuations, we set $ \nu=3/2 $ in the expression for the mode
functions eq.(\ref{gnu}). This simplification results in
considering the scalar field fluctuations as massless and minimally
coupled to gravity. 

The energy density in the vacuum state defined by the new initial
conditions is 
\be 
\rho = {}_\mathcal{S}\langle 0| T_{00}|0
\rangle_\mathcal{S} \label{rho}\,. 
\ee 
The renormalized energy
density from the fluctuations of the inflaton field is found to
be \cite{motto2,effpot,quant,quadru1}
\be \label{rhoren} 
\rho = \rho^{BD}+ I_1 + I_2 \; ,
\ee
where $ \rho^{BD} $ corresponds to the Bunch-Davies vacuum initial
conditions $ N_k=0 $ and 
\bea 
I_1  & = & \frac1{2 \, \pi^2}
\int_0^\infty dk\,k^2\, \left\{N_\phi(k)\, |\dot{F}(k,\eta)|^2+
\sqrt{N_\phi(k)[1+N_\phi(k)]}\,
\mathrm{Re}\left[e^{-i\theta_k}\left(\dot{F}(k,\eta)\right)^2\right]
\right\} \label{I1}\\
I_2  & = & \frac1{2 \, \pi^2} \int_0^\infty dk\,k^2\,\frac{k^2}{a^2}
\left\{N_\phi(k) \,| {F}(k,\eta)|^2+ \sqrt{N_\phi(k)[1+N_\phi(k)]}\,
\mathrm{Re}\left[e^{-i\theta_k}\left( {F}(k,\eta)\right)^2\right]
\right\} \label{I2} 
\eea 
where $ F(k,\eta) $ is given in terms of
the Bunch-Davis mode function eq.(\ref{g32}) for $ \nu=3/2 $ as 
\be
F(k,\eta) =  (-H\,\eta) \; g_{\frac32}(k,\eta)   = \frac{H}{\sqrt{2
\; k^3}} \; e^{-i \; k \; \eta} \; (i-k \; \eta) \label{BDmode} \; \,. 
\ee
The power spectrum of the inflaton fluctuations is given by
\cite{hu}, 
\be 
P_{\phi}(k,t) =
{}_\mathcal{S}\langle0||\phi_k(\eta)|^2|0\rangle_\mathcal{S} =
P^{BD}_{\phi}(k,t)+ \frac{k^3}{\pi^2}\left\{ N_\phi(k) \; |
{F}(k,\eta)|^2+ \sqrt{N_\phi(k)[1+N_\phi(k)]} \;
\mathrm{Re}\left[e^{-i\theta_k}\left({F}(k,\eta)\right)^2\right]\right\}
\; , 
\ee 
where we used eq.(\ref{expa}) and 
\be 
P^{BD}_{\phi}(k,t) =
\frac{k^3}{2\pi^2} \; | {F}(k,\eta)|^2 \; . 
\ee 
We find, 
\bea \label{I1fin} 
&&I_1= \frac{(H\eta)^4}{(2\pi)^2} \int_0^\infty dk\,
k^3 \left\{ N_\phi(k)-\sqrt{N_\phi(k)[1+N_\phi(k)]}\cos[2 \, k \,
\eta+\theta_k] \right\} \\ \cr 
&& I_2 =\frac{(H^2\eta)^2}{(2\pi)^2}
\int_0^\infty dk \, k  \left\{ N_\phi(k)\,(1+k^2 \,
\eta^2)-\sqrt{N_\phi(k)[1+N_\phi(k)]}\left[(1-k^2 \, \eta^2) \cos[2
\, k \, \eta+\theta_k]+2k \; \eta \sin[2 \, k \,
\eta+\theta_k]\right]\right\}\label{I2fin}  \; , \\ \cr 
&& P_{\phi}(k,t)= \left(\frac{H}{2\pi}\right)^2 \left\{(1+k^2 \,
\eta^2)[1+2 \; N_\phi(k)]-2 \; \sqrt{N_\phi(k)[1+N_\phi(k)]}
\left[(1-k^2 \, \eta^2)\cos[2 \, k \, \eta+\theta_k]+2 \, k \, \eta
\sin[2 \, k \, \eta+\theta_k]\right] \right\} \nonumber  \; . 
\eea
Evaluating the power spectrum after horizon crossing $ |k \, \eta|
\ll 1 $, yields 
\be \label{Powerratio} 
\frac{P_\phi}{P^{BD}_\phi}\Bigg|_{|k \; \eta| \ll 1}
= 1+ D_\phi(k) \; ,
\ee 
where we have introduced the
\emph{transfer function for initial conditions} 
\be \label{Dofk}
D_\phi(k)= 2 \;| {B}_\phi(k)|^2 -2 \; 
\mathrm{Re}\left[A_\phi(k)B^*_\phi(k)\right]
= 2 \; N_\phi(k)-2 \; \sqrt{N_\phi(k)[1+N_\phi(k)]} \; \cos
\theta_k \; .
\ee 
The integrals $ I_{1,2} $ are finite provided that
asymptotically for $ k\rightarrow \infty $ the occupation numbers
behave as 
\be \label{adiabatic}
N_\phi(k)=\mathcal{O}\left(\frac1{k^{4+\delta}}\right) \; , 
\ee 
with $ \delta >0 $. Namely, the finiteness of the energy momentum tensor
constrains the asymptotic behaviour of the occupation numbers to
vanish faster than $ 1/k^4 $ for $ k \to \infty $
\cite{motto2}. Of course, this asymptotic condition leaves a large
freedom on the occupation numbers $ N_k $.

We systematically impose the constraint eq.(\ref{adiabatic}) which
guarantees the finiteness of energy momentum tensor 
\cite{effpot,quant,quadru1}. This is not
always the case for initial conditions considered in the literature
(see ref.\cite{otros}).

Let us establish a bound on the large momentum behavior of $ N_k $
inserting the asymptotic behavior 
\be \label{ocupa} 
N_k = N_\mu
\left(\frac{\mu}{k}\right)^{4+\delta}  \; , 
\ee 
with $ 0< \delta \ll 1 $ in the integrals $ I_{1,2} $. Namely, assuming
that the integrals are dominated by the region of high momenta $ k/H
\gg 1 $ and that the occupation number attains the largest possible
values consistent with ultraviolet finiteness
[eq.(\ref{adiabatic})]. We observe that  $ k \; |\eta| \gg 1 $ in
the early stages of inflation for large $ k $, and that the maximum
contribution from these integrals are at early time $ \eta \sim -1/H
$. Hence, the oscillatory terms in $ I_1, \; I_2 $ average out and
we have from eqs.(\ref{I1fin})-(\ref{I2fin}) \cite{quadru1}, 
\be \label{integ} I_1
\sim I_2 \sim \frac{N_\mu}{(2\pi)^2} \,\frac{\mu^4}{\delta}  \; .
\ee 
The contribution from the fluctuations to the energy momentum
tensor does not lead to large back reaction effects affecting the
inflationary dynamics provided that $ I_1, \; I_2 \ll  H^2 \;
M^2_{Pl} $, which yields 
\be  \label{Nmu} 
{N_\mu} \ll 2 \, \pi^2 \;
\frac{H^2 \; M^2_{Pl}}{\mu^4} \; \delta \; . 
\ee
Eq.(\ref{Nmu}) provides an occupation number distribution exhibiting
the largest asymptotic value compatible with an UV finite energy
momentum tensor. This maximal occupation number distribution falls
off for $ k \to \infty $ with the minimal acceptable power tail
exponent $ 4 + \delta $ with $ \delta \ll 1 $ \cite{quadru1}.

\medskip

Gravitons are massless particles with two independent polarizations,
therefore their energy momentum tensor is given by eq.(\ref{T00sub})
times a factor two. The first term in the energy momentum
pseudotensor for gravitational waves eq.(\ref{tensorT00}) features
only one time derivative, which results in only one factor $ k $ for
large momenta, whereas the terms with two time or spatial
derivatives yield $ k^2 $. Therefore, the first term is subdominant
in the ultraviolet and the short wavelength  contribution to the
energy momentum (pseudo) tensor of gravitational waves is the same
as that for a free massless scalar field, up to a factor 2 from the
physical polarization states \cite{giova2,quant}. Therefore, we
can directly extend the results obtained above for scalar
fluctuations to the case of tensor fluctuations.

\medskip

Small backreaction effects from the fluctuations is a necessary
consistency condition for the validity of the inflationary scenario.
In addition, the condition that different initial states
\emph{should not affect} the renormalization aspects of the energy
momentum tensor is a consistency condition on the renormalizability
of the effective field theory of inflation: the theory should be
insensitive to the short distance physics for \emph{any} initial
conditions. These criteria lead to the following important
consequences \cite{quadru1}:

\begin{itemize}
\item{If $ \mu \sim M_{Pl} $ then $ N_\mu  \lesssim H^2/M^2_{Pl}   \ll 1 $
because $ H/M_{Pl} \ll 1 $  in the effective field theory expansion
and the effect of initial conditions becomes negligible. }

\item{For $ \mu \sim \sqrt{H\,M_{Pl}} \sim
N^{\frac14} \; M $, namely  $ \mu $ of the order of the scale of
inflation during the slow-roll stage, then $ N_{\mu} \lesssim 1 $.
For example for $ \delta \sim 0.01 $ one obtains   $ N_\mu \sim 0.1 $.
If $ \mu \ll \sqrt{H\,M_{Pl}} $, for example $ \mu \sim H $, the
bound eq.(\ref{Nmu}) is rather loose allowing a wide range of $
N_\mu $ with potentially appreciable effects. }

\item{The condition that the occupation number falls off faster
than $ 1/k^4 $ for large wavevector, implies that the possible
effects from different initial conditions are more prominent for the
smaller wavevectors, those that exited the Hubble radius the
\emph{earliest}. For cosmologically relevant wavevectors, these are
those that crossed about 60 efolds before the end of inflation.
Today those wavevectors correspond to the present Hubble scale,
hence the low multipoles in the CMB.  }
\end{itemize}

We conclude that consistent with renormalizability and small back
reaction there may be a substantial effect from the initial
conditions when the characteristic scale is $ \mu \leq \sqrt{H\;
M_{Pl}} $. The rapid fall-off of the occupation numbers $ N_\phi(k)
$ for subhorizon wavelengths and the back-reaction constraint
eq.(\ref{Nmu}) entails that for these modes the transfer function
eq.(\ref{Dofk}) for initial conditions simplifies to 
\be \label{simpleD} 
D_\phi(k) \buildrel{N_\phi(k) \ll 1}\over=
 -2 \; \sqrt{N_\phi(k)} \; \cos\theta_k \; ,
\ee 
and that the \emph{smaller values} of $ k $ yield the
\emph{larger corrections  from initial conditions}. The result
eq.(\ref{simpleD}) implies a {\bf suppression} of the power
spectrum for $ \cos\theta_k >0 $.  These observations will be
crucial below when we study the effect of initial conditions on the
multipoles of the CMB \cite{quadru1}.

While this discussion focused on the fluctuations of the inflaton,
they are directly applicable to the case of gauge invariant
perturbations.

The contribution from gravitational waves to the
energy momentum (pseudo) tensor is gauge invariant and up to a
factor of two from the polarizations is exactly of the form
eq.(\ref{T00sub}) with $ \phi $ replaced by $ h $
\cite{quant}. Thus, the constraint on the occupation number
eq.(\ref{ocupa})-(\ref{Nmu}) from the analysis of the backreaction
and renormalizability translate \emph{directly} to the
case of gravitational waves for the occupation number $ N_T(k) $. 

This implies that corrections to the power spectrum of tensor modes
from initial conditions are substantial if $ \mu $, the asymptotic $ k $
scale of $ N_T(k) $, is $ \mu \leq  \sqrt{H \; M_{Pl}} \sim N^{\frac14} 
\; M $, [see discussion above for $ N_{\mu} $]. 
We get from eq.(\ref{Dofkh}) for $ N_T(k) \ll 1 $ and to leading
order in slow-roll, 
\be \label{Dh2} 
D_T(k) \buildrel{N_T(k)\ll
1}\over= -2 \; \sqrt{N_T(k)} \; \cos\theta_k \; . 
\ee 
Again, for a
positive $ \cos \theta_k $, we have a {\bf negative} $ D_T(k) $.
That is, the initial conditions {\bf suppress} the tensor power
spectrum in such case.

\medskip

We have focused on the backreaction effects from
initial conditions beginning with the gauge invariant energy
momentum tensor for scalar and tensor perturbations. Since the
fluctuation modes are initialized on a fixed time hypersurface while
their wavelength are well inside the Hubble radius, we established a
correspondence with ref.\cite{motto2} which refer solely to the
quantum fluctuations of the inflaton field. The effect of different
initial conditions is encoded in the Bogoliubov coefficients, and in
particular in the occupation numbers $ N_k $ and the phases $
\theta_k $. Ultraviolet allowed initial conditions require that $
N_k $ diminishes faster than $ 1/k^4 $ for asymptotically large
momenta. Small backreaction effects require in general that $ N_k
\ll 1$.

This analysis applies to UV allowed initial conditions on the
quantum fluctuations associated with gauge invariant variables, 
both scalar and tensor perturbations, studied in sec. \ref{gistf}.

\subsubsection{The effect of initial conditions on the low multipoles of the 
CMB}\label{eiclm}

We have shown above that the fast fall off of the occupation number
$ N(k) $ (for the corresponding perturbation) entails that initial
conditions can only provide substantial corrections for perturbation
modes whose wavevectors crossed out of the Hubble radius
\emph{early} during inflation. In turn, today  these wavevectors
correspond to scales of the order of the  Hubble radius, namely to
the low multipoles in the CMB.

In the region of the Sachs-Wolfe plateau for $ l \lesssim 30 $, the
matter-radiation transfer function can be set equal to unity and the
$ C_l's $ are given by \cite{hu,CMBgiova} 
\be 
C_l =\frac{4\pi}{9} \int_0^\infty \frac{dk}{k}\,  {P}_X(k) \left\{
j_l[k(\eta_0-\eta_{LSS})]\right\}^2 \label{Cls} \; , 
\ee 
where $ P_X $ is the power spectrum of the corresponding perturbation, $
X=\mathcal{R} $ for curvature perturbations and $ X=T $ for tensor
perturbations,  $ j_l(x) $ are spherical Bessel functions \cite{abra} 
and $ \eta_0-\eta_{LSS} $ is the comoving
distance between today and the last scattering surface (LSS)
given by eqs.(\ref{etaLSS}) and (\ref{329}).

Notice that $ k/H_0 \sim d_H/\lambda_{phys}(t_0) $ is the ratio
between today's Hubble radius and the physical wavelength. The power
spectra for curvature ($ \mathcal R $) or gravitational wave (T)
perturbations are of the form given by eqs.(\ref{powR}),
(\ref{potBD}), (\ref{potT}), (\ref{ptbd}), 
\be 
{P}_X = |\Delta_k^X|^2 \left( \frac{k}{k_0}\right)^{n_s-1}
\left[1+D_X(k)\right]\label{power} \; , 
\ee 
with $ n_s=n_\mathcal{R}
$ for curvature perturbations, $ n_s = 1+n_T $ for tensor
perturbations, and $ k_0 \sim  H_0 $ is a pivot scale. Then,
from eqs.(\ref{Powerratio}) and (\ref{Cls}), the relative change 
$ \Delta C_l $ in the $ C_l's $ due to the effect of generic initial 
conditions (generic vacua), is given by 
\be \label{DelC}
C_l \equiv C ^{BD}_l + \Delta C_l
\quad , \quad \frac{\Delta C_l}{C_l} = \frac{\int^\infty_0
D_X(\kappa \; x)~ f_l(x) \; dx}{\int^\infty_0 f_l(x) \; dx} 
\ee 
where $ x= k(\eta_0-\eta_{LSS}) = k/\kappa $ and from eq.(\ref{329}), $ \kappa \equiv   
H_0/3.296\ldots . \; D(\kappa\,x) $ is the transfer function of
initial conditions for the corresponding perturbation and 
\be 
f_l(x)= x^{n_s-2}[j_l(x)]^2 \,. \label{fls} 
\ee 
\noindent We now focus on
curvature perturbations since these are directly related to the
temperature fluctuations \cite{WMAP1,WMAP3,WMAP5}. Since $ n_s \sim 1 $, 
the functions $ f_l(x) $ are strongly peaked at $ x
\sim l $ as shown by eq.(\ref{maxjl}). Therefore, $ \Delta C_l/C_l
$ is dominated by wavenumbers $ k \sim l \; \kappa $.

Low multipoles $ l $ correspond to  wavelengths \emph{today} of the
order of the Hubble radius. These wavelengths crossed the Hubble
radius about 60  efolds before the end of inflation. Therefore,
since inflation lasted a total number of efolds $ N_{tot} \gtrsim
64 $, these wavevectors were subhorizon during the first few
efolds, namely during the slow-roll stage $ k\gg H $. As already
discussed, let us take for these wavevectors the occupation number $
N_k \ll 1 $  as given by the asymptotic expression eq.(\ref{ocupa}) 
and assume that the angles $ \theta_k $ are slowly varying functions of 
$ k $ in the region of $ k $ corresponding to \emph{today's Hubble 
radius} so that $ \cos\theta_k \approx \overline{\cos \theta} $. Then, we 
find that the fractional change in the coefficients $ C_l $ is given 
by \cite{quadru1}
\be
\frac{\Delta C_l}{C_l} \approx - 2 \; \sqrt{N_{\mu}} \;
\left(\frac{3.3 \; \mu}{H_0} \right)^{2+\frac{\delta}2} \;
\overline{\cos \theta}~~ \frac{I_l(n_s-2-\frac{\delta}2)}{I_l(n_s)}
\label{Claps} \ee where \be I_l(n_s) = \frac1{2^{3-n_s}} \;
\frac{\Gamma\left(3-n_s \right) \; \Gamma\left(\frac12 \; (2l+1-3+n_s+1)
\right)}{\Gamma^2\left(2-\frac12 \; n_s) \right) \;
\Gamma\left(\frac12 \; (2l+1+3-n_s+1) \right)} \label{Ilns}  \; . 
\ee 
To obtain an estimate of the corrections, we take the values $ n_s=1,
\; \delta=0 $ and find 
\be 
\frac{\Delta C_l}{C_l} \approx - \frac43
\; \sqrt{N_{\mu}} \;  \left(\frac{3.3\,\mu}{H_0} \right)^ 2
\, \frac{\overline{\cos\theta}}{(l-1)(l+2)} \; . \label{Cl10} 
\ee
The $ \sim 1/l^2 $ behavior is a result of the $ 1/k^2 $ fall off of
$ D(k) $, a consequence of the renormalizability condition on the
occupation number.

When the scale $ \mu $ in the asymptotic form of the
occupation number eq.(\ref {ocupa}) is of the order of the
largest scale of cosmological relevance \emph{today}, one has 
$ \mu \sim H_0 $ and for example with $ N_\mu \sim 0.1 $ we find that 
the fractional change in the quadrupole is given by: 
\be  \label{quad} 
\frac{\Delta C_2}{C_2}\sim - {\overline {\cos \theta}} \; ,
\ee 
namely a {\it suppression} of the
order of $ \sim 1 $ in the quadrupole provided that $
\overline{\cos\theta} \sim 1 $. This corresponds to $ \mu $ of the
order of the Hubble parameter during the slow-roll stage 
\cite{quadru1}. Namely, changing the initial
conditions in such a way {\bf can} explain the observed suppression
of the CMB quadrupole \cite{cobe,WMAP1,WMAP3,WMAP5}.

We emphasize that these are general arguments based on the criteria
of renormalizability and small backreaction which initial conditions
must fulfill \cite{quadru1}.

In ref.\cite{quadru2} (see sec. \ref{fastroll}) we showed that these 
initial conditions are effectively equivalent to the presence of a 
fast-roll stage before the slow-roll regime. We show in ref. 
\cite{quadru2} that a short stage just prior to the onset of slow-roll
inflation and in which the inflaton field evolves \emph{fast},
imprints by the beginning of slow-roll a behaviour on the curvature 
perturbations which is similar to  the non-Bunch-Davis initial 
conditions considered above.

\subsection{The early fast-roll inflationary stage  and the CMB quadrupole 
suppression}\label{fastroll}

Although there are no statistically significant departures from the
slow-roll inflationary scenario at small angular scales ($ l\gtrsim
100 $), the WMAP data again confirm the surprisingly
low quadrupoles $ C_2^{TT} $ and   $ C_2^{TE} $
\cite{WMAP3,WMAP5} and suggest that
it cannot be completely explained by galactic foreground
contamination. The low value of the quadrupole has been an
intriguing feature on large angular scales since first observed by
COBE/DMR \cite{cobe}, and confirmed by the WMAP data 
\cite{WMAP3,WMAP5}.

\medskip

In order to asses the statistical relevance of the observed
quadrupole suppression, we studied the best fit to the $\Lambda$CDM model
using the WMAP5 data. We find that the probability that the quadrupole is as low or lower 
than the observed value is just 0.031. Even if one does not care about the 
specific multipole and looks for any multipole as low or lower than the 
observed quadrupole with respect to the $\Lambda$CDM model value, then
the probability remains smaller than 5\%. Therefore, it is relevant to 
find a cosmological explanation of the quadrupole supression beyond the 
$\Lambda$CDM model.

\medskip

Generically, the classical evolution of the inflaton has a brief {\bf
fast-roll stage} that precedes the slow-roll regime. The fast-roll
stage leads to a purely attractive potential in the wave equations of
curvature and tensor perturbations. Such potential is a \emph{generic} feature of this brief
\emph{fast-roll} stage that merges smoothly with   slow-roll
inflation. This stage is a consequence of generic initial conditions
\emph{for the classical inflaton dynamics} in which the kinetic and
potential energy of the inflaton are of the same order, namely, the
energy scale of slow-roll inflation. During the early fast-roll
stage the inflaton evolves rapidly during a brief period,  but slows
down by the cosmological expansion settling in  the slow-roll stage
in which the kinetic energy of the inflaton is much smaller than its
potential energy.

As shown in refs. \cite{quadru1,quadru2,lasenby} the attractive potential 
in the wave equations of
curvature and tensor perturbations during the fast-roll stage 
leads to a {\bf suppression} of the
quadrupole moment for CMB and B-mode angular power spectra.
Both scalar and tensor low multipoles are suppressed. However,
the potential for tensor perturbations is about an order of magnitude
smaller than the one for scalar fluctuations and hence the suppression
of the low $ \ell $ tensor perturbations is much less significant 
\cite{quadru1,quadru2}.

The observation of a low quadrupole \cite{cobe,WMAP3,WMAP5} 
and the surprising alignment of quadrupole
and octupole \cite{tegmark,virgo} sparked many different proposals
for their explanation \cite{expla}.

The fast-roll explanation of the quadrupole does not require to introduce new
physics neither modifications of the inflationary potential.
The only new feature is that the quadrupole mode should exit the horizon during
the generic fast-roll stage that precedes slow-roll inflation.

A single {\bf new} parameter emerges dynamically due to the fast-roll stage: the
comoving wave number $ k_{tran} $, characteristic scale of the attractive potential
felt by the fluctuations during fast-roll.  The fast-roll stage modifies the
initial power spectrum by a transfer function $ D(k) $ that we compute solving
the classical inflaton evolution equations (see fig. \ref{dkflin4}). $ D(k) $ 
effectively suppresses the primordial power for $ k < k_{tran} $ and possesses the 
scaling property $ D(k) = \Psi(k/k_{tran}) $ where $ \Psi(x) $ is an universal function.
$ D(k) $ has a main peak around $ k_M \simeq 1.9 \; k_{tran} $ and oscillates around
zero with decreasing amplitude as a function of $ k $ for $ k > k_M $.  $ D(k) $
vanishes asymptotically for large $ k $, as expected.

\medskip 

We reported in ref.\cite{quamc} the results of a MCMC analysis of the 
WMAP-3, small--scale CMB and SDSS data including the fast-roll stage and 
find the value $ k_{tran} = 0.290 \; {\rm Gpc}^{-1} $. This mode 
$ k_{tran} $ happens to exit the horizon precisely 
at the transition from the fast-roll
to the slow-roll stage. The quadrupole mode $ k_Q = 0.238 \; 
{\rm Gpc}^{-1} $ exits the horizon {\bf during} the fast-roll stage 
approximately {\bf 0.2 efolds} earlier than $ k_{tran} $. We compare in
ref.\cite{quamc} the fast-roll fit with a fit without fast-roll but 
including a sharp lower cutoff on the primordial power. That is, setting 
the the primordial power to zero for $ k < k_{tran} $. 

We analyze with MCMC and compare three classes of cosmological models:
\begin{itemize}
\item The usual slow-roll $\Lambda$CDM, the $\Lambda$CDM$+r$ and the
$\Lambda$CDM+$r$+BNI models.  BNI stands for {\em 
Binomial New Inflation}. In this last model we {\bf enforce} the 
theoretical functional relation between $ n_s $ and $ r $ valid in BNI.
\item The slow-roll $\Lambda$CDM+$r$+BNI model with a sharp 
cutoff for $ k < k_{tran} $.
\item The $\Lambda$CDM+$r$+BNI model including both fast and
slow-roll stages.
\end{itemize}

We observe that the oscillatory form of the fast-roll transfer function $
D_\mathcal{R}(k) $, by {\bf depressing as well as enhancing} the 
primordial power spectrum at long wavelengths (see fig. \ref{dkflin4}),
leads also to new superimposed {\bf
oscillatory corrections} on the low multipoles. As far as fitting to 
current data is concerned, such corrections are more effective than the 
pure reduction caused by a sharp cutoff. The fast-roll oscillations
yield better gains in likelihood than the sharp cutoff case \cite{quamc}. 

\medskip

The quadrupole suppression by the early fast-roll stage can be simply 
understood by causality in a qualitative way. During fast-roll inflation
the Hubble parameter decreases fast (much faster than during slow-roll 
inflation) as shown in fig. \ref{fR1}. Therefore, the Hubble radius 
$ d_H $ grows with time and the $ d_H $ curve in fig. \ref{inflation} is 
not horizontal but down bended during the fast-roll stage. Hence, the 
fluctuation modes cannot exit the horizon before slow-roll and therefore
the primordial power gets suppressed for $ k < k_{tran} $.

\subsubsection{The Effect of Fast-roll on the Inflationary 
Fluctuations.}\label{frcua}

Both scalar curvature and tensor fluctuations obey the Schr\"odinger-type 
equation
\be \label{Wgen}
\left[\frac{d^2}{d\eta^2}+k^2-W(\eta)\right]S(k;\eta) = 0 \; . 
\ee
with $ \eta $ as the coordinate, $ k^2 $ as the
energy and $ W(\eta) $ the potential determined by the classical inflaton dynamics.
$ W_\mathcal{R}(\eta) $ and $ W_T(\eta) $ are given by eqs.(\ref{Wb}) and 
(\ref{WRT}). The Schr\"odinger form of the mode equations (\ref{Wgen})
in one dimension suggests to consider them more generally
as a \emph{scattering problem} by a potential \cite{quadru1,quadru2}. 

During slow-roll both potentials $ W_\mathcal{R}(\eta) $ and $ W_T(\eta) $ 
are {\bf repulsive} and have the shape of a centrifugal barrier:
$$ 
W(\eta) \simeq \frac{\nu^2-1/4}{\eta^2} \quad {\rm slow-roll},
$$
where
\be
\nu =  \Bigg\{ \begin{array}{l}
\nu_{\mathcal{R}}=\frac32+ 3 \, \epsilon_v -\eta_v +  {\cal O}\left(\frac1{N^2}\right)
\quad \mathrm{for~curvature~perturbations}        \\ \\
\nu_T = \frac32+ \epsilon_v + {\cal O}\left(\frac1{N^2}\right) \quad 
 \mathrm{for~tensor~perturbations} \; .  \\
\end{array} \label{defnu}
\ee
We choose as initial condition the Bunch-Davies asymptotic condition
\be \label{BD}
S(k;\eta \rightarrow -\infty) = \frac{e^{-i \; k \; \eta}}{\sqrt{2k}} \; . 
\ee
We formally consider here inflation and the conformal time starting at $
\eta = -\infty $. However, it is natural to consider that the
inflationary evolution of the universe starts at some negative value
$ \eta_i < {\bar \eta} $, where $ {\bar \eta} $ is  the conformal
time when fast-roll ends and slow-roll begins.

It is convenient to explicitly separate the behavior of $ W(\eta) $
during the slow-roll stage by writing
\be\label{defV}
W(\eta)=  \mathcal{V}(\eta) +\frac{\nu^2-\frac14}{\eta^2} \; ,
\ee
The potential $ \mathcal{V}(\eta) $ is localized in the fast-roll stage \emph{prior} 
to slow-roll (during which cosmologically relevant modes cross out of the Hubble radius),
$ \mathcal{V}(\eta) $ vanishes during slow-roll. In terms of the potential
$ \mathcal{V}(\eta) $ the equations for the quantum fluctuations read,
\be 
\left[\frac{d^2}{d\eta^2}+k^2-\frac{\nu^2-\frac14}{\eta^2}-
\mathcal{V}(\eta)  \right]S(k;\eta) = 0 \; . \label{eqnpsr2} 
\ee
During the slow-roll stage, the fluctuations equations (\ref{Wgen}) can be solved 
in close form in terms of Bessel functions as discussed in secs. 
\ref{4A}-\ref{sec:tensor}. This does not apply during the fast-roll stage. 

As discussed above, the potential $ \mathcal{V}(\eta) $ describes the
deviation from the slow-roll dynamics during the (brief) fast-roll
stage prior
to slow-roll and is vanishingly small for $ \eta > {\bar \eta} $,
where $ {\bar \eta} $ denotes the beginning of the slow-roll stage
during which modes of cosmological relevance today exit the Hubble
radius.

The retarded Green's function $ G_k(\eta,\eta') $ of
eq.(\ref{eqnpsr2}) for $ \mathcal{V}(\eta) \equiv 0 $   obeys 
\be
\label{green}
\left[\frac{d^2}{d\eta^2}+k^2-\frac{\nu^2-\frac14}{\eta^2}
 \right]G_k(\eta,\eta') = \delta(\eta-\eta') ~~;~~ G_k(\eta,\eta') =0~
\mathrm{for}~\eta'>\eta \quad , 
\ee 
it  is given by 
\be \label{Gret}
G_k(\eta,\eta') = i \left[g_\nu(k;\eta) \; g^*_\nu(k;\eta')-
g_\nu(k;\eta') \; g^*_\nu(k;\eta) \right] \Theta(\eta-\eta') \quad ,
\ee 
where $ g_\nu(k;\eta) $ is given by eq.(\ref{gnu}).

The solution of the mode equation (\ref{eqnpsr2}) can be written as
an integral equation using the  Green's function eq.(\ref{Gret}) 
\be
\label{sol} S(k;\eta)= g_\nu(k;\eta) + \int^{0}_{-\infty}
G_k(\eta,\eta') \; \mathcal{V}(\eta') \; S(k;\eta') \; d\eta' \;.
\ee 
This is the Lippmann-Schwinger equation familiar in potential
scattering theory. Inserting eq.(\ref{Gret}) into eq.(\ref{sol})
yields, 
\be \label{solu} 
S(k;\eta)= g_\nu(k;\eta) + i \;
g_\nu(k;\eta)\,\int^{\eta}_{-\infty}
 g^*_\nu(k;\eta') \; \mathcal{V}(\eta') \;  S(k;\eta') \; d\eta'-i  \;
g^*_\nu(k;\eta)\,\int^{\eta}_{-\infty}
 g_\nu(k;\eta') \; \mathcal{V}(\eta') \;  S(k;\eta') \; d\eta'\quad .
\ee
 This solution has the Bunch-Davies asymptotic condition
\be 
S(k;\eta \rightarrow -\infty) = \frac{e^{-i \; k \; \eta}}{\sqrt{2k}} \; . 
\ee
Since $ \mathcal{V}(\eta) $ vanishes for  $ \eta > {\bar \eta} $,  the
mode functions $ S(k;\eta) $ for  $ \eta > {\bar \eta} $ can be written 
as linear combinations of the mode functions $ g_\nu(k;\eta) $ and $
g^*_\nu(k;\eta) $,
\be \label{solSR}
S(k;\eta) = A(k) \; g_\nu(k;\eta) + B(k) \; g^*_\nu(k;\eta)
\quad , \quad \eta > {\bar \eta} \quad ,
\ee
where the coefficients $ A(k) $ and $ B(k) $ can be read from eq.(\ref{solu}),
\bea
A(k) & = &  1+ i\int^{0}_{-\infty} g^*_\nu(k;\eta) \; \mathcal{V}(\eta) \;  S(k;\eta) \;
d\eta\label{aofk} \cr \cr
 B(k) & = & -i \int^{0}_{-\infty}  g_\nu(k;\eta) \; \mathcal{V}(\eta)  \;
S(k;\eta) \;  d\eta\label{bofk}  \; .
\eea
The coefficients $ A(k) $ and $ B(k) $ are therefore {\bf calculated}
from the {\bf dynamics  before} slow-roll, that is, during fast-roll.
[recall that $ \mathcal{V}(\eta)  = 0 $ for  $ \eta > {\bar \eta} $ during slow-roll.]
$ A(k) $ and $ B(k) $ fulfil the constraint eq.(\ref{constraint}) and can therefore
be represented as in eq.(\ref{bogonum}).

\medskip

Starting with Bunch-Davies initial conditions for $ \eta \to
-\infty $, the action of the fast-roll potential $ \mathcal{V}(\eta) $ generates a mixture 
(Bogoliubov transformation) of the two linearly independent mode functions that result in the mode
functions eq.(\ref{solSR}) for $ \eta > {\bar \eta} $ when the potential 
$ \mathcal{V}(\eta) $ vanishes. This is clearly equivalent to starting the evolution
of the fluctuations at the \emph{beginning} of slow-roll $ \eta =
{\bar \eta} $ with initial conditions defined by the fast-roll Bogoliubov
coefficients $ A(k) $ and $ B(k) $ given by eq.(\ref{bofk}) as stressed in 
ref.\cite{quadru2}. Namely, one can obtain a similar suppression
on the quadrupole either taking into account the fast-roll stage
 {\bf or} changing the initial conditions of the
mode functions at the beginning of slow-roll.

\medskip

\begin{figure}[h]
\begin{center}
\begin{turn}{-90}
\psfrag{"chi.dat"}{$ {\cal H}(\tau) $ vs. $ \ln a $}
\psfrag{"pe.dat"}{$ p/\rho $ vs. $ \ln a $}
\includegraphics[height=12cm,width=8cm,keepaspectratio=true]{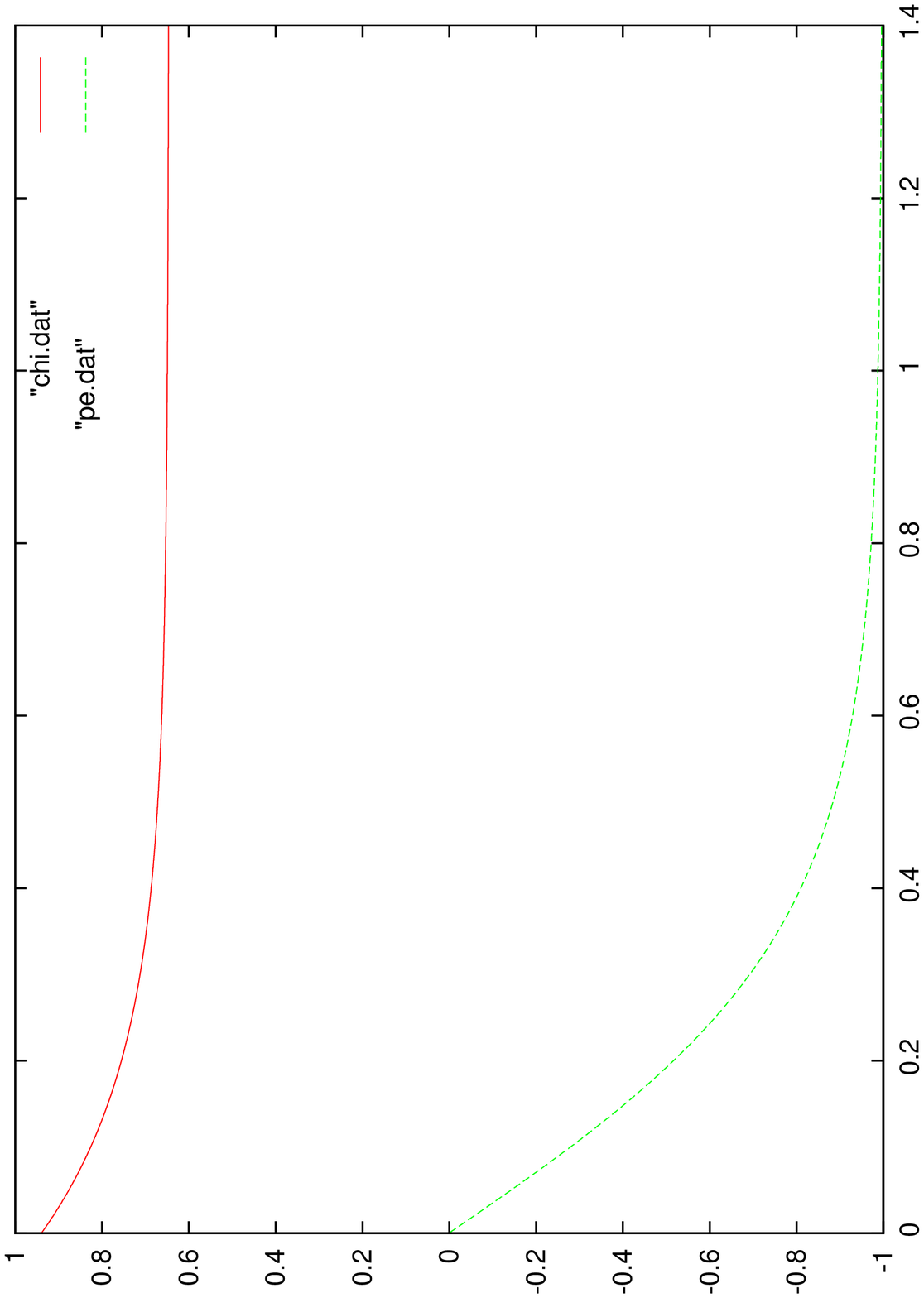}
\psfrag{"fi.dat"}{$ \chi(\tau) $ vs. $ \ln a $}
\psfrag{"fip.dat"}{$ {\dot \chi}(\tau) $ vs. $ \ln a $}
\psfrag{"epv.dat"}{$ \log [N \; \epsilon_v(\tau)] $ vs. $ \ln a $}
\includegraphics[height=12cm,width=8cm,keepaspectratio=true]{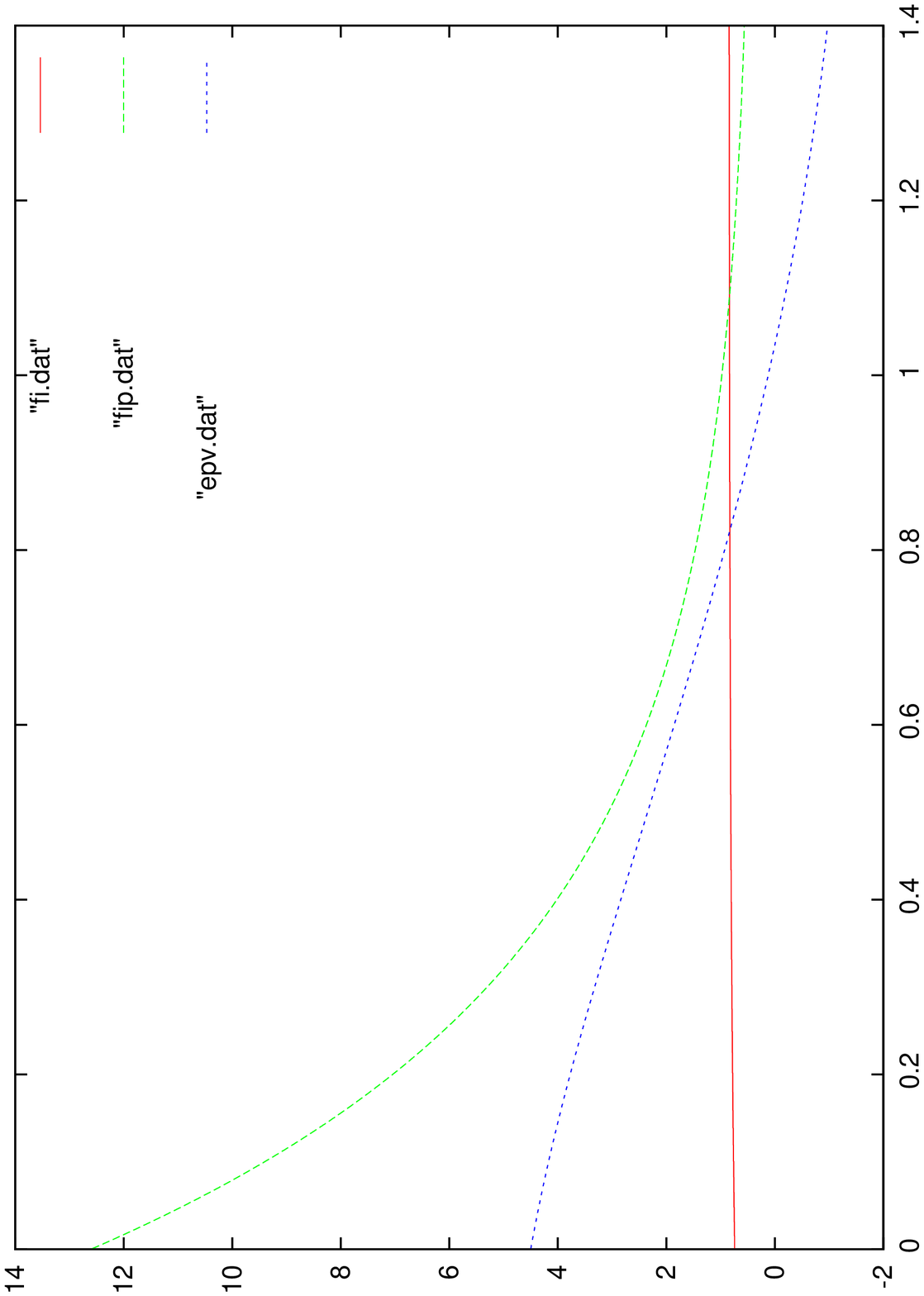}
\end{turn}
\caption{Upper panel: the Hubble parameter $ {\cal H}(\tau) $ and the equation of state
$ p/\rho $ as a function of $ \ln a $ during fast-roll inflation.
Lower panel: the dimensionless fields $ \chi(\tau), \; {\dot \chi}(\tau) $
and the parameter $ \log[\epsilon_v(\tau) \; N] $ as function of $ \ln a $ during fast-roll.
Notice that fast-roll ends when $ \epsilon_v(\tau) \times N = 1 $ at 
$ \ln a(\tau) = 1.0347\ldots $. 
Both figures are for the inflaton potential
eq.(\ref{wnue}) with $ y = 1.26 $ and $ N_{tot} \simeq 64 $ efolds of inflation.}
\label{fR1}
\end{center}
\end{figure}

The integral equation (\ref{solu}) can be solved iteratively in a perturbative
expansion if the potential $ \mathcal{V}(\eta) $
is small when compared to 
$$ 
k^2 - \frac{\nu^2-1/4}{\eta^2} \; .
$$
In such case, we can use for the coefficients $ A(k), \; B(k) $ the first
approximation obtained by replacing $ S (k;\eta') $ by
$ g_\nu(k;\eta') $ in the integrals eqs.(\ref{aofk})-(\ref{bofk}).
This is the Born approximation, in which
\be
A(k)   =    1+  i \int^{0}_{-\infty}
   \mathcal{V}(\eta)\,|g_\nu(k;\eta)|^2 \, d\eta   \quad  , \quad
 B(k)  =  - i  \int^{0}_{-\infty}
   \mathcal{V}(\eta)\, g^2_\nu(k;\eta)  \, d\eta \; .\label{bofk0}
\ee
That is, the standard slow-roll power spectrum $  P^{sr}(k) $ results modified by the
fast-roll stage as displayed in eqs.(\ref{curvapot}) and (\ref{powR}).

The simple expressions (\ref{bofk0}) are very illuminating. For
asymptotically large $ k $ the form eq.(\ref{fnuasy}) for the mode
functions can be used, and if the potential $ \mathcal{V}(\eta) $ is
differentiable and of compact support, an integration by parts
yields
 \be 
B(k) \buildrel{k \to \infty}\over=
-\frac1{4 \, k^2} \int^{0}_{-\infty}
 e^{-2 \, i \, k \; \eta} \; \mathcal{V}'(\eta) \, d\eta \label{bofk01} \; ,
\ee 
where the prime stands for derivative with respect to $ \eta $.
Therefore, according to the Riemann-Lebesgue lemma, $ N_k  =
|B(k)|^2 \lesssim 1/k^4 $ for large $ k $ and UV convergence in the
integrals for the energy momentum tensor is guaranteed. Hence, an
immediate consequence of the explanation of the initial conditions
as a scattering problem with a localized potential is that these
initial conditions are \emph{automatically} ultraviolet allowed.

\medskip

The transfer function of initial conditions given by eq.(\ref{DofkR}) can be computed
in the Born approximation, which is indeed appropriate in this situation, using
eqs.(\ref{bofk0}) for the Bogoliubov coefficients $ A(k) $ and $ B(k) $ to dominant
order in $ 1/N $  \cite{quadru2}
\be \label{Dborn2} 
D(k) = \frac1{k} \int^0_{-\infty} d\eta \;
\mathcal{V}(\eta) \left[\sin(2\, k \; \eta) \left(1  - \frac1{k^2 \,
\eta^2} \right)+ \frac2{k \, \eta} \, \cos(2\, k \, \eta) \right] \; .
\ee 
The potential $ \mathcal{V}(\eta) $ is obtained from eq.(\ref{defV}) as
$$
\mathcal{V}(\eta) = W(\eta) - \frac{\nu^2-1/4}{\eta^2} \; .
$$
In the integral eq.(\ref{DelC}) that yields the coefficients $
\Delta C_l/C_l $, the transfer function $ D(k) $ multiplies a
function that is strongly peaked at $ x \sim l $, namely, for
momenta $ k \sim l \; \kappa $. Therefore, if  $ k \; |\eta_0| \sim
l \; \kappa \;  |\eta_0| \gg 1 $, the rapid oscillations in $ D(k) $
average out in the integrand, resulting in a vanishing contribution
to the $ \Delta C_l/C_l's $. Hence, there are significant
contributions to $ \Delta C_l/C_l $ only  when $ l \; \kappa \;
|\eta_0| \sim 1 $. For the quadrupole this corresponds to, $ a_0 \; 
H_0 \; |\eta_0|  \sim 1 $.

Inserting the expression (\ref{Dborn2}) for $ D(k) $ into eq.(\ref{DelC}) yields
\be\label{cuaB}
\frac{\Delta C_\ell}{C_\ell} = \frac1{\kappa}
\int^0_{-\infty} d\eta \;  \mathcal{V}(\eta) \; \Psi_\ell(\kappa \; \eta)
\ee 
where
\be \label{defPsi}
\Psi_\ell(x) \equiv 2 \; \ell(\ell+1) \; 
 \int_0^{\infty} \frac{dy}{y^4} \left[ j_\ell(y) \right]^2
\left[ \left(y^2 -\frac1{x^2} \right)\sin(2\, y \; x )+ \frac{2 \,
y}{x} \; \cos(2\, y \; x ) \right]
\ee
$ \Psi_\ell(x) $ is an odd function of $ x $. $ \Psi_2(x) $ turns to
be {\bf positive} for $ x < 0 $ as shown in ref.\cite{quadru2}.
Since the potential $ \mathcal{V}(\eta) $ turns to be an {\bf
attractive} potential $ \mathcal{V}(\eta) < 0 $, the correction $
\Delta C_2/C_2 $ given by eq.(\ref{cuaB}) in the Born approximation 
is  {\bf negative} clearly revealing a {\bf suppression} in the CMB quadrupole.

The functions $ \Psi_\ell(x) $ exhibit oscillations for $ \ell > 2 $
and take both positive as well as negative values for $ x < 0 $.
Therefore, we see from eq.(\ref{cuaB}) that the $ \ell > 2 $ CMB multipoles 
can be suppressed as well as enhanced by the effect 
of fast-roll as explicitly shown in ref.\cite{quamc}.

The correction $ \Delta C_l/C_l $ of higher  multipoles is smaller, 
falling off as $ 1/l^2 $ \cite{quadru2}.

\medskip  

To explicitly compute $ \mathcal{V}_\mathcal{R}(\eta) $ as a function of $
\eta $ for the curvature fluctuations we solve numerically the equations of
motion (\ref{evol}) for new inflation [eq.(\ref{wnue})] and insert the
solution for the inflaton $ \chi(\eta) $ in eqs.(\ref{Wb})-(\ref{slrsd}).
No large $ N $ approximation is used in this numerical calculation since we
cover in the evolution the fast-roll region where slow-roll obviously does
not apply \cite{quamc}.

It is illuminating to study $ \chi(\tau) , \;  {\dot \chi}(\tau) , \; 
{\cal H}(\tau), \; p(\tau)/\rho(\tau) $ and $ \log[N \; \epsilon_v(\tau)] $
[see eq.(\ref{slrsd})] during the short fast-roll stage for new inflation 
[eq.(\ref{wnue})]. They are plotted 
in figs. \ref{fR1} during the fast-roll stage (see figs. \ref{infla1}, 
\ref{infla2} and \ref{hubZ} for the full inflationary stage).
We choose the best fit coupling $ y = 1.26 $ [see table 6] and a
total number of efolds equal to 64. We choose the initial values
of $ \chi $ and $ \dot \chi $ such that their initial kinetic and potential
energies are equal [eq.(\ref{cifr})] and therefore $ p/\rho = 0 $,
 initially.

We see from figs. \ref{fR1} that the equation of state $ p/\rho(\tau) $ 
goes down from the initial value $ p/\rho = 0 $ to $ p/\rho \simeq -1 $ 
by the end of fast-roll. Both $ {\dot \chi}(\tau) $ and 
$ \epsilon_v(\tau) $ decrease very fast during fast-roll. We see that the 
fast-roll stage ends by $ \tau \simeq 0.0247 $ and 
$ \log a(\tau) = 1.0347 $ when $ \epsilon_v(\tau)\times N = 1 $. Notice 
that $ \epsilon_v(0) = 1.5 \gg 1/N $ for the initial conditions 
eq.(\ref{cifr}).

\begin{figure}[h]
 \includegraphics[height=13cm,width=8cm,keepaspectratio=true]{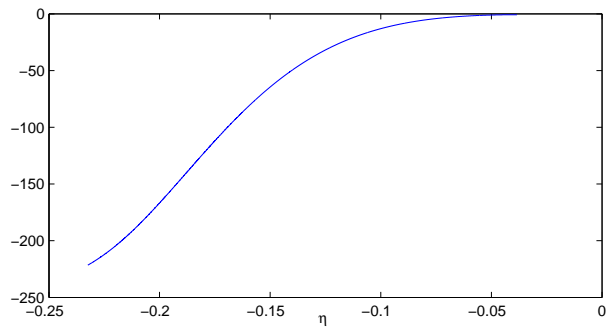}
  \caption{The potential $ \mathcal{V}_\mathcal{R}(\eta) $ felt by the
fluctuations vs. $ \eta $ for new inflation with $ y = 1.26 $.
$ \mathcal{V}_\mathcal{R}(\eta) $ is {\bf attractive} during
    fast-roll and vanishes by the end of fast-roll ($ \eta \sim -0.04 $).}
  \label{veta}
\end{figure}

In fig. \ref{veta} we plot $ \mathcal{V}_\mathcal{R}(\eta) $ vs. $ \eta $
for new inflation and the same initial conditions. We see that the 
potential $ \mathcal{V}_\mathcal{R}(\eta) $ is {\bf attractive} in the 
fast-roll stage and asymptotically vanishes by the end of fast-roll 
$ \eta \sim -0.04 $ \cite{quamc}.

\begin{figure}[h]
  \includegraphics[scale=0.8]{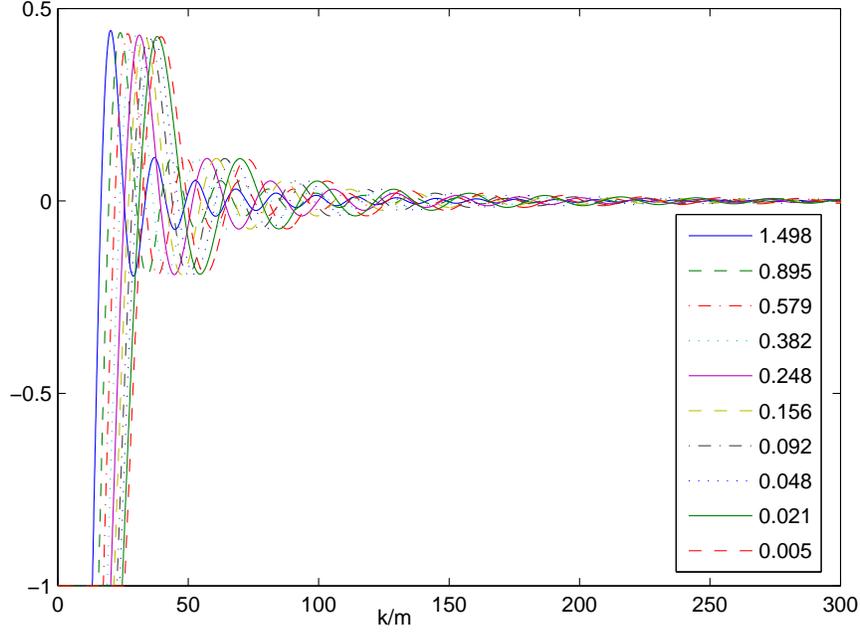}
  \caption{The fast-roll transfer function $ D_\mathcal{R}(k) $ vs. $ k/m $ 
for new inflation and ten different couplings $ 0.00536 < y < 1.498 $.
We see that the plots of $ D_\mathcal{R}(k) $ for different couplings 
follow from each other by changing the scale in
the variable $ k $ as summarized by eq.(\ref{Qk1}).}
  \label{dkflin4}
\end{figure}

We obtain the transfer function $ D_\mathcal{R}(k) $ by inserting $
\mathcal{V}_\mathcal{R}(\eta) $ into eq.(\ref{Dborn2}) and computing the
integral over $ \eta $ numerically. In fig.  \ref{dkflin4} we plot $
D_\mathcal{R}(k) $ vs. $ k/m $ for new inflation [eq.(\ref{wnue})] and ten
different couplings $ 0.00536 < y < 1.498 $ with a total number of efolds
equal to 64. We see that $ D_\mathcal{R}(k) $ oscillates around zero and
therefore produces {\bf suppressions as well as enhancements} in the low
multipoles [see eq.(\ref{curvapot})]. $ D_\mathcal{R}(k)
$ vanishes asymptotically for large $ k $ as expected.

The first peak in $ D_\mathcal{R}(k) $ is clearly its dominant feature.
The $ k $ of this peak corresponds to $k$-modes which are today horizon
size and affect the lowest CMB multipoles (see below and table 2)
\cite{quadru1,quadru2}.

For small $ k $ the Born approximation to $ D_\mathcal{R}(k) $ yields large
negative values indicating that this approximation cannot be used in this
particular small $ k $ regime.  We introduce the scale $k_{tran}$ by the
condition $ D_\mathcal{R}(k_{tran}) = -1 $ and then just take $ D_\mathcal{R}(k) = -1 $ 
for $ k \leq k_{tran} $. This corresponds to vanishing primordial power for the lowest
values of $ k $ (see fig. \ref{dkflin4}).

\medskip 

From fig. \ref{dkflin4} we also see that the plots of $ D_\mathcal{R}(k) $
for different couplings follow from each other almost entirely by changing
the scale in the variable $ k $ as summarized by eq.(\ref{Qk1}).  Indeed,
the characteristic scale $ k_{tran} $ plays here a further important role.

\begin{figure}[h]
  \includegraphics[scale=0.8]{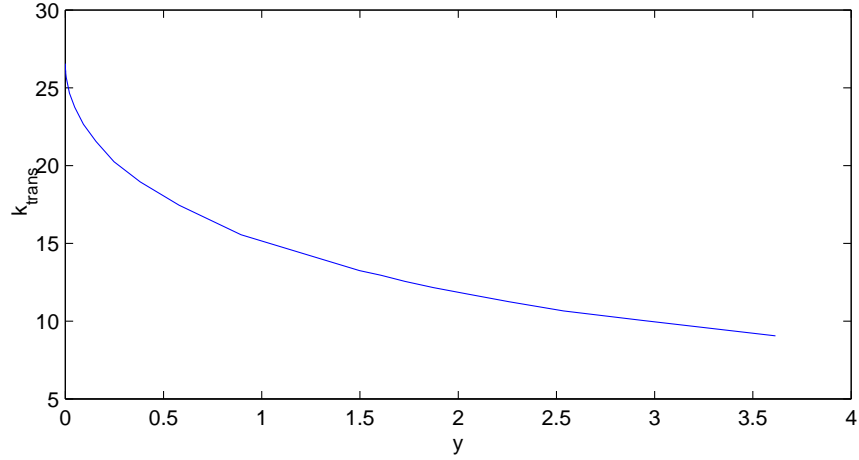}
  \caption{$ k_{tran}/m $ vs. $ y $ for new inflation.}
\label{k1y}
\end{figure}

Analysing $ \mathcal{V}_\mathcal{R}(\eta) $ and $ D_\mathcal{R}(k) $ for
different couplings $ y $ we find that they {\bf scale} with $ k_{tran} $.
Namely, 
\be\label{Qk1} 
\mathcal{V}_\mathcal{R}(\eta) = k_{tran}^2 \; Q(k_{tran} \;
\eta) \quad , \quad D_\mathcal{R}(k) = \Psi\left(\frac{k}{k_{tran}}\right) \; ,
\ee 
where $ Q(x) $ and $ \Psi(x) $ are universal functions. That is, $ Q(x) $
and $ \Psi(x) $ do not depend on the coupling $ y $ while $ k_{tran}/m $ is a
function of $ y $. We display $ k_{tran}/m $ vs. $ y $ in fig. \ref{k1y}.
These scaling properties arise from the fact that
the scale of the potential $ \mathcal{V}_\mathcal{R}(\eta) $ in $ k^2 $ does
determine the scale of variation of the transfer function  $ D_\mathcal{R}(k) $
with $ k $.

\begin{figure}[h]
  \includegraphics[scale=0.8]{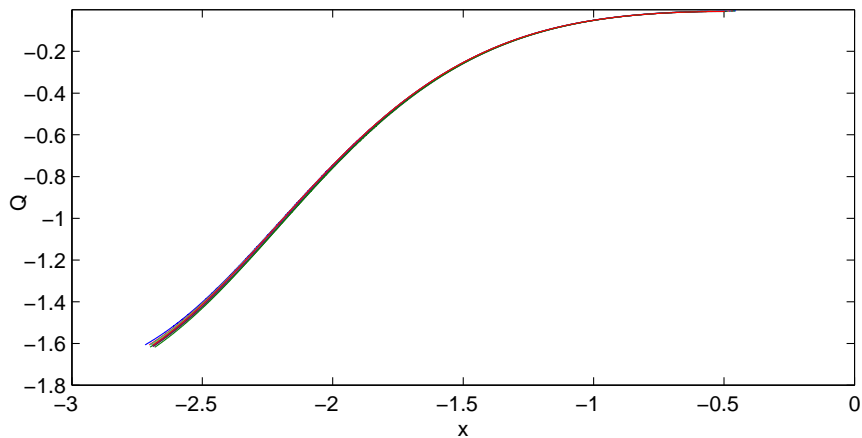}
  \caption{$ Q(x) $ for the ten values of $ y $ of fig.~\ref{dkflin4},
    according to eq.(\ref{Qx}).  All curves collapse to a common one
    proving the scaling properties eq.(\ref{Qk1}).}
  \label{Qxf}
\end{figure}

We obtain the function $ Q(x) $ from eq.(\ref{Qk1}) as,
\be\label{Qx}
Q(x) = \frac1{k_{tran}^2} \; \mathcal{V}_\mathcal{R}\left(\frac{x}{k_{tran}}\right)
\ee
We plot $ Q(x) $ in fig. \ref{Qxf} as follows from the r. h. s.
of eq.(\ref{Qx}) for ten different values of $ y $. We see that all the curves collapse
on a common curve proving the validity of the quasi-scaling properties eq.(\ref{Qk1}).

\subsubsection{MCMC analysis of CMB and LSS data including the early 
fast-roll inflationary stage}\label{mcmcsec}

In order to test the theoretical quadrupole supression predicted by the fast-roll
inflationary stage against the current experimental data we performed in
ref.~\cite{quamc} a Monte Carlo Markov Chains (MCMC) analysis of the commonly
available cosmological data using the {\em CosmoMC} program \cite{lewis}.

For LSS we considered SDSS (DR4). For CMB we first considered the three--years
WMAP data (with the second release of WMAP likelihood code) and the small scale data
(ACBAR-2003, CBI2, BOOMERANG03). While the work in ref.~\cite{quamc} was in
progress, the five--years WMAP data were released, and we repeated our MCMC
analysis almost completely with these new data, using also the newer 2007 ACBAR
release (commonly denoted as ACBAR08) \cite{acbar08}. Actually WMAP3 or WMAP5
provide by far the dominant contribution and small scale experiments have very
little relevance for the quadrupole supression issue. In this review we only
report MCMC results based on the WMAP5 data, in combination with the SDSS data
and either ACBAR08 or the most recent supernovae compilation (SN for
short) \cite{SN}.

\medskip 

In our MCMC analysis we modified the {\em CosmoMC} code and
introduced explicitly the transfer function $ D_\mathcal{R}(k) $ in the
primordial power spectrum according to eq.~(\ref{curvapot}).

We ran {\em CosmoMC} on pc clusters with Message Passing Interface (MPI),
producing from 10 to 24 parallel chains, with the `R-1' stopping criterion
set equal to 0.03 (this criterion looks at the fluctuations among parallel 
chains to decide when to stop the run). The statistical convergence was 
also verified a posteriori with the help of the {\em getdist} program of 
{\em CosmoMC}.

As discussed in section~\ref{mcmc}, the preferred reference model for
slow--roll inflation cosmology is the $\Lambda$CDM+$r $ model, that is the
standard six-parameters $\Lambda$CDM model augmented by the tensor-scalar 
ratio $ r $. Indeed, the current experimental accuracy provides sensible 
bounds for the index $ n_s $ and the ratio $ r $ eqs.(\ref{ns})
and (\ref{defr}). Specific slow--roll scenarios, such as those based on new
(small--field) or chaotic (large--field) inflation lead to specific 
theoretical constraints in the $ (n_s,\,r) $ plane presented in 
sec.~\ref{mcmc} \cite{mcmc}.

\medskip

We point out that we used the default {\em CosmoMC} pivot scale $ k_0 = 
0.05 ~ {\rm Mpc}^{-1}$ rather than the customary WMAP choice of 
$ k_0 = 0.002\,{\rm Mpc}^{-1}$. As evident from eq.~(\ref{potBD}) this 
leads to a small difference with respect to the WMAP choice in the 
definition itself of the tensor to scalar ratio
$ r $. In particular, the CosmoMC $ r $ is roughly 10\% larger than the 
WMAP one.

\begin{figure}[h]
\includegraphics[height=9.cm,width=12.cm]{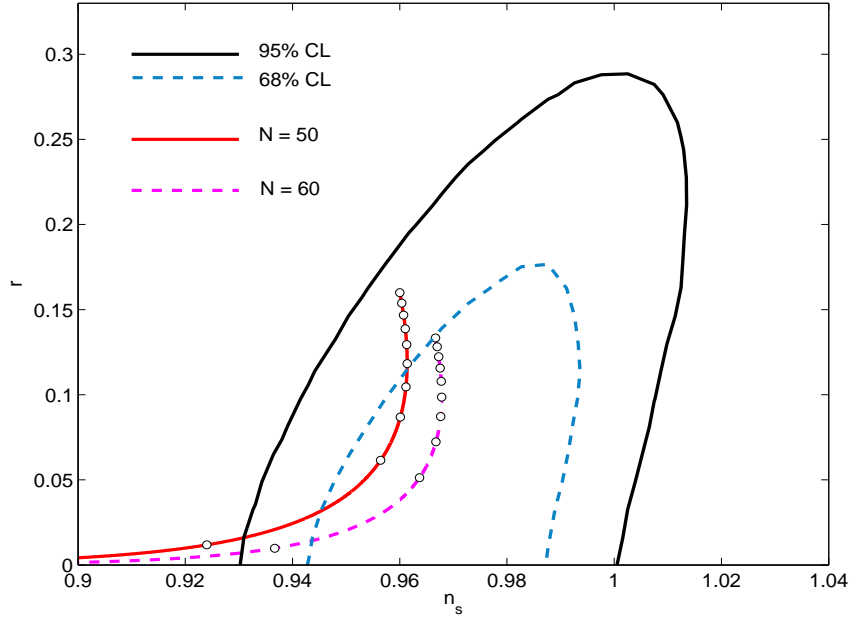}
\caption{Binomial New Inflation compared to $\Lambda$CDM+$r $ model in the
$(n_s,\,r)$ plane. The contour plots correspond to $68\% $ and $ 95 \% $
confidence levels for $\Lambda$CDM+$r$ according to WMAP5, SN and SDSS 
data. $C_\mathrm{BNI}$ is the solid red curve for $ N=50 $ or the dashed 
magenta curve for $ N=60 $. The white dots correspond to the values $ 0.01 
+ 0.11*n , \; n=0,1, \ldots, 9 $, of the variable $ z $ in eq.~\eqref{BNI},
starting from the leftmost ones. The quartic coupling $ y $ increases 
monotonically starting from the uppermost dots, corresponding to the
free-field, purely quadratic inflaton potential $ y = 0, \; z = 1 $ till
the strong coupling region $ y \gg 1, \; z \ll 1 $ in the 
lower part of the $C_\mathrm{BNI}$ curve.
We see that very small values of $ r $ {\bf are excluded} since they 
correspond to $ n_s < 0.92 $ outside the $ 95 \% $ confidence level 
contour. That is, we obtain a {\bf lower bound} for $ r : \;
r > 0.027 $ at 95\% C. L.}
\label{nsr}
\end{figure}

\subsubsection{MCMC analysis with Binomial New Inflation
without the fast-roll stage: $ D_\mathcal{R}(k) = 0$.}\label{BNIMCMC}

\begin{figure}[h]
\includegraphics[height=8cm]{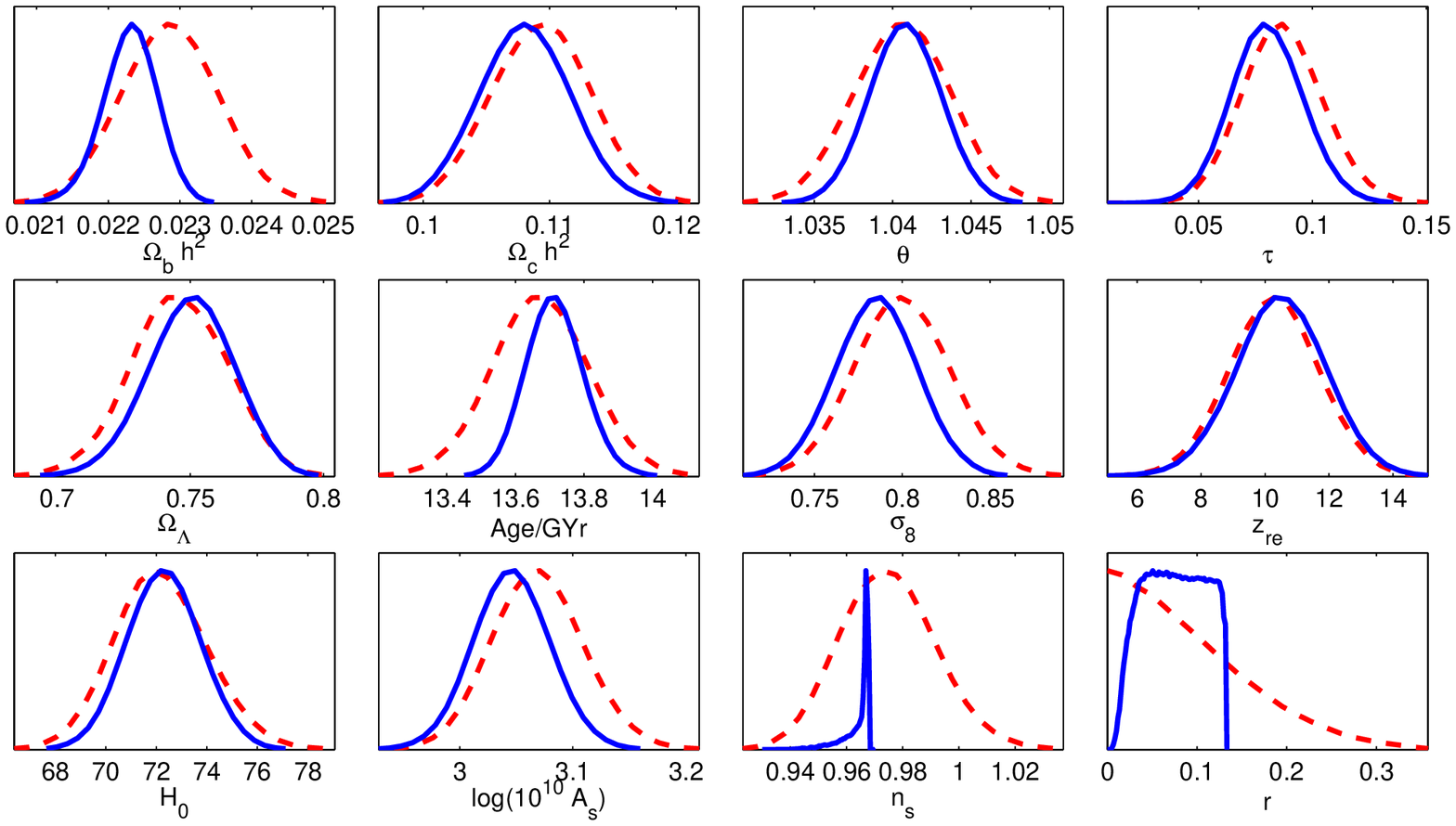}
\caption{The $\Lambda$CDM+BNI model (solid blue lines) compared to the
$\Lambda$CDM+$r$ model (dashed red lines) in the probability distributions
of relevant cosmological parameters, for $ N=60 $ with WMAP5+SN+SDSS data.}
\label{bnivsstd}
\end{figure}

Let us first present, to fix the reference, our MCMC analysis with the 
standard slow-roll primordial power eq.(\ref{potBD}). That is, without 
including the early fast-roll stage and therefore with a vanishing 
transfer function $ D_\mathcal{R}(k) $.

For instance, in the simplest binomial realization of new inflation 
described by the inflaton potential eq.~\eqref{wnue}, $ n_s $ and $
r $ are constrained to the curve $ C_\mathrm{BNI} $ (BNI stands
for {\em Binomial New Inflation}) parametrized by the quartic coupling 
$ y $ as [see eq.~\eqref{nsbinew}]:
\begin{equation}\label{BNI}
  n_s=1- \frac{y}{N}\, \frac{3 \,z + 1}{(1-z)^2}  \;,\qquad
  r=\frac{16 \; y}{N} \frac{z}{(1-z)^2} \;, \qquad
  y=z-1-\log z \;,\qquad z = \frac{y}8 \; \chi^2  \;, \qquad 0<z<1 \;.
\end{equation}
This situation is clearly displayed in fig.~\ref{nsr} in which the curve $
C_\mathrm{BNI} $, for the two choices $ N=50 $ and $ N=60 $, is drawn over the
contour plots of the likelihood distribution for $ n_s $ and $ r $ in the $
\Lambda$CDM+$r $ model obtained with CosmoMC, using the WMAP5, SN and SDSS data.
Very similar contour plots apply when ACBAR08 is used in place of SN.

All the MCMC results reported below refer to the case $ N=60 $. 

The likelihood $ L $, as function of the whole set of parameters, provides a
quantitative measure of the power of a given model to fit the multipoles $
C_{\ell}^\gamma $. As customary, we set $ -2\log L = \chi_L^2 $, although it is
well known that, due particularly to cosmic variance, the shape of $ L $ as
function of the $ C_{\ell}^\gamma $ is not Gaussian especially for low $ \ell$.

Now, as evident from eq.~(\ref{BNI}) and fig.~\ref{nsr}, one could expect from
the $ \Lambda\mathrm{CDM} $ model constrained to $ C_\mathrm{BNI} $ a fit to the
data not as good as in the $\Lambda$CDM+$r$ model since the current data seem
to favor smaller values for $ r $.  However, we find very small increase of
min $ \chi_L^2 $  \cite{quamc}
\begin{equation}\label{dchi21}
  {\rm min}\,\chi_L^2(\Lambda{\rm CDM+}r{\rm +BNI}) 
  -{\rm min}\,\chi_L^2(\Lambda\mathrm{CDM+}r) \simeq 0.2 \;.
\end{equation}
This shows that the BNI constraint is naturally compatible with the data.
This result was obtained for $ N=60 $ by direct minimization of $ \chi_L^2 $ in
the neighbourhood of $ C_\mathrm{BNI} $, using the data of a large collection of
long chain runs (with a grandtotal of more than one million steps) for the $
\Lambda$CDM+$r $ model with the WMAP5, SN and SDSS data. The flat
priors on the cosmological parameters were the standard ones of CosmoMC, that is
\begin{center}
  \begin{tabular}{l l l}
    $0.005<\omega_b<0.1 \quad ,$ & $\;0.01<\omega_c<0.99 \quad ,$ &$\,0.5<\theta <10$ \\
    $0.01<\tau<0.8 \quad , $ & $\;2.7 < \log(10^{10} A_s)<4 \quad , $ & $\;0.5<n_s<1.5$ \\
  \end{tabular} 
\end{center}
while for the tensor-scalar ratio we imposed as prior
\begin{equation*}
  0 < r < 0.35 \;.
\end{equation*}
Another parametrization, that unlike the direct minimization of $ \chi_L^2 $ over $
C_\mathrm{BNI} $, does take advantage of the explicit analytic parametrizations
in eq.~\eqref{BNI}, is to use the single variable $ z $ as MCMC parameter,
instead of the constrained pair $ (n_s\,r) $, with a flat prior over all the
allowed range $ 0<z<1 $. Let us call $\Lambda$CDM+BNI the
$6$-parameter model $ \Lambda {\rm CDM} $ constrained on $C_{\rm BNI}$ using the
variable $ z $. Then, we find that taking into account the natural fluctuations
due to the large number of data (which make the likelihood landscape over the
MCMC parameters quite complex) and the various approximations and numerical
errors in the theoretical calculation of the multipoles, the increasing in $
\chi_L^2 $ due to the $ C_{\rm BNI} $  constraint eq.~\eqref{BNI} compared to the $
\Lambda$CDM  model essentially vanishes (see Table IV below).

For completeness and reference, we report in Table IV our best fit (or 
most likely) values for the MCMC cosmological parameters, as well as the 
variations in $\chi_L^2$ with respect to the $\Lambda$CDM$+r$ model. The 
dataset includes WMAP5, SN and SDSS. We report in the first row of the 
table also our best fit for the standard $\Lambda$CDM model, which has six
free parameters as the $\Lambda$CDM+BNI model since $ r $ is set to zero by fiat.
\begin{table}\label{T1}
  \begin{tabular}{|c|c|c|c|c|c|c|c|c|}
 \hline   & $~~10\,\omega_b~~$ & $~~~~\omega_c~~~~$ & $~~10\,\theta~~$ & $~~~~\tau~~~~$ 
    & $~~10^{9}A_s~~$ & $~~~~n_s~~~~$ & $~~~~r~~~~$ & $~~\Delta\chi_L^2/2~~$ \\ \hline
    $\Lambda$CDM
    & 0.226 & 0.110 & 1.040 & 0.896 & 2.164 & 0.966 & 0 &  -0.2\\ \hline
    $\Lambda$CDM$+r$ 
    & 0.225 & 0.110 & 1.039 & 0.785 & 2.108 & 0.962 & 0.010 & --\\\hline
    $\Lambda$CDM+$r$+BNI
    & 0.225 & 0.110 & 1.040 & 0.807 & 2.115 & 0.964 & 0.052 & +0.2 \\\hline
    $\Lambda$CDM+BNI 
    & 0.226 & 0.111 & 1.040 & 0.867 & 2.159 & 0.965 & 0.057 & -0.3 \\\hline
  \end{tabular} 
  \caption{Best fit values for the MCMC cosmological parameters without
    quadrupole suppression, using WMAP5, SN and SDDS. 
$ C_\mathrm{BNI} $ means the curve on which $ n_s $ and $ r $ are 
constrained in Binomial New Inflation (BNI), eq.~\eqref{BNI} with $ N=60 $.
$\Lambda$CDM+$r$+BNI means the $\Lambda$CDM+$r$ model constrained on 
$ C_\mathrm{BNI} $ using the constrained pair of variables $ (n_s,r) $. 
$\Lambda$CDM+BNI denotes the $\Lambda$CDM model constrained on 
$ C_\mathrm{BNI} $ using the single variable $ z $ eq.~\eqref{BNI} as MCMC 
variable instead of the constrained pair $ (n_s,r) $.}
\end{table}
\begin{figure}[h]
\includegraphics[height=6.8cm]{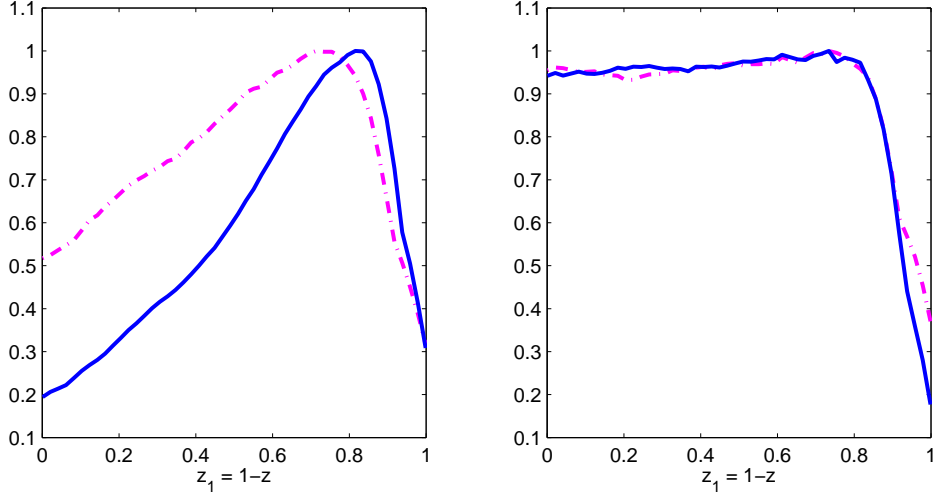}
\caption{Comparison of probability distributions (solid blue lines) and 
mean likelihoods (magenta dashed-dotted lines) in the $\Lambda$CDM+TNI 
model (left) and in the $\Lambda$CDM+BNI model (right). Here 
$ N=60 $ and WMAP5, SN and SDSS data are used.}
\label{ztvszb}
\end{figure}

The sligthly better performance of $\Lambda$CDM+BNI over
$\Lambda$CDM+$r$+BNI (and also over $\Lambda$CDM+$r$ itself) is not very
relevant since is most likely due to the finer search possibility in a 
space with 6 parameters by compared to one in a space with 7 parameters.
We want to draw the attention to the fact that including the constraint 
over $ C_\mathrm{BNI} $ does not produce any statistically significant 
change on the most likely values of the cosmological parameters, except 
of course on $ n_s $ and $ r $ themselves. This shows how natural turns to 
be to constraint the CMB+LSS data by the BNI model.
In particular, comparing the $\Lambda$CDM+BNI model
with respect to the $\Lambda$CDM+$r$ model, 
the most likely value of $ n_s $ is practically unchanged, while that of 
$ r $ improves in $\Lambda$CDM+BNI: $ r $
changes from values of order $ 10^{-2} $ (or just $0$) in $\Lambda$CDM
to values such as $ 0.052 $ and $ 0.057 $ in $\Lambda$CDM+BNI.

\begin{figure}[ht]
\includegraphics[height=8cm]{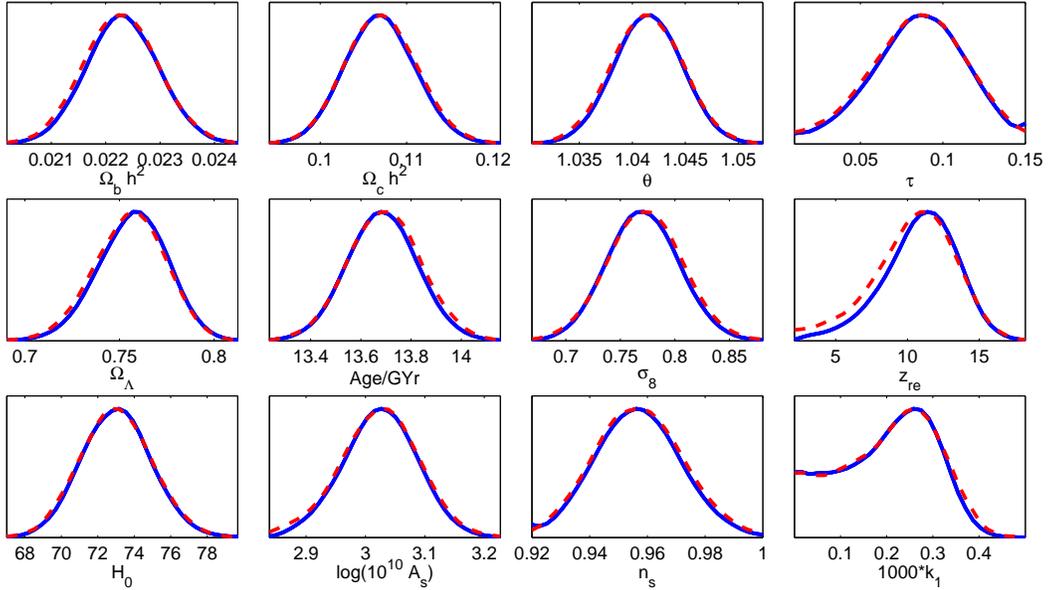}
\caption{Marginalized distributions (solid blue lines) and mean likelihoods
(red dotted lines) for the parameters of the $\Lambda$CDM+sharp 
cutoff model, using the WMAP3+small scale\_CMB+SDSS data.
$ k_{tran} $ is in $ ({\rm Gpc})^{-1} $.}
\label{sharpcut}
\end{figure}

Concerning marginalized distributions, we find no significant changes but 
for $ n_s $ and $ r $. These results are very close to those in section 
\ref{mcmc}, where TNI was considered, as can be appreciated from 
fig.~\ref{new_vs_std5} and
\ref{bnivsstd}. The only minor change is in the broader shape of the peak 
in the marginalized distribution for $ r $, which for BNI extends almost 
flatly up to
the theoretical limit $ r=2/15 = 0.133\ldots $. This is due to the absence 
in BNI of the asymmetry $ h $ which in the TNI case allows for the 
broadening of the allowed $n_s-r$ region as $ z_1\to 1 $ ({\em i.e.} as 
$ z=1-z_1\to 0 $, which is the strong coupling $ y $ regime). This effect 
may be directly observed in
the probability distributions and mean likelihoods for $ z_1 $, which we 
compare in fig.~\ref{ztvszb}. As an obvious
consequence, with respect to the TNI case, we obtain a higher bound on 
$ r $ in the $\Lambda$CDM+BNI model: $ r>0.027 $ at 95\% CL.

\begin{figure}[ht]
\includegraphics[height=10cm]{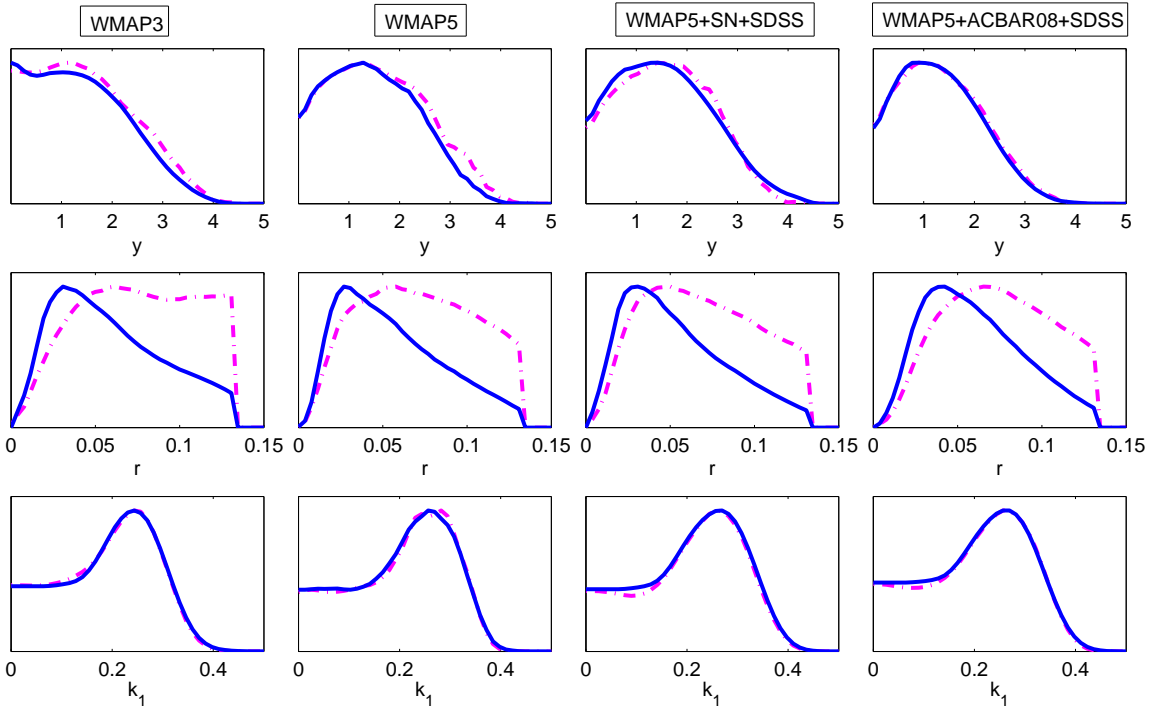}
\caption{Marginalized distributions (solid blue lines) and mean likelihoods
(magenta dash-dotted lines) for the $\Lambda$CDM+fast-roll model, 
with flat prior on the coupling $ y $ and using the dataset shown on top
of the columns. The slightly better fitting power of fast--roll over 
sharp--cutoff may be also seen from fig.~\ref{clsfig} where the best fit 
for the $ C_{\ell}^{\rm TT} $ multipoles is compared to the experimental 
data at low $ \ell $.}
\label{yrk1}
\end{figure}

\medskip 

Our inflation--based 6-parameter $\Lambda$CDM+BNI model
predicting a value of $ r $ well below 0.2 is {\bf as likely} as
the $\Lambda$CDM model itself. Recall that the  current CMB+LSS data 
analysis, without any theoretical constraint, put only an upper bound on 
$ r $ (namely $ r < 0.22 $ with 95\% C. L. in the most recent WMAP5 
analysis \cite{WMAP5}). This means that the theoretical grounds of a
given model take a more important role in the analysis and interpretation 
of the CMB+LSS data. For instance, from an inflationary viewpoint, the 
choice that $ r $ exactly vanishes, appears {\bf unlikely and unphysical}. 
Notice that $ n_s - 1 = 0 = r $ corresponds to a singular limit and a 
critical (massless) point where the inflaton potential vanishes \cite{mcmc}
while the CMB$+$LSS data analysis for both the $\Lambda$CDM+BNI and 
$\Lambda$CDM+TNI models yield {\bf lower bounds} for $ r $ as discussed 
here and in sec. \ref{mcmc:trino}
\cite{mcmc,quamc}. All together, our data analysis shows that the 
6-parameter $\Lambda$CDM+BNI model is {\bf better} than the standard 
6-parameter $\Lambda$CDM model.

\begin{figure}
\includegraphics[height=8cm]{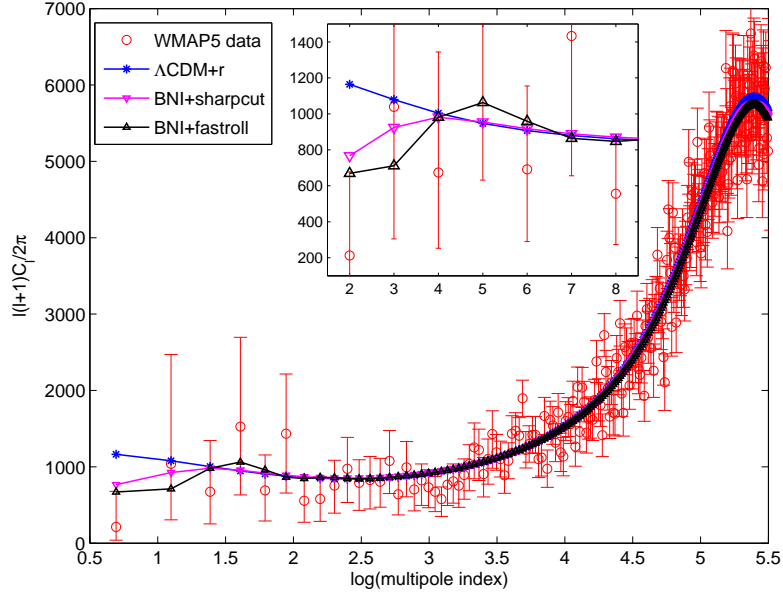}
\caption{Comparison, with the experimental WMAP5 data, of the theoretical 
$ C_{\ell}^{\rm TT} $ multipoles computed in the best fit point of the 
various models of the main text, using the WMAP5+SN+SDSS data. The insert 
contains an enlargement in linear scale of the first seven multipoles. 
The  $ C_{\ell}^{\rm TT} $ units are $ [\mu \; K^2] $. The
error bars in the plotted range of $ \ell $ are mostly due to cosmic 
variance. Error bars of the WMAP5 data are one-sigma ($68\%$C.L.).}
\label{clsfig}
\end{figure}

\subsubsection{MCMC analysis with Binomial New Inflation
including the fast-roll stage: $ D_\mathcal{R}(k) \neq 0 $.}\label{BNIFR}

Let us now further develop this argument by considering the quadrupole
supression, avoiding the a priori dismissal based on the simple invocation
of cosmic variance or experimental inaccuracy. In the standard $\Lambda$CDM
model the simplest, purely phenomenological way to decrease the low
multipoles is to introduce a infrared sharp cutoff in the primordial power
spectrum of the curvature fluctuations. That is, one assumes that
$ P_\mathcal{R}(k)=0 $ for $ k<k_{tran} $ and treats 
$ k_{tran} $ as a new MCMC 
parameter to be fitted against the data.  It is actually not necessary for 
the moment to include also a cut on the tensor power spectrum, since it 
would lead to changes certainly not appreciable within the current 
experimental accuracy.

With this procedure we obtained in ref.\cite{quamc}, 
using either the WMAP3 data alone or the WMAP3+small scale\_CMB+SDSS 
data:
\begin{equation*}
   {\rm min}\,\chi_L^2(\Lambda\mathrm{CDM}+{\rm sharp cutoff}) 
   - {\rm min}\,\chi_L^2(\Lambda\mathrm{CDM}) \simeq -1.4 
\end{equation*} 
This result is slightly better than the one reported in ref.\cite{WMAP3}, 
but still the likelihood gain hardly compensates the price of a new 
parameter, especially because its nature appears quite {\em ad hoc}.  
Furthermore, the gain in likelihood is sizable smaller when WMAP5+SDSS+SN 
data are used:
\begin{equation*}
   {\rm min}\,\chi_L^2(\Lambda\mathrm{CDM}+{\rm sharp cutoff}) 
   - {\rm min}\,\chi_L^2(\Lambda\mathrm{CDM}) \simeq -0.9 
\end{equation*}
In fig.~\ref{sharpcut} we plot the marginalized probabilities and mean
likelihoods of the seven MCMC parameters plus other standard derived 
parameters in the WMAP3+small scale\_CMB+SDSS case. If any other 
combination of more recent
data such as WMAP5+SN are used, the plots are almost identical. The main
point is that there are no significant changes from $\Lambda$CDM to
$\Lambda$CDM+sharp cutoff in their common parameters, in either most
likely values or marginalized distributions. The distribution of the new
cutoff parameter $ k_{tran} $ shows a well defined peak centered on 
its most likely value (ML), which corresponds to today's physical 
wavelength
 \begin{equation*}
   (k_{tran})_{\rm ML} = \begin{cases} 
     0.291 \; ({\rm Gpc})^{-1} \quad \textrm{(WMAP3 only)}\\
     0.245 \; ({\rm Gpc})^{-1} \quad \textrm{(WMAP5+ACBAR08+SDSS)}\\
     0.273 \; ({\rm Gpc})^{-1} \quad \textrm{(WMAP5+SN+SDSS)} 
   \end{cases}  \qquad   (\Lambda{\rm CDM} + {\rm sharp cutoff}) \;,
 \end{equation*}
that is of the order of today's inverse Hubble radius, as expected.

Introducing the infrared sharp cutoff on $ P_\mathcal{R}(k) $ in the
$\Lambda$CDM+BNI model we find similar gains
\begin{equation*}
  {\rm min}\,\chi_L^2(\Lambda\mathrm{CDM+BNI}+{\rm sharp cutoff}) 
  - {\rm min}\,\chi_L^2(\Lambda\mathrm{CDM+BNI}) =
  \begin{cases}
    -1.4 \quad \textrm{(WMAP3)} \\
    -1.1 \quad \textrm{(WMAP5)}\\
    -1.0 \quad \textrm{(WMAP5+SN+SDSS)}\\
    -0.7 \quad \textrm{(WMAP5+ACBAR08+SDSS)}
   \end{cases}
\end{equation*}
The differences in these gains are partly due to the tighter bound on 
$ r $ provided by the new WMAP5 data and by the SDSS data.

Let us now consider the fast-roll case, when the fast-roll transfer function $
D_\mathcal{R}(k) $ eq.(\ref{Dborn2}) and fig. \ref{dkflin4} is used, treating
the scale $ k_{tran} $ in eq.(\ref{Qk1}) as a MCMC parameter.  That is, in the MCMC
analysis we use the initial power spectrum
eq.(\ref{curvapot}) modified by the fast-roll transfer function $
D_\mathcal{R}(k) $.  We computed once and forever $ D_\mathcal{R}(k) $ from eq.
(\ref{Dborn2}) (see fig. \ref{dkflin4}).  $ D_\mathcal{R}(k) $ is a function of
$ k $ and $ k_{tran} $ with the scaling form eq.(\ref{Qk1}), $ \Psi(x) $ being an
universal function.

\begin{table}[ht]
\begin{tabular}{|l|c|c|c|c|}
     \hline $\Lambda$CDM+BNI+sharp cutoff & $~~k_{tran}$ (best fit)~~ &
     $~n_s $ (best fit)~~&$~~r$ (best fit)~~ & $~~r$ (95\% CL)~~\\ \hline
     WMAP3 &  0.275  $ {\rm Gpc}^{-1}$ & 0.960 & 0.150 & $> 0.028$ \\ 
\hline
     WMAP5  &  0.257  $ {\rm Gpc}^{-1}$ & 0.961 & 0.040 & $> 0.025$ \\ 
\hline
     WMAP5+SDSS+SN &  0.245 $ {\rm Gpc}^{-1}$ & 0.963 & 0.048 & $> 0.022$ 
\\ \hline
     WMAP5+SDSS+ACBAR08 &  0.260 $ {\rm Gpc}^{-1}$ & 0.965 & 0.060 & $> 
0.029$ \\ \hline
\end{tabular}
\caption{The most likely values of $ k_{tran} , \; n_s, \; r $ and the lower 
bound on $ r $ in the $\Lambda$CDM+BNI+sharp cutoff model model. We used
$ N=50 $ in the run with the WMAP3 dataset, while we used $ N=60 $ in the 
other three runs.}
\end{table}

\begin{figure}
\includegraphics[width=12cm,height=8cm]{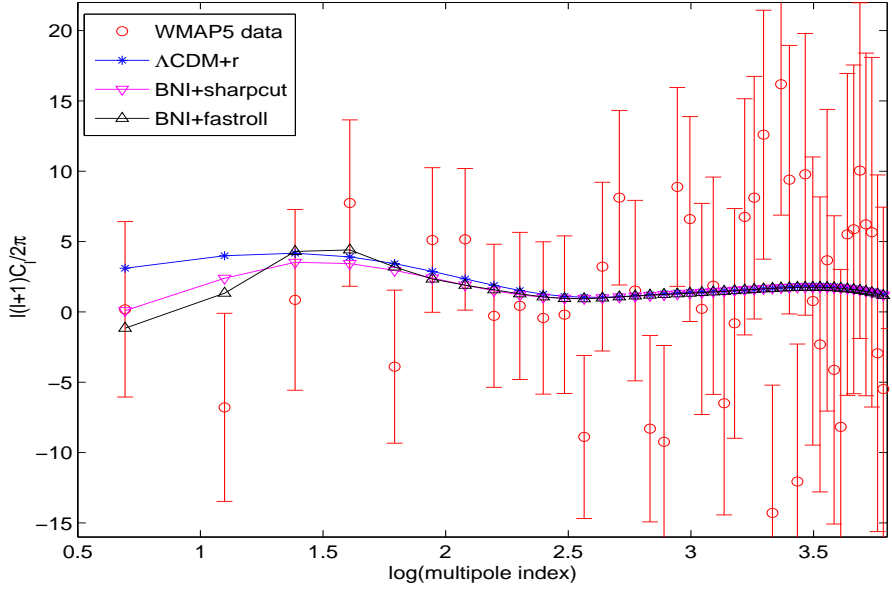}
\caption{Comparison, with the experimental WMAP5 data, of the theoretical 
$ C_{\ell}^{\rm TE} $ multipoles computed in the best fit point of the
$\Lambda$CDM+$r $ model, fast-roll and sharp cutoff models. The
$ C_{\ell}^{\rm TE} $ units are $ [\mu \; K^2] $. Error bars of the WMAP5 
data are one-sigma ($68\%$c.l.).}
\label{clsTE}
\end{figure}

We then find
\begin{equation*}
  {\rm min}\,\chi_L^2(\Lambda{\rm CDM+BNI+fastroll}) 
  - {\rm min}\,\chi_L^2(\Lambda\mathrm{CDM+BNI}) =
  \begin{cases}
    -1.8 \quad \textrm{(WMAP3)} \\
    -1.2 \quad \textrm{(WMAP5)}\\
    -1.4 \quad \textrm{(WMAP5+SN+SDSS)}\\
    -1.0 \quad \textrm{(WMAP5+ACBAR08+SDSS)}
   \end{cases} \;
\end{equation*}
We see that the gains in likelihood are {\em more significant in the fast-roll
  case} than in the sharp cutoff case \cite{quamc}.  Clearly, this fit
improvement through power modification by fast-roll over power reduction by sharp
cutoff is too small to constitute a real experimental evidence. But still, it is
very interesting that the theoretically well founded approach based on fast-roll
inflation works {\bf better} than the purely phenomenological cutoff.

In Table V we report the most likely values (ML) of $ k_{tran} , \; n_s $ and $ r $ 
as well as the 95\% CL lower bound on $ r $ for the $\Lambda$CDM+BNI
model with sharp cutoff denoted $\Lambda$CDM+BNI+sharp cutoff.
In Table VI we do the same when fast--roll is 
used instead of sharp cutoff. In this case we report also the best fit for the
quartic coupling $ y $ for future use.
\begin{table}[ht]
   \begin{tabular}{|l|c|c|c|c|c|}
\hline    $\Lambda$CDM+BNI+fastroll& $~~k_{tran}$ (best fit)~~ &
     $~n_s$ (best fit)~~& $~~r$ (best fit)~~ & $~~r$ (95\% CL)~~ & $~~y$ 
(best fit)~~  \\ \hline
     WMAP3 &  0.249 $ {\rm Gpc}^{-1} $ & 0.961 & 0.146 &$>0.025$ & 0.03\\ 
\hline
     WMAP5 &  0.298 $ {\rm Gpc}^{-1} $ & 0.965 & 0.056 &$>0.024$ & 1.11\\ 
\hline
     WMAP5+SDSS+SN  &  0.290  $ {\rm Gpc}^{-1} $ & 0.964 & 0.051 & $>0.023$ 
& 1.26 \\ \hline
     WMAP5+SDSS+ACBAR08  &  0.284  $ {\rm Gpc}^{-1} $ & 0.963 & 0.047 & 
$>0.029$ & 1.38 \\ \hline
   \end{tabular}
   \caption{The most likely values of $ k_{tran} , \; n_s, \;  r $ and the quartic 
coupling $ y $ and the lower bound on $ r $ in the $\Lambda$CDM+BNI+fast roll model. 
We set $ N=50 $ in the run with the WMAP3 dataset, while we set $ N=60 $ in the other 
three runs.}
\end{table}
In fig.~\ref{yrk1} we plot the probability and mean likelihood 
distributions of $ y, \; r $ and $ k_{tran} $, with flat prior on the coupling 
$ y $, in the case of the $\Lambda$CDM+BNI+fast-roll model
(the plot when a sharp
cut is used is almost identical). It is worth noticing that the 
introduction of the new MCMC parameter $ k_{tran} $ typically restores the peak 
in $ z $ (or equivalently in $ y=1-z-\log z $) that was lost upon setting 
the asymmetry $ h $ to zero in the restriction of the $\Lambda$CDM+TNI 
model to the $\Lambda$CDM+BNI model (see the upper row in fig.~\ref{yrk1}).
(The only exception to this rule is when the WMAP3 data alone are used, 
most likely because WMAP3 does not provide a tight enough upper
bound for $ r $, as can be appreciated also from Tables V and 

\begin{figure}
\includegraphics[width=12cm,height=8cm]{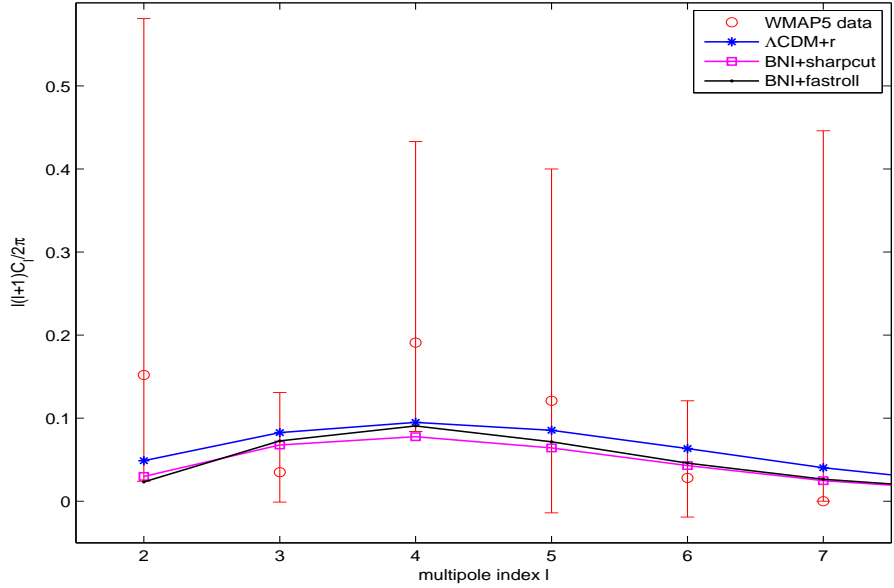}
\caption{Comparison, with the experimental WMAP-5 data, of the 
theoretical $ C_{\ell}^{\rm EE} $ multipoles computed in the best fit 
point of the $\Lambda$CDM+$r$ model, fast-roll and sharpcut models
as functions of the natural logarithm of $ \ell $.
Error bars of the WMAP-5 data are one-sigma ($68\%$c.l.).
Notice that both fast-roll and sharpcut models produce
a suppression of the low EE multipoles including the EE quadrupole.
The  $ C_{\ell}^{\rm EE} $ units are $ [\mu \; K^2] $.}
\label{clsEE}
\end{figure}

\medskip 

It must be noticed that in all cases (except for the obsolete WMAP3 alone 
case), we have $ w''(\chi) < 0 $ at horizon exit for the best fit values in
tables V and VI as follows from eq.(\ref{sigwseg}). That is, the best fit 
corresponds to an inflaton field exiting the horizon in the {\bf negative 
concavity region} $ w''(\chi) < 0 $ intrinsic to {\it new inflation}. The 
inflaton field at horizon exit takes the value $ \varphi_{exit} \sim 7 \; 
M_{Pl} $ as stated in eq.~(\ref{fiexit}). Notice that the energy density 
then is $ \sim N \; M^4 \ll M_{Pl}^4 $ well below the Planck energy scale 
guaranteeing the validity of the effective theory of inflation as discussed
in sec.~\ref{validez}.

\medskip

The slightly better fitting power of fast--roll over sharp--cutoff may be
appreciated also from fig.~\ref{clsfig} where the best fit for the $
C_{\ell}^{\rm TT} $ multipoles is compared to the experimental data at low 
$ \ell $. We see that the oscillatory form of the fast-roll transfer 
function $ D_\mathcal{R}(k) $, by {\bf depressing as well as enhancing} the
primordial power spectrum at long wavelengths, leads also to new 
superimposed {\bf oscillatory corrections} on the multipoles. As far as 
fitting to current data is concerned, such corrections are more effective 
than the pure reduction caused by a sharp cutoff.

We plot in fig. \ref{clsTE} the best fit for the $ C_{\ell}^{\rm TE} $
multipoles compared to the experimental data at low $ \ell $.  
We plot in fig. \ref{clsEE} the best fit for the $ C_{\ell}^{\rm EE} $
multipoles compared to the experimental WMAP-5 data at low $ \ell $. 
We see that both fast-roll and sharpcut models produce
a suppression of the low EE multipoles including the EE quadrupole.

We did not display in figs. \ref{clsfig} and \ref{clsTE} 
the  $\Lambda$CDM+sharp cutoff results since they are indistinguishable 
from the $\Lambda$CDM+BNI+sharp cutoff values.

\medskip

Our MCMC analysis with the $\Lambda$CDM+BNI model predict {\bf non-zero} 
lower bounds on $ r $: see Tables V and VI. The best fit values of the 
other cosmological parameters remain practically unchanged as compared to 
the $\Lambda$CDM model. Similarly, their 
marginalized probability distributions are almost unchanged, 
with the natural exception of $ n_s $, which in BNI has a
theoretical upper limit [see eq.~(\ref{cotsup})]. 

In the case of the $\Lambda$CDM+BNI+fast-roll model with the 
CMB+LSS datasets, the most likely value of the quartic coupling $ y $ is 
slightly larger than unity. Then from fig.~\ref{k1y} we read a value $
\sim 14 $ for the ratio $ k_{tran}/m $ at the beginning of inflation.

It is important to notice that the value of $ k_Q = 0.238 \; 
({\rm Gpc})^{-1} $ is {\bf slightly smaller} than the characteristic scale
$ k_{tran} $ as displayed in Table VI.

\medskip

A $k$-mode crosses the horizon when 
$ k = H(t) \; a(t) = a(\tau) \; m \; \sqrt{N} \; {\cal H}(\tau) $.
We compute $ \ln a $ and $ \epsilon_v $ at horizon crossing for relevants 
wavenumbers $k$ using the numerical solution of eqs.(\ref{evol}) and 
(\ref{wnue}) for $ \chi(\tau), \; {\dot\chi}(\tau), \; {\cal H}(\tau) $ and
$ \ln a(\tau) $ plus eq.(\ref{slrN}). 
We display in Table VII the relevant wavenumbers: 
$ k_Q, \; k_{tran} , \; k_0 $ 
and the number of efolds since the beginning of inflation when they
exit the horizon. 
We see that the quadrupole modes {\bf exit} the horizon during the
fast-roll stage, approximately $ 2/10 $ of an efold before the end of 
fast-roll.  The mode $ k_{tran} $ exit the horizon by $ \ln a = 1.024 $, 
very close to the point $ \ln a = 1.035 $ where $ \epsilon_v = 1/N $ 
(see fig. \ref{fR1}). That is, $ k_{tran} $ exits the horizon precisely
when {\bf fast-roll ends and becomes slow-roll}.

\begin{figure}
\includegraphics[height=8cm]{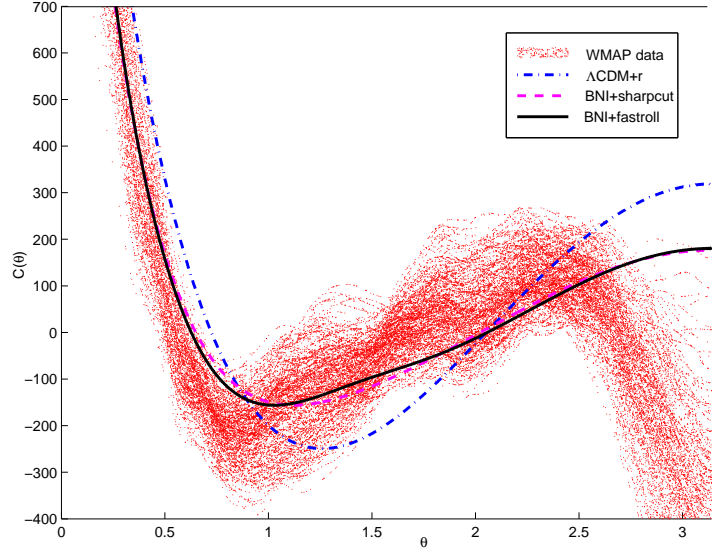}
\caption{The real space two point TT correlation function $ C^{TT}(\theta) 
$ for $\Lambda$CDM, sharp cutoff and fast-roll models vs. the angle 
$ \theta $.  The $\Lambda$CDM correlator differs from the two others only 
for large angles $ \theta \gtrsim 1 $. Since all $l$-modes except the 
lowest ones are
practically identical in the three cases, this shows how important are the
low multipoles in the large angle correlations. Also shown are the WMAP 
data. The truly observed correlator runs approximately in the middle of 
the red band. The width of the data band is mostly due to the cosmic 
variance. The WMAP $ C^{TT}(\theta) $ plotted here may not coincide, 
especially for the largest values $ \theta \sim \pi $, with the correlator
directly measured from sky maps due to the pixel weighting in the WMAP 
data analysis. The $ C^{TT}(\theta) $ units are $ [\mu \; K^2] $.}
\label{corrR}
\end{figure}

\begin{table}
  \begin{tabular}{|c|c|c|c|c|}
 \hline   $~~~~~ k ~~~~~ $ (today) & $ \ln a $ at horizon exit 
    &  $ \epsilon_v $  at horizon exit & $k^{init}$ & $ \ell $ \\ \hline
$ k_Q = 0.238 \; {\rm Gpc}^{-1} $ &  $ 0.822 $ & $ 0.038 \gtrsim 1/N $
& $ 1.39 \; 10^{14} $  GeV & $ 2 $ \\ \hline
$  k_{tran} = 0.290  \; {\rm Gpc}^{-1} $ & $ 1.024 $ & $ 0.017 \simeq 1/N $
& $ 1.69 \; 10^{14} $ GeV  & $ \sim 2 $ \\ \hline
    $  k_{0} = 2 \; {\rm Gpc}^{-1} $ (WMAP)  &  $ 2.97 $ 
    & $ 0.0031\lesssim 1/N $  & $ 1.17 \; 10^{15} $ GeV & 25 \\ \hline
$ k = 16.1 \; {\rm Gpc}^{-1} $ & $ 5.06 $ &  $ 0.0033\lesssim 1/N $  &
$ 9.41 \; 10^{15} $ GeV & 220 \\ \hline
$  k_{0} = 50 \; {\rm Gpc}^{-1} $ (CosmoMC)  &  $ 6.19 $ 
& $ 0.0034\lesssim 1/N $  & $ 2.91 \; 10^{16} $ GeV & 510 \\ \hline
$ k = 0.71 \; {\rm Mpc}^{-1} $ & $ 8.86 $ & $ 0.0037\lesssim 1/N $ & $ 4.2 
\; 10^{17} $ GeV & $ 10^4 $\\ \hline
  \end{tabular}   \;.
\caption{The number of efolds $ \ln a $ since the beginning of inflation 
till relevant wavevectors exit the horizon and the parameter $ \epsilon_v $
at horizon exit. $ k^{init} $ stands for the value at the beginning of 
inflation. $  k_{0} $'s are pivot wavenumbers.
$ \ell $ stands for the $ C_\ell $ to which the corresponding $k$-mode 
contributes most reentering  the horizon. The quadrupole mode $ k_Q $ 
{\bf exits} the horizon during the fast-roll stage, about $ 0.2 $ efolds 
before fast-roll ends. $ k_{tran} $ exits the horizon precisely
at the {\bf transition} 
from the fast-roll to the slow-roll stage. We used here eqs.(\ref{qini}), 
(\ref{rangk}), (\ref{maxjl}), $ N_{tot} = 64 $ and $ \beta \sim 2.5 $. 
Recall that $ 1/N =1/60=0.0166\ldots$.}
\end{table}

In Table VII we denote by $ k_{0} $ the pivot scale
in the WMAP \cite{WMAP3,WMAP5} and CosmoMC codes \cite{lewis},
where the indices $ n_s , \; r $ and the running of $ n_s $ are computed. 
Both $ k_{0}$'s exit the horizon well inside the slow-roll regime.

\subsubsection{Real Space Two Point TT-Correlator}

We display in fig. \ref{corrR} the real space two point TT-correlation 
function $ C^{TT}(\theta) $ for $\Lambda$CDM, sharp cutoff and fast-roll 
models \cite{quamc},
$$
C^{TT}(\theta) = \frac1{4 \, \pi} \sum_{l=2}^{\infty}(2\,l+1) \; C_l^{TT} 
\; P_l(\cos \theta) \; .
$$
We see that the $\Lambda$CDM correlator becomes really different from the 
two others only for large angles $ \theta \gtrsim 1 $. Since all $l$-modes 
except the lowest ones are practically identical in the three cases, 
this shows how dominant  are the low multipoles in the large angle 
correlations. We also show the WMAP data, the width of the data is mostly 
due to the cosmic variance.

As is clear from fig. \ref{corrR}, both fast-roll and sharp cutoff models 
reproduce the two point correlator  $ C^{TT}(\theta) $ better than the 
pure slow-roll  $\Lambda$CDM$+r$ model.

\subsubsection{Conclusions} 

The best values for the ratio and spectral index including the early
fast-roll stage are given by the WMAP5+SN+SDSS
line of Table VI: $ r = 0.051, \; n_s = 0.964 $ and $ y = 1.26 $.

Fast-roll provides a slightly better fit than a sharp cutoff both for the 
$ C_{\ell}^{\rm TT} $ and the $ C_{\ell}^{\rm TE} $ coefficients. Besides 
reproducing the quadrupole supression, the 
fast-roll fit accounts for the oscillations of the lower multipole data.

We get the following picture of the inflationary universe 
explaining the quadrupole suppression from the effective 
(Ginsburg-Landau) theory of inflation combined with MCMC simulations 
of CMB$+$LSS data. A fast-roll stage lasting 
about one efold is followed by a slow-roll stage lasting $ \sim 63 $ 
efolds [see sec. \ref{ntot64}].  After these $ \sim 63+1 = 64 $ inflation efolds, 
the radiation dominated era follows. The quadrupole modes exit the horizon {\bf during}
the fast-roll stage about 0.8 efold after the beginning of inflation
and are therefore {\bf suppressed} compared with the modes exiting 
the horizon later during the slow-roll stage.

The fast-roll stage explains the quadrupole suppression and {\bf fixes the
total number of efolds} of inflation to be $ N_{tot} \simeq 64 $
as shown in sec. \ref{ntot64}.

\section{Quantum Loop Corrections}
\subsection{The domain of validity of the effective theory of 
inflation.}\label{validez}

We discuss in the subsections below the validity domain of the effective 
theory of inflation focusing on three regimes where corrections could be 
expected: a) large (transplanckian) values of the inflaton field, b) 
large $k$ fluctuation modes (transplanckian modes), c) parametric and 
spinodal resonant fluctuation modes.

\medskip

We consider in this review the inflaton with the canonical kinetic term
$$
X = \frac{\dot{\varphi}^2}2 -\frac{(\nabla\varphi)^2}{2 \; a^2(t)} \; .
$$
Lagrangeans with general functions of $ X $ has been considered in the literature
\cite{cine}. Since $ X $ has mass dimension fourth, these Lagrangeans introduce
at least one new dimensionful parameter $ \mu $. The corrections to the 
canonical kinetic case will be in the Ginsburg-Landau context of the order 
$$
{\cal O} \left(\frac{M^4}{\mu^4}\right) \; ,
$$
since $ X \sim M^4 $ during inflation (see sec. \ref{potuniv}). 
Moreover, it must be $ \mu \sim M_{Pl} $ in the Ginsburg-Landau context
since $ M_{Pl} $ is the only available dimensionful parameter here.
[$ \mu \sim M $ is excluded since the dominant kinetic term is always canonical
in Ginsburg-Landau effective theory.] Therefore, the observable effects of 
non-canonical kinetic terms in the Ginsburg-Landau effective theory are of the order
$$
{\cal O} \left(\frac{M^4}{ M_{Pl}^4}\right) \sim 10^{-11} \; ,
$$
and can be safely neglected.

\subsubsection{Large Inflaton Fields}

It is sometimes stated that the validity of the
effective field theory of inflation entails that
\be\label{val}
\frac{\varphi}{M_{Pl}} \ll 1   \; .
\ee
From this stringent constraint on the validity of an
effective field theory bounds on $ r $ have been proposed \cite{lyth}.

The validity of the constraint eq.(\ref{val}) suggested 
that effective field theory predicts values of $ r \ll 1 $  probably
too small to be observed in the next generation of CMB
experiments \cite{vpj}. Alternatively if larger values of $r$ are
measured then this would entail a breakdown of the effective field
theory approach to inflation.

This line of reasoning is incorrect for three different but
complementary reasons:

\begin{itemize}
\item{The validity of an  effective field theory expansion {\bf does
not} rely on  $ \varphi/M_{Pl} \ll 1 $, instead it relies on a
wide separation between the energy scale of inflation and the higher
energy scale of the underlying microscopic theory. If the
effective field theory emerges from integrating out degrees of
freedom at the GUT scale, then $ H/M_{GUT} \sim 10^{-2} $, if such
scale is instead the Planck scale then $ H/M_{Pl} \sim 10^{-5} $, and in 
either case an effective field theory description is valid. Indeed, 
detailed calculations discussed in secs. \ref{QLC1} and \ref{trece}
reveal that the quantum corrections to
slow-roll inflation are of the order $ \left(H/M_{Pl}\right)^2 $ 
\cite{pardec,effpot,quant}.
Any breakdown of an effective field theory is typically manifest
in large quantum corrections but the results of 
\cite{pardec,effpot,quant} presented in secs. \ref{QLC1} and \ref{trece}
unambiguously point out that quantum corrections are well under
control for $ H/M_{Pl} \ll 1 $. This provides a reassuring
confirmation of the validity of the effective field theory for a
scale of inflation consistent with the WMAP data. }

\item{Three simple examples highlight that the criterion
$ \varphi/M_{Pl} \ll 1 $ \emph{cannot} be the deciding factor for the
reliability of the effective field theory: consider the following
series
\bea 
& & \sum_{p=0}^{\infty}(-1)^p \left(\frac{\varphi}{M_{Pl}}
\right)^{2 \; p} = \frac1{1+ \left(\frac{\varphi}{M_{Pl}}\right)^2}
\label{seriesA} \\
& & \sum_{p=0}^{\infty}\frac{(-1 )^p}{(2p)!}\,\left(\frac{\varphi}{M_{Pl}}
\right)^{2 \; p} = \cos\left(\frac{  \varphi}{M_{Pl}}
\right)\label{seriesB}\\
& & \sum_{p=0}^{\infty} \left(\frac{\varphi}{M_{Pl}} \right)^{2 \; p}
=  \frac1{1- \left(\frac{\varphi}{M_{Pl}}\right)^2} \label{seriesC}
\; . 
\eea
The sum of both series (\ref{seriesA}) and
(\ref{seriesB}) is perfectly well defined for $ \varphi > M_{Pl} $. In
particular, the series (\ref{seriesB}) is a prototype for an
axion-type potential \cite{natu}, while certainly the series
(\ref{seriesC}) breaks down for $ \varphi \sim M_{Pl} $. The series
(\ref{seriesA})-(\ref{seriesB}) do not feature any real
singularity in the variable $ \varphi/M_{Pl} $ whereas
eq.(\ref{seriesC}) has a singularity at $ \varphi = M_{Pl} $. These
elementary examples highlight that what constraints the
reliability of the effective field theory description of the
inflationary potential is \emph{not} the value of the ratio
$ \varphi/M_{Pl} $ but the position of the \emph{singularities} as a
function of this variable. These singularities are determined by
the large order behavior of the \emph{coefficients} in the series.
If the series has a non-zero radius of convergence or if it is
just Borel summable, it defines the potential $ V(\varphi) $ {\bf
uniquely}. Clearly, a thorough knowledge of the radius of
convergence of the series in the effective field theory requires a
detailed knowledge of the underlying microscopic theory. However,
it should also be clear that the requirement $ \varphi/M_{Pl}\ll 1$ is
overly restrictive in general. } \item{One of the main results of
ref.\cite{1sN} (see sec. \ref{potuniv})
is that the combination of WMAP data and slow-roll
expansion suggest the universal form of the inflation potential eqs.(\ref{V})-(\ref{chifla}).
Even in the case when
the coefficients in a $\chi$-series expansion of $ w(\chi) $ lead
to a breakdown of the series at $ \chi \sim 1 $, namely at $ \varphi
\sim \sqrt{N} \; M_{Pl}$, there is still room for values of $
M_{Pl} < \varphi < \sqrt{N} \;  M_{Pl} $ for which the series would
be reliable. However, no \emph{ a priori} physical reason for such
a breakdown can be inferred without a reliable calculation of the
effective field theory from a microscopic fundamental theory.
Therefore, we expect that the effective field theory potential
$ V(\varphi)=N \; M^4 \; w(\chi) $ would be reliable \emph{at least} up
to $ \varphi \sim \sqrt{N} \; M_{Pl} $ and most generally for values $
\chi \gg 1 $ and hence $ \varphi \gg M_{Pl} $. }
\end{itemize}

As mentioned above, the studies in ref.\cite{pardec,effpot,quant} reveal
that  quantum corrections in the effective field theory yields an
expansion in $ (H/ M_{Pl})^2 $ for
\emph{general inflaton potentials}. This indicates
that the use of the inflaton potential $ V(\varphi) $ from effective
field theory is consistent for
$$
\left(\frac{H}{M_{Pl}}\right)^2 \ll 1 \quad {\rm and~hence} \quad
V(\varphi) \ll  M_{Pl}^4 \; .
$$
We  find using  eqs.(\ref{V})-(\ref{chifla}):
\be 
w(\chi) \ll \frac1{N} \;\left(\frac{M_{Pl}}{M}\right)^4 \sim \frac1{N} \; 10^{12} \; . 
\ee
This condition yields an  {\bf upper bound} in the inflaton field
$ \varphi $ depending on the large field behaviour of $ w(\chi) $. We
find for relevant behaviours of the inflaton potential the
following upper bounds on $ \chi $ and $ \varphi$:
\bea \label{cotxi}
&&  w(\chi) \buildrel{\chi \gg 1}\over \simeq \chi^2 \quad :
 \quad \chi \ll \frac{10^6}{\sqrt{N}} \quad , \quad
\varphi \ll  10^6 \; M_{Pl} \cr \cr
&& w(\chi) \buildrel{\chi \gg 1}\over \simeq \chi^4 \quad :
 \quad \chi \ll \frac{10^3}{N^{\frac14} }
 \quad , \quad  \varphi \ll 2783 \; M_{Pl} \quad {\rm for} \quad N\simeq 60
\cr \cr
&& w(\chi) \buildrel{\chi \gg 1}\over \simeq \chi^n \quad :
 \quad \chi \ll \left(\frac{10^{12}}{N}\right)^{\frac1{n}}
 \quad , \quad  \varphi \ll 10^{\frac{12}{n}} \; N^{\frac12 - \frac1{n}}
\; M_{Pl} \cr \cr
&& w(\chi) \buildrel{\chi \gg 1}\over \simeq e^{\chi}\quad :
 \quad \chi \ll 12 \; \ln 10 - \ln N = 23.54
 \quad , \quad  \varphi \ll 182 \;  M_{Pl}\quad {\rm for} \quad N\simeq 60 \; .
\eea 
We see that the effective field theory is {\bf consistent}
for values of the inflaton field well {\bf beyond} the Planck mass
even for very steep potentials, such as the exponential function $
e^{\chi} $.

\medskip

We remark that the arguments presented above suggest that the
reluctance to use the inflaton potential $ V(\varphi) $ for $ \varphi
\gtrsim M_{Pl} $ arises from a prejudice which is unwarranted
under the most general circumstances, unless of course, the
inflaton potential features singularities. The true upper bound
for the validity  of the effective field theory description of
inflation is $ V(\varphi) \ll M_{Pl}^4 $ which in fairly general cases
allows large values of $ \frac{\varphi}{M_{Pl}} $ as emphasized by
eqs.(\ref{cotxi}).

\subsubsection{Transplanckian and Resonant 
Fluctuation Modes.}\label{transPyR}

Fluctuation modes can have large wavenumbers going beyond the Planck mass.
The question is whether such large $k$ modes can have observable effects
through the primordial power on CMB anisotropies. 

The analysis presented in sec. \ref{tres} gives for the cosmologically 
relevant modes the window eq.(\ref{rangok}) where $ \beta \sim 2 $
according to the best fit eq.(\ref{myM}).
Hence, the upper limit in eq.(\ref{rangok}) becomes 
$ \simeq 7 \times  10^{18} \; e^{N_{tot}-64} $ GeV just
above the Planck mass for $ N_{tot} \sim 64 $.

Moreover, the CMB multipoles $ \ell < 200 $ where features have been 
observed \cite{WMAP1,WMAP3,WMAP5}, are definitely sub-Planckian for 
$ N_{tot} \sim 64 $ since they have $ k < 4 \times  \; 10^{16} \; 
e^{N_{tot}-64}$ GeV according to eqs.(\ref{rangk}) and (\ref{maxjl}).

\medskip

Inflaton modes with $ k \sim m $ often exhibit resonant exponential growth 
just after the beginning of inflation \cite{prehea,chalo,chalo2}. This 
phenomena arises from spinodal or
parametric instabilities in new and chaotic inflation, respectively.
One can wonder whether such exponential mode growth can have observable
consequences through the CMB spectrum. Since $ m \simeq  10^{13}$ GeV
[eq.(\ref{myM})], these potentially growing modes are {\bf below} the lower
end $ k^{init} \simeq 30 \; m $ of the observable window
eq.(\ref{rangok}) for $ N_{tot} > 61 $. It should be recalled that the 
nonlinear backreaction soon shuts off these instabilities 
\cite{chalo,chalo2}. In addition, these instabilities are unimportant in 
the thermalization process in classical field theory \cite{fi4}.

\subsection{Quantum Loop Corrections to the Inflaton Dynamics}\label{QLC1}

The study of quantum corrections to the inflaton dynamics is 
necessary to establish the validity of slow-roll inflation as a
reliable effective field theory description and in addition, it is
important to understand novel phenomena associated with quantum loop 
corrections in an expanding cosmology.

We study several novel aspects of quantum fluctuations that {\bf do not
have counterpart} in Minkowski space-time. This study also leads to
the conclusion that there is a small parameter $ (H/M_{Pl})^2 \sim
10^{-9} $ (where $ H $ is the Hubble parameter during inflation), which
emerges: this small ratio determines the magnitude of the 
quantum corrections. The reliability of the effective field theory 
approach is then a consequence of the smallness of this ratio during slow 
roll inflation.

\subsubsection{Quantum Decay Rate}\label{13A}

Particle production and decay are of
fundamental importance in cosmology. Most of the treatments of
particle decay in cosmology rely on the concept of the decay rate of
a particle in Minkowski space time. The framework for obtaining the
decay rate in Minkowski space time is the usual S-matrix approach or
at a more fundamental level Fermi's golden rule which manifestly
makes use of the energy-momentum conservation resulting in reaction
thresholds \cite{particulas}. In an expanding cosmology there is no global 
time-like Killing vector and energy is not conserved, although in an 
isotropic and homogeneous cosmology momentum is conserved. 
The lack of energy conservation implies major modifications in the 
kinematics for decay since there are no longer reaction thresholds. 
Furthermore, the S-matrix theory based on the in-out formulation of 
quantum field theory is ill-suited to study time dependent phenomena in an 
expanding cosmology and a completely different approach is needed to
understand particle production and decay. The non-equilibrium
quantum field theory provides a systematic set of tools to study
precisely such phenomena. Linear response is specially suited to
study relaxation phenomena in general and is tailored to study
particle decay in particular. The strategy is to study the effective
equations of motion for the expectation value of the fields in
presence of an external source as an initial value problem. An
external source prepares the initial expectation value, which
upon switching off the source relaxes towards equilibrium. 
According whether the field  expectation value decays exponentially
with time or not, the dynamics of relaxation is described by a
decay rate or by some alternative concept.

\medskip

In ref.\cite{pardec} we study the quantum decay of
a particle into other particles during inflation. The decaying
particle could be the inflaton but our study is more
general. In this section we provide an
understanding of the concept of decay of a particle in a rapidly
expanding cosmology. We introduce and implement a method that
allows a systematic and unambiguous study of the relaxational
dynamics of quantum fields and in particular allows to extract the
decay law resulting from interactions.  

In Minkowski space-time
there are two alternative but equivalent manners to define the
decay rate of a particle: (i) the total decay rate as the inclusive
transition probability \emph{per unit time} from an initial 'in'
state to final 'out' states, (ii) the total decay rate as the
imaginary part of the space-time Fourier transform of the
self-energy of the particle evaluated on the particle's mass shell
and divided by its mass-shell energy. Both definitions are
equivalent by dint of the optical theorem, or alternatively,
unitarity. The calculation of a total decay rate from definition
(i) involves calculating the transition amplitude from some initial
time $ t_i \rightarrow -\infty $ to a final time $ t_f \rightarrow
+\infty $ and multiplying by its complex conjugate. In Minkowski
space-time the transition amplitude from an asymptotic state in
the past to an asymptotic state in the future is proportional to
an energy conserving delta function. In squaring the amplitude,
the square of this delta function is interpreted as the total time ($T$)
elapsed in the reaction multiplying an energy conserving
delta function. Dividing by the total time ($T$) of the reaction
one extracts the decay rate. 

The calculation of the decay rate
from the total width via definition (ii) requires that the
self-energy be a function of the time difference and invokes
energy-momentum conservation at each interaction vertex. The
space-time Fourier transform of the self-energy features branch
cut singularities in the complex frequency plane and the imaginary
part across these cuts at the position of the particle mass shell
gives the decay width or decay rate \cite{particulas}. 

The important point in this discussion is that in the two cases above 
referred the concept of a decay rate relies heavily on energy 
(and momentum) conservation. Therefore, the concept of a decay rate 
(an inclusive transition probability per unit time)
may not have a translation to rapidly expanding cosmologies where 
there is no global
timelike Killing vector associated with conservation of energy
even when there may be space-like Killing vectors associated with
spatial translational symmetries and momentum conservation. Such
is the case for spatially flat FRW
cosmologies. The manifest lack of energy conservation in an
expanding cosmology makes possible processes that are
forbidden in static space-times by energy conservation.
In addition, contrary to Minkowski spacetime, cosmological modes in general
do not decay exponentially with time, therefore the definition of the 
decay rate requires the kind of analysis we provide here.

\vspace{1mm}

{\bf The method:}

Particle decay in de Sitter space-time was studied
in reference \cite{prem} for some very special cases that allowed a
solution of the equation of motion. In ref.\cite{pardec} we introduced a
method that allows to study the relaxation of quantum fields  and
particle decay  in great generality. The main strategy is to study
the effective equations of motion of the expectation values of
fields as an initial value problem in linear response including
the self-energy corrections. The solution of the equations of
motion lead to an unambiguous identification of the decay law from
the relaxation of the amplitude of the field as a consequence of
the self-energy corrections (interactions). When self-energy
corrections  are included the equations of motion become non-local
(non-Markovian) and cannot be solved in general in closed form.

When a perturbative solution of the equations of motion is
attempted there emerge \emph{secular terms}, namely terms that
grow in time and invalidate the perturbative expansion. These
secular terms indicate precisely the relaxation (or production)
time scales. We implement the dynamical renormalization group
introduced in \cite{DRG} to provide a systematic resummation of
these secular terms leading to the correct description of
relaxation and decay. Such program has been successfully applied
to a wide variety of non-equilibrium situations in Minkowski space
time (see \cite{DRG} and references therein).

In ref.\cite{DRG} we study the relaxation of
the expectation value of quantum fields as an initial value
problem. In ref.\cite{pardec}, we illustrate its application
and study the decay of a massive particle (it could be the inflaton) 
coupled to conformally coupled massless particles via a trilinear 
vertex in de Sitter space time.

 This simple setting allows to present the main aspects of the
 program and reveal the important features associated with
 the expansion in a clear manner. The relaxation and decay law is
 studied to lowest order in the coupling both for wavelengths that
 are inside and outside the Hubble radius during inflation. The
 decay constant for superhorizon modes have an interesting
 interpretation in terms of the Hawking temperature of de
Sitter space-time. After extracting the decay law to lowest order
in the loop expansion for the self-energy, we show that the limit of
Minkowski space-time reproduces the known Minkowski space-time
decay rates. In all cases we find that the decay
is \emph{enhanced} during inflation as compared to the Minkowski
space-time result. The decay law for modes deep within the horizon
feature a wavevector dependence that leads to a larger suppression
of the amplitude for longer wavelengths \cite{pardec}.

 \medskip

We begin by studying in refs. \cite{pardec} 
the general case of a cubic interaction of
scalar particles minimally coupled to gravity,  allowing the decay
of one field  into two others during de Sitter inflation. The
masses of all particles are much smaller than the Hubble constant,
which leads to a strong infrared behavior in the self-energy
loops. We introduce an expansion in terms of the small parameter
$ M^2/H^2 $, where $ M \ll H $ is the mass of the particle in the quantum 
loop, which naturally regulates the infrared behavior.

Long-time divergences associated with secular terms in the solutions of the
equations of motion are systematically resummed by implementing
the DRG introduced in ref.\cite{DRG} and lead to the
decay law. We then apply these general results to the case of
quasi-de Sitter slow-roll inflation. We show that in this case a
similar small parameter emerges which is a simple
function of slow-roll parameters and regulates the infrared
behavior {\it even} for massless particles. We study the decay of
superhorizon fluctuations as well as of fluctuations with
wavelengths deep inside the horizon. A rather striking aspect is
that a particle {\bf can decay} into
 \emph{itself} precisely as a consequence of the lack of energy
 conservation in a rapidly expanding cosmology. We then focus on
 studying the decay of the  inflaton quantum fluctuations into
 their \emph{own quanta}, namely the {\it self-decay} of the inflaton
 fluctuations, discussing the potential implications on the power
 spectra of primordial perturbations and on non-gaussianity.

 \vspace{2mm}

Our main results on quantum decay rate go as follows \cite{pardec}:

\begin{itemize}
\item{In the case of de Sitter inflation for particles with mass
$ M \ll H $ a small parameter $ \sim M^2/H^2 $ regulates the
infrared regime. We introduce an expansion in this small parameter 
$ M^2/H^2 $ akin to the $ \varepsilon $  expansion in dimensionally 
regularized critical theories \cite{wilkog}. After implementing the DRG 
resummation, we obtain the decay laws in a $ M^2/H^2 $ expansion.}

\item{Minimally coupled particles decay \emph{faster} than those
conformally coupled to gravity due to the strong infrared behavior
both for superhorizon modes as well as for modes with wavelengths
well inside the Hubble radius.}

\item{The decay of short wavelength modes, those inside the
horizon during inflation, is \emph{enhanced} by soft collinear
\emph{bremsstrahlung radiation of superhorizon quanta} which
becomes the dominant decay channel for the physical wave vectors which
obey, 
\be 
k_{ph}(\eta) \lesssim \frac{H}{\eta_v-\epsilon_v}= \frac{2 \; H}{n_s - 1
+\frac14 \; r} \; ,
\ee 
where $ \eta_v, \; \epsilon_v $ are the slow-roll parameters 
eq.(\ref{slrsd}).}

\item{Expanding cosmologies allow processes that are forbidden
in Minkowski space-time by energy conservation \cite{pardec,woo}: 
in particular, for particle masses $ \ll H $, 
\emph{kinematic thresholds} are absent allowing a particle to decay into 
\emph{itself}. Namely, the \emph{self-decay} of quantum fluctuations is a 
feature of an interacting theory in a rapidly expanding cosmology. A
self-coupling of the inflaton leads to the self-decay of its
quantum fluctuations both for modes inside as well as
\emph{outside} the Hubble radius. }

\item{The results obtained in de Sitter space-time directly apply to the
\emph{self decay} of the  quantum fluctuations of the inflaton
during slow-roll (quasi de Sitter) expansion. In this case,
a simple linear combination of slow-roll parameters regulates the 
infrared. For superhorizon modes we find that the amplitude of the  
inflaton quantum fluctuations relaxes as a power law $ \eta^{\Gamma} $
in conformal time $ \eta $ where $ \Gamma $ is the decay rate. To lowest 
order in slow-roll, we find $ \Gamma $ completely determined by the slow 
roll parameters and the amplitude of the power spectrum of curvature 
perturbations $ \triangle^2_{\mathcal{R}} $ eq.(\ref{ampliI}): 
\be 
\Gamma = \frac{8 \; \xi^2_v \; 
\triangle^2_{\mathcal{R}}}{(\epsilon_v-\eta_v)^2}
\left[1+\mathcal{O}(\epsilon_v,\eta_v)\right]
\ee
\noindent where $ \xi_v, \; \eta_v, \; \epsilon_v $ are the slow-roll 
parameters eqs.(\ref{slrsd}), (\ref{xisig}) and (\ref{sigmv}). As a 
consequence, the growing mode which dominates at late time evolves as 
\be 
\frac{\eta^{\eta_v-\epsilon_v +\Gamma}}{\eta} \;.
\ee 
featuring an {\it anomalous dimension} $ \Gamma $ slowing down
the growth of the dominant mode.

 The decay of the inflaton quantum fluctuations with wavelengths
 deep within the Hubble radius during slow-roll inflation is
 {\bf enhanced} by the infrared behavior associated with the collinear
 emission of ultrasoft  quanta, namely \emph{ bremsstrahlung
radiation of superhorizon fluctuations}. The decay  hastens as the
physical wavelength approaches the horizon because the phase space
for the emission of superhorizon quanta opens up as the wavelength
nears horizon crossing.}

\item{We discuss the implications of these results for scalar and
tensor perturbations, and establish a connection with previous
calculations of non-gaussian correlations.}
\end{itemize}

\subsubsection{Particle decay in inflationary cosmology}\label{partdecay} 

Let us first consider a simple model of two quantum scalar fields 
$ \phi $ and $ \varphi $ with masses $ M $ and $ m $,
respectively, in an expanding FRW background 
with a cubic coupling $ g \; \phi \; \varphi^2 $.
This is a paradigm to study the decay 
$ \phi \rightarrow \varphi \; \varphi $ and the
action for the model in cosmic time is 
\be
A= \int d^3x \; dt \;  a^3(t) \Bigg\{ \frac12 \; 
{\dot{\phi}^2}-\frac{(\nabla
\phi)^2}{2 \; a^2}-\frac12 \Big(M^2+\xi_\chi \; \mathcal{R}\Big)\phi^2 +
\frac12 \; {\dot{\varphi}^2}-\frac{(\nabla
\varphi)^2}{2 \; a^2}-\frac12 \Big(m^2+\xi_\delta \;
\mathcal{R}\Big)\varphi^2 - g \;  \phi \, \varphi^2 +J(t) \;
\phi\Bigg\}
\ee
\noindent with 
\be 
\mathcal{R} = 6 \;  \Big(\frac{\ddot{a}}{a}+\frac{\dot{a}^2}{a^2}\Big) 
\ee 
being the Ricci scalar and $ \xi $ an arbitrary coupling to the Ricci 
scalar: $ \xi= 0 $ corresponds to minimal coupling  and $ \xi= 1/6 $ 
corresponds to conformal coupling. The source $ J(t) $ induces an 
expectation value of the field $ \phi $ and allows to set up an initial 
value problem for its dynamics. In this case, neither $ \phi $ nor 
$ \varphi $ are the inflaton.

It is convenient to pass to conformal time $ \eta $ [eq.(\ref{confo})]
and introduce a conformal rescaling of the fields,
$$
a(t) \; \phi(\vx,t) = \chi(\vx,\eta) \quad , \quad
a(t) \; \varphi(\vx,t)=\delta(\vx,\eta)\;.
$$
The action becomes (after discarding surface terms that will not
change the equations of motion) 
\be 
A\Big[\chi,\delta\Big]= 
\int d^3x \; d\eta \; \Bigg\{\frac12\left[ {\chi'}^2-(\nabla
\chi)^2-\mathcal{M}^2_{\chi}(\eta) \; \chi^2 + {\delta'}^2 -(\nabla
\delta)^2-\mathcal{M}^2_{\delta}(\eta) \; \delta^2 \right] -g
a(\eta) \;  \chi \; \delta^2 - a^3(\eta) \; J(\eta) \;  \chi \Bigg\}
\; , 
\ee 
\noindent with primes denoting derivatives with respect to
conformal time $ \eta $ and 
\be 
\mathcal{M}^2_{\chi}(\eta) =
\Big(M^2+\xi_\chi \; \mathcal{R}\Big)
a^2(\eta)-\frac{a''(\eta)}{a(\eta)} \quad ,  \quad
 \mathcal{M}^2_{\delta}(\eta) =  \Big(m^2+\xi_\delta \; \mathcal{R}\Big)
a^2(\eta)-\frac{a''(\eta)}{a(\eta)}  \; , 
\ee 
For inflationary cosmology the scale factor describes a de Sitter
space-time, namely 
\be \label{scalefactor} 
a(t)= e^{Ht} ~~;~~ a(\eta) = -\frac1{H \; \eta} \quad , \quad t> 0 \quad , \quad \eta< 0
\; ,
\ee 
\noindent with $ H $ the Hubble
constant. The Heisenberg equations of motion for the spatial Fourier
modes of wavevector $ k $  of the fields in the non-interacting
($ g=0 $) and source-free [$ J(\eta)=0 $] theory are given by 
\bea
&&\chi''_{\vk}(\eta)+
\Big[k^2-\frac1{\eta^2}\Big(\nu^2_{\chi}-\frac14 \Big)
\Big]\chi_{\vk}(\eta)  =   0 ~~;~~ \delta''_{\vk}(\eta)+
\Big[k^2-\frac1{\eta^2}\Big(\nu^2_{\delta}-\frac14 \Big)
\Big]\delta_{\vk}(\eta)  =   0 \label{modes}  \; , \cr \cr
&& \nu^2_{\chi}  =  \frac{9}{4}-\Big( \frac{M^2}{H^2}+12 \; \xi_\chi
\Big) ~~;~~ \nu^2_{\delta}  = \frac{9}{4}-\Big( \frac{m^2}{H^2}+12
\; \xi_\delta \Big) \; . \label{nus}
\eea
The spatial Fourier
transform of the free field Heisenberg operators $ \chi_{\vk}(\eta),
\; \delta_{\vk}(\eta) $ are therefore written as 
$$
\chi_{\vk}(\eta)  =  \alpha_{\vk} \; g_{\nu_\chi}(k;\eta)+
\alpha^\dagger_{-\vk} \; g^*_{\nu_\chi}(k;\eta) \quad ,  \quad
\delta_{\vk}(\eta)  =  \beta_{\vk} \;
g_{\nu_\delta}(k;\eta)+\beta^\dagger_{-\vk} \;
g^*_{\nu_\delta}(k;\eta) \; , 
$$
\noindent where the Heisenberg operators 
$ \alpha_{\vk},  \; \alpha^\dagger_{\vk} $
and $ \beta_{\vk},  \; \beta^\dagger_{\vk} $ obey the usual canonical
commutation relations eq.(\ref{ccr}). The mode functions $ g_{\nu}(k,\eta) $ with
Bunch-Davis boundary conditions are given by eq.(\ref{gnu}).

\begin{figure}
\includegraphics[height=3 in,width=3 in,keepaspectratio=true]{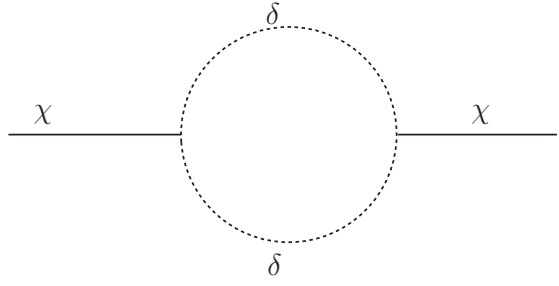}
\caption{Self energy loop of  $ \delta $ particles (dashed lines). }
\label{fig:selfenergy}
\end{figure}

The equation of motion for the expectation value of the field $\chi$
is obtained by writing 
\be \label{shift} 
\chi_{\vk}(\eta) =
X_{\vk}(\eta)+ \sigma_{\vk}(\eta)~~;~~\langle
\chi_{\vk}(\eta)\rangle = X_{\vk}(\eta)~~;~~\langle
\sigma_{\vk}(\eta)\rangle =0 \; . 
\ee 
\noindent in the above
expressions $ \langle \cdots \rangle $ stand for expectation values in
the initial state which is prepared by switching on the external
source term $ J $ to displace the field and switching the source off
to let the field evolve. This is the usual method to prepare an
initial value problem in linear response. Implementing the condition
$ \langle \sigma_{\vk}(\eta)\rangle =0 $ order by order in an
expansion in the coupling $ g $ yields the equation of motion
including the self-energy to the given order. Up to one loop order
we obtain 
\be \label{eqnofmot2}
X''_{\vk}(\eta)+\left[k^2+\mathcal{M}^2_{\chi}(\eta)\right]
X_{\vk}(\eta)+ 2 \, g^2 \; a(\eta) \; \int_{\eta_0}^{\eta}
d\eta'~a(\eta') \; \mathcal{K}_k(\eta,\eta') \; X_{\vk}(\eta')=0\;.
\ee 
The Feynman diagram for the self energy kernel $ \mathcal{K}_k $ is
depicted in fig. \ref{fig:selfenergy}. In the general case the
self-energy kernel is a complicated non-local function of
$ \eta,\eta' $, but it simplifies in the case when the $ \varphi $
particles are massless and conformally coupled to gravity, in which
case $ \nu_{\delta} = 1/2 $ and
\be\label{confmode} 
g_{\nu_{\delta}}(k;\eta)= \frac1{\sqrt{2k}} \;
e^{-i \; k \; \eta} \; ,
\ee 
\noindent and the non-local kernel is given by 
\be\label{kernelconf} 
\mathcal{K}_k(\eta,\eta') = \int\frac{d^3q}{(2\pi)^3}
\frac{\sin\left[(q+|\vec{k}+\vec{q}|)(\eta-\eta')\right]}{2 \, q \; |\vec{k}+\vec{q}|}
=-\frac1{8\pi^2}\cos[k(\eta-\eta')]~\mathcal{P}
\left(\frac1{\eta-\eta'} \right) 
\ee 
\noindent where $ \mathcal{P} $
stands for the principal part. We \emph{define} the principal part
prescription as follows 
\be\label{PP} 
\mathcal{P} \left(\frac1{\eta-\eta'} \right) \equiv
\frac{\eta-\eta'}{\left(\eta-\eta'\right)^2 + (\epsilon \;
\eta')^2}= \frac12\left[\frac1{\eta-\eta'+i\epsilon \; \eta'}+
\frac1{\eta-\eta'-i\epsilon \; \eta'}\right] ~~;~~ \epsilon
\rightarrow 0^+
\ee 
This prescription for the principal part regulates
the short distance divergence in the operator product expansion with
a dimensionless infinitesimal quantity $ \epsilon $ independent of
time. Although the equation of motion (\ref{eqnofmot2}) is linear, it
is non-local and a general solution is unavailable. However for weak
coupling $ g^2 $ a perturbative solution can be found by writing 
\be
X_{\vk}(\eta)= \sum_{n=0}^{\infty}(g^2)^n~X_{n,\vk}(\eta) \; , 
\ee
\noindent 
leading to the hierarchy of coupled equations
\begin{eqnarray}\label{perteqn}
&&X''_{0,\vk}(\eta)+\left[k^2-\frac1{\eta^2}\Big(\nu^2_{\chi}-
\frac14 \Big) \right]X_{0,\vk}(\eta)
 =  0  \; ,\\
&&X''_{n,\vk}(\eta)+\left[k^2-\frac1{\eta^2}\Big(\nu^2_{\chi}-
\frac14 \Big) \right]
X_{n,\vk}(\eta) = \mathcal{R}_n(k;\eta) ~~;~~n=1,2\cdots\\
&&  \mathcal{R}_n(k;\eta) =   - 2 \; a(\eta)\int_{\eta_0}^{\eta}
d\eta'~a(\eta') \; \mathcal{K}_k(\eta,\eta') \; X_{n-1,\vk}(\eta')
\; . \label{rn}
\end{eqnarray}
The hierarchy can be solved order by order by introducing the
retarded Green's function 
\be\label{GF} 
\mathcal{G}(k;\eta,\eta')=i
\left[g_{\nu_\chi}(k;\eta) \; g^*_{\nu_\chi}(k;\eta')-
g^*_{\nu_\chi}(k;\eta) \; g_{\nu_\chi}(k;\eta')
\right]\Theta(\eta-\eta')\;. 
\ee 
Hence the solution of the hierarchy
of equations for $n\geq 1$ is given by 
\be \label{forsol}
X_{n,\vk}(\eta)= \int_{\eta_0}^{0} d\eta' \;
\mathcal{G}(k;\eta,\eta') \; \mathcal{R}_{n}(k;\eta') \; . 
\ee

{\bf Superhorizon modes: $k=0$ } In this case the solution to the
equation of motion at zeroth order is 
\be\label{Xcero}
X_{0,\vec{0}}(\eta) = a \; (-\eta)^{\beta_+}+ b \; (-\eta)^{\beta_-}
\; , \ee where $ a $ and $ b $ are constant coefficients and   \be
\beta_{\pm}  = \frac12\pm \nu_{\chi}~~,~~\nu_{\chi}=
\sqrt{\frac94-\frac{M^2}{H^2}}\label{betapm}  \; . 
\ee
The retarded Green's function Eq.(\ref{GF}) necessary to solve the
hierarchy of coupled equations, for $k=0$ is given by
\be\label{GFk0} 
\mathcal{G}(\eta,\eta')=  \frac{\sqrt{\eta \;
\eta'}}{2 \; \nu_{\chi}}
\left[\left(\frac{\eta'}{\eta}\right)^{\nu_{\chi}}-
\left(\frac{\eta}{\eta'}\right)^{\nu_{\chi}}\right]\Theta(\eta-\eta')
\; . 
\ee 
The first order correction $X_{1,\vec{0}}(\eta)$ is given
by 
\bea &&X_{1,\vec{0}}(\eta)=- \frac1{2 \; (2 \; \pi \; H)^2 \;
\nu_{\chi}} \left( a \; (-\eta)^{\beta_+} \left\{ \left[
\log\epsilon + \gamma + \psi(1-\beta_+) \right] \log
\frac{\eta}{\eta_0} + \mbox{non-secular}\right\} \right. \cr \cr
&&\left. - b \; (-\eta)^{\beta_-}  \left\{ \left[ \log\epsilon +
\gamma + \psi(1-\beta_-) \right] \log \frac{\eta}{\eta_0} +
\mbox{non-secular}\right\} \right)\;, 
\eea 
\noindent where the
non-secular terms are terms {\bf bounded} in the limit $ \eta
\rightarrow 0 $. Therefore, to first  order in the coupling we find
the solution of the equation of motion for the $ k= 0$ mode to
be given by 
\bea\label{solX0} 
&&X_{0}(\eta) = a \; (-\eta)^{\beta_+}\left(1-\frac{g^2}{2 \; (2 \; \pi \; H)^2 \;
\nu_{\chi}} \left\{ \left[ \log\epsilon + \gamma +  \psi(\beta_-)
\right] \log \frac{\eta}{\eta_0} + \mbox{non-secular}\right\}
\right) +\cr \cr 
&&+b \; (-\eta)^{\beta_-}\left(1+\frac{g^2}{2 \; (2
\; \pi \; H)^2 \; \nu_{\chi}} \left\{ \left[ \log\epsilon + \gamma +
\psi(\beta_+) \right] \log \frac{\eta}{\eta_0} +
\mbox{non-secular}\right\} \right) \; . 
\eea 
where we used that $ \beta_+ + \beta_- = 1 $ [eq.(\ref{betapm})]. Obviously the
\textbf{secular} terms, namely the terms $ \propto \log \eta $ grow in
the limit $ \eta \rightarrow 0 $ and would lead to a breakdown of the
perturbative expansion. The \emph{dynamical renormalization group}
furnishes a systematic \emph{resummation} of the perturbative
expansion that yields an asymptotically convergent result. After a
mass renormalization, the solution of the \emph{dynamical
renormalization group equation} yields 
\be \label{resumX0ren}
X_{0}(\eta)=
\left[\frac{\eta}{ {\eta}_0}\right]^{\Gamma_1}\Bigg\{a( {\eta}_0) \;
(-\eta)^{\beta_+} [1+\mathcal{O}(g^2)]+b( {\eta}_0) \;
(-\eta)^{\beta_-}[1+ \mathcal{O}(g^2)]\Bigg\} \; ,
\ee  
where $ \eta_0 $ is an arbitrary renormalization scale. We now
clearly see that a decay rate $ \Gamma_1 $ can be unambiguously
identified from the contribution that multiplies \emph{both}
solutions, it is given by 
\be\label{Mgra} 
\Gamma_1 = \frac{g^2 \;
\tanh\left[\pi\sqrt{\frac{M^2}{H^2} - \frac{9}{4}}\right]}{16 \; \pi
\;  H^2 \; \sqrt{\frac{M^2}{H^2} - \frac{9}{4}}}\; . 
\ee 
The decay
rate in de Sitter space-time in cosmic time can be read off from the
above expression since  \be
\left[\frac{\eta}{\eta_0}\right]^{\Gamma_1} = e^{-\Gamma_1 \; H \;
t} \equiv e^{-\Gamma_{dS} \; t}. \label{GammaDS} \ee and $
\Gamma_{dS} = H \; \Gamma_1 $.

{\bf Minkowski space time limit:}

\vspace{1mm}

Minkowski space time is recovered in
the limit $H\rightarrow 0$. In such limit we find from
  eqs.(\ref{betapm}) and (\ref{Mgra})
 \bea
&&\nu_{\chi}\buildrel{H \rightarrow 0} \over= i \; \frac{M}{H} \cr
\cr
 &&\Gamma_1 \buildrel{H \rightarrow 0} \over=\frac{g^2}{16\pi M H}\,\tanh\left[
\frac{\pi\,M}{H}\right] \buildrel{H \rightarrow 0} \over=
\frac{g^2}{16\pi M H}\nonumber \\
&&\beta_{\pm} \buildrel{H \rightarrow 0} \over= \pm i \; \frac{M}{H}
\; . \eea 
Therefore, we find in this limit 
\be\label{limitzeromom}   
\left[\frac{\eta}{\eta_0}\right]^{\Gamma_1} = e^{-\frac{g^2}{16\pi M
}t} = e^{-\Gamma_M \;  t}\; , 
\ee 
\noindent which displays the exponential decay of the amplitude with the correct
decay rate $ \Gamma_{M}= g^2/(16 \pi M)$ for long-wavelength
excitations with mass $ M $ in Minkowski space-time. Since $ -\eta =
e^{-Ht}/H $,  up to an overall normalization of the field, the
dynamical renormalization group improved solution in this limit is
given by 
\be 
X_0(t) = e^{-\frac{g^2}{16\pi M }t}\left[A \; e^{i M
t}+ B \; e^{-iMt}\right] 
\ee 
\noindent which is the correct solution
for a zero momentum excitation in Minkowski space-time \cite{pardec}.

{\bf Modes that exit the horizon during inflation $ k \neq
0~~,~~ \eta \rightarrow 0^-$. }

\vspace{1mm}

For arbitrary $k$ the integral equation (\ref{eqnofmot2}) takes the
form \be \label{kgen}
X''_{\vk}(\eta)+\left[k^2-\frac1{\eta^2}\Big(\nu^2_{\chi}-
\frac14 \Big) \right] X_{\vk}(\eta)- \left(\frac{g}{2 \; \pi \;
H}\right)^2 \; \frac1{\eta} \; \int_{\eta_0}^{\eta}
\frac{d\eta'}{\eta'} \; \frac{(\eta-\eta') \; \cos k(\eta-\eta')}{
(\eta-\eta')^2 + ( \epsilon \; \eta')^2 } \; X_{\vk}(\eta')=0 \ee
where we used eqs.(\ref{kernelconf} )-(\ref{PP}).

To first order in $ g^2 $ the solution $ X_{1,\vk}(\eta) $ given by
Eq.(\ref{forsol}) becomes 
\be\label{x1kg}
 X_{1,\vk}(\eta) = \frac1{(2\pi H)^2} \int_{\eta_0}^{\eta}
\frac{d\eta'}{\eta'} \;  \mathcal{G}(k;\eta,\eta')
\int_{\eta_0}^{\eta'}\frac{d\eta''}{\eta''} \; \frac{(\eta'-\eta'')
\; \cos k(\eta'-\eta'')}{ (\eta'-\eta'')^2 + ( \epsilon \; \eta'')^2
} \; X_{0,\vk}(\eta'') \;, 
\ee 
where $ \mathcal{G}(k;\eta,\eta') $ is
given by eq.(\ref{GF}) and for simplicity we consider the solution
$$
 X_{0,\vk}(\eta) = A_{\vk}\; g_{\nu_{\chi}}(k;\eta) \; .
$$
with $ g_{\nu_{\chi}}(k;\eta) $ the mode function with Bunch-Davies
initial condition eq.(\ref{gnu}). Eq.(\ref{x1kg}) can be written in the following
form 
\be \label{x1u}
 X_{1,\vk}(\eta) = A_{\vk} \int_{\eta_0}^{\eta} \frac{d\eta'}{\eta'} \;
 g_{\nu_{\chi}}(k;\eta') \; J(\eta,\eta')  \; ,
\ee where \be \label{jota} J(\eta,\eta') = \frac{\sqrt{\eta}}{8\pi^2
H^2}
 \; \int_{\eta'}^{\eta} \frac{d\eta''}{\sqrt\eta''}
\frac{(\eta'-\eta'') \; \cos k(\eta'-\eta'')}{ (\eta'-\eta'')^2 + (
\epsilon \; \eta')^2 } \; \mbox{Im}\left[
H^{(1)}_{\nu_{\chi}}(k \; \eta) \; H^{(2)}_{\nu_{\chi}}(k \; \eta'')
\right]\;. 
\ee
Our goal is to evaluate $  X_{1,\vk}(\eta) $ for $ \eta \to 0^- $.
In order to achieve this we need $ g_{\nu_{\chi}}(k;\eta) $ and the
integrand in Eq.(\ref{jota}) for small arguments: 
\bea\label{gHH} 
&&
g_{\nu}(k;\eta) \buildrel{\eta\rightarrow 0^- }\over=
\frac{\sqrt{\pi \; \eta}}2 \; i^{-\nu-\frac12} \left\{ \frac{i
\; \Gamma(\nu)}{\pi} \; \left( \frac2{k \; \eta} \right)^{\nu}
\left[1 + {\cal O}\left(k^2 \; \eta^2 \right) \right]
+ \frac{1 - i \; \cot\pi\nu }{ \Gamma(\nu+1)} \left( \frac{k \;
\eta}{2}\right)^{\nu}\left[1 + {\cal O}\left(k^2 \; \eta^2 \right)
\right]\right\} \; , \cr \cr &&\mbox{Im}\left[
H^{(1)}_{\nu_{\chi}}(k \; \eta) \; H^{(2)}_{\nu_{\chi}}(k \; \eta')\right]
\buildrel{\eta,\eta'\rightarrow 0^- }\over=
 \frac1{\pi \; \nu} \left[ \left( \frac{\eta}{\eta'} \right)^{\nu} -
\left( \frac{\eta'}{\eta} \right)^{\nu}\right] \left[1 + {\cal
O}\left(\eta^2 , {\eta'}^2 \right) \right]\; . 
\eea 
Inserting Eq.(\ref{gHH}) in Eq.(\ref{x1u}) and (\ref{jota}) yields, 
\be \label{x1m}
 X_{1,\vk}(\eta)\buildrel{\eta \rightarrow 0^-} \over= A_{\vk}\;
 \frac{2^{\nu-2} \; \Gamma(\nu)}{\sqrt{\pi} \; \nu
\;  i^{\nu+\frac32} \; k^{\nu}} \; \eta^{\frac12-\nu} \; S\left(
\frac{\eta_0}{\eta}\right) \; , 
\ee 
where 
\be
S\left(\frac{\eta_0}{\eta}\right) \buildrel{\eta \rightarrow 0^-}
\over= - \frac12 \int_1^{\frac{\eta_0}{\eta}} \frac{dy}{y}
\left(y^{-2\,\nu} - 1 \right) \int_1^{\frac{\eta_0}{y \, \eta}}
\frac1{(1+i\epsilon)t - 1} \; \frac{dt}{t^{\nu+\frac12}} +
(\epsilon \to - \epsilon) \; . 
\ee 
Carrying out the integrations leads to the following result 
\be 
\eta^{\frac12-\nu} \; S\left(\frac{\eta_0}{\eta}\right) 
\buildrel{\eta \rightarrow 0^-}
\over=\eta^{\frac12-\nu}\left\{ \left[ \psi\left(\nu + \frac12
\right) + \gamma + \ln\epsilon \right] \log\frac{\eta}{\eta_0} +
{\cal O}\left(\eta_0\right)\right\} \; . 
\ee 
Notice that this $ \eta \rightarrow 0^- $ behavior turns out to be $k$-independent. This is
due to the fact that the term $k^2$ in Eq.(\ref{kgen}) becomes
negligible compared to the $ \frac1{\eta^2} $ term for  $ \eta
\rightarrow 0^- $ after the modes cross the horizon.

Mass renormalization  cancels the logarithmic short distance
singularity $ \ln\epsilon $ leading to the following result up to
order $ g^2 $, 
\be
 X_{0,\vk}(\eta) + g^2 \;  X_{1,\vk}(\eta)
\buildrel{\eta \rightarrow 0^-}\over= A_{\vk}\;
 \frac{2^{\nu-1} \; \Gamma(\nu)}{\sqrt{\pi}
\;  i^{\nu-\frac12} \; k^{\nu}} \; \eta^{\frac12-\nu} \left[ 1 +
\frac{g^2 \; \tan\pi\nu_{\chi}}{16 \; \pi \; \nu_{\chi}
 \; H^2} \; \log\frac{\eta}{\eta_0} + {\cal O}\left(\eta_0\right)\right]
\; . 
\ee 
This result features  the secular term $ \log \eta $ which
is resummed by implementing the DRG as in sec.IV-A with the result \cite{pardec},
\be
 X_{\vk}(\eta)\buildrel{\eta \rightarrow 0^-}\over=
\frac{2^{\nu-1} \; \Gamma(\nu)\;{\eta}^{\frac12-\nu}}{\sqrt{\pi} \;
i^{\nu-\frac12} \; k^{\nu}} \; A_{\vk}(\tilde{\eta}_0)
\left[\frac{\eta}{\tilde{\eta}_0}\right]^{\Gamma_1}\left[1 +
 {\cal O}\left(g^2\right)\right]\;,
\ee 
\noindent  where $ \Gamma_1 $ was defined in eq.(\ref{gamma}). It
is clear from this result that the  behaviour for  $ \eta
\rightarrow 0^- $ and  $ k \neq 0 $ and fixed, is
the same as that for the case  $ k=0 $ [see eq.(\ref{resumX0ren})].
This due to the fact that the physical wavenumbers $ k \; \eta $ become
so small for $ \eta \rightarrow 0^- $ that they bear no relevance on
the late time dynamics. While this result could be expected on
physical grounds, it is important to see it emerge from the
systematic implementation of the DRG method.

{\bf Wavelengths much smaller than the Hubble radius: $|k \; \eta| \gg
1$:}

\vspace{1mm}

For $ |k\; \eta| \gg 1 $ corresponding to modes with
 wavelengths much smaller than the Hubble radius during inflation,
the hierarchy of equations  of motion up to $ \mathcal{O}(g^2) $ is
given by
\begin{eqnarray}\label{perteqnKgran}
  X''_{0,\vk}(\eta)+k^2\,X_{0,\vk}(\eta) & =  &  0  \\
   X''_{1,\vk}(\eta)+k^2\,X_{1,\vk}(\eta)  & =  &
   \mathcal{R}_1(k;\eta)  \; ,
\end{eqnarray}
   \noindent with the inhomogeneity now given by
\begin{equation}\label{inho}
  \mathcal{R}_1(k;\eta) =  -\frac{\delta M^2_1}{H^2\eta^2} \;
X_{0,\vk}(\eta)  - 2 \; a(\eta)\int_{\eta_0}^{\eta} d\eta'~a(\eta')
\; \mathcal{K}_k(\eta,\eta') \; X_{0,\vk}(\eta') \; .
\end{equation}
where $ \mathcal{K}_k(\eta,\eta') $ is given by Eq.(\ref{kernelconf}) 
and we have explicitly introduced a mass renormalization
counterterm $ \delta M^2_1 $.

The solution of the zeroth order equation is 
\be 
X_{0,\vk}(\eta)= A_k\, e^{-i \; k \; \eta}+B_k \, e^{i \; k \; \eta}  \; . 
\ee 
The counterterm $ \delta M^2_1 $ is chosen to cancel the short distance 
divergence proportional to $ (\ln\epsilon)/\eta^2 $. After straightforward 
but lengthy algebra we find the following expression for the
inhomogeneity in the limit $ |k \; \eta_0| \gg |k \; \eta| \gg 1 $ 
\cite{pardec}
\be
\mathcal{R}_1(k;\eta)= \frac1{8\pi^2 \eta^2 H^2} \Bigg\{
A_k\,e^{-i \; k \; \eta} \left[\ln \frac{\eta}{\eta_0} + i \; \frac{\pi}2
\right] + B_k\, e^{i \; k \; \eta} \left[\ln \frac{\eta}{\eta_0} - i \;
\frac{\pi}2 \right] + \cdots\Bigg\}\;, 
\ee 
\noindent where the
dots stand for terms that are subleading in the limit $ |k \; \eta| \gg
1$. The inhomogeneous equation for the first order correction can
now be solved in terms of the retarded Green's function 
\be
\mathcal{G}(\eta,\eta') = \frac1{k} \; \sin[k(\eta-\eta')] \;
\theta(\eta-\eta') \; .
\ee
To leading order in the limit $ |k \; \eta_0| \gg |k \; \eta| \gg 1 $ 
we find
\be 
X_{1,\vk}(\eta) = - \frac{A_k \; e^{-i \; k \; \eta}}{32 \; \pi \;  H \;
k }\left\{ a(\eta)-a(\eta_0) - \frac{2 \; i}{\pi \;  H \; \eta}
\ln\frac{\eta}{\eta_0} +\cdots  \right\}- \frac{B_k \;
e^{i \; k \; \eta}}{32 \; \pi \;  H \;  k }\left\{ a(\eta)-a(\eta_0) +
\frac{2 \; i}{\pi \;  H \; \eta} \ln\frac{\eta}{\eta_0} +\cdots
\right\} \label{X1largek} 
\ee 
\noindent where again the dots stand
for terms that are subleading in the $ |k \; \eta_0| \gg |k \; \eta| \gg 1 $
limit and $ a(\eta)=-1/H\eta $ is the scale factor. The term
$ a(\eta)-a(\eta_0) $ in the above expression is truly a
\emph{secular} term, since it grows by a factor larger than $e^{64}$
during inflation. The validity of the perturbative expansion for
this term is determined by the requirement that $|k/Ha(\eta)|=
|k \; \eta| \gg1$, namely that the wavelengths are much smaller than the
Hubble radius all throughout inflation.

Thus the solution up to this order is given by 
\bea 
&&X_{\vk}(\eta) =   A_k\, e^{-i \; k \; \eta}\left\{ 1- \frac{g^2}{32\pi H k}\left[
a(\eta)-a(\eta_0) \right] + i \; \frac{g^2}{16\pi^2 H^2}
\frac{\ln\frac{\eta}{\eta_0} }{k \; \eta}+\cdots\right\}+\nonumber \\
& & +B_k\,e^{i \; k \; \eta}\left\{ 1- \frac{g^2}{32\pi H k}\left[
a(\eta)-a(\eta_0) \right] - i \; \frac{g^2}{16\pi^2 H^2}
\frac{\ln\frac{\eta}{\eta_0}}{k \; \eta}+\cdots\right\} \; , 
\eea
\noindent where the dots stand for terms that are of higher order in
$ g^2 $ and subleading in the limit $ |k \; \eta_0|\gg |k \; \eta| \gg 1 $. The
dynamical renormalization group resummation leads to the following
  improved solution 
\be\label{kgransolRG} 
X_{\vk}(\eta)=
e^{-\frac{g^2}{32\pi H k}\left[ a(\eta)-a(\eta_0) \right]}\Bigg\{
A_k\,e^{-i[k \; \eta+\varphi_k(\eta)]}\left[1+\mathcal{O}(g^4)\right]+
B_k\,e^{i[k \; \eta+\varphi_k(\eta)]}\left[1+\mathcal{O}(g^4)\right]\Bigg\}\;,
\ee 
\noindent where $ \varphi_k(\eta) $ is a logarithmic  phase that
is not relevant for the  decay of the amplitude, and the terms in
the brackets are truly perturbative in the long time limit for
wavelengths much smaller than the Hubble radius. In
eq.(\ref{kgransolRG}) we have chosen the renormalization scale
$ \tilde{\eta}_0 $ to coincide with $ \eta_0 $. The DRG improved
solution (\ref{kgransolRG}) reveals the decay of the amplitude with
the scale factor. The result above has the correct limit in
Minkowski space-time as can be seen from the following argument. In
cosmic time, the difference $ a(\eta)-a(\eta_0)= e^{H \, t}- e^{H \, t_0} $,
therefore in the limit $ H \rightarrow 0 $ 
\be \label{Kgranlim} 
\frac{g^2}{32\pi H k}\left[ a(\eta)-a(\eta_0) \right] 
\buildrel{H \to 0} \over= \frac{g^2}{32\pi  k} \; (t-t_0) \;,
\ee 
\noindent which gives
the correct exponential relaxation of the amplitude of the field for
large momentum in Minkowski space-time as shown in ref.\cite{pardec}.

The results for the decay laws reproduce the decay rates in
Minkowski space time in the limit $H\rightarrow 0$ [see eqs.
(\ref{limitzeromom}) and (\ref{Kgranlim})] thus confirming the
reliability of the DRG approach.

The asymptotic behavior of the power spectrum (here we do not
include the $k^3$ normalization)  of the unperturbed solution for
modes deep inside the horizon $|k \; \eta|\gg 1$ and those well outside
the horizon $|k \; \eta| \rightarrow 0$ is given by 
$$
|X_{0,\vk}(\eta)|^2 \buildrel{ |k \; \eta| \gg 1}\over= 
\frac1{2k} \quad ,  \quad
|X_{0,\vk}(\eta)|^2 \buildrel{ |k \; \eta| \rightarrow 0 }\over= 
\frac{2^{2\nu-2} \; \Gamma^2(\nu)\;{\eta}}{{\pi} \;
 \; (k \; \eta)^{2\nu}}
$$
Particle decay modifies the amplitude of the solution and
consequently the power spectrum, which after the DRG resummation is
given by 
\bea 
|X_{\vk}(\eta)|^2 \buildrel{ |k \; \eta| \gg 1}\over   = &
& \frac1{2k}\; e^{-\frac{g^2}{16\pi
H k}\left[ a(\eta)-a(\eta_0) \right]} \label{in}\\
|X_{\vk}(\eta)|^2 \buildrel{ |k \; \eta| \rightarrow 0 }\over  = & &
\frac{2^{2\nu-2} \; \Gamma^2(\nu)\;{\eta}}{{\pi} \;
 \; (k \; \eta)^{2\nu}} \; A^2_{\vk}(\tilde{\eta_0})
\left[\frac{\eta}{\tilde{\eta_0}}\right]^{2\Gamma_1}\label{out} 
\eea
\noindent where we have normalized the mode functions to
Bunch-Davies initial conditions at the beginning of inflation
$ \eta=\eta_0 $ in eq.(\ref{in}). The solution for wavelengths larger
than the Hubble radius is independent of the scale $ \tilde{\eta}_0 $
because the amplitude $ A_{\vk}(\tilde{\eta}_0) $ obeys the DRG
equation. This amplitude at a given scale $ \tilde{\eta}_0 $ is
obtained by matching the asymptotic forms of the DRG improved
solution at a scale $ \tilde{\eta}_0 $. Clearly this amplitude will
depend on the decay law of modes deep inside the horizon, which
reflects a larger suppression of the amplitude for long wavelength
modes.

These results are general, hence they are also valid for the decay of
the quantum fluctuations of the inflaton field. Since the quantum
 fluctuations of the inflaton field seed scalar density
 perturbations the result obtained above implies 
 that the process of particle decay can lead to
 modifications of the power spectrum of superhorizon density
 perturbations which is obtained when the fluctuations freeze after
 horizon crossing as $ \eta \rightarrow  0^- $. The new renormalization 
scale $ \tilde{\eta}_0 $ will lead to violations of scale
 invariance much in the same way as in the renormalization group
 approach to deep inelastic scattering.

Clearly in order to derive the corrections to the power spectrum of density
perturbations from decay of quantum fluctuations, the full gauge invariant 
treatment of the density perturbations presented in sec. \ref{trece} must 
be used.
 
\subsubsection{Quantum corrections to the equations of
motion for the inflaton}\label{13C}

As mentioned above the rapid cosmological
expansion and the lack of a global time-like Killing vector leads to
new decay processes as a consequence of the lack of reaction
thresholds. This new aspect of a rapidly expanding cosmology entails
that in a non-linear field theory, the quanta of a field can decay
into other quanta of the \emph{same} field. This phenomenon is
obviously not available in Minkowski space time by energy and
momentum conservation. In the previous subsection we studied the
decay of superhorizon fluctuations of one scalar field into the
quanta of another field, but following the same method we can obtain
the \emph{self-decay } of modes that are inside the horizon during
inflation.

Our approach relies on two distinct and fundamentally different
expansions: (i) the effective field theory (EFT) expansion and  (ii)
the slow-roll expansion.

As mentioned above, the effective field
theory approach  relies on the separation between the energy scale
of inflation and the cutoff scale, which here is the Planck scale.
The scale of inflation is determined by the Hubble parameter during
the relevant stage of inflation when wavelengths of cosmological
relevance cross the horizon.  Therefore, the dimensionless ratio
that defines the EFT approximation is the ratio $ [H(\Phi_0)/M_{Pl}]^2 $,
where $ H(\Phi_0) $ is the Hubble parameter during the relevant
inflationary stage. In scalar field driven inflation the reliability
of this approximation \emph{improves} upon dynamical evolution since
the scale of inflation {\it diminishes} with  time as shown in
sec. \ref{3C}. Phenomenologically, the EFT approximation is an excellent 
one since during inflation $ (H/M_{Pl})^2 \sim 10^{-9} $ according 
to eq.(\ref{myH}).

We consider single field inflationary models described by a general
self-interacting scalar field theory in a spatially flat
FRW cosmological space time with scale factor
$ a(t) $. In comoving coordinates the action  is given by
\begin{equation}\label{action}
S= \int d^3x \; dt \;  a^3(t) \Bigg[ \frac12 \;
{\dot{\phi}^2}-\frac{(\nabla \phi)^2}{2a^2}-V(\phi) \Bigg] \;.
\end{equation}
We consider a \emph{generic} potential $ V(\phi) $, the only
requirement is that its \emph{derivatives} be small in order to
justify the slow-roll expansion as in the class of potentials
eq.(\ref{V}). In order to study the corrections
from the quantum fluctuations we separate the classical homogeneous
expectation value of the scalar field from the quantum fluctuations
by writing 
\be\label{tad} 
\phi(\vec{x},t)=\Phi_0(t)+\varphi(\vec{x},t)\;, 
\ee 
\noindent with 
\be\label{exp}
\Phi_0(t)=\langle \phi(\vx,t) \rangle~~;~~ \langle
\varphi(\vx,t)\rangle =0 \;, 
\ee 
where the expectation value is in
the non-equilibrium quantum state. Expanding the Lagrangian density
and integrating by parts, the action becomes 
\be\label{Split} 
S= \int d^3x \; dt \;  a^3(t) \; \mathcal{L}[\Phi_0(t),\varphi(\vx,t)]\;, 
\ee 
\noindent with
\bea\label{lagra} 
&&\mathcal{L}[\Phi_0(t),\varphi(\vx,t)]  =
\frac12 \; {\dot{\Phi}^2_0}-V(\Phi_0)+\frac12 \;
{\dot{\varphi}^2}-\frac{(\nabla \varphi)^2}{2 \, a^2} -\frac12 \;
V^{''}(\Phi_0)\; \varphi^2 \cr \cr &&- \varphi\;
\left[\ddot{\Phi}_0+3 \, H \,\dot{\Phi}_0+V^{'}(\Phi_0)\right] -
\frac16\; V^{'''}(\Phi_0)\; \varphi^3 - \frac1{24}\;
V^{(IV)}(\Phi_0)\; \varphi^4+ \textmd{higher orders in}\, \varphi \; . 
\eea 
We will obtain the equation of motion for the homogeneous
expectation value of the inflaton field  by   requiring the
condition $ \langle \varphi(\vx,t)\rangle =0 $ consistently in a
perturbative expansion by treating the \emph{linear}, cubic, quartic
(and higher order) terms in the Lagrangian density eq.(\ref{lagra})
as perturbations.

The Friedmann equation and the classical equation of motion for
$ \Phi_0 $ are 
\bea 
&& H_0^2 = \frac1{3 \, M^2_{Pl}}\left[\frac12(\dot{\Phi}_0)^2+
V(\Phi_0)\right]\;, \label{hub2} \\
\label{claseq} 
&&{\ddot\Phi}_0 +3\,H_0\,\dot{\Phi}_0+V'(\Phi_0) =0
\;. 
\eea
[$ H_0 $ stands in secs. \ref{QLC1} and \ref{trece} for the classical
Hubble parameter during inflation. In the rest of the article
$ H_0 $ stands for the Hubble constant today.]

We now introduce the effective mass of the fluctuations $ M^2 $  and
the cubic and quartic self-couplings $ g, \; \lambda $ respectively as
\bea 
&&M^2 \equiv M^2(\Phi_0)  =   V''(\Phi_0) = 3 \; H_0^2 \;
\eta_v + \mathcal{O}(\epsilon_v \; \eta_v)\,, \label{flucmass}\\
&& g\equiv g(\Phi_0)  =  \frac12 \;  V^{'''}(\Phi_0)\quad , \quad
\lambda \equiv \lambda (\Phi_0)  =  \frac16 \;
 V^{(IV)}(\Phi_0) \; . \label{lambda}
\eea 
In particular, the dimensionless ratio of the cubic coupling and the 
Hubble parameter is given to leading order in slow-roll by 
\be\label{gcoup} 
\frac{g}{H_0} = \frac{3\,\xi_v}{2\sqrt{2\;\epsilon_v}} \; 
\frac{H_0}{M_{Pl}} =
\frac32 \, \pi \;  \Delta_{\mathcal{R}}\left[ \frac{r}2 \left(n_s -
1 + \frac{3 \, r}{16} \right) - \frac{dn_s}{d \ln k}\right] =
{\cal O} \left(\frac{M^2}{N \; M_{Pl}^2}\right) \; , 
\ee
\noindent where the slow-roll parameters $ \epsilon_v $ and $ \xi_v $ are 
given by eqs.(\ref{slrN}) and (\ref{xisig}), respectively and 
$ \Delta_{\mathcal{R}} $ is given by eq.(\ref{ampliI}).

The quartic coupling $ \lambda $ can be conveniently
written in terms of slow-roll and effective-field theory parameters as 
\be\label{lam} 
\lambda = \frac{\sigma_v}{4 \,
\epsilon_v}\left(\frac{H_0}{M_{Pl}}\right)^2 = 2 \, \pi^2 \;
\Delta_{\mathcal{R}}^2 \; \sigma_v = {\cal O} 
\left(\frac{M^4}{N \; M_{Pl}^4}\right) \; ,
\ee 
where $ \sigma_v $ is given by eq.(\ref{sigmv}). Moreover, $ \lambda $ can
be written solely in terms of CMB observables inserting the
expression eq.(\ref{gorda}) for $ \sigma_v $ into eq.(\ref{lam}).

During slow-roll the effective mass and couplings are not constants
but  \emph{very slowly varying functions of time}. 
The time dependence of these couplings is implicit through their
dependence on $ \Phi_0 $ but is slow in the slow-roll stage.

Eqs.(\ref{gcoup}) and (\ref{lam}) show that $ (g/H_0)^2 $ and
$ \lambda/N $ are of the \emph{same order} in the EFT expansion, namely
$ \mathcal{O}(H^2_0/[N^3 \; M^2_{Pl}]) $. That is, 
both the cubic and quartic self
couplings are small, the quartic being of higher order in slow-roll
than the cubic etc, namely
\be\label{jerar}
1\gg \frac{g}{H_0} \gg \lambda \gg \cdots
\ee
\noindent where the dots stand for self-couplings arising from
higher derivative of the potential as displayed in eq.(\ref{lagra}).
This observation will be important in the calculation of the self-energy 
correction for the quantum fluctuations below. Eqs.(\ref{gcoup}) and 
(\ref{lam}) for $ g $ and $ \lambda $ must be compared with eqs.(\ref{hub})
and (\ref{lambdaG4}) for the couplings at zero inflaton field. We see that 
the hierarchy eq.(\ref{jerar}) is fulfilled by eqs.(\ref{lambdaG4}).

\medskip

The condition $ \langle \varphi (\vec{x},t) \rangle =0 $ up to leading
order ($ \mathcal{O}(g) $) leads to the equation of motion
\be\label{1lupeqn} 
\ddot{\Phi}_0(t)+3 \, H \; \dot{\Phi}_0(t)+V'(\Phi_0)+g(\Phi_0) \; \langle
[\varphi(\vx,t)]^2\rangle =0 \;. 
\ee 
The first three terms here are the familiar ones for the classical 
equation of motion of the inflaton.

The last term in eq.(\ref{1lupeqn}) is the one-loop correction to the 
equations of motion of purely quantum mechanical origin. 
The one-loop quantum correction
to the equations of motion is completely determined by the power
spectrum of inflaton fluctuations, 
\be\label{lupPS} 
\langle [\varphi(\vx,t)]^2\rangle = \int \frac{d^3 k}{(2\pi)^3} \; \langle
|\varphi_{\vk}(t)|^2 \rangle = \int_0^{\infty} \frac{dk}{k} \;
\mathcal{P}_{\varphi}(k,t)\,. 
\ee   
In order to compute the one-loop
contribution, it is convenient to work in conformal time and to
conformally rescale the field 
\be\label{rescale} 
\varphi(\vx,t) =\frac{\chi(\vx,\eta)}{a(\eta)} \quad , 
\ee 
\noindent $ a(\eta) $
being the scale factor in conformal time. The spatial Fourier
transform of the free field Heisenberg operators $ \chi(\vx,\eta) $,
$ \chi_{\vk}(\eta) $ are written in terms
of annihilation and creation operators that act on Fock states  as
\be\label{ope} 
\chi_{\vk}(\eta) = a_{\vk} \; g_{\nu}(k,\eta)+
a^{\dagger}_{-\vk} \; g^{*}_{\nu}(k,\eta) 
\ee 
\noindent where the mode functions $ g_{\nu}(k,\eta) $ with  
Bunch-Davis boundary conditions are given by eq.(\ref{gnu})
and the index $ \nu $ is given by
\be\label{nu} 
\nu = \frac32 + \epsilon_v-\eta_v +
\mathcal{O}\left(\frac1{N^2}\right) \; . 
\ee
Notice that the index $ \nu $ for inflaton fluctuations is similar but not 
identical to the index $ \nu_\mathcal{R} $ for scalar curvature 
fluctuations eq.(\ref{eqnz}).

In the Bunch-Davis vacuum, defined so that $ a_{\vk} |0>_{BD} =0 $
we find   
\be\label{PSSR} 
\mathcal{P}_{\varphi}(k,t) = \frac{H^2_0}{8 \, \pi} \; (-k \; \eta)^3 \; 
|H^{(1)}_\nu(- k \eta)|^2 \; . 
\ee
The momentum integral in eq.(\ref{lupPS}) features ultraviolet and
infrared divergences. The ultraviolet divergences are absorbed into
renormalization of the parameters in the tree level potential. The
infrared divergences are a reflection of the \emph{near} scale
invariance of the inflaton fluctuations, and are regularized by a
particular linear combination of slow-roll parameters 
\be \label{delta}
\Delta \equiv \eta_v-\epsilon_v = \frac12\left(n_s - 1
+\frac14 \; r\right) = \mathcal{O}\left(\frac1{N}\right) \; . 
\ee 
The infrared regularization by
the slow-roll combination $ \Delta $ is akin to the regularization of
infrared divergences in critical phenomena through the expansion
around four dimensions, namely the $\epsilon-$expansion \cite{wilkog}. 
We find,
\be\label{fineq}
\ddot{\Phi}_0(t)+3\,H_0\,\dot{\Phi}_0(t)+V^{'}_R(\Phi_0)+
\left(\frac{H_0}{4 \, \pi}\right)^2
\frac{V^{'''}_R(\Phi_0)}{\Delta}=0 \;. 
\ee 
An important aspect of
this equation  is the following: naively, the quantum correction is
of order $ V^{'''}_R(\Phi_0) $, therefore  of second order in slow
roll, but the strong infrared enhancement arising from the quasi
scale invariance of inflationary fluctuations brings about a denominator
which is of first order in slow-roll $ \mathcal{O}(1/N) $. Hence, the 
lowest order quantum correction in the slow-roll expansion, is actually of 
the same order as $ V^{'}_R(\Phi_0) $. To highlight this observation, it
proves convenient to write eq.(\ref{fineq}) in terms of the EFT and
slow-roll parameters, 
\be\label{fineqsr}
\ddot{\Phi}_0(t)+3\,H_0\,\dot{\Phi}_0(t)+
V^{'}_R(\Phi_0)\left[1+\left(\frac{H_0}{2\pi \, M_{Pl}}\right)^2
\frac{\xi_v}{2\,\epsilon_v\,\Delta}\right]=0 \;. 
\ee 
Since $ \xi_v \sim {\cal O}\left(1/N^2\right) $ and 
$ \Delta \sim \eta_v \sim \epsilon_v \sim {\cal O}\left(1/N\right) $ 
[see eqs.(\ref{slrsd}), (\ref{xisig}) and (\ref{delta})], the leading 
quantum corrections are of zeroth order in slow-roll ($ \sim N^0 $). 
This is a consequence of the infrared enhancement resulting from the 
nearly scale invariance of the power spectrum of scalar fluctuations. 
The quantum correction is suppressed by an EFT factor
$ H^2_0/M^2_{Pl} \ll 1 $.

\subsubsection{Quantum corrections to the Friedmann equation: the one
loop effective potential}\label{13D}

The zero temperature effective potential in Minkowski space-time is
often used to describe the scalar field dynamics during inflation,
but this ignores the true nature of the quantum fluctuations in the
nearly de-Sitter cosmology.

Since the fluctuations of the inflaton field are quantized, the
interpretation of the `scalar condensate' $ \Phi_0 $ is that of the
expectation value of the full quantum field $ \phi $ in a homogeneous
coherent quantum state. Consistently with this, the Friedmann
equation must necessarily be understood in terms of the
\emph{expectation} value of the field energy momentum tensor, namely
\be\label{FRW2} 
H^2= \frac1{3 \, M^2_{Pl}}\left\langle \frac12
\; \dot{\phi}^2+\frac12 \; \left(\frac{\nabla
\phi}{a(t)}\right)^2+V(\phi) \right\rangle  \;. 
\ee 
Separating the
homogeneous condensate from the fluctuations as in eq.(\ref{tad})
with the condition that the expectation value of the quantum
fluctuation vanishes eq.(\ref{exp}), the Friedmann equation becomes
\be\label{FRexp} 
H^2= \frac1{3 \, M^2_{Pl}}\left[ \frac12 \;
{\dot{\Phi_0}}^2 + V_R(\Phi_0)+\delta V_R(\Phi_0)\right]+ \frac1{3
\, M^2_{Pl}}\left\langle \frac12 \; \dot{\varphi}^2+\frac12
\; \left(\frac{\nabla \varphi}{a(t)}\right)^2+\frac12 \;
V^{''}(\Phi_0)\; \varphi^2 +\cdots\right\rangle 
\ee 
where $ V_R $ refers to the renormalized tree level potential and 
$ \delta V_R $ contains the counterterms that cancel the ultraviolet 
divergences. The dots inside the angular brackets correspond to terms with 
higher derivatives of the potential which are smaller in the slow-roll
expansion. The quadratic term $ \langle \varphi^2 \rangle $ has been
calculated above to leading order in slow-roll and given by eq.
(\ref{lupPS}). Calculating the expectation value in eq.(\ref{FRexp})
in free field theory  corresponds to obtaining the corrections to
the energy momentum tensor by integrating the fluctuations \emph{up
to one loop}.

The first two terms of the expectation value in  eq.(\ref{FRexp})
feature ultraviolet divergences that are absorbed into the
renormalization of the tree-level potential, but \emph{do not}
feature infrared divergences for $ \nu=3/2 $ because the time and
spatial derivatives introduce  two extra powers of the loop momentum
in the corresponding integrals. The last term does feature the
infrared enhancement and leads to the following expression
\be\label{FRren} 
H^2 = \frac1{3 \, M^2_{Pl}}\left[ \frac12 \;
{\dot{\Phi_0}}^2 + V_R(\Phi_0) + \left(\frac{H_0}{4 \,
\pi}\right)^2\frac{V^{''}_R(\Phi_0)}{\Delta} +\textmd{higher orders
in slow-roll}\right] \equiv H^2_0 + \delta H^2 \;, 
\ee 
\noindent
where $ H_0 $ is the  Hubble parameter in absence of quantum
fluctuations:
$$
 H^2_0 = \frac{V_R(\Phi_0)}{3 \, M^2_{Pl}} \left[1+\frac{\epsilon_v}3+
\mathcal{O}\left(\frac1{N^2}\right) \right] \; .
$$
Using the lowest order slow-roll relation eq.(\ref{flucmass}), the
last term in eq.(\ref{FRren}) can be written as follows
\be\label{delH} 
\frac{\delta H^2}{H^2_0} = \left(\frac{H_0}{4 \,
\pi\,M_{Pl}}\right)^2 \frac{\eta_v}{\Delta}\; . 
\ee
This equation defines the back-reaction correction to the scale
factor arising from the quantum fluctuations of the inflaton.

Hence, while the ratio $ \eta_v/\Delta $ is of order zero in slow
roll, the one loop correction to the Friedmann equation is of the
order $H^2_0/M^2_{Pl} \ll 1$ consistently with the effective field
theory expansion. The Friedmann equation suggests the identification
of the effective potential 
\bea\label{Veff} 
&&V_{eff}(\Phi_0) = V_R(\Phi_0)+ \left(\frac{H_0}{4 \,
\pi}\right)^2\frac{V^{''}_R(\Phi_0)}{\Delta} +\textmd{higher orders
in slow-roll} = \\ \cr &&= V_R(\Phi_0)\left[1+ \left(\frac{H_0}{4 \,
\pi\,M_{Pl}}\right)^2\frac{\eta_v}{ \Delta} + \textmd{higher orders
in slow-roll} \right]= \cr \cr && =
V_R(\Phi_0)\left[1+\frac{\Delta^2_{\mathcal{T}}}{32} \;
 \frac{n_s -1 + \frac38 \; r}{n_s -1 + \frac14 \; r}
+\textmd{higher orders in slow-roll}\right] \; . \label{Vefsr} 
\eea
We see that the equation of motion for the inflaton eq.(\ref{fineq})
takes the natural form
$$
\ddot{\Phi}_0(t)+3\,H_0\,\dot{\Phi}_0(t)+ \frac{\partial
V_{eff}}{\partial\Phi_0}(\Phi_0) = 0 \;.
$$
where the derivative of $ V_{eff} $ with respect to $ \Phi_0 $ is
taken at fixed Hubble and slow-roll parameters. The quantum
corrections to the effective potential lead to loop corrections to
the slow-roll parameters, for example defining the \emph{effective}
slow-roll parameters as \be\label{effSR} \epsilon_{eff} =
\frac{M^2_{Pl}}2\;
\left[\frac{V^{'}_{eff}(\Phi_0)}{V_{eff}(\Phi_0)}\right]^{\!
2}~~;~~\eta_{eff} =  M^2_{Pl} \;
\frac{V^{''}_{eff}(\Phi_0)}{V_{eff}(\Phi_0)} \; , \ee \noindent
eq.(\ref{Veff}) yields to leading order in EFT and slow-roll
expansions: \bea &&\epsilon_{eff}  =  \epsilon_v \left[ 1+
\left(\frac{H_0}{4 \, \pi \, M_{Pl}} \right)^2 \, \frac{4 \, \eta_v
\Delta - \xi_v}{ \Delta^2}
\right] \label{epseff}\\
&&\eta_{eff}  =  \eta_v  \left\{1+ \left(\frac{H_0}{4 \, \pi \,
M_{Pl}} \right)^2 \, \frac1{\Delta^2 } \left[\frac{\xi_v^2}{\eta_v
\Delta} - \frac{\sigma_v}{2\,\eta_v} - \frac{\xi_v}{\eta_v}
\left(\eta_v+ 6 \, \epsilon_v \right) + 4 \, \eta_v \, \left( \eta_v
+ 4 \, \epsilon_v\right) - 20 \, \epsilon_v^2 \right]\right\} \; .
\nonumber \eea
 A remarkable feature of the quantum corrections to the slow
 roll parameters is that they are of \emph{zeroth} order in slow
 roll $ {\cal O}( N^0) $. Again, this is a consequence of the infrared 
enhancement of the loop diagrams for a nearly scale invariant spectrum of
 fluctuations. Higher order slow-roll parameters can be obtained
 similarly. The smallness of the quantum corrections are determined
 by the EFT ratio $H_0/M_{Pl}$. Thus the validity of the EFT
 approach guarantees that the quantum corrections are small.

 There is a striking difference between the effective potential in
 quasi de-Sitter space-time and the Minkowski case, in the latter it
 is given by 
\be \label{potefM}
V_{eff}^{Minkowski}(\Phi_0) = V_R(\Phi_0)+\frac{[V^{''}_R(\Phi_0)]^2}{64 \,
\pi^2} \ln\left[\frac{V^{''}_R(\Phi_0)}{M^2} \right]\; , 
\ee
\noindent where $ M $ is a renormalization scale. Therefore the naive
use of the Minkowski space-time effective potential in the dynamics
of the inflaton field during slow-roll inflation is a gross and
misleading approximation that cannot describe the infrared
properties of inflationary dynamics.

Indeed, early papers (as refs. \cite{infnue} and \cite{astw}) used 
effective potentials like eq.(\ref{potefM}) during inflationary expansion.
It will be interesting to revisit these early works at the light
of our present knowledge on the  effective potential during inflation.

\subsubsection{Quantum corrections to superhorizon modes and their scaling
dimensions}\label{anomdim}

In order to study the equations of motion for the \emph{quantum
fluctuations} of the inflaton including self-energy corrections, it
is convenient to first pass to conformal time and to implement the
conformal rescaling of the field as in eq.(\ref{rescale}). The
action is now given by 
\be 
S= \int d^3x \; d\eta \;
\mathcal{L}_c[\chi,\Phi_0] \;, 
\ee \noindent 
where the Lagrangian density $ \mathcal{L}_c[\chi,\Phi_0] $ is given by 
\bea\label{lagconf}
&&\mathcal{L}_c[\chi,\Phi_0] = a^4(\eta)\left[ \frac12 \;
{\dot\Phi}^2_0-V_R(\Phi_0)-\delta V_R(\Phi_0) \right] +
\frac{{\chi'}^2}2-\frac{(\nabla
\chi)^2}2-\frac12 \; {\mathcal{M}^2(\eta)} \; \chi^2 - \nonumber \\
&& - a^3(\eta) \; \chi \; \left[ \ddot{\Phi}_0+3 \, H \,
\dot{\Phi}_0+V^{'}_R(\Phi_0)+\mathcal{C}_2[\Lambda,H] \;
V^{'''}_R(\Phi_0)+\cdots\right] - \frac12 \; \delta
\mathcal{M}^2(\eta) \; \chi^2 -\frac{g}3 \; a(\eta) \;
\chi^3-\frac{\lambda}{4} \; \chi^4 +\cdots 
\eea 
\noindent where
$ a(\eta) $ is the scale factor in conformal time, the dots on
$ \Phi_0 $ stand for derivatives with respect to cosmic time, the
primes on $ \chi $ stand for derivatives with respect to conformal
time, we have used the definitions given in eqs.(\ref{lambda})
and the dots inside the angular brackets correspond to terms with higher
derivatives of the potential which are smaller in the slow-roll
expansion.

The effective (time dependent) mass term is given by
\be \label{renmass}   
\mathcal{M}^2(\eta) = V^{''}_R(\Phi_0) \; a^2(\eta)-
\frac{a''(\eta)}{a(\eta)} = -\frac1{\eta^2}\left(\nu^2-\frac14\right) \; ,
\ee 
where $ \nu $ is given by eq.(\ref{nu}) and
$ \delta \mathcal{M}^2(\eta) $ and $ \mathcal{C}_2 $ are counterterms that
will cancel the ultraviolet divergences in the one-loop self-energy.
$ -\mathcal{M}^2(\eta) $ is  similar but not identical to the potential 
$ W_\mathcal{R}(\eta) $ felt by the scalar curvature fluctuations 
eq.(\ref{eqnz}).

The effective equation of motion for the fluctuations is obtained in
the linear response approach by introducing an external source that
induces an expectation value for the field $ \chi(\vx,\eta) $,
switching off the source this expectation value will evolve in time
through the effective equation of motion of the fluctuations. The
first step is to take the spatial Fourier transform of
$ \chi(\vec{x},\eta) $ and write 
\be\label{split} 
\chi_{\vk}(\eta) = X_{\vk}(\eta) +
 \sigma_{\vk}(\eta)~~;~~ \langle \chi_{\vk}(\eta)\rangle
 =X_{\vk}(\eta) ~~;~~\langle \sigma_{\vk}(\eta)\rangle =0
 \;,
 \ee
 \noindent were $ X_{\vk}(\eta) $ is the  spatial Fourier transform of
 the expectation value of the fluctuation field $ \chi $ induced by
 the external source term. Implementing the   condition
 $ \langle \sigma\rangle=0 $ up to one loop we obtain the
 effective equation of motion
\be \label{eqnofmotfluc}
X''_{\vk}(\eta)+\left[k^2-\frac{\nu^2-\frac14 }{\eta^2}\right]
X_{\vk}(\eta)+ \int_{\eta_0}^{\eta} \Sigma(k,\eta,\eta') \;
X_{\vk}(\eta') \; d\eta' = 0 \;. 
\ee 
The counterterms cancel the UV
divergent parts of the self-energy which yield a local term
($ \propto \delta(\eta-\eta') $), after renormalization, the
self-energy kernel is given by 
\be\label{Sigma}
\Sigma(k,\eta,\eta')=  \frac{2\,g^2}{H^2_0 \, \eta \, \eta'} \;
\mathcal{K}_{\nu}(k;\eta,\eta') \;. 
\ee 
The one-loop kernel $ \mathcal{K}_{\nu}(k;\eta,\eta') $ is given by 
\be\label{kernel}
\mathcal{K}_{\nu}(k;\eta,\eta') = 2 \int \frac{d^3q}{(2\pi)^3} \;
\mathrm{Im}\left[ g_{\nu}(q,\eta)
g^*_{\nu}(q,\eta')g_{\nu}(|\vq-\vk|,\eta) \;
g^*_{\nu}(|\vq-\vk|,\eta')\right]\;,
\ee 
\noindent where the mode functions $ g_{\nu}(k,\eta) $ are given
by eq.(\ref{gnu}). We are primarily interested in obtaining the
superhorizon behavior of the fluctuations ($ |k \, \eta| \ll 1 $) to
obtain the scaling behavior in this limit, therefore we set  $k=0$.
To leading order in the slow-roll
expansion and leading logarithmic order the kernel is given by
\bea\label{Knu} 
\mathcal{K}_{\nu}(0;\eta,\eta')= &&-\frac1{8 \,
\pi^2} \; \frac{\eta-\eta'}{\left(\eta-\eta'\right)^2 + (\epsilon \;
\eta')^2}+ \nonumber \\ 
&& + \frac1{6 \, \pi^2} \left[
\left(\frac1{2 \, \Delta}+\frac23 \right)
\left(\frac{\eta'}{\eta^2}-\frac{\eta}{\eta^{'2}}\right) -
\frac{\eta'}{\eta^2} \; \ln\frac{\eta'}{\eta}+
\left(\frac{\eta}{\eta^{'2}}- \frac{\eta'}{\eta^2} \right) \,
\ln\left(1-\frac{\eta}{\eta'} \right) +
\frac1{\eta'}-\frac1{\eta} \right] \;,
\eea 
where $ \epsilon
\rightarrow 0 $ furnishes a regularization.

The equation of motion (\ref{eqnofmotfluc}) is solved in a
perturbative loop expansion as follows
\be\label{pertsol}
X_{\vk}(\eta)=X_{0,\vk}(\eta)+X_{1,\vk}(\eta)+\textmd{higher loop
corrections} \;, 
\ee 
\noindent where $X_{0,\vk}(\eta)$ is the free
field solution, $X_{1,\vk}(\eta)$ is the one-loop correction, etc.
This expansion to one loop order leads to the following hierarchy of
coupled equations [see eqs.(\ref{perteqn})-(\ref{rn})],
\bea
&&X''_{0,\vk}(\eta)+\left[k^2-\frac1{\eta^2}\Big(\nu^2-
\frac14 \Big) \right]X_{0,\vk}(\eta)
 =  0  \; ,\label{X0}\\
&&X''_{1,\vk}(\eta)+\left[k^2-\frac1{\eta^2}\Big(\nu^2-
\frac14 \Big) \right] X_{1,\vk}(\eta) = \mathcal{R}_1(k,\eta)
\label{X1}\;.
\eea
The superhorizon solution of the zeroth order equation is 
\be\label{X00} 
X_{0,0}(\eta)  =  A\;\eta^{\beta_+}+B\;\eta^{\beta_-} \quad ;
\quad \beta_{\pm} \equiv \frac12\pm \nu \; . 
\ee
and after renormalization we find 
\be\label{source0}
\mathcal{R}_1(0,\eta)= A \; \frac{\eta^{\beta_+}}{\eta^2}\; (2 \,
\nu \, d^+)\,+B \; \frac{\eta^{\beta_-}}{\eta^2} \; (2 \, \nu \,
d^-)\,+F[\eta,\eta_0]\;, 
\ee 
\noindent 
where   the coefficients
$d^{\pm}$ are entirely of quantum origin (one-loop) and given by
\bea 
d^+ & = & - \frac1{2 \, \nu}\left[\frac{3 \, \lambda}{8 \,
\pi^2\,\Delta}+ \frac1{6 \, \pi^2 }
\left(\frac{g}{ H_0  \, \Delta}\right)^2 \right]\label{DP} \\
d^- & = & - \frac1{2 \, \nu}\left[\frac{3 \, \lambda}{8 \,
\pi^2\,\Delta}+ \frac1{2 \, \pi^2 } \left(\frac{g}{ H_0 \,
\Delta}\right)^2\right]\;, \label{DM} 
\eea 
and  $ F[\eta,\eta_0] $ refers to the contribution of the lower 
integration limit and does
not produce secular terms in the limit $ \eta \rightarrow 0 $.

The solution of the inhomogeneous eq.(\ref{X1}) is given by 
\be 
X_{1,\vk}(\eta)= \int_{\eta_0}^{0} d\eta' \;
\mathcal{G}_\nu(k;\eta,\eta') \; \mathcal{R}_1(k,\eta') \; . 
\ee
\noindent $ \mathcal{G}_\nu(k;\eta,\eta') $ is the retarded Green's
function  (\ref{GF}). Up to one loop order we find the superhorizon
solution 
\be 
X_0(\eta)= A\;\eta^{\beta_+}\left[1+d^+
\ln\left(\frac{\eta}{\eta_0}
\right)\right]+B\;\eta^{\beta_-}\left[1- d^-
\ln\left(\frac{\eta}{\eta_0}\right)\right]+\textmd{non-secular
terms} \;.
\ee 
The resummation of the logarithmic secular terms is
performed by implementing the dynamical renormalization group
resummation, leading to the following result 
\be\label{DRGsol}
X_0(\eta)=A_{\overline{\eta}} \;
\left(\frac{\eta}{\overline{\eta}}\right)^{\beta_++d^+} +
B_{\overline{\eta}} \;
\left(\frac{\eta}{\overline{\eta}}\right)^{\beta_--d^-}
=\left(\frac{\eta}{\overline{\eta}}\right)^{\Gamma}
\left[A_{\overline{\eta}} \;
\left(\frac{\eta}{\overline{\eta}}\right)^{\beta_++\gamma}+
B_{\overline{\eta}} \;
\left(\frac{\eta}{\overline{\eta}}\right)^{\beta_--\gamma}\right]\;,
\ee \noindent where $ \overline{\eta} $ is a renormalization scale;
the amplitudes $ A_{\overline{\eta}}, \; B_{\overline{\eta}} $ are
given at this renormalization scale and obey a renormalization group
equation, so that the full solution $X_0(\eta)$ is independent of
the renormalization scale, as it must be. The exponents are given by
\bea 
\gamma & = & \frac12 \left(d^++d^- \right)= -\frac1{2 \,
\nu}\left[ \frac{3 \, \lambda}{8 \, \pi^2\,\Delta}+ \frac1{3 \,
\pi^2 }
\left(\frac{g}{ H_0 \,  \Delta}\right)^2 \right]\;, \label{gamma}\\
\Gamma & = & \frac12 \left(d^+-d^- \right) = \frac1{12 \,
\pi^2\,\nu } \left(\frac{g}{ H_0 \, \Delta}\right)^2 \;.
\label{Gamma} 
\eea  
Since $ \eta = -e^{-H_0\,t}/H_0$, in cosmic
time the amplitude of superhorizon fluctuations decays exponentially
with the decay rate
 \be\label{decrat}
\Gamma_{\varphi \rightarrow  \varphi \varphi} = H_0 \, \Gamma =
\frac{H_0}{12 \, \pi^2\,\nu} \left(\frac{g}{ H_0 \, \Delta}\right)^2
\; , \ee \noindent where the subscript ${\varphi \rightarrow \varphi
\varphi}$ emphasizes that this is the rate of \emph{self-decay} of
inflaton fluctuations, a novel phenomenon which is a consequence of
the inflationary expansion and the fact that in the expanding
cosmology there is no global time-like Killing vector that would
lead to reaction thresholds.

In addition in the limit $ \eta \rightarrow 0^- $ the growing mode
features a \emph{novel} scaling dimension $ -d^- $ namely 
\be
X_0(\eta) \buildrel{\eta \to 0}\over= B_{\overline{\eta}} \;
\left(\frac{\eta}{\overline{\eta}}\right)^{\frac12-\nu-d^-} \;.
\ee 
This correction to scaling is related to the decay rate $ \Gamma
$ of superhorizon fluctuations eq.(\ref{Gamma}). From
eqs.(\ref{gcoup}), (\ref{lam}) and (\ref{DM}), to leading order in
slow-roll and EFT expansions, $ d^- $  and the cosmic time decay
rate $ \Gamma_{\varphi \rightarrow \varphi\varphi}$ of superhorizon
inflaton fluctuations are given by 
\bea 
-d^- = && \left(
\frac{H_0}{4\pi\,M_{Pl}} \right)^{\! \! 2} \frac{\sigma_v \;
(\eta_v-\epsilon_v)+6 \, \xi^2_v}{2 \,
\epsilon_v\,(\eta_v-\epsilon_v)^2}= \Delta_{\mathcal{R}}^2 \; \frac{
\sigma_v \; (\eta_v-\epsilon_v) + 6 \, \xi^2_v}{4 \,
(\eta_v-\epsilon_v)^2}
\;, \label{diman}\\
\Gamma_{\varphi \rightarrow  \varphi\varphi} = &&  \left(
\frac{H_0}{4 \, \pi \; M_{Pl}} \right)^{\! \! 2} \frac{H_0 \;
\xi^2_v}{\epsilon_v\,(\eta_v-\epsilon_v)^2} = \frac12 \;
\Delta_{\mathcal{R}}^2 \; \frac{H_0 \;
\xi^2_v}{(\eta_v-\epsilon_v)^2} \label{gamslow}
 \eea
\noindent where the slow-roll parameters are given by eqs.
(\ref{slrsd})-(\ref{xisig}). Whereas the exponent $ \nu=\frac32-\Delta=
\frac32+\epsilon_v -\eta_v $ is determined by eq.(\ref{X0}) for the
free mode functions, the novel scaling exponent $-d^-$ is determined
by the quantum corrections arising from the  interactions. Again
eq.(\ref{diman}) highlights an important aspect of  the effective
field theory approach. These expressions are of \emph{zero order} in slow 
roll $ N^0 $. This is a consequence of the {\em infrared enhancement} 
of the self-energy for $ \nu = 3/2 + {\cal O}(1/N) $ manifest as 
$ \Delta^{-2}= (\eta_v-\epsilon_v)^{-2} ={\cal O}(N^2) $.
However, the novel dimension is perturbatively small precisely because of 
the effective field theory factor $ H^2_0/M^2_{Pl} \ll 1 $.

\medskip

We get from eqs.(\ref{ampR}), (\ref{ampliI}), (\ref{xisig}) and 
(\ref{gcoup}) that
\be\label{gsobH}
\frac{g}{H_0} = 3 \, \pi \; \xi_v \; \Delta_{\mathcal{R}} = 
\frac{\sqrt3}{2 \; N} \; 
\left(\frac{M}{M_{Pl}}\right)^2 \; \frac{|w'''(\chi)|}{\sqrt{w(\chi)}}
\ee
Using eq.(\ref{valorM}) we get as estimate on the cubic coupling
$ g/H_0 \sim 10^{-6} $.

We can analogously estimate the rate 
$ \Gamma_{\varphi \rightarrow  \varphi\varphi} $ [Eq.(\ref{gamslow})],
\be
\frac{\Gamma_{\varphi \rightarrow  \varphi\varphi}}{H_0} =
\frac1{3 \; \pi^2} \left(\frac{M}{M_{Pl}}\right)^4 \; 
\frac{w(\chi) \; [w'''(\chi)]^2}
{2 \; w(\chi) \; w''(\chi) - [w''(\chi)]^2} \sim \frac1{3 \; \pi^2} 
\left(\frac{M}{M_{Pl}}\right)^4 \sim 10^{-11} \; .
\ee
This gives in cosmic time for a typical value $ H \simeq 10^{14} $ GeV,
$ \Gamma_{\varphi \rightarrow  \varphi\varphi} \sim 10^3$ GeV .

\subsubsection{Connection with non-gaussianity}

Non-gaussianity of the spectrum of fluctuations is associated with
three (and higher) point correlation functions. An early assessment
of non-gaussian features  of temperature fluctuations in an
interacting  field theory  was given in ref.\cite{allen}. In
ref.\cite{srednicki} the simplest inflationary potential with a
cubic self-interaction for the inflaton field was proposed as a
prototype theory to study possible departures from gaussianity.
The three point correlation function of a scalar field in a theory with
cubic interaction as well as the four point correlation function in
a theory with quartic interaction in de Sitter space-time were calculated 
in ref.\cite{bartolo}.

\begin{figure}[ht!]
\begin{center}
\includegraphics[height=5cm,width=6cm]{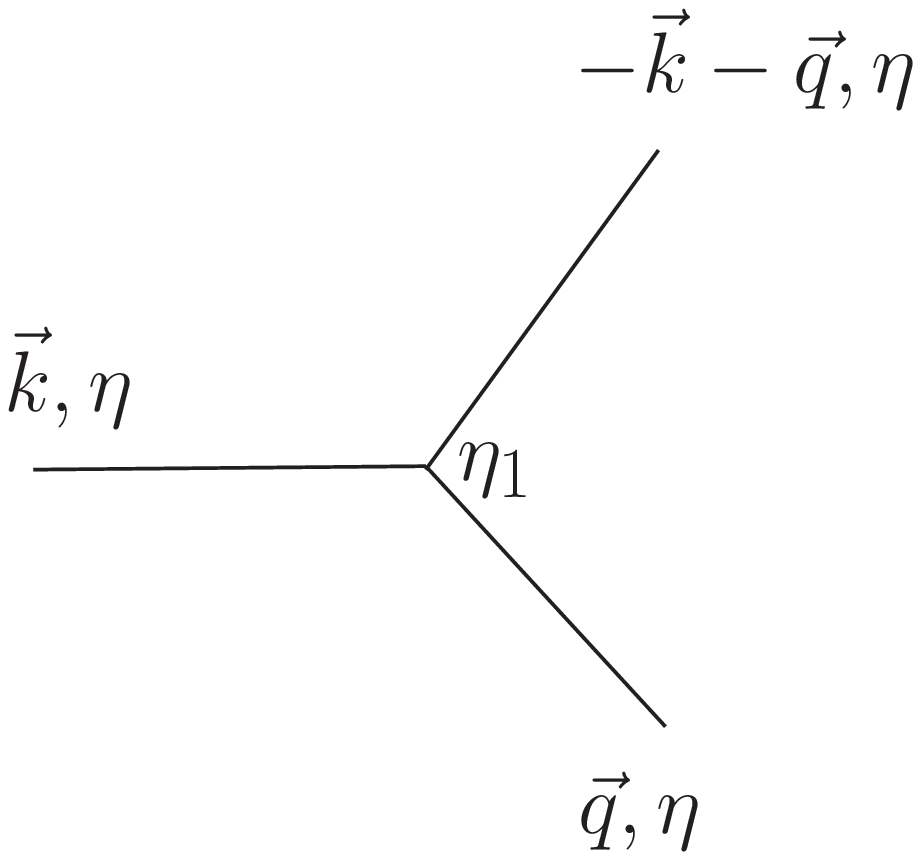}
\includegraphics[height=5cm,width=6cm]{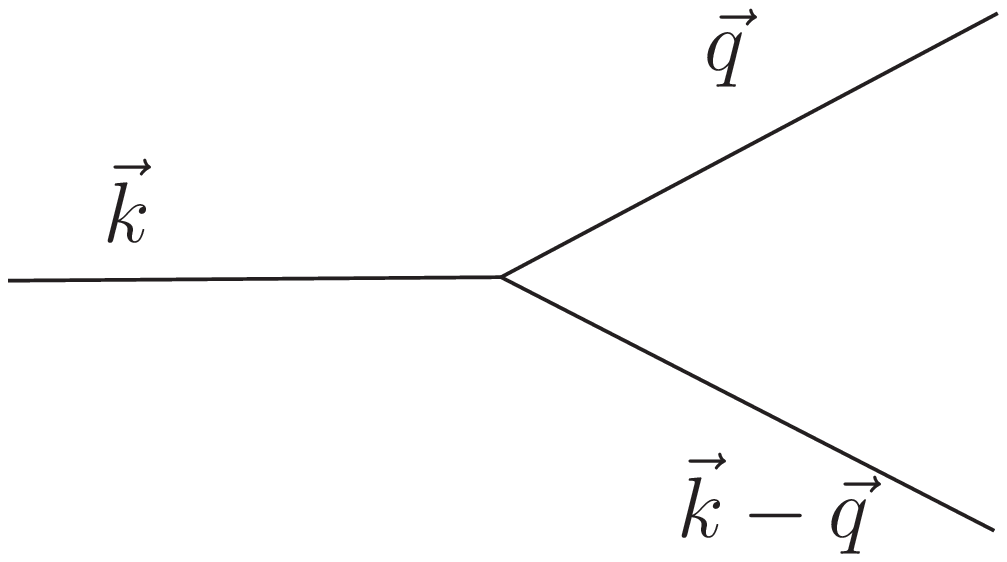}
\caption{Left panel: Equal-time three point function 
$ \langle \chi(\vk,\eta) \; \chi(\vq,\eta) \; \chi(-\vk-\vq,\eta)\rangle $
in de Sitter space-time in the Born approximation. The conformal time 
coordinate $ \eta_1 $ of the vertex is integrated out. Right panel: 
Self-decay of quantum fluctuations of the inflaton. All
lines correspond to the field $ \phi $, i.e, the quantum fluctuations
of the inflaton.} \label{fig:expecval}
\end{center}
\end{figure}

The long time ($ \eta\rightarrow 0 $) behavior of the equal time
three point correlation function in the scalar field theory defined by
Eq.(\ref{lagra})  for $ M = 0 $ (and hence $ \nu = \frac32 $),
is given by \cite{srednicki}
\be \label{3point}\langle
\chi(\vk,\eta) \; \chi(\vq,\eta) \; \chi(-\vk-\vq,\eta)\rangle =
\frac{2\pi^3}3 \; a^3(\eta) \; g \;  H^2
\frac{F(\vk,\vq;\eta)}{\left[ k \;  q \;  |\vk+\vq| \right]^3}
\ee
\noindent where
\be \label{ef}
F(\vk,\vq;\eta)= \left[k^3+q^3+|\vk+\vq|^3
\right]\left[\ln (k_T\eta) +\gamma\right]-(k^2+q^2+|\vk+\vq|^2
)k_T+k\,q\,|\vk+\vq|~~;~~k_T= k+q+|\vk+\vq|
\ee
A diagrammatic interpretation of the equal time expectation
value Eq.(\ref{3point}) is depicted  in fig. \ref{fig:expecval}, which
illustrates the {\it similarity} with the {\it decay process} depicted in 
fig. \ref{fig:expecval}.

Furthermore, the logarithmic secular term in Eq.(\ref{ef}) indicates
that the three point function features
secular divergences {\bf even at the tree level}. It is argued in
refs. \cite{srednicki,bartolo}
that $ -\ln [k_T\eta]\sim 60 $  which is the
number of efolds from the time when fluctuations of wavenumber
$ k_T $ first crossed the horizon till the end of inflation. However,
such infrared logarithms are {\bf secular terms} and have precisely
the {\bf same} origin as in the self-energy kernel and in the inflaton
fluctuations discussed here (secs. IIIC and IIID). The same holds for the
 infrared logarithms in the three graviton scattering
vertex \cite{IRcosmo,dolgov}.

In particular, the self-energy computation
corresponds to a further integration over the loop momentum $ q $. In a
fairly loose manner, the self-energy is basically the square of the
three point correlation function integrated over the loop momentum.
This is akin to the unitarity relation between the imaginary part of
the forward scattering amplitude and the square of the transition
amplitude in S-matrix theory.

In summary, the {\it interaction} between the quantum fluctuations gives 
rise to  {\it non-gaussian} correlations which are determined by the three
point function which is  precisely related to the
{\it self-energy}  and  the {\it decay} of the quantum fluctuations.
Therefore, the decay of the quantum fluctuations of the scalar field
will also lead to non-gaussian correlations and
non-gaussianity in the power spectrum.

Then, there is a direct relationship between the self-energy, quantum
decay and non-gaussian features of the power spectrum.

\subsection{Quantum Loop Corrections to the inflaton potential and the 
power spectra from superhorizon modes and trace anomalies}\label{trece}

Our goal is to obtain the \emph{effective} potential that includes
the one loop quantum  corrections from fields that are
\emph{light} during the relevant inflationary stage.

Our program of study focuses on the understanding of quantum
aspects of the basic inflationary paradigm. We previously
addressed the decay of inflaton fluctuations in secs. 
\ref{13A}-\ref{partdecay} \cite{pardec}. In secs. 
\ref{13C}-\ref{13D} \cite{effpot} we focused on the quantum
corrections to the equations of motion of the inflaton and the
scalar fluctuations during slow-roll inflation. We integrated
out not only the inflaton fluctuations but also the excitations
associated with another scalar field. Since the power spectra of
fields with masses $ m \ll H $ are nearly scale invariant, strong
infrared enhancements appear as revealed in these
studies described in sec. \ref{QLC1} \cite{pardec,effpot}. In addition, 
we find that a particular linear combination $ \Delta $ of slow-roll 
parameters eq.(\ref{delta}) which measures the
departure from scale invariance of the fluctuations provides a
{\bf natural} infrared regularization.

The small parameter that determines the validity of inflation as an 
effective \emph{quantum field theory} below the Planck scale is 
$ (H/M_{Pl})^2 $ where $ H $ is the Hubble parameter during inflation and 
therefore the scale at which inflation occurs. The slow-roll expansion is 
in a very well defined sense an \emph{adiabatic} approximation since the
time evolution of the inflaton field is slow on the expansion scale.
Thus the small dimensionless ratio $ (H/M_{Pl})^2 \sim  10^{-9} $
[see eq.(\ref{myH})], which is required for
the validity of an effective field theory (EFT) is logically
\emph{independent} from the small dimensionless combinations of
derivatives of the potential which ensure the validity of the
slow-roll expansion. 

Therefore, in this review article we invoke \emph{two independent}
approximations, the effective field theory (EFT) and the slow-roll
approximation. The former is defined in terms of an expansion in
the ratio $ (H/M_{Pl})^2 $, whereas the latter corresponds to an
expansion in the (small) slow-roll parameters which is
an expansion in $ 1/N $ (see sec. \ref{potuniv}) \cite{1sN}.
Quantum loop corrections during inflation are considered also in
ref.\cite{otrosq}.

It is important to highlight the main differences between slow-roll
inflation and the post-inflationary stage. During slow-roll
inflation the dynamics of the scalar field is slow on the time scale
of the expansion and consequently the change in the amplitude
of the inflaton is small and quantified by the slow-roll parameters.
The slow-roll approximation is indeed an \emph{adiabatic approximation}.
In striking contrast to this situation, during the post-inflationary stage 
of reheating the scalar field undergoes rapid and large amplitude 
oscillations that cannot be studied in a perturbative expansion 
\cite{chalo,chalo2,baacke,ramsey}.

\medskip

{\bf Brief summary of results in this section:}

We obtain the quantum corrections to the inflaton potential
up to one loop by including the contributions from scalar and
tensor perturbations of the metric as well as one scalar and
one fermion field coupled generically to the inflaton.
Therefore, this study provides the most complete assessment of the
general backreaction problem up to one loop that includes not only
metric perturbations, but also the contributions from fluctuations
of other light fields with a generic treatment of both bosonic and
fermionic degrees of freedom. Motivated by an assessment of the
quantum fluctuations that \emph{could} be of observational
interest, we focus on studying the effective inflaton potential
during the cosmologically relevant stage of slow-roll inflation.

\vspace{2mm}

Both light bosonic fields as well
as scalar density perturbations feature an infrared enhancement of
their quantum corrections which is naturally regularized by the slow-roll
parameters. 

Fermionic contributions as expected do not feature any
infrared enhancement and neither does the graviton contribution to
the energy momentum tensor.

We find that in slow-roll and for light bosonic and fermionic
fields there is a clean separation between the super and
subhorizon contributions to the quantum corrections from scalar
density metric and light bosonic field perturbations. For these
fields the superhorizon contribution is of zero order in slow
roll $ \sim N^0 $ as a consequence of the infrared enhancement regularized 
by slow-roll parameters. The subhorizon contribution to the energy
momentum tensor from all the fields is completely determined by
the trace anomaly of minimally coupled scalars, gravitons and
fermionic fields.  

We find the one loop effective potential to be
\be
V_{eff}(\Phi_0) = V(\Phi_0)\Bigg[1+ \frac{H^2_0}{3 \; (4 \pi)^2 \;
M^2_{Pl}}\Bigg( \frac{\eta_v-4\,\epsilon_v}{\eta_v-3 \,
\epsilon_v}+\frac{3\,\eta_\sigma}{\eta_\sigma-\epsilon_v}+\mathcal{T}\Bigg)
  \Bigg] \label{Veffinsum}
\ee
\noindent where $ V(\Phi_0) $ is the \emph{classical} inflaton
potential, $ \eta_v, \epsilon_v, \eta_\sigma $ slow-roll parameters
and $ \mathcal{T} = \mathcal{T}_{\Phi}+ \mathcal{T}_s+\mathcal{T}_t
+ \mathcal{T}_{\Psi} =-\frac{2903}{20} =- 145.15 $ is the total trace 
anomaly from the scalar metric, tensor, light scalar and fermion 
contributions.

  The terms that feature ratios of slow-roll parameters arise from
  superhorizon contributions from curvature and scalar field
  perturbations. The last term in eq.(\ref{Veffinsum}) is
independent of slow-roll parameters and is completely determined by the 
trace anomalies of the different fields. The term $ \mathcal{T} $
is the hallmark of the subhorizon contributions.

In the case when the mass of the light bosonic scalar field is
much smaller than the mass of the inflaton fluctuations, we find
the following result for the scalar curvature and tensor
fluctuations including the one-loop quantum corrections, 
\bea &&
|{\Delta}_{k,eff}^{\mathcal{R}}|^2 = |{\Delta}_{k}^{\mathcal{R}}|^2 \left\{ 1+
\frac23 \left(\frac{H_0}{4 \; \pi \; M_{Pl}}\right)^2
\left[1+\frac{\frac38 \; r \; (n_s - 1) + 2 \; \frac{dn_s}{d \ln
k}}{(n_s - 1)^2} + \frac{2903}{40} \right] \right\} \cr \cr &&
|{\Delta}_{k,eff}^{(T)}|^2 =|{\Delta}_{k}^{(T)}|^2  \left\{ 1
-\frac13 \left(\frac{H_0}{4 \; \pi \; M_{Pl}}\right)^2
\left[-1+\frac18 \; \frac{r}{n_s - 1}+ \frac{2903}{20}
\right]\right\} \; , \cr \cr 
&& r_{eff} \equiv 
\frac{|{\Delta}_{k,eff}^{(T)}|^2}{|{\Delta}_{k,eff}^{\mathcal{R}}|^2}
= r \;  \left\{ 1-\frac13 \left(\frac{H_0}{4 \; \pi \; M_{Pl}}\right)^2
\left[1+\frac{\frac38 \; r \; (n_s - 1) +
\frac{dn_s}{d \ln k}}{(n_s - 1)^2} + \frac{8709}{20} \right] \right\} \; .
\eea 
The quantum corrections turn out to {\bf enhance} the scalar curvature fluctuations 
$ |{\Delta}_{k,eff}^{\mathcal{R}}|^2 $
and to {\bf reduce} the tensor fluctuations $ |{\Delta}_{k,eff}^{(T)}|^2 $
as well as their ratio $ r $. The quantum corrections are always small, 
of the order $ (H_0/ M_{Pl})^2 $, but it is interesting 
to see that these quantum effects are dominated by the trace anomalies
and they correct both scalar and tensor fluctuations in a definite
direction. Moreover, it is the tensor part of the trace anomaly
which numerically yields the largest contribution.

\medskip

Quantum trace (conformal) anomalies of the energy momentum tensor
in gravitational fields constitute an important aspect of quantum
field theory in curved backgrounds, (see for example \cite{BD} and
references therein). In black hole backgrounds they are
related to the Hawking radiation. It is interesting to see here
that the trace anomalies appear in a relevant cosmological problem
and dominate the quantum corrections to the primordial spectrum of
curvature and tensor fluctuations.

\subsubsection{Statement of the problem}

Our goal is to obtain the one loop quantum corrections from fields with 
masses well below the energy scale of inflation $ M $.
Therefore, we consider the inflaton coupled to a
scalar field $ \sigma $ and to Fermi fields with a generic
Yukawa-type coupling. We take the fermions to be Dirac fields but it
is straightforward to generalize to Weyl or Majorana fermions. We
also include the contribution to the effective potential from scalar
and tensor metric perturbations, thereby considering their
\emph{backreaction} up to one loop.

The Lagrangian density is taken to be 
\be 
\mathcal{L} = \sqrt{-g} \; \Bigg\{ \frac12 \;  \dot{\varphi}^2 -\frac12 \; 
\left(\frac{\vec{\nabla}\varphi}{a}\right)^2-V(\varphi)+\frac12
\; \dot{\sigma}^2 - \frac12 \; \left(\frac{\vec{\nabla}\sigma}{a}\right)^2 -
\frac12 \;  m^2_\sigma \;  \sigma^2 -G(\varphi)  \;  \sigma^2+
\overline{\Psi}\Big[i\,\gamma^\mu \;  \mathcal{D}_\mu  -m_f -
Y(\varphi)\Big]\Psi \Bigg\} \label{lagrangian} 
\ee 
\noindent where $ m_\sigma \ll M , \; m_f \ll M $, 
$ G(\Phi) $ and $ Y(\Phi) $ are generic interaction terms between the
inflaton and the scalar and fermionic fields. Obviously this
Lagrangian can be further generalized to include a multiplet of
scalar and fermionic fields and such case can be analyzed as a
straightforward generalization. For simplicity we consider one
bosonic and one fermionic Dirac field.

The Dirac $ \gamma^\mu $ are the curved space-time $ \gamma $ matrices
and the fermionic covariant derivative is given by  
\be
\mathcal{D}_\mu  = 
\partial_\mu + \frac18 \; [\gamma^c,\gamma^d] \;  V^\nu_c  \;
\left(D_\mu V_{d \nu} \right) \quad , \quad
D_\mu V_{d \nu} =  \partial_\mu V_{d \nu} -\Gamma^\lambda_{\mu
\nu} \;  V_{d \lambda} \nonumber 
\ee \noindent 
where the vierbein field is defined as
$$
g^{\mu\,\nu} = V^\mu_a \;  V^\nu_b \;  \eta^{a b} \; ,
$$
\noindent $ \eta_{a b} $ is the Minkowski space-time metric and the
curved space-time  matrices $ \gamma^\mu $ are given in terms of the
Minkowski space-time ones $ \gamma^a $  by (greek indices refer to
curved space time coordinates and latin indices to the local
Minkowski space time coordinates)
$$
\gamma^\mu = \gamma^a V^\mu_a \quad , \quad
\{\gamma^\mu,\gamma^\nu\}=2 \; g^{\mu \nu}  \; .
$$
We will consider that the light scalar field $ \sigma $ has vanishing
expectation value at all times, therefore inflationary dynamics is
driven by one single scalar field, the inflaton $ \phi $. We now
separate the homogeneous expectation value of the inflaton field
from its quantum fluctuations as usual by writing
$$
\varphi(\vec{x},t) = \Phi_0(t) +\delta\varphi (\vec{x},t) \; .
$$
We will consider the contributions from the quadratic fluctuations
to the energy momentum tensor. There are \emph{four} distinct
contributions: i) scalar metric (density) perturbations, ii) tensor
(gravitational waves) perturbations, iii) fluctuations of the light
bosonic scalar field $ \sigma $, iv) fluctuations of the light
fermionic field $ \Psi $.

Fluctuations in the metric are studied as usual, writing the metric
as
$$
g_{\mu\nu}= g^0_{\mu\nu}+\delta^s g_{\mu\nu}+\delta^t g_{\mu\nu}
$$
\noindent where $ g^0_{\mu\nu} $ is the spatially flat FRW background
metric which in conformal time is given by
$$
g^0_{\mu\nu}= a^2(\eta) \;  \eta_{\mu\nu} 
$$
\noindent and $ \eta_{\mu\nu}=\textrm{diag}(1,-1,-1,-1) $ is the flat
Minkowski space-time metric.  $ \delta^{s,t} g_{\mu\nu} $ correspond
to the scalar and tensor perturbations respectively. In longitudinal
gauge 
\be
\delta^{s} g_{00}  =  a^2(\eta) \; 2  \; \phi \quad , \quad
\delta^{s} g_{ij}  =  a^2(\eta) \;  2 \;  \psi \;  \delta_{ij} 
\quad , \quad 
\delta^{t} g_{ij}  =  -a^2(\eta) \;  h_{ij} \nonumber
\ee \noindent 
where $ h_{ij} $ is transverse and traceless and we
neglect vector modes since they are not generated in single field
inflation.

Gauge invariant variables associated with the fluctuations of the
scalar field and the potentials $ \phi, \; \psi $ are constructed
explicitly in ref.\cite{hu} where the reader can find their
expressions. Expanding up to quadratic order in the scalar fields,
fermionic fields and metric perturbations the part of the Lagrangian
density that is quadratic in these fields is given by
$$
\mathcal{L}_Q
=\mathcal{L}_s[\delta\varphi^{gi},\phi^{gi},\psi^{gi}]+
\mathcal{L}_t[h]+\mathcal{L}_\sigma[\sigma]+
\mathcal{L}_\Psi[\overline{\Psi},\Psi] \; ,
$$
\noindent where
\bea 
&&\mathcal{L}_t[h] = \frac{M^2_{Pl}}8 \; a^2(\eta) \;
\partial_\alpha h^j_i  \; \partial_\beta h^i_j \;  \eta^{\alpha \beta} \; ,
\cr \cr
&& \mathcal{L}_\sigma[\sigma]=
 a^4(\eta) \; \Bigg\{\frac12 \;
\left(\frac{\sigma'}{a}\right)^2-\frac12 \; \left(\frac{\nabla
\sigma}{a}\right)^2 -\frac12 \; M^2_\sigma[\Phi_0]\;
\sigma^2\Bigg\}  \; , \cr \cr
&& \mathcal{L}_\Psi[\overline{\Psi},\Psi]= \overline{\Psi}\Big[i \;
\gamma^\mu  \; \mathcal{D}_\mu -M_\Psi[\Phi_0]\Big]\Psi \; , \nonumber 
\eea 
\noindent where the prime stands for derivatives with
respect to conformal time and the labels (gi) stand for the gauge
invariant quantities \cite{hu}. The explicit expression for
$ \mathcal{L}[\delta\varphi^{gi},\phi^{gi},\psi^{gi}] $ is given in
ref.\cite{hu}. The effective masses for the bosonic and fermionic fields are given by 
\be\label{sigmamass}
M^2_\sigma[\Phi_0]  =  m^2_\sigma + G(\Phi_0) \quad , \quad M_\Psi[\Phi_0]  = m_f+Y(\Phi_0) \; . 
\ee
We will focus on the study of the quantum corrections to the Friedmann equation, for the case in
which both the scalar and fermionic fields are light in the sense
that during slow-roll inflation, 
\be 
M_\sigma[\Phi_0], \; M_\Psi[\Phi_0] \ll H_0 \; , 
\ee 
\noindent at least during the cosmologically relevant stage corresponding to the 60 or so efolds
before the end of inflation.

In conformal time the vierbeins $V^\mu_a$ are particularly simple
\be 
V^\mu_a = a(\eta) \; \delta^\mu_a 
\ee 
\noindent and the Dirac
Lagrangian density simplifies to the following expression 
\be \label{ecdi} 
\sqrt{-g} \; \overline{\Psi}\Bigg(i \; \gamma^\mu \;
\mathcal{D}_\mu -M_\Psi[\Phi_0]\Bigg)\Psi  =
a^{\frac32}\overline{\Psi} \;  \Bigg[i \;
{\not\!{\partial}}-M_\Psi[\Phi_0] \; a(\eta) \Bigg]
\left(a^{\frac32}{\Psi}\right) 
\ee 
\noindent where $ i {\not\!{\partial}} $ is the usual Dirac differential operator in
Minkowski space-time in terms of flat space time $ \gamma $ matrices.

From the quadratic Lagrangian given above the  quadratic quantum
fluctuations to the energy momentum tensor can be extracted.

The effective potential is identified with the expectation value of
the $00$ component of the energy momentum tensor in a state in which
the expectation value of the inflaton field is $ \Phi_0 $. During slow
roll inflation the expectation value $ \Phi_0 $ evolves very slowly in
time, the slow-roll approximation is indeed an adiabatic
approximation, which justifies treating $ \Phi_0 $ as a constant in
order to obtain the effective potential. The time variation of
$ \Phi_0 $ only contributes to higher order corrections in slow-roll.
This is standard in \emph{any} calculation of an effective
potential. The energy momentum tensor is computed in the FRW
inflationary background determined by the \emph{classical}
inflationary potential $ V(\Phi_0) $, and the slow-roll parameters are
also explicit functions of $ \Phi_0 $. Therefore the energy momentum
tensor depends \emph{implicitly} on $ \Phi_0 $ through the background
and \emph{explicitly} on the masses for the light bosonic and
fermionic fields given above.

Therefore the effective potential is given by 
\be 
V_{eff}(\Phi_0) = V(\Phi_0)+ \delta V(\Phi_0)
\ee 
\noindent where 
\be\label{dVeff}
\delta V(\Phi_0)= \langle T^{0}_{0}[\Phi_0] \rangle_s + \langle
T^{0}_{0}[\Phi_0] \rangle_t + \langle T^{0}_{0}[\Phi_0]
\rangle_\sigma +\langle T^{0}_{0}[\Phi_0] \rangle_\Psi 
\ee 
\noindent $ (s,t,\sigma,\Psi) $ correspond to the energy momentum
tensors of the quadratic fluctuations of the scalar metric, tensor
fluctuations (gravitational waves), light boson field $ \sigma $ and
light fermion field $ \Psi $ fluctuations respectively. Since these
are the expectation values of a quadratic energy momentum tensor,
$ \delta V(\Phi_0) $ corresponds to the \emph{one loop correction} to
the effective potential.

\subsubsection{Light scalar field coupled to the inflaton}

We begin by analyzing the contribution to the effective potential
from the light bosonic scalar field $ \sigma $ because this study
highlights the main  aspects which are relevant in the case of
scalar metric (density) perturbations.

The bosonic  Heisenberg field operators are expanded as follows 
\be\label{sigm} 
\sigma(\vec{x},\eta) = \frac1{a(\eta)} \int 
\frac{d^3k}{(2 \; \pi)^\frac32}\left[
 e^{i \vk \cdot \vx} a_{\sigma,\vec{k}}
\, S_\sigma(k,\eta)+ e^{-i\vk\cdot\vx} \;  
a^\dagger_{\sigma,\vec{k}} \,S^*_\sigma(k,\eta)
\right] \; .
\ee 
During slow-roll inflation the effective mass of the $ \sigma $ field
is given by eq.(\ref{sigmamass}), just as for the inflaton, it is
convenient to introduce a parameter $ \eta_\sigma $ defined to be
\be\label{etasig} 
\eta_\sigma = \frac{M^2_\sigma[\Phi_0]}{3 \;
H^2_0 } \; . 
\ee 
Hence, the statement that the $ \sigma $ field is light corresponds to the 
condition 
\be \label{light}
\eta_\sigma \ll 1 \; . 
\ee 
This dimensionless parameter plays the same
role for the $ \sigma $ field as the parameter $ \eta_v $ given by eq.
(\ref{slrsd}) does for the inflaton fluctuation. It is indeed a slow
roll parameter for the $ \sigma $ field.

The mode functions $ S_\sigma(k,\eta) $ in eq.(\ref{sigm}) obey the
following equations up to quadratic order
$$
 S^{''}_{\sigma}(k,\eta)+ \left[k^2 + M^2_{\sigma}(\Phi_0) \;  a^2(\eta)-
\frac{a^{''}(\eta)}{a(\eta)} \right]S_{\sigma}(k,\eta)=0 \,.
$$
Using the slow-roll expressions eq.(\ref{adeta}) and in terms of
$ \eta_\sigma $, these mode equations become
$$ 
S^{''}_{\sigma}(k,\eta)+\left[k^2-
\frac{\nu^2_{\sigma}-\frac14}{\eta^2} \right]S_{\sigma}(k,\eta)
= 0 ~~;~~ \nu_\sigma = \frac32+\epsilon_v-\eta_{\sigma} +
\mathcal{O}(\frac1{N^2},\eta^2_{\sigma},\frac{\eta_{\sigma}}{N}) \; .
$$
During slow-roll inflation $ \Phi_0 $ is approximately constant, and
the slow-roll expansion is an \emph{adiabatic} expansion. As usual
in the slow-roll approximation, the above equation for the mode
functions is solved by assuming that $ \Phi_0 $, hence $ \nu_\sigma $
are \emph{constant}. This is also the same type of approximation
entailed in \emph{every} calculation of the effective potential.
Therefore during slow-roll, the solution of the mode functions above
are given by eq.(\ref{gnu}),
$$
S_{\sigma}(k,\eta) = \frac12 \; \sqrt{-\pi\eta} \;
\; i^{\nu_{\sigma} +\frac12} \; H^{(1)}_{\nu_\sigma}(-k \; \eta) \;  .
$$
This choice of mode functions  defines the Bunch-Davis vacuum, which
obeys $ a_{\vec{k}}|0>_{BD}=0 $. It is important to highlight that
there is no unique choice of vacuum or initial state, a recognition
that has received considerable attention in the literature, see for
example \cite{inistate,motto2} and references therein. In this study
we focus on  Bunch-Davis initial conditions since this has been the
standard choice to study the power spectra and metric perturbations,
hence we can compare our results to the standard ones in the
literature. The assessment for initial states others than Bunch-Davis
is given in ref.\cite{quadru1}. 

The contribution to the effective potential from the light scalar
field $ \sigma $ is given by
$$ 
\langle T_{00} \rangle_\sigma =
\frac12\,\Bigg\langle\dot{\sigma}^2+ \left(\frac{\nabla
\sigma}{a(\eta)}\right)^2+M^2_\sigma[\Phi_0] \;  \sigma^2
\Bigg\rangle \; ,
$$
\noindent where the dot stands for derivative with respect to cosmic
time. The expectation values are in the Bunch-Davis vacuum state and
yield the following contributions 
\bea 
\frac12 \; \big\langle\left(\dot{\sigma}\right)^2 \big\rangle & = &
\frac{H^4_0}{16\pi}\int_0^\infty \frac{dz}{z} \;  z^2 \;
\Big|\frac{d}{dz} \left[z^\frac32 \; H^{(1)}_{\nu_\sigma}(z) \right]
\Big|^2 \label{sigdot2} \\
\frac12\,\big\langle\ \left(\frac{\nabla
\sigma}{a^2(\eta)}\right)^2 \big \rangle & = &
\frac{H^4_0}{16\pi}\int_0^\infty \frac{dz}{z} \;  z^5 \;
\Big|H^{(1)}_{\nu_\sigma}(z)  \Big|^2 \label{gradsig2} \\
\frac{M^2_\sigma[\Phi_0]}2 \; \big \langle \sigma^2(\vec{x},t)
\big \rangle & = & \frac{3\,H^2_0\,\eta_\sigma}2  \int_0^\infty
\frac{dk}{k} \;  \mathcal{P}_\sigma(k,t) \label{sig2} \; , 
\eea
\noindent where $ \mathcal{P}_\sigma(k,t) $ is the power spectrum of
the $ \sigma $ field, which in terms of the spatial Fourier transform
of the field $ \sigma_{\vec{k}}(t) $ is given by
$$
\mathcal{P}_\sigma(k,t) = \frac{k^3}{2\pi^2} \; \big \langle
\left|\sigma^2_{\vec{k}}(t)\right| \big \rangle = \frac{H^2_0}{8\pi}
\;  (-k \; \eta)^3 \; \left|H^{(1)}_{\nu_\sigma}(-k \; \eta)\right|^2  \; .
$$
For a light scalar field during slow-roll the power spectrum of the
scalar field $ \sigma$  is nearly scale invariant and the index
$ \nu_\sigma \sim 3/2 $. In the exact scale invariant case $ \nu_\sigma
= 3/2 $,
$$
z^3 \left|H^{(1)}_\frac32(z)\right|^2 = \frac2{\pi} [1 + z^2]
$$
\noindent and  the integral of the power spectrum in eq.
(\ref{sig2}) not only features logarithmic and quadratic
\emph{ultraviolet} divergences but also a logarithmic
\emph{infrared} divergence. During slow-roll and for a light but
massive scalar field the quantity
$$ 
\Delta_\sigma = \frac32 - \nu_\sigma = \eta_\sigma-\epsilon_v +
\mathcal{O}\left(\frac1{N^2},\eta^2_{\sigma},\frac{\eta_{\sigma}}{N}\right)
 \ll 1 \; ,
$$
\noindent is a measure of the departure from scale invariance and
provides a natural \emph{infrared regulator}. We note that the
contribution from eq.(\ref{sig2}) to the effective potential, which
can be written as
$$
\frac{3\,H^4_0\,\eta_\sigma}{16 \; \pi} \int_0^\infty \frac{dz}{z}
\; z^3 \;  \left|H^{(1)}_{\nu_\sigma}(z)\right|^2  \; ,
$$
\noindent is \emph{formally} smaller than the contributions from
eqs.(\ref{sigdot2})-(\ref{gradsig2}) by a factor $ \eta_\sigma \ll
1 $. However, the logarithmic infrared divergence in the exact scale
invariant case, leads to a single \emph{pole} in the variable
$ \Delta_\sigma $ as described in refs. \cite{pardec,effpot}. To
see this feature in detail, it proves convenient to separate the
infrared contribution by writing the integral above in the following
form
$$
\int_0^\infty \frac{dz}{z} \; z^3 \;
\left|H^{(1)}_{\nu_\sigma}(z)\right|^2 = \int_0^{\mu_p} \frac{dz}{z}
\; z^3 \;  \left|H^{(1)}_{\nu_\sigma}(z)\right|^2 +
\int_{\mu_p}^\infty \frac{dz}{z} \; z^3 \;
\left|H^{(1)}_{\nu_\sigma}(z)\right|^2  \; .
$$
In the first integral we obtain the leading order contribution in
the slow-roll expansion, namely the pole in $ \Delta_\sigma $, by
using the small argument limit of the Hankel functions
$$
z^3 \, \left|H^{(1)}_{\nu_{\sigma}} (z)\right|^2 \buildrel{z \to
0}\over=\left[ \frac{2^{\nu_\sigma} \; \Gamma(\nu_\sigma)}{\pi}
\right]^2 \; z^{2 \, \Delta_\sigma}
$$
\noindent which yields
$$
\int^{\mu_p}_0 \frac{dz}{z} \; z^3 \, \left|H^{(1)}_{\nu_{\sigma}}
(z)\right|^2 = \frac2{\pi}\left[\frac1{2 \, \Delta_\sigma}+
\frac{\mu^2_p}2 + \gamma - 2 + \ln(2 \; \mu_p)
+\mathcal{O}(\Delta_\sigma)\right]\,,
$$
\noindent In the second integral for small but fixed  $ \mu_p $, we can
safely set $ \Delta_\sigma=0 $ and by introducing an upper momentum
(ultraviolet) cutoff $ \Lambda_p $, we finally find
$$
\int_0^{\Lambda_p} \frac{dz}{z} \; z^3 \;
\left|H^{(1)}_{\nu_\sigma}(z)\right|^2 = \frac1{\pi}
\left[\frac1{\Delta_\sigma} + {\Lambda_p}^2 + \ln \Lambda_p^2
 + 2 \, \gamma - 4 + \mathcal{O}(\Delta_\sigma) \right]
$$
 The simple pole in $ \Delta_\sigma $ reflects the infrared enhancement
 arising from a nearly scale invariant power spectrum. While the
 terms that depend on $ \Lambda_p $ are of purely ultraviolet origin
and correspond to the specific regularization scheme, the
simple pole in $ \Delta_\sigma $ originates in the \emph{infrared} behavior
 and is therefore independent of the regularization scheme. A
 covariant regularization of the expectation value $ \langle
 \sigma^2(\vec x,t)\rangle $ yields a result which
feature the simple pole in $ \Delta_\sigma $ plus terms which are
ultraviolet finite and regular in the limit $ \Delta_\sigma
\rightarrow 0 $. Such regular terms
 yield a contribution  $ \mathcal{O}(H^4 \; \eta_\sigma) $
to eq.(\ref{sig2}) and are subleading in the limit of light scalar
fields because they do not feature a denominator $ \Delta_\sigma $.

 Therefore, to leading order in the slow-roll expansion and in
 $ \eta_\sigma \ll 1 $, the contribution from eq.(\ref{sig2}) is given by,
$$
\frac{M^2_\sigma[\Phi_0]}2 \; \big \langle \sigma^2(\vec{x},t)
\big \rangle =
\frac{3\,H^4_0}{(4 \; \pi)^2}\; \frac{\eta_\sigma}{\eta_\sigma-\epsilon_v}+
\mathrm{subleading  ~in ~ slow ~ roll}.
$$
 In the first two contributions given by 
eqs.(\ref{sigdot2})-(\ref{gradsig2})
 extra powers of momentum arising either from the time or spatial
 derivatives, prevent the logarithmic infrared enhancements. These
 terms are infrared finite in the limit $ \Delta_\sigma \rightarrow 0 $ and
 their leading contribution during slow-roll can be obtained by
 simply setting $ \nu_\sigma = 3/2 $ in these integrals, which feature
 quartic, quadratic and logarithmic ultraviolet divergences. A
 covariant  renormalization of these two terms leads to an
 ultraviolet and an infrared finite contribution to the energy momentum 
tensor of $ \mathcal{O}(H^4_0) $, respectively. For the term given by 
eq.(\ref{sig2}), the infrared contribution that yields the pole in 
$ \Delta_\sigma $ compensates for the $ \eta_\sigma \ll 1 $ in the 
numerator, after renormalization of the ultraviolet divergence, the 
ultraviolet and infrared finite contributions to this term yields a
 contribution to the energy momentum tensor of order
 $ \mathcal{O}(H^4_0 \; \eta_\sigma) $, without the small denominator,
and therefore subleading. This analysis
 indicates that the leading order contributions to the energy
 momentum tensor for light scalar fields is determined by the
 infrared pole $ \sim 1/\Delta_\sigma $ from eq.(\ref{sig2}) and the fully
 renormalized contributions from (\ref{sigdot2})-(\ref{gradsig2}),
 namely to leading order in slow-roll and $ \eta_\sigma $
 \be
\langle T_{00} \rangle_\sigma =
 \frac{3\,H^4_0}{(4 \; \pi)^2}\frac{\eta_\sigma}{ \frac32-\nu_\sigma}+
\frac12\,\Bigg\langle\dot{\sigma}^2+ \left(\frac{\nabla
\sigma}{a(\eta)}\right)^2 \Bigg\rangle_{ren} \label{T00siglead} \ee
In the expression above we have displayed explicitly the pole at
$3/2-\nu_\sigma = \eta_\sigma-\epsilon_v$.

In calculating the second term (renormalized expectation value) to
leading order in eq.(\ref{T00siglead}) we can set to zero the slow
roll parameters $ \epsilon_v, \; \eta_v $ as well as the mass of the
light scalar, namely $ \eta_\sigma=0 $. Hence, to leading order, the
second term is identified with the $00$ component of the
renormalized energy momentum tensor for a free massless minimally
coupled scalar field in exact de Sitter space time. Therefore we can
extract this term from the known result for the renormalized energy
momentum tensor for a minimally coupled free scalar boson of mass
$ m_\sigma $ in de Sitter space time with a Hubble constant $ H_0 $
given by \cite{BD,fordbunch,sanchez} 
\bea 
\langle T_{\mu \nu}\rangle_{ren} &=& \frac{g_{\mu \nu}}{(4 \,
\pi)^2}\Bigg\{m^2_\sigma \; H^2_0 \left(1-\frac{m^2_\sigma}{2 \,
H^2_0}\right)\left[-\psi\left(\frac32+
\nu\right)-\psi\left(\frac32-\nu\right)+\ln\frac{m^2_\sigma}{H^2_0}
\right]+ \frac23 \; m^2_\sigma  \; H^2_0-\frac{29}{30} \; H^4_0
\Bigg\} \; , \cr \cr
 \nu &\equiv&  \sqrt{\frac94-\frac{m^2_\sigma}{H^2_0}}\label{TmunudS} \; .
\eea 
where $ \psi(z) $ stands for the digamma function. This
expression corrects a factor of two in ref.\cite{BD,dowker}. In eq.
(6.177) in \cite{BD} the D'Alambertian acting on $G^1(x,x')$ was
neglected. However, in computing this term, the D'Alambertian must
be calculated \emph{before} taking the coincidence limit. Using the
equation of motion yields the extra factor 2 and the expression
eq.(\ref{TmunudS}). This result  eq.(\ref{TmunudS}) for the
renormalized energy momentum tensor was obtained by several
different methods: covariant point splitting, zeta-function and
Schwinger's proper time regularizations \cite{BD,dowker}.

The simple pole at $ \nu=3/2 $ manifest in eq.(\ref{TmunudS})
$$
\psi\left(\frac32-\nu\right) \buildrel{\nu \to \frac32}\over= 
\frac1{\nu-\frac32} \; .
$$
coincides precisely with the similar simple pole in eq.
(\ref{T00siglead}) as can be gleaned by recognizing that $ m^2_\sigma
= 3 \; H^2 \;  \eta_\sigma $ as stated by eq.(\ref{etasig}). 
This pole originates in the term $ m_\sigma^2  \; <\sigma^2> $, which
features an infrared divergence in the scaling limit
$ \nu_\sigma=3/2 $. All the terms that contribute to the energy
momentum tensor with space-time derivatives are infrared finite in
this limit. Therefore, from the energy momentum tensor
eq.(\ref{TmunudS}) we can extract straightforwardly the leading
contribution to the renormalized expectation value in
eq.(\ref{T00siglead}) in the limit $ H_0 \gg m_\sigma $, and
neglecting the slow-roll corrections to the scale factor. It is
given by the last term in the bracket in eq.(\ref{TmunudS}). Hence,
we find the leading order contribution for $ m_\sigma^2 \ll H_0^2 $,
\be \label{Tsiglea} 
\langle T_{00} \rangle_\sigma =
 \frac{H^4_0}{(4 \; \pi)^2}\left[\frac{3\,\eta_\sigma}{
 \eta_\sigma-\epsilon_v}-\frac{29}{30}+
\mathcal{O}(\epsilon_v,\eta_\sigma,\eta_v)\right] \; .
\ee
The last term is completely determined by the trace
anomaly \cite{BD,sanchez,fordbunch,duff,dowker,hartle,fujikawa} which
is in turn determined by the short distance correlation function of
the field and the background geometry.

\subsubsection{Heavy scalar field coupled to the inflaton}

We can study now the quantum corrections induced by {\bf heavy} scalar particles
with mass  $ m_\sigma $ in the range
$$
H^2_0 \ll  m_\sigma^2 \ll M^2 \; ,
$$
where the effective theory approach is valid. We obtain from eq.(\ref{TmunudS}) for
$  m_\sigma^2 \gg H^2_0 $,
\be
\langle T_{00} \rangle_\sigma = - \frac{ m_\sigma^2 \;  H^2_0}{12 \; (4 \; \pi)^2}
\left[1 - \frac{32 \; H^2_0}{5 \;  m_\sigma^2} + {\cal O}\left(\frac{H^4_0}{ m_\sigma^4}\right)
\right] \; .
\ee
This behavior can be compared with eq.(\ref{Tsiglea}) valid for  $ m_\sigma^2 \ll H_0^2 $.

In any case the relative change of the effective potential induced by the quantum
corrections is small:
$$
\frac{\delta V_{eff}}{ V_{eff}} = \frac{\langle T_{00} \rangle_\sigma}{3 \; M_{Pl}^2 \;  H^2_0}=
- \frac1{36 \;  (4 \; \pi)^2} \frac{ m_\sigma^2}{ M_{Pl}^2}
\left[1 - \frac{32 \; H^2_0}{5 \;  m_\sigma^2} + {\cal O}\left(\frac{H^4_0}{ m_\sigma^4}\right)
\right] \; .
$$
Therefore, we emphasize that in the slow-roll approximation there is
a clean and unambiguous separation between the contribution from
superhorizon modes, which give rise to simple poles in slow-roll
parameters and that of subhorizon modes whose leading contribution
is determined by the trace anomaly and the short distance behavior
of the field.

\subsubsection{Scalar curvature perturbations}

The gauge invariant energy momentum tensor for quadratic scalar
metric fluctuations has been obtained in ref.\cite{abramo,quant,quadru1} where the
reader is referred to for details. 

We discussed in detail the zero-zero component of the gauge invariant energy momentum tensor
in sec. \ref{sec:scalar}. Just as in the case of the $ \sigma $ field, we
expect an infrared enhancement arising from superhorizon modes,
therefore, following ref.\cite{abramo,quant,quadru1}  we split the contributions
to the energy momentum tensor as those from superhorizon modes,
which yields the infrared enhancement, and the subhorizon modes
for which we can set all slow-roll parameters to zero. Just as
discussed above for the case of the $ \sigma $ field, since
spatio-temporal derivatives bring higher powers of the momenta, we
can neglect all derivative terms for the contribution from the
superhorizon modes. Therefore, the contribution from superhorizon
modes which reflects the infrared enhancement is extracted
from \cite{abramo,quant,quadru1} 
\be 
\langle T_{00} \rangle_{IR} \approx
\frac12 \;  V''[\Phi_0] \; \langle \left(\delta \varphi
(\vec{x},t)\right)^2 \rangle + 2 \; V'[\Phi_0] \;  \langle
\phi(\vec{x},t)\,\delta \varphi(\vec{x},t) \rangle  \; . 
\ee 
The analysis of the solution of eq.(\ref{phieq}) for superhorizon
wavelengths in ref.\cite{hu} shows that in exact de Sitter
space time $ \phi_{\vec k} \sim \mathrm{constant} $, hence it follows
that during quasi-de Sitter slow-roll inflation for superhorizon
modes 
\be \label{fipun}
 \dot{\phi}_{\vec k} \sim
(\mathrm{slow~roll}) \times H_0 \; \phi_{\vec k} 
\ee 
Therefore, for superhorizon modes, the constraint equation (\ref{vin})
yields 
\be \label{rela} 
\phi_{\vec k} = -\, \frac{V'(\Phi_0)}{2 \;
V(\Phi_0)}  \;  \delta \varphi_{\vec k} + {\rm higher ~ orders ~ in
~ slow ~ roll} \; . 
\ee 
Inserting this relation in eq.(\ref{delfieqn}) and consistently neglecting the term
$ \dot{\phi}_{\vec k} $ according to eq.(\ref{fipun}), we find the
following equation of motion for the gauge invariant scalar field
fluctuation in longitudinal gauge 
\be \label{delfieqn3} 
\delta {\ddot\varphi}_{\vec k}+3 \; H_0  \; \delta{\dot \varphi}_{\vec
k}+\left[\frac{k^2}{a^2(\eta)}+3 \;  H^2_0 \;  \,\eta_\mathcal{R}
\right]\delta \varphi_{\vec k}=0  \; , 
\ee 
\noindent where we have used the definition of the slow-roll parameters 
$ \epsilon_v; \; \eta_v $ given in eq.(\ref{slrsd}), and introduced 
\be\label{etadelta}
\eta_\mathcal{R} \equiv \eta_v-2 \; \epsilon_v 
\ee 
Eq.(\ref{delfieqn3} ) is the equation of motion for a minimally coupled scalar 
field with mass squared $ 3 \;  H^2_0  \; \eta_\mathcal{R} $ and we can 
use the 
results obtained in the case of the scalar field $ \sigma $ above. 
These superhorizon fluctuations $ \delta \varphi_{\vec k}(t) $ coincide 
with the scalar curvature fluctuations $ S_\mathcal{R}(k;\eta) $ for 
superhorizon modes studied in sec. \ref{4A}, as it must be.

The quantum field $ \delta \varphi(\vec x,t) $ is expanded as in 
eq.(\ref{curvau})
\be  \label{deltaexp} 
\delta \varphi(\vec{x},\eta) \simeq \frac{u(\vx,\eta)}{a(\eta)}
=\frac1{a(\eta)} \int \frac{d^3k}{(2 \; \pi)^\frac32}\left[
\alpha_\mathcal{R}(\vk) \; S_\mathcal{R}(k;\eta) \; e^{i\vk\cdot\vx}  +
\alpha^\dagger_\mathcal{R}(\vk) \; S^*_\mathcal{R}(k;\eta)  
\; e^{-i\vk\cdot\vx}\right] \; ,
\ee 
\noindent where the mode functions $ S_\mathcal{R}(k;\eta) $
are given by eq.(\ref{gnu}) with [see eq.(\ref{eqnz})]
\be \label{delmod} 
\nu = \nu_\mathcal{R} = \frac32+\epsilon_v-\eta_\mathcal{R} = \frac32+3 \;
\epsilon_v-\eta_v + {\cal O}\left(\frac1{N^2}\right) = \frac32
- \frac12(n_s - 1) + {\cal O}\left(\frac1{N^2}\right) \; .
\ee 
In this case, the slow-roll quantity that regulates the infrared
behavior is $ \Delta_\mathcal{R} \equiv \eta_v-3 \; \epsilon_v =
\frac12(n_s - 1) $.

Again we choose the Bunch-Davies vacuum state annihilated by the
operators $ a_{\delta,\vec{k}} $. Therefore, the  contribution to
$ \langle T_{00}\rangle $ from superhorizon modes to lowest order in
slow-roll is given by 
\be 
\langle T_{00} \rangle_{IR} = 3 \; H^2_0
\left(\frac{\eta_v}2 -2\,\epsilon_v\right) \left[\int_0^\infty
\frac{dk}{k} \mathcal{P}_\mathcal{R}(k,\eta) \right]_{IR} 
\ee 
\noindent
where the power spectrum of scalar fluctuations is given by
\be\label{powspdel} 
\mathcal{P}_\mathcal{R}(k,\eta) = \frac{k^3}{2\pi^2}
\; \big \langle \left|\delta \varphi_{\vec{k}}(t)\right|^2 \big
\rangle = \frac{H^2_0}{8\pi} \; (-k \; \eta)^3 \;
\left|H^{(1)}_{\nu_\mathcal{R}}(-k \; \eta)\right|^2 
\ee 
\noindent and the
subscript $IR$ in the integral refers only to the infrared pole
contribution to $ \Delta_\mathcal{R} $. Repeating the analysis presented in
the case of the scalar field $ \sigma $ above, we finally find 
\be
\langle T_{00} \rangle_{IR} = \frac{3 \;  H^4_0}{(4 \; \pi)^2} \left[
\frac{\eta_v-4\,\epsilon_v}{\eta_v-3 \, \epsilon_v} +
{\cal O}\left(\frac1{N}\right)\right]=\frac{3 \;  H^4_0}{(4 \; \pi)^2} 
\left[1 - \frac{r}{8 \, (n_s - 1)} + {\cal O}\left(\frac1{N}\right)
\right] \; .
\ee 
Since both $ n_s - 1 $ and $ r $ are $ {\cal O}(1/N) , \;  
\langle T_{00} \rangle_{IR} $ is generically infrared finite. 
The denominator $ n_s - 1 $ indicates that a near scale invariant 
primordial power produces infrared enhancement.

\medskip

For subhorizon modes with
wavevectors $ k \gg a(t) \;  H_0 $, the solutions of the equation
(\ref{phieq})  are \cite{hu} 
\be \label{apro2} 
\phi_{\vec k}(t) \approx e^{\pm i k \eta} \Rightarrow 
\dot{\phi}_{\vec k}(t) \sim
\frac{i \, k}{a(t)} \,{\phi}_{\vec k}(t) 
\ee 
For $ k \gg a(t) \; H_0 $ the constraint equation (\ref{vin}) entails
that \cite{abramo,quant,quadru1} 
\be \label{conss2} 
\phi_{\vec k}(t) \approx \frac{i
\, a(t)}{2\,M_{Pl}\,k} \;  \dot{\Phi}_0  \;  \delta\varphi_{\vec k}.
\ee 
Replacing the expressions eqs.(\ref{apro})-(\ref{conss}) in
eq.(\ref{Too}) yields that all the terms featuring the
gravitational potential $ \phi $ are suppressed with respect to those
featuring the scalar field fluctuation $ \delta\varphi $ by powers of
$ H_0 \; a(t)/k \ll 1 $ as observed in ref.\cite{abramo}. 
Therefore, the contribution from subhorizon modes to $ \langle T_{00}
\rangle $ is given by 
\be 
\langle T_{00}\rangle_{UV} \simeq
\frac12\langle (\dot{\delta \varphi})^2 \rangle + \frac{\langle
(\nabla \delta \varphi)^2\rangle}{2\,a^2} 
\ee 
\noindent where we
have also neglected the term with $ V''[\Phi_0] = 3 \; H^2_0 \;
\eta_v $ since  $ k^2/a^2 \gg H^2_0 $ for subhorizon modes.
Therefore, to leading order in slow-roll we find the renormalized
expectation value of $ T_{00} $ given by 
\be \label{T00sfin}
\langle T_{00}\rangle_{ren} \simeq \frac{3 H^4_0}{(4 \; \pi)^2}
\frac{\eta_v-4\,\epsilon_v}{\eta_v-3 \, \epsilon_v} +
\frac12\Bigg\langle (\dot{\delta \varphi})^2 + \left(\frac{\nabla
\delta \varphi}{a(\eta)}\right)^2 \Bigg\rangle_{ren} 
\ee 
To obtain the renormalized expectation value in eq.(\ref{T00sfin}) one can set
all slow-roll parameters to zero to leading order and simply
consider a massless scalar field minimally coupled in de Sitter
space time. This is precisely what we have already calculated in the
case of the scalar field $ \sigma $ above by using the known results
in the literature for the covariantly renormalized energy momentum
tensor of a massive minimally coupled
field \cite{BD,fordbunch,sanchez,dowker}, and we can just borrow the
result from eq.(\ref{Tsiglea}). We find the following final result
to leading order in slow-roll 
\be \label{T00sfinal2}
\langle T_{00} \rangle_{ren} =
 \frac{H^4_0}{(4 \; \pi)^2}\left[\frac{\eta_v-4\,\epsilon_v}{\eta_v-3 \,
\epsilon_v}-\frac{29}{30}+
\mathcal{O}(\epsilon_v,\eta_\sigma,\eta_v)\right] \; .
\ee 
The last term in eq.(\ref{T00sfinal2}) is completely determined
by the trace anomaly of a minimally coupled scalar field in de
Sitter space time\cite{BD,sanchez,duff,dowker}.

\subsubsection{Tensor perturbations}

Tensor perturbations correspond to massless fields with two physical
polarizations. They are treated in detail in sec. \ref{sec:tensor}.

The energy momentum tensor for gravitons only depends on
derivatives of the field $ h^i_j $ therefore its  expectation value
in the Bunch Davies (BD) vacuum does not feature infrared
singularities in the limit $ \epsilon_v \rightarrow 0 $. The absence
of infrared singularities in the limit of exact de Sitter space
time entails that we can extract the leading contribution to the
effective potential from tensor perturbations by evaluating the
expectation value of $T_{00}$ in the BD vacuum in exact de Sitter
space time, namely by setting all slow-roll parameters to zero. This
 yields the leading order in the slow-roll expansion.

Because de Sitter space time is maximally symmetric, the expectation
value of the energy momentum tensor is given by\cite{weinberg,BD}
\be \label{trace} 
\langle T_{\mu \nu} \rangle_{BD} = \frac{g_{\mu
\nu}}{4} \; \langle T^{ \alpha}_{\alpha } \rangle_{BD} 
\ee 
\noindent
and $ T^{ \alpha}_{\alpha }$ is a space-time constant, therefore the
energy momentum tensor is manifestly covariantly conserved. Of
course, in a quantum field theory there emerge ultraviolet
divergences and the regularization procedure must be compatible with
the maximal symmetry. A large body of work has been devoted to study
the trace anomaly in de Sitter space time implementing a variety of
powerful covariant regularization methods that preserve the
symmetry\cite{duff,BD,dowker,hartle,fujikawa,sanchez} yielding a
renormalized value of the expectation value of the $ \langle T_{\mu
\nu} \rangle_{BD} $ given by eq.(\ref{trace}). Therefore, the full
energy momentum tensor is completely determined by the trace anomaly
\cite{BD,sanchez,duff}.

The contribution to the trace anomaly from gravitons has been given
in refs.\cite{duff,sanchez,BD}, it is 
\be 
\langle T^{\alpha}_{\alpha } \rangle_t = -\frac{717}{80 \; \pi^2} \;  H^4_0
\label{traza} 
\ee 
From this result, we conclude that 
\be 
\langle T_{00} \rangle_t = -\frac{717}{320 \;  \pi^2} H^4_0 \label{T00grav}
\ee 
This result differs by a numerical factor from that obtained in
ref.\cite{finelli}, presumably the difference is a result of a
different regularization scheme.

\subsubsection{Spinor fields}

The Dirac equation in the FRW geometry is given by [see
eq.(\ref{ecdi})], 
\be 
\Bigg[i \; {\not\!{\partial}}-M_\Psi[\Phi_0]
\; a(\eta) \Bigg] \left(a^{\frac32}{\Psi({\vec x},\eta)}\right) = 0 \; . 
\ee 
The solution $ \Psi({\vec x},\eta) $ can be expanded in spinor mode 
functions as 
\be \label{psiex} 
\Psi(\vec{x},\eta) = \frac1{a^{\frac32}(\eta)}  \int
\frac{d^3k}{(2 \; \pi)^{\frac32}} \sum_{\lambda}\,  e^{i \vec{k}\cdot
\vec{x}}\left[b_{\vec{k},\lambda}\, U_{\lambda}(\vec{k},\eta)+
d^{\dagger}_{-\vec{k},\lambda}\, V_{\lambda}(-\vec{k},\eta)\right] \; ,
\ee 
where the spinor mode functions $U,V$ obey
the  Dirac equations 
\bea 
\Bigg[i \; \gamma^0 \;  \partial_\eta - \vec{\gamma}\cdot \vec{k}
-M(\eta) \Bigg]U_\lambda(\vec{k},\eta) & = & 0 \label{Uspinor} \\
\Bigg[i \; \gamma^0 \;  \partial_\eta + \vec{\gamma}\cdot \vec{k}
-M(\eta) \Bigg]V_\lambda(\vec{k},\eta) & = & 0 \label{Vspinor} \eea
\noindent and \be M(\eta) \equiv M_\Psi[\Phi_0]  \;
a(\eta)\label{Fmass} 
\ee 
Following the method of refs.\cite{boyarel,baacke}, it proves convenient to write \bea
U_\lambda(\vec{k},\eta) & = & \Bigg[i \; \gamma^0 \;  \partial_\eta
- \vec{\gamma}\cdot \vec{k} +M(\eta)
\Bigg]f_k(\eta)\, \mathcal{U}_\lambda \label{Us}\\
V_\lambda(\vec{k},\eta) & = & \Bigg[i \; \gamma^0 \;  \partial_\eta
+ \vec{\gamma}\cdot \vec{k} +M( \eta)
\Bigg]g_k(\eta)\,\mathcal{V}_\lambda \label{Vs} 
\eea 
\noindent with
$ \mathcal{U}_\lambda;\mathcal{V}_\lambda $ being constant
spinors \cite{boyarel,baacke} obeying 
\be \gamma^0 \;
\mathcal{U}_\lambda  =  \mathcal{U}_\lambda \label{Up} \qquad ,
\qquad \gamma^0 \;  \mathcal{V}_\lambda  =  -\mathcal{V}_\lambda 
\ee
The mode functions $f_k(\eta);g_k(\eta)$ obey the following
equations of motion 
\bea 
\left[\frac{d^2}{d\eta^2} +
k^2+M^2(\eta)-i \; M'(\eta)\right]f_k(\eta) & = & 0 \label{feq}\\
\left[\frac{d^2}{d\eta^2} + k^2+M^2(\eta)+i \;
M'(\eta)\right]g_k(\eta) & = & 0 \label{geq} 
\eea 
Neglecting the derivative of $ \Phi_0 $ with respect to time, namely terms of order
$ 1/\sqrt{N} $ and higher, the equations of motion for the mode
functions are given by 
\bea 
\left[\frac{d^2}{d\eta^2} +
k^2-\frac{\nu^2_+
-\frac14}{\eta^2}\right]f_k(\eta) & = & 0 \label{fplu}\\
\left[\frac{d^2}{d\eta^2} + k^2-\frac{\nu^2_-
-\frac14}{\eta^2}\right]g_k(\eta) & = & 0 \label{gmin} 
\eea
\noindent where
$$
\nu_{\pm} = \frac12\pm i  \; \frac{M_\Psi[\Phi_0]}{H_0} 
$$
The scalar product of the spinors $ U_\lambda(\vec k,\eta), \;
V_\lambda(\vec k,\eta) $ yields 
\be
U^\dagger_\lambda(\vec k,\eta)
\; U_{\lambda'}(\vec k,\eta) = \mathcal{C}^+(k) \; 
\delta_{\lambda,\lambda'} \quad , \quad
V^\dagger_\lambda(\vec k,\eta) \; V_{\lambda'}(\vec k,\eta) = 
\mathcal{C}^-(k) \; \delta_{\lambda,\lambda'} \nonumber \; ,
\ee
\noindent where 
\bea 
\mathcal{C}^+(k) & = & f^{*'}_k(\eta) \;
f'_k(\eta)+\left(k^2+M^2(\eta)\right) f^*_k(\eta) \; f_k(\eta)+i \;
M(\eta)\left(f'_k(\eta) \; f^*_k(\eta)- f_k(\eta) \;
f^{*'}_k(\eta)\right)\; , \cr \cr
\mathcal{C}^-(k) & = & g^{*'}_k(\eta) \;
g'_k(\eta)+\left(k^2+M^2(\eta)\right) g^*_k(\eta) \; g_k(\eta)-i \;
M(\eta)\left(g'_k(\eta) \; g^*_k(\eta)-
g_k(\eta) \; g^{*'}_k(\eta)\right) \; , \nonumber 
\eea 
\noindent are constants of motion by dint of the equations of
motion for the mode functions $f_k(\eta), \; g_k(\eta)$. The
normalized spinor solutions of the Dirac equation are therefore
given by 
\bea 
&& U_\lambda(\vec k,\eta) =
\frac1{\sqrt{\mathcal{C}^+(k)}}\left[i \;
f'_k(\eta)-\vec{\gamma}\cdot\vec{k} \; f_k(\eta)+M(\eta) \;
f_k(\eta) \right] \,\mathcal{U}_\lambda  \cr \cr 
&& V_\lambda(\vec k,\eta) =
\frac1{\sqrt{\mathcal{C}^-(k)}}\left[-i \;
g'_k(\eta)+\vec{\gamma}\cdot\vec{k} \; g_k(\eta)+M(\eta) \;
g_k(\eta)
\right] \,\mathcal{U}_\lambda \; . \nonumber 
\eea 
We choose the solutions of the mode equations
(\ref{fplu})-(\ref{gmin}) to be 
\be 
f_k(\eta) = \sqrt{\frac{-\pi k \eta}2} \; i^{\nu_+ + \frac12}
\; H^{(1)}_{\nu_+}(-k \; \eta) \label{fksol} \qquad , \qquad
g_k(\eta) = \sqrt{\frac{-\pi k \eta}2} \; i^{-\nu_- - \frac12}
\; H^{(2)}_{\nu_-}(-k \; \eta) \; .
\ee 
We also choose the Bunch-Davies vacuum state such that 
$$ 
b_{\vec k,\lambda}|0>_{BD}=0 \quad,\quad d_{\vec k,\lambda}|0>_{BD}=0 
\quad .
$$ 
The choice of the mode functions eq.(\ref{fksol}) yield the following
normalization factors
$$
\mathcal{C}^+(k) = \mathcal{C}^-(k) = 2 \; k^2 \; . 
$$
The energy momentum tensor for a spin $1/2$ field is given
by \cite{BD}
$$ 
T_{\mu \nu} = \frac{i}2 \left[\overline{\Psi} \gamma_{(\mu}
\stackrel{\leftrightarrow}{\mathcal{D}}_{\nu)}\Psi \right]\,
$$ 
\noindent and its expectation value in the Bunch-Davis vacuum is
equal to
$$
\langle T_{00} \rangle_{BD} = \frac2{a^4(\eta)}\int
\frac{d^3k}{(2\pi)^3}\Bigg\{M(\eta)+
\mathrm{Im}\left[g'_k(\eta)g^*_k(\eta)\right] \Bigg\}
$$
where $ M(\eta) $ and $ g_k(\eta) $ are given by eqs.(\ref{Fmass})
and (\ref{fksol}), respectively. It is clear that this energy
momentum tensor does not feature any infrared sensitivity because
the real part of the Bessel functions index is Re$ \nu_\pm = 1/2 $. Of
course this is expected since fermionic fields cannot feature large
amplitudes due to the Pauli principle.

A lengthy computation using covariant point splitting regularization
yields the following result 
\be 
\langle T_{00} \rangle_{\Psi} =
\frac{11 \, H^4_0}{960 \; \pi^2}+ \frac12 \; \mu^2 \; \Phi_0^2 + 
 \frac14 \; \lambda \; \Phi_0^4 + \frac1{8 \; \pi^2} \; M_\Psi^2[\Phi_0]
\left(M_\Psi^2[\Phi_0] +  H^2_0\right) \left[\mathrm{Re}\,
\psi\left(1+i \frac{M_\Psi[\Phi_0]}{H_0}\right) + \gamma + 
\ln \frac{H_0}{\mu}  \right] \quad , 
\label{T00fermi} 
\ee 
We included here the mass and coupling constant renormalization
which make $ \mu^2 $ and $ \lambda $ free and finite parameters.
Eq.(\ref{T00fermi} ) in the $ H_0 \to 0 $ limit becomes
the effective potential for fermions in Minkowski space-time.

The first term in the bracket in eq.(\ref{T00fermi}) is recognized as
the trace anomaly for fermions and is the only term that survives in
the massless limit \cite{duff,hartle,BD,fujikawa,dowker,sanchez}. For
light fermion fields, $  M_\Psi[\Phi_0] \ll H_0 $, and the leading
contribution to the energy momentum tensor is completely determined
by the trace anomaly, hence in this limit the contribution to the
covariantly regularized effective potential from (Dirac) fermions is
given by
$$ 
\langle T_{00} \rangle_{\Psi} = \frac{11 \, H^4_0}{960 \;
\pi^2}\left[ 1 + \mathcal{O}\left(\frac{M_\Psi[\Phi_0]}{H_0} \right)^2
\right] \; .
$$
This result is valid for Dirac fermions and it must be divided by a
factor 2 for Weyl or Majorana fermions.

Gathering all the contributions we find that the effective potential
at one-loop is given by,
$$
\delta V(\Phi_0)  = \frac{H^4_0}{(4 \; \pi)^2} \Bigg[\frac{\eta_v-4
\; \epsilon_v}{\eta_v-3 \; \epsilon_v}+
\frac{3\,\eta_\sigma}{\eta_\sigma-\epsilon_v}+ \mathcal{T}_{\Phi}+
\mathcal{T}_s+\mathcal{T}_t +\mathcal{T}_{\Psi}+
\mathcal{O}(\epsilon_v,\eta_v,\eta_\sigma,\mathcal{M}^2) \Bigg] \; ,
$$
\noindent where $ (s,t,\sigma,\Psi) $ stand for the contributions of
the scalar metric, tensor fluctuations, light boson field $ \sigma $
and light fermion field $ \Psi $, respectively, where  
\be\label{Tt}
\mathcal{T}_{\Phi}   =   \mathcal{T}_s = -\frac{29}{30} ~~;~~
\mathcal{T}_t   =   -\frac{717}{5} ~~;~~ \mathcal{T}_{\Psi}  =
\frac{11}{60}  
\ee  
The terms that feature the \emph{ratios} of
combinations of slow-roll parameters arise from the infrared or
superhorizon contribution from the scalar density perturbations and
scalar fields $ \sigma $ respectively. The terms
$ \mathcal{T}_{s,t,\Psi} $ are completely determined by the trace
anomalies of scalar, graviton and fermion fields respectively.
Writing $ H^4_0 = V(\Phi_0) \;  H^2_0/[3 \; M^2_{Pl}] $ we can finally
write the effective potential to leading order in slow-roll 
\be\label{Veffin}
V_{eff}(\Phi_0) = V(\Phi_0)\Bigg[1+ \frac{H^2_0}{3 \; (4 \; \pi)^2
\; M^2_{Pl}}\Bigg( \frac{\eta_v-4\,\epsilon_v}{\eta_v-3 \;
\epsilon_v}+ \frac{3\,\eta_\sigma}{
\eta_\sigma-\epsilon_v}-\frac{2903}{20}\Bigg) \Bigg] 
\ee 
There are several remarkable aspects of this result:

(i) the infrared
enhancement as a result of the near scale invariance of scalar
field fluctuations, both from scalar density perturbations as
well as from a light scalar field, yield corrections of \emph{zeroth
order in slow-roll}. This is a consequence of the fact that
during slow-roll the particular combination $ \Delta_\sigma = \eta_\sigma-
\epsilon_v $ of slow-roll parameters yield a natural infrared
cutoff.

(ii) the final one   loop contribution to the effective potential
displays the effective field theory dimensionless parameter $ H^2_0/M^2_{Pl} $
confirming our previous study in sec. \ref{QLC1} \cite{pardec,effpot},

(iii) the last term is completely determined by the trace anomaly, 
a purely geometric result of the short distance properties of the theory.

\subsubsection{Quantum Corrections to the Scalar Curvature and Tensor
power spectra}

The quantum corrections to the effective potential lead to quantum
corrections to the amplitude of scalar and tensor fluctuations.
The scalar curvature and tensor fluctuations
are given by eqs.(\ref{ampR}) and (\ref{ampliT}), respectively.

We can include the leading quantum corrections in eqs.(\ref{ampR}) and (\ref{ampliT})
replacing in it $ H $ and $ \epsilon_v $ by the corrected parameters
$ H_{eff} $ and $ \epsilon_{eff} $. That is, 
\be\label{hyep}
H_{eff}^2 = H_0^2 + \delta H^2 \quad ,  \quad \epsilon_{eff} =
\epsilon_v +  \delta \epsilon_v \ee with \be\label{efe} H_{eff}^2 =
\frac{V_{eff}(\Phi_0)}{3 \;  M_{Pl}^2} \quad , \quad \epsilon_{eff}
=\frac{M^2_{Pl}}2 \;
\left[\frac{V_{eff}^{'}(\Phi_0)}{V_{eff}(\Phi_0)} \right]^2  \; ,
\ee 
and where $ V_{eff}(\Phi_0) $ is given by eq.(\ref{Veffin}). We
thus obtain, 
\bea\label{dhe} 
&&\frac{\delta H^2}{H_0^2} =
\frac13\left(\frac{H_0}{4 \; \pi \; M_{Pl}}\right)^2 \Bigg[
\frac{\eta_v-4\,\epsilon_v}{\eta_v-3 \, \epsilon_v}+
\frac{3\,\eta_\sigma}{\eta_\sigma-\epsilon_v}-\frac{2903}{20}\Bigg]
\; ,
\\ \cr
&&\frac{\delta\epsilon_v}{\epsilon_v} =  \frac23\left(\frac{H_0}{4
\; \pi \; M_{Pl}}\right)^2 \left\{ \frac{\xi_v + 12 \; \epsilon_v
\left( 2 \; \epsilon_v - \eta_v \right)}{2 \; \left( \eta_v - 3 \;
\epsilon_v\right)^2} + \frac{3 \; \eta_\sigma}{\left(\eta_{\sigma} -
\epsilon_v\right)^2} \left[\eta_\sigma + \eta_v - 2 \; \epsilon_v
 - \sqrt{2 \; \epsilon_v} \; M_{Pl} \;
\frac{d \log M_{\sigma}[\Phi_0]}{d \Phi_0}  \right] -\frac{2903}{20}
\right\} \nonumber 
\eea
Inserting eq.(\ref{dhe}) into eqs.(\ref{hyep}) and (\ref{efe})
yields after calculation, for the scalar perturbations,
\bea\label{delref}  
&& |{\Delta}_{k,eff}^{\mathcal{R}}|^2  =
|{\Delta}_{k}^{\mathcal{R}}|^2  \left[ 1 -
\frac{\delta\epsilon_v}{\epsilon_v} + \frac{\delta H^2}{H^2} \right]
= |{\Delta}_{k}^{\mathcal{R}}|^2  \left\{ 1 - \frac13 \left(\frac{H_0}{4 \;
\pi \; M_{Pl}}\right)^2 \left[\frac{\xi_v + 12 \; \epsilon_v^2 -
\eta_v^2 - 5 \;  \epsilon_v \; \eta_v}{\left(\eta_v - 3 \;
\epsilon_v\right)^2} + \right. \right. \cr \cr &&\left.  \left.
+\frac{3 \; \eta_\sigma}{\left(\eta_{\sigma} - \epsilon_v\right)^2}
\left[\eta_\sigma - 3 \; \epsilon_v + 2 \; \eta_v - 2\;  \sqrt{2 \;
\epsilon_v} \; M_{Pl} \; \frac{d \log M_{\sigma}[\Phi_0]}{d \Phi_0}
\right] - \frac{2903}{20}\right] \right\} \; , \eea and for the
tensor perturbations, \be\label{delref2} |{\Delta}_{k,eff}^{(T)}|^2
=|{\Delta}_{k}^{(T)}|^2  \left[ 1 + \frac{\delta H^2}{H^2} \right]=
|{\Delta}_{k}^{(T)}|^2 \left\{ 1 + \frac13 \left(\frac{H_0}{4 \; \pi
\; M_{Pl}}\right)^2 \Bigg[ \frac{\eta_v-4\,\epsilon_v}{\eta_v-3 \,
\epsilon_v}+
\frac{3\,\eta_\sigma}{\eta_\sigma-\epsilon_v}-\frac{2903}{20}\Bigg]
 \right\} \; .
\ee  
where $ M_{\sigma}[\Phi_0] $ and $ \eta_\sigma $ are given by
eqs.(\ref{sigmamass}) and (\ref{etasig}), respectively.

\medskip

Since the field $ \sigma $ is assumed much lighter than the inflaton,
and since
$$
\eta_\sigma \sim \left(\frac{m_\sigma}{m_{inflaton}} \right)^2
\;\eta_v \; ,
$$
for $ m_\sigma^2 \ll m_{inflaton}^2 $, we can neglect terms
proportional to  $ \eta_\sigma $. in the expressions for  $
|{\Delta}_{k}^{\mathcal{R}}|^2 $ and $  |{\Delta}_{k,eff}^{(T)}|^2 $.
Moreover, it is particularly illuminating to express the slow-roll
parameters in eqs.(\ref{delref})-(\ref{delref2}) in terms of the
observables $ n_s, \; r $ and the spectral running of the scalar
index using 
\bea\label{gorda} 
&&\epsilon_v  = \frac{r}{16} \quad , \quad \eta_v =
\frac12\left( n_s - 1 +  
\frac38 \, r \right) \quad , \cr \cr 
&& \xi_v = \frac{r}4 \left(n_s - 1 +  \frac3{16}
\, r \right) - \frac12\frac{dn_s}{d \ln k} \quad , \quad \eta_v - 3
\; \epsilon_v = \frac12 (n_s - 1) \quad , \cr \cr 
&&\sigma_v = - \frac{r}8 \left[ \left(n_s-1+\frac{r}{32}\right)^2 -
\frac{9 \, r^2}{1024} \right] + \frac14\left(n_s-1-\frac{9 \, r}8 +
\frac{r^2}{16} \right) \frac{dn_s}{d \ln k} \cr \cr &&-
\frac14\left(1 - \frac{r}6
 \right)  \left(n_s-1+\frac{3 \, r}8\right) \frac{dr}{d \ln k} +
\frac12 \left(1 - \frac{r}6  \right) \frac{d^2n_s}{d (\ln k)^2} \; . 
\eea 
We find from eqs.(\ref{delref})-(\ref{delref2}), 
\bea \label{deltcorr}
&& |{\Delta}_{k,eff}^{\mathcal{R}}|^2  =  
|{\Delta}_{k}^{\mathcal{R}}|^2  \left\{ 1+
\frac23 \left(\frac{H_0}{4 \; \pi \; M_{Pl}}\right)^2
\left[1+\frac{\frac38 \; r \; (n_s - 1) + 2 \; \frac{dn_s}{d \ln
k}}{(n_s - 1)^2} + \frac{2903}{40} \right] \right\} \cr \cr &&
|{\Delta}_{k,eff}^{(T)}|^2  =|{\Delta}_{k}^{(T)}|^2  \left\{ 1
-\frac13 \left(\frac{H_0}{4 \; \pi \; M_{Pl}}\right)^2
\left[-1+\frac18 \; \frac{r}{n_s - 1}+ \frac{2903}{20}
\right]\right\} \; . 
\eea 
The denominators in $ n_s - 1 $ indicate that a near scale invariant 
primordial power produces infrared enhancement.

We see that the anomalies contribution $
\frac{2903}{40} = 72.575 $ and  $ \frac{2903}{20} = 145.15 $
presumably dominate both quantum corrections. The other terms are
expected to be of order one and anyway smaller than these large
anomalies contribution. The anomalies contribution is dominated in
turn by the tensor part $ \mathcal{T}_t $ [see eq.(\ref{Tt})]. 
Only fermions give
contributions with the opposite sign. However, one needs at least
$783$ species of fermions to compensate the tensor part.

These quantum corrections also affect the ratio $ r $ of
tensor/scalar fluctuations as follows, 
\be \label{rcorr}
r_{eff} \equiv 
\frac{|{\Delta}_{k,eff}^{(T)}|^2}{|{\Delta}_{k,eff}^{\mathcal{R}}|^2} 
= r \; \left\{ 1-\frac13 \left(\frac{H_0}{4 \; \pi \; M_{Pl}}\right)^2
\left[1+\frac{\frac38 \; r \; (n_s - 1) + \frac{dn_s}{d \ln k}}{(n_s
- 1)^2} + \frac{8709}{20} \right] \right\} 
\ee 
We expect this
quantum correction to the ratio to be negative as the anomaly
contributions  dominates: $ \frac{8709}{20} = 435.45 $.

The brackets in eqs.(\ref{deltcorr})-(\ref{rcorr}) are of order $ N^0 $ 
since the numerators are of the same order in $ {\cal O}(1/N) $ than the 
denominators.

\medskip

Therefore, the quantum corrections {\bf enhance} the scalar
curvature fluctuations while they {\bf reduce} the tensor
fluctuations as well as their ratio $r$. The quantum corrections are
small, of the order $ \left(H_0/M_{Pl}\right)^2 $, but it is
interesting to see that the quantum effects are dominated by the
trace anomalies and they correct both fluctuations in a definite
direction.

\subsubsection{Conclusions and further questions}

The results of our study bring about several questions and implications:
\begin{itemize}
\item{The effective theory of inflation indicates that the quantum inflaton
decay rate into itself is of the order $ \Gamma_{\varphi \rightarrow  
\varphi\varphi}/H_0 \sim 10^{-11} $ as shown in
sec. \ref{anomdim}. This corresponds to  a decay rate in cosmic time 
$ \Gamma_{\varphi \rightarrow  \varphi\varphi} \sim 10^3 $ GeV.
Although these values may seem small, it must be noticed that the
decay is a secular, namely cumulative effect.}
\item{The generation of superhorizon fluctuations during inflation is
usually referred to as `acausal' since the superhorizon region is
causally disconnected from the observer. We have found that fluctuation 
modes deep inside the horizon decay into superhorizon modes, and therefore 
there is a coupling between modes inside and outside the horizon in 
quantum theory. The phase space for this process opens up as the physical 
wavelength approaches the horizon. This process that couples modes inside
and outside the horizon with a coupling that effectively depends on
the wave vector leads to {\it distortions} in the power
spectrum. These distortions are of the order $ (M/M_{Pl})^2 \ll 1 $ as  
shown in refs. \cite{effpot,quant} (see sec.\ref{trece}).}

\item{In the non-interacting theory, the equation of motion for the gauge
invariant Newtonian potential (equal to the curvature
perturbation) features a constant of motion for superhorizon
wavelengths \cite{hu,stamuk}. This is used to
estimate the spectrum of density perturbations in inflationary
universe models. It is conceivable
that this conservation law will no longer hold in higher orders in
slow-roll when interactions are included.
We expect this to be the case for two reasons: the
coupling between modes inside and outside the horizon as well as the
decay of superhorizon modes. Clearly the violation of the
conservation law, if present, will be small in slow-roll, but this
non-conservation may also lead to small {\it distortions} in the power 
spectrum.}

\item{While we have focused on the decay process during inflation,
our results, in particular the decay of superhorizon
fluctuations and the coupling between modes inside and outside the
Hubble radius, raise the possibility of similar processes being
available during the radiation dominated phase. If this would be the case,
the decay of short wavelength modes into superhorizon modes can
serve as an active process for seeding superhorizon fluctuations.  }

\end{itemize}

Forthcoming  observations of CMB anisotropies as well as large
scale surveys with ever greater precision will provide a
substantial body of high precision observational data which may
hint at corrections to  the generic and robust predictions of slow-roll
and fast-roll inflation. Such observations will pave 
the way for a better determination of inflaton potential. Studying
the possible observational consequences of the quantum phenomena
presented in this review article will therefore prove a worthwhile 
endeavor.

The quantum loop corrections turn to be of the order 
$ (M/M_{Pl})^2 \sim 10^{-9} $ which validates the tree level results
and the effective field theory approach to inflationary dynamics. 

\subsection{Outlook and future perspectives}

This review presents the state of the art of the effective theory of inflation and 
its successful confrontation with the CMB and LSS data.

We can highlight as perspectives for a foreeseable future:

\begin{itemize}
\item{Measurement of the tensor/scalar ratio $ r $ by the forthcoming CMB
experiments. This would be the {\bf first} detection of (linearized)
gravitational waves as predicted by Einstein's General Relativity.
In addition, since such primordial gravitational waves were born as quantum
fluctuations, this would be the {\bf first} detection 
of gravitons, namely, {\bf quantized} gravitational waves at tree level.
Such detection of the primordial gravitational waves will test
our prediction $ r \simeq 0.05 $ based on the effective theory of 
slow-roll inflation
(broken symmetric binomial and trinomial potentials) \cite{mcmc,quamc}.}
\item{The running of the spectral index $ dn_s/d \ln k $.
Since the range of the cosmologically relevant modes is
$ \Delta \ln k < 9 $, we have $  \Delta n_s < 9/N^2 \sim 0.0025 $,
where we use the generic estimate eq.(\ref{run}). Therefore, 
the effective theory of slow-roll inflation indicates that 
the detection of the running calls for measurements of $ n_s $
with a one per thousand precision on a wide range of wavenumbers.}
\item{Non-gaussianity measurements. Although this subject is beyond the 
scope of this review, let us recall that primordial non-gaussianity is of 
the order $ f_{NL} \sim 1/N $ in single-field slow-roll inflation \cite{bartolo}. Such 
small primordial non-gaussianity is hardly expected to be measured in a 
near future.}
\item{More precise measurements of $ n_s $ together with better data on 
$ r $ and $ dn_s/d \ln k $ will permit to better select the correct
inflationary model. This will test our prediction that a broken symmetric 
inflaton potential with moderate nonlinearity (new inflation) best 
describes the data \cite{mcmc,quamc}.}
\end{itemize}

\begin{acknowledgments}
We thank M. Giovannini, E. Komatsu, A. Lasenby, 
L. Page, R. Rebolo, D. Spergel for fruitful discussions.
D. B. acknowledges support from the U.S. National Science Foundation 
through grant No:  PHY-0553418.
\end{acknowledgments}

\end{document}